\documentclass[twoside,a4paper,12pt,openright]{book}
      \usepackage{setspace}
\usepackage{cite}
 \usepackage{tabu}
        \usepackage{graphicx} 
        \usepackage{rotating} 
        \usepackage{multirow} 
        \usepackage{amsfonts}
        \usepackage{amsmath}
\usepackage[font=avantgarde]{quoting}
	\usepackage{float}
\usepackage{array}
\usepackage{booktabs}
        \usepackage{colortbl}
        \usepackage{fancyhdr}
        \usepackage{textcomp}
        \usepackage{subfigure}
\definecolor{dullpurple}{rgb}{0.431,0.188,0.534}
\definecolor{darkgreen}{rgb}{0.133,0.545,0.133}
\definecolor{dullred}{rgb}{0.706,0.208,0.192}
\RequirePackage[colorlinks=true
,urlcolor=dullpurple
,anchorcolor=blue
,citecolor=dullred
,filecolor=blue
,linkcolor=darkgreen
,menucolor=blue
,pagecolor=blue
,linktocpage=true
,pdfproducer=medialab
]{hyperref}        
\newcommand\note[2][]{%
\if!#1!%
\stepcounter{footnote}\footnotetext{#2}%
\else%
{\renewcommand\thefootnote{#1}%
\footnotetext{#2}}%
\fi}
\usepackage{notoccite}
                \usepackage{picinpar}  
\usepackage{bm}
\usepackage{latexsym}
\usepackage{colortbl}
\usepackage{multirow}
\usepackage{array}
\usepackage{booktabs}
\usepackage{graphicx}
        \definecolor{dullpurple}{rgb}{0.431,0.188,0.534}
\hypersetup{urlcolor=dullpurple}
\hypersetup{citecolor=dullpurple}

\usepackage{bm}
\usepackage{amsopn}
\usepackage{latexsym}
\usepackage{colortbl}
\usepackage{multirow}
\usepackage{array}
\usepackage{booktabs}
        \usepackage{fancyhdr}
\usepackage{graphicx}
        \usepackage[charter]{mathdesign}
        \usepackage{enumerate}
\usepackage{cancel}
\usepackage{amsfonts}
   \usepackage{bm}
\DeclareBoolOption{reuseMathAlphabets}
\DeclareBoolOption{OMLmathrm}
\DeclareBoolOption{OMLmathbf}
\DeclareBoolOption{OMLmathsf}
\DeclareBoolOption{OMLmathsfit}
\DeclareBoolOption{OMLmathtt}
\DeclareVoidOption{OMLmathsans}{\isomath@OMLmathsfittrue}

\let\oldsqrt\sqrt
\def\sqrt{\mathpalette\DHLhksqrt}
\def\DHLhksqrt#1#2{%
\setbox0=\hbox{$#1\oldsqrt{#2\,}$}\dimen0=\ht0
\advance\dimen0-0.2\ht0
\setbox2=\hbox{\vrule height\ht0 depth -\dimen0}%
{\box0\lower0.4pt\box2}}

\def\clap#1{\hbox to 0pt{\hss#1\hss}}

\newcommand{\A}{\mathcal{A}}

\def\ignorecitefornumbering#1{%
     \begingroup
         \@fileswfalse
         #1
    \endgroup
}

\DeclareMathOperator{\RePart}{Re}

\renewcommand{\Re}{\RePart}

\renewcommand{\vec}[1]{\bm{\mathrm{{#1}}}}

\newcolumntype{Q}{>{$\displaystyle}l<{$}}
\newcolumntype{q}{>{\columncolor[gray]{0.9}$\displaystyle}l<{$}}
\newcolumntype{R}{>{$\displaystyle}r<{$}}
\newcolumntype{S}{>{$\displaystyle}c<{$}}
\newcolumntype{s}{>{\columncolor[gray]{0.9}$\displaystyle}c<{$}}
\newcolumntype{T}{>{\columncolor[gray]{0.9}}c<{}}

\newsavebox{\tableA}
\newsavebox{\tableB}

\newsavebox{\boxplot}
\newsavebox{\boxplota}

\makeatother

\renewcommand{\geq}{\geqslant}
\renewcommand{\leq}{\leqslant}

\newcommand{\HRule}{\rule{\linewidth}{0.5mm}}
\usepackage[avantgarde]{quotchap}

		\makeindex

\newcommand{\be}{\begin{eqnarray}}
\newcommand{\ee}{\end{eqnarray}}
\newcommand{\mpl}{{M_{\rm {pl}}}}

\newcommand{\dd}{\, {\rm d}}
\newcommand{\gsim}{\;\mbox{\raisebox{-0.5ex}{$\stackrel{>}{\scriptstyle{\sim}}$}
}\;}
\newcommand{\lsim}{\;\mbox{\raisebox{-0.5ex}{$\stackrel{<}{\scriptstyle{\sim}}$}
}\;}
\newcommand{\rs}{r_{\rm s}}
\newcommand{\ys}{y_{\rm s}}
\newcommand{\pn}{\Phi_{\rm N}}
\newcommand{\pno}{\Phi_{\rm N,0}}
\newcommand{\ver}{\vec{r}}

\newcommand{\gao}{\Gamma_{1,0}}
\newcommand{\thn}{{\theta}}
\newcommand{\omr}{{\omega_R}}
\newcommand{\oms}{\omega_{\rm s}}
\newcommand{\iii}{_{\rm i}}
\newcommand{\mmm}{_{\rm m}}
\newcommand{\eff}{_{\rm eff}}

\newcommand{\dc}{\delta_{\rm c}}
\newcommand{\tdc}{\delta_{\rm c}}
\newcommand{\dpp}{\delta\phi}
\newcommand{\bp}{\bar{\phi}}
\newcommand{\tg}{\tilde{g}}
\newcommand{\pr}{^\prime}
\newcommand{\pmi}{\phi_{\rm min}}
\newcommand{\gp}{{g^\prime}}
\newcommand{\rc}{\rho_{\rm c}}
\newcommand{\bv}{\beta_\varphi}

\newcommand{\h}{t_{\rm H}}

\newcommand{\pdp}{ {\Phi^\dagger\Phi}}

\newcommand{\Dc}{\Delta_{\rm c}}

\begin{document}
 
\begin{titlepage}
\hspace*{-0,8cm}
\includegraphics[width=2.5cm]{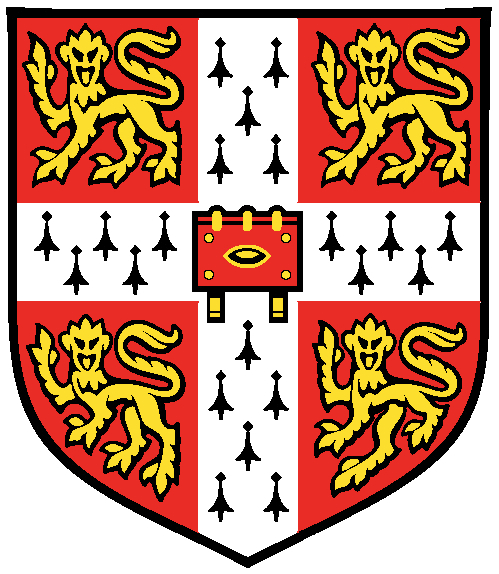}
\vspace*{0.1cm}

\noindent\begin{Large}University of Cambridge
\end{Large}\\
\begin{large}
Department of Applied Mathematics 

\vspace*{-0.1cm}

\noindent and Theoretical Physics
\end{large}

\vspace*{2cm}

\noindent\HRule   \HRule
\vspace*{0,5cm}
{\Huge \bfseries Astrophysical Tests of Modified Gravity\\}
\vspace*{0,3cm}

\begin{Large}
\noindent\textit{Jeremy Aaron Sakstein}
\end{Large}

\noindent\HRule \HRule
\vspace*{0.5cm}

\begin{flushright} 
\begin{small}
\noindent This thesis is submitted to the University of Cambridge\\
	for the degree of Doctor of Philosophy
\end{small}
\end{flushright}

\vspace*{1.3cm}

\begin{flushright} 
		
\hspace*{-0,7cm}  
\includegraphics[width=2cm]{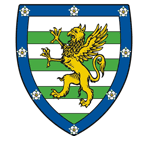}
						   
\noindent\begin{large}Downing College
\end{large}\\
\noindent\begin{normalsize} Cambridge, United Kingdom\\ March 2014
\end{normalsize}
\end{flushright}
	
\end{titlepage}

\newpage
		\pagestyle{fancy}

	\thispagestyle{empty}
	\begin{center}
	\vspace*{\fill}
	Ph.D. Dissertation, March 2014 \\
	 \textbf{\textit{Jeremy Aaron Sakstein}}\\ Department of Applied Mathematics and Theoretical Physics\\ \& \\ Downing
College\\
	University of Cambridge, United Kingdom
	\end{center}

	\newpage
\thispagestyle{empty}

\newpage

\chapter*{Abstract}
\addcontentsline{toc}{chapter}{Abstract} \markboth{Abstract}{}
Einstein's theory of general relativity has been the accepted theory of gravity for nearly a century but how well have we really
tested it? The laws of gravity have been probed in our solar system to extremely high precision using several different tests and
general relativity has passed each one with flying colours. Despite this, there are still some mysteries it cannot account for,
one of which being the recently discovered acceleration of the universe and this has prompted a theoretical study of modified
theories of gravity that can self-accelerate on large scales. Indeed, the next decade will be an exciting era where several
satellites will probe the structure of gravity on cosmological scales and put these theoretical predictions to the test. Despite
this, one must still worry about the behaviour of gravity on smaller scales and the vast majority of these theories are
rendered cosmologically uninteresting when confronted with solar system tests of gravity. This has motivated the study of theories
that differ from general relativity on large scales but include screening mechanisms which act to hide any modifications in our
own solar system. This then presents the problem of being able to distinguish these theories from general relativity. In the last
few years, astrophysical scales have emerged as a new and novel way of probing these theories. These scales encompass the mildly
non-linear regime between galactic and cosmological scales where the astrophysical objects have not yet joined the Hubble flow.
For this reason, the screening mechanism is active but not overly efficient and novel effects may be present.
Furthermore, these tests do not require a large sample of galaxies and hence do not require dedicated surveys; instead they can
piggyback on other experiments.

This thesis explores a class of theories of screened modified gravity which are scalar-tensor theories where the field is
conformally coupled to matter via the metric and includes chameleon and symmetron models as well as those that screen using the
environment-dependent Damour-Polyakov effect. The thesis is split into two parts.

The first is aimed at searching for new and novel astrophysical probes and using them to place new constraints on the model
parameters. In particular, we derive the equations governing hydrodynamics in the presence of an external gravitational field
that includes the modifications of general relativity. Using this, we derive the equations governing the equilibrium structure of
stars and show that unscreened stars are brighter and hotter than their screened counterparts owing to the larger nuclear burning
rate in the core needed to combat the additional inward force. These theories have the property that the laws of gravity are
different in unscreened galaxies from our own. This means that the inferred distance to an unscreened galaxy using a stellar
effect that depends on the law gravity will not agree with a measurement using a different method that is insensitive
gravitational physics. We exploit this property by comparing the distances inferred using pulsating Cepheid variable stars, tip of
the red giant branch stars and water masers to place new constraints on the model parameters that are three orders of magnitude
stronger than those previously reported. Finally, we perturb the equations of modified gravity hydrodynamics to first order and
derive the equations governing the oscillations of stars about their equilibrium structure. By solving these equations we show
that unscreened stars are more stable to small perturbations than screened stars. Furthermore, we find that the oscillation
period is far shorter than was previously estimated and this means that the current constraints can potentially be improved using
previous data-sets. We discuss these new results in light of current and future astrophysical tests of modified gravity.

The final part of this thesis is dedicated to the search for supersymmetric completions of modified theories of gravity. There
have been recent investigations into the quantum stability of these models and there is evidence that they may suffer from
quantum instabilities. Supersymmetric theories enjoy powerful non-renormalisation theories that may help to avoid these issues.
For this reason, we construct a framework for embedding these models into global supersymmetry and investigate the new features
this introduces. We show how supersymmetry is broken at a scale set by the ambient density and that, with the exception of
no-scale models, supergravity corrections already constrain the model parameters to levels where it is not possible to probe the
theories with astrophysics or laboratory experiments. Next, we construct a class of supersymmetric chameleon models and
investigate their cosmology. In particular, we find that they are indistinguishable from the $\Lambda$CDM model at the background
level but that they may show deviations in the cold dark matter power spectrum that can be probed using upcoming experiments.
Finally, we introduce a novel mechanism where a cosmological constant in the form of a Fayet-Illiopoulos term can appear at late
times and investigate the constraints this imposes on the model parameter space.
\clearpage



\tableofcontents
\listoffigures
\listoftables

	\newpage
\thispagestyle{empty}

\newpage

	\renewcommand\chapterheadstartvskip{\vspace*{-5\baselineskip}}
\mainmatter  
\clearpage

\newpage
\thispagestyle{empty}
\newpage
\clearpage

\newpage

\thispagestyle{empty}

 \vspace*{\fill}

\begin{quote}
An Englishman's self-assurance is founded on his being a citizen of the best organised state in the world and on the fact that, as
an Englishman, he always knows what to do, and that whatever he does as an Englishman is unquestionably correct.
\qauthor{Leo Tolstoy, \textit{War and Peace} }
\end{quote}
 \vspace*{\fill}

\newpage
\thispagestyle{empty}

\newpage

\begin{savequote}[29pc]
Were the succession of stars endless, then the background of the sky would present us a uniform luminosity, like that displayed by
the Galaxy --- since there could be absolutely no point, in all that background, at which would not exist a star. The only mode,
therefore, in which, under such a state of affairs, we could comprehend the voids which our telescopes find in innumerable
directions, would be by supposing the distance of the invisible background so immense that no ray from it has yet been able to
reach us at all.
\qauthor{Edgar Allen Poe}
\end{savequote}
\chapter{Introduction}\label{chap:one}

Physics has always been an experimentally driven science. Whenever a new and
unaccountable phenomena has been observed in nature there are two
possible explanations: either there is new, previously unknown physics or the
current theory is not as complete as we believe. In 1854 Hermann von Helmholtz
proposed that the energy source powering the Sun was its own gravitational
contraction. This presented a problem since this time-scale for this process is
of the order of millions of years yet there was geological evidence at the time
that the Earth is older than this. It was not until 1938 that Hans Bethe
provided the solution in the form of new physics. Using the theory of nuclear
physics he was able to show that nuclear reactions in the Sun's core could
provide the relevant energy and the correct lifetime. Similarly, by 1846
astronomers had discovered several irregularities in the orbit of Uranus
compared to the prediction found using Newtonian gravity. Urbain Le Verrier
inferred that there must be another planet orbiting further than Uranus if
the current laws of gravity were correct at this distance. Neptune was indeed
discovered in 1846, another instance of new physics. By 1859, a similar problem
had arisen with Mercury, its orbit was precessing and this could not be
accounted for by either the finite extent of the Sun or perturbations due to
any known planets. After his success with Neptune, Le Verrier again inferred the
existence of a new planet orbiting closer to the Sun: Vulcan. This time
however, no such planet could be found and new physics was not the answer. It
was not until the advent of general relativity that the solution was found. In
some sense, general relativity is a modified theory of Newtonian gravity. In
non-relativistic and low-density environments it reproduces the predictions of
Newton's laws but away from this regime its predictions are drastically
different.  

It has been nearly 100 years since Einstein first published his theory of
general relativity \cite{Einstein:1916vd}. Since then, it has been
remarkably successful at predicting and explaining all known phenomena in the
solar system from the perihelion of Mercury to the bending of light by the Sun.
So much so that modern experiments\footnote{See \cite{Will:2004nx} for a review
of experimental tests of general relativity.} constrain any other theory to be
indistinguishable on solar system scales. The radius of the Earth differs from
the size of the observable universe by 18 orders of magnitude and the density
differs by 41 and so it is natural to ask: is the structure of gravity the
same across all of these scales?

Indeed, there are several reasons why this question is well-motivated. It is
well-known that general relativity is not a renormalisable quantum field theory,
which suggests that it may require ultra-violet (UV) modifications. These would
be indistinguishable from general relativity on cosmological and solar system
scales but could show drastic differences in the strong-field
regime\footnote{One may n\"{a}ively think that we have already probed this
regime in the form of binary pulsars. In fact, whereas neutron stars themselves
are highly relativistic, the Newtonian potential $GM/Rc^2$ of binary pulsar
systems lie in the range 10$^{-5}$--10$^{-4}$ and so these systems only probe
the post-Newtonian structure of gravity.}. There are also hints that the
structure of gravity may be different in the infra-red (IR). The recent discovery
that the universe is accelerating
\cite{Riess:1998cb,Perlmutter:1998np} is one of the biggest
unsolved problems in modern physics. One possible explanation (see
\cite{Copeland:2006wr,Clifton:2011jh} for reviews) is that gravity is altered
on cosmological scales such that it allows for self-accelerating solutions. For
this reason, infra-red modifications of gravity have received a renewed
interest in the last few years.

We are living in an age of precision cosmology and the standard
cosmological model predicated on a cosmological constant and dark matter is the
best-fit to the cosmic microwave background (CMB) as measured by Planck \cite{Ade:2013ktc}. Inherent in this analysis is the
assumption that
density perturbations evolve according to general relativity. When this
assumption is relaxed and the theory of gravity is allowed to vary in
conjunction with the cosmology the current cosmological data constrains the
theory with far less precision than solar system tests \cite{Bean:2010zq}.

The situation will change drastically in the next decade or so with a series of
ambitious programs aimed at probing the structure of gravity on new scales.
Lensing missions such as Euclid \cite{Amendola:2012ys} will
measure the redshift-distance relation and baryon acoustic oscillation (BAO)
scale out to redshift 2 and will hence probe dark energy over much of the
period where it has dominated the expansion of the universe. Imaging surveys
such as the \textit{Large Synoptic Survey Telescope} (LSST)
\cite{Ivezic:2008fe} and the \textit{Dark Energy Survey} (DES)
\cite{Albrecht:2006um} will provide galaxy clustering, supernova and lensing
data that can be combined to constrain the dark energy equation of state and
distinguish dark energy from modified gravity. LSST will also provide optical
data pertaining to variable stars in the nearby universe which can be used to
constrain certain modified gravity models directly. The \textit{Wide-Field
Infra-red Survey Telescope} (WFIRST) \cite{Spergel:2013tha} will
provide near-infra-red data on galactic properties and stellar populations
which can be useful in placing astrophysical constraints on certain modified
theories of gravity. Finally, gravitational wave detectors such as the Advanced
Laser Interferometer Gravitational-Wave Observatory (Advanced-LIGO) \cite{Abbott:2007kv} and eLISA
\cite{AmaroSeoane:2012km}
will probe black hole and neutron star mergers and in-spirals and will thus
provide a window into the strong-field regime of gravity.

One may wonder how such drastic deviations from general relativity are
compatible with the solar system tests that have stood the test of time for
nearly a century. Indeed, general relativity (and a cosmological constant) is
the unique Lorentz-invariant theory of a massless spin-2 particle
\cite{Weinberg:1965rz} and any modification necessarily introduces new degrees of freedom. In the absence of any
symmetry, we would then generically expect these to couple to
matter with order-one strength, which gives rise to additional or
\textit{fifth}- forces. In general, these forces are present at both the
Newtonian and post-Newtonian level and hence must be fine-tuned to be
negligible in order to be compatible with the solar system bounds. This
generally renders them cosmologically uninteresting\footnote{The same is not
necessarily true in the strong-field regime which can show interesting effects
such as spontaneous scalarisation \cite{Damour:1993hw}.}. 

Many theories of modified gravity do indeed fall at this
first hurdle but in the last 10 years or so a new and
interesting class of theories which include \textit{screening mechanisms} have
emerged. Screening mechanisms allow for order-one (or larger) deviations
from general relativity on large, galactic or cosmological scales whilst
\textit{screening} these modifications on smaller scales so that they satisfy the
current solar system bounds. They thus represent both viable and interesting
modifications of general relativity and it is this class of theory that this thesis is concerned with.

Screened modified gravity generally falls into two classes. The chameleon
\cite{Khoury:2003aq,Khoury:2003rn} effect, the symmetron mechanism
\cite{Hinterbichler:2010es} and the environment-dependent Damour-Polyakov effect (EDDP) \cite{Brax:2010gi} all screen by
suppressing the scalar charge to mass
ratio of dense objects and the Vainshtein mechanism \cite{Vainshtein:1972sx}
screens by suppressing the scalar-gradients sourced by dense objects\footnote{We do not consider theories that screen linearly \cite{Sakstein:2014isa,Sakstein:2014aca} in this thesis.}. This thesis investigates several aspects of the first
class\footnote{Reviews of different classes of screened modified gravity can be
found in \cite{Hui:2009kc,Jain:2010ka,Khoury:2010xi,Davis:2011qf,Brax:2012gr,
Sakstein:2013pda,Sakstein:2014nfa}.}, which are referred to collectively as chameleon-like
theories when necessary. 

These mechanisms all rely on dense environments to screen any fifth-forces.
In particular, our own galaxy must be screened and so one must generally look
for new and novel probes in more under-dense environments such as dwarf galaxies,
inter-galactic scales and the cosmological background and to date, experimental
tests of chameleon-like theories have either focused on laboratory experiments
~\cite{Mota:2006ed,Mota:2006fz,Brax:2007hi,Gies:2007su,Brax:2007vm,Brax:2009bk,
Brax:2010xx, Brax:2010gp, Brax:2011wp, Upadhye:2012fz} or cosmological probes\footnote{Recently, binary pulsar tests have also
been considered \cite{Brax:2013uh}.}
\cite{Brax:2004qh,Brax:2005ew,Hu:2007nk,Davis:2009vk,Hinterbichler:2011ca,
Li:2011pj, Li:2011qda,Brax:2012gr, Brax:2012sy,
Li:2012by,Brax:2012nk,Llinares:2012ds,Jennings:2012pt,Lee:2012bm,Brax:2013mua}.
In the last few years, astrophysical effects have emerged as a new and novel
probe of these theories. Astrophysical tests probe scales between the Milky Way
and the Hubble flow and provide smoking-gun signals. This regime is mildly
non-linear so that some degree of screening is present but not so much as to
destroy all observable consequences. Cosmology on the other hand only probes the
linear regime and so any deviations from general relativity are degenerate with other theories of dark energy and
modified gravity. Fully non-linear scales show only minimal deviations
from general relativity. Thus, astrophysical tests have the potential to
constrain these theories to levels unattainable using other methods.

The first part of this thesis is dedicated to finding new and novel
astrophysical probes of chameleon-like theories and using current data to place
new constraints. In particular, we investigate the structure and evolution of
stars in chameleon-like theories of gravity, first using a simple, semi-analytic
method and then by implementing the modified equations of stellar structure into
an existing stellar structure code that is precise enough to allow a
comparison with observational data. Subsequently, we do just this; by examining
the effects on tip of the red giant branch stars we obtain a new and independent
constraint that the Milky way is self-screening\footnote{This had previously
been debated in the field.} and by examining the properties of Cepheid stars we
place new constraints on the model parameters two orders of magnitude stronger
than the previous bounds. At the time of writing, these are currently the
strongest constraints in the literature.

Going beyond equilibrium, we derive the equations governing hydrodynamics in the
presence of modified gravity and perturb them to first-order to find
the new equations governing the radial oscillations of stars about their
equilibrium configurations. We identify two new effects: the period of
oscillations is shorter than one might n\"{a}ively expect using previous
approximations and stars are more stable in modified gravity. Next, we solve the
modified equations numerically using both semi-analytic convective models and
realistic Cepheid models and investigate the size of these effects. We show that
the Cepheid oscillation period can be up to three times shorter than previously
predicted and therefore that the current constraints can be improved using the
same data-sets. Finally, we estimate the possible improvement and discuss the
results in light of future and upcoming astrophysical tests of gravity.

The remainder of this thesis is dedicated to the search for supersymmetric completions of
chameleon-like theories. Recently, the quantum stability of chameleon theories
have come into question after it has emerged that certain models suffer from
strong-coupling issues \cite{Upadhye:2012vh} and that it is difficult to evolve
them through the cosmological era of radiation domination
\cite{Erickcek:2013dea}. One might hope to circumvent these issues by imposing
some sort of symmetry that protects the structure of the theory. Unfortunately,
this is not possible since any Lie symmetry would imply that any
fields in the system are Goldstone bosons with a technically-natural light
mass. This is incompatible with the screening mechanism. The exception is supersymmetry, which transfers this property to a
Goldstone fermion instead. 

An investigation into possible supersymmetric completions is therefore
worthwhile, both to avoid these issues and to aid in the search for an
ultra-violet completion. We will present a framework for embedding
chameleon-like theories into global supersymmetry\footnote{Such a bottom-up
approach is sensible for low-energy infra-red modifications of general
relativity such as these.} and investigate the new features that arise. At finite density, supersymmetry is broken at a scale
that depends on the ambient density and not the TeV scale associated with supersymmetry breaking in the observable sector. We
prove a general no-go theorem showing that when supergravity corrections are accounted for, the model-independent parameters are
already so constrained that it is not possible to probe the theories with laboratory or astrophysical tests\footnote{The
exception to this are no-scale models, which we discuss thoroughly.} and that only the background cosmology and linear
perturbations can show deviations from general relativity. For this reason, we construct a class of supersymmetric chameleon
models and investigate their cosmological behaviour. Like all
chameleon-like models, these ultimately require a cosmological constant to
account for the observed acceleration of the universe
\cite{Wang:2012kj}\footnote{Technically, these models do not satisfy the authors
assumptions, however the conclusions are the same.}. Unlike regular models, the introduction of a cosmological constant at the
level of the action
breaks supersymmetry and is forbidden. Hence, we next introduce a mechanism
where a small cosmological constant can be generated at late times in
the form of a Fayet-Illiopoulos (FI) term and explore the associated parameter
space. Finally, we prove a general no-go theorem showing that, with the exception
of no-scale models, which, to date, have been unsuccessful at incorporating the
chameleon mechanism, every object in the universe is self-screened and the theory's predictions are identical to those of general
relativity in all environments.


This thesis is organised as follows: The rest of this chapter presents some background material. A basic introduction to the
salient features of general relativity, cosmology, dark energy, stellar structure and supersymmetry is given. This is by no means
a comprehensive review and only the aspects relevant to the original work presented in this thesis are included, often without
derivation. In chapter \ref{chap:two} we give a detailed description of modified gravity with screening mechanisms, focussing on
chameleon-like models since these are the main focus of this thesis. In chapters \ref{chap:three}, \ref{chap:four} and
\ref{chap:five} we turn our attention to astrophysical probes of chameleon-like models. In chapter \ref{chap:three} we present
the equations of modified gravity hydrodynamics and use them to derive and solve the new equations governing equilibrium stellar
structure both analytically and numerically. We then discuss potential observational probes using the new features we will
discern. In chapter \ref{chap:four} we will use distance indicator measurements to place new constraints on the model-independent
parameters appearing in these theories, which at the time of writing are the strongest in the literature by two orders of
magnitude. In chapter \ref{chap:five} we perturb the equations of modified gravity hydrodynamics to first order and obtain the
new equations governing the radial, adiabatic oscillations of stars. We find that stars are more stable in modified theories of
gravity than general relativity and furthermore that the oscillation period is far shorter than has previously been estimated. We
discuss these results in the context of current and upcoming astrophysical tests of modified gravity. Next, we change focus and
investigate supersymmetric theories of chameleon-like models in chapters \ref{chap:six} and \ref{chap:seven}. In chapter
\ref{chap:six} we present a general framework for embedding these models into global supersymmetry and discuss the new features
that this construction predicts. In chapter \ref{chap:seven} we present a class of supersymmetric chameleon models and
investigate their cosmology. We conclude in chapter \ref{chap:eight} with a discussion of the future of astrophysical probes of
modified gravity in light of the results presented here and discuss the prospects for finding viable supersymmetric completions.

\section{Conventions}

Throughout this thesis we will use the metric convention
$\eta_{\mu\nu}=\textrm{diag}\left(-,+,+,+\right)$ for the Minkowski metric and
will use the same signature convention when dealing with curved space-times.
Greek letters refer to four-dimensional Lorentzian coordinates when used as
indices and Roman letters likewise for three-dimensional Euclidean coordinates.
When describing identical theories of gravity in different frames the Jordan
frame quantities are distinguished from their Einstein frame counterparts using
tildes, for example, the Jordan frame metric is $\tilde{g}_{\mu\nu}$. $\nabla$
denotes a covariant derivative and $\partial$ a partial derivative. We will work
in units where $\hbar=c=1$. The Planck mass is $\mpl^2=1/8\pi G$. We will often use abbreviations for cumbersome expressions and,
for convenience, a complete list is given in table \ref{tab:abrv}.

\begin{table}
\heavyrulewidth=.08em
	\lightrulewidth=.05em
	\cmidrulewidth=.03em
	\belowrulesep=.65ex
	\belowbottomsep=0pt
	\aboverulesep=.4ex
	\abovetopsep=0pt
	\cmidrulesep=\doublerulesep
	\cmidrulekern=.5em
	\defaultaddspace=.5em
	\renewcommand{\arraystretch}{1.6}
\centering
\begin{tabu}{l|[2pt] l}
Abbreviation & Expression   \\
\tabucline[2pt]{-}
AGB & asymptotic giant branch  \\
CDM & cold dark matter \\
CGS & centimeter-grams-seconds \\
CMB & cosmic microwave background \\
EDDP & environment-dependent Damour-Polyakov effect\\
FI & Fayet-Illiopoulos \\ 
FLRW & Friedman-Lema\^{i}tre-Robertson-Walker \\
LAWE & linear adiabatic wave equation\\
LOS & line of sight \\
MLAWE & modified linear adiabatic wave equation  \\
PNLF & planetary nebula luminosity function\\
TRGB & tip of the red giant branch \\
VEV & vacuum expectation value
\end{tabu}
\caption{Abbreviations used throughout this thesis.}
\label{tab:abrv}
\end{table}

\section{General Relativity}
\label{sec:chap1GR}

Before discussing any modified theory of gravity one must first discuss general relativity. This is governed by the
Einstein-Hilbert action
action\footnote{Up to boundary terms that are not relevant for this thesis.} 
\begin{equation}\label{eq:EHaction}
 S_{\rm EH}=\int \dd^4 x\sqrt{-g}\frac{\mpl^2}{2}R+S_{\rm
m}[g_{\mu\nu};\psi_{\rm i}],
\end{equation}
where $\psi_{\rm i}$ represent the various matter fields. Varying this with
respect to the metric yields the Einstein field equations
\begin{equation}\label{eq:einsteinE}
 G_{\mu\nu}=8\pi G T_{\mu\nu},
\end{equation}
where $T_{\mu\nu}$ is the energy-momentum tensor for matter and the Einstein
tensor $G_{\mu\nu}=R_{\mu\nu}+1/2Rg_{\mu\nu}$ with $R_{\mu\nu}$ and $R$ the
Ricci tensor and scalar respectively. In this theory of
gravity, the energy-momentum tensor is conserved, $\nabla_\mu T^{\mu\nu}=0$ and
this implies that particles move on geodesics of the metric. To see this, one
can consider the energy-momentum tensor for a pressureless fluid of density
$\rho$:
\begin{equation}\label{eq:fluidTP=0}
 T^{\mu\nu}=\rho u^{\mu}u^{\nu},
\end{equation}
where $u^\mu={x^\prime}^\mu$ is the 4-velocity and a dash denotes a derivative
with
respect to the affine parameter $\lambda$. The conservation of the
energy-momentum tensor then implies that (recall $\nabla_\mu(\rho u^\mu)=0$)
\begin{equation}\label{eq:geode1}
 u^\mu\nabla_\mu u^\nu=u^\mu(\partial_\mu
u^\nu+\Gamma^\nu_{\mu\alpha}u^\alpha)={x^{\prime\prime}}^\nu+\Gamma^\nu_{
\mu\alpha } { x^\prime }
^\alpha{x^\prime}^\mu=0,
\end{equation}
which is the geodesic equation. This can also be found by extremising the action
for a point particle
\begin{equation}\label{eq:pointparticleGR}
 S=-m\int\dd s ,
\end{equation}
where $m$ is the mass of the
particle and $s=\sqrt{-g_{\mu\nu}{x^\prime}^\mu{x^\prime}^\nu}$ is the proper
time. One may then find the non-relativistic limit of this theory by
considering perturbations about Minkowski space-time in the \textit{conformal
Newtonian gauge}
\begin{equation}\label{eq:minkpertconfnewt}
 \dd s^2=-(1+2\Phi)\dd t^2+(1-2\Psi)\dd x^2,
\end{equation}
where the metric potentials are time-independent. Using this metric along with
the non-relativistic condition $\dd x^i/\dd x^0=v/c\ll 1$ in the geodesic
equation (\ref{eq:geode1}) we find that the particles evolve according to
\begin{equation}\label{eq:Newton1}
 \ddot{\vec{x}}=-\nabla\Phi,
\end{equation}
where a dot denotes a derivative with respect to the time coordinate. Equation (\ref{eq:Newton1}) is simply Newton's second law
with $\Phi$ identified with the Newtonian
potential $\pn$. Using the Einstein equations (\ref{eq:einsteinE}) with a non-relativistic density source
$T^{\mu\nu}=\mathrm{diag}(\rho,P,P,P)$ such that $P\ll\rho c^2$ we find the Poisson equation
\begin{equation}\label{eq:poisson1}
\nabla^2\pn=4\pi G \rho. 
\end{equation}
One can then see that general relativity reproduces Newton's law of gravitation
exactly in the non-relativistic limit.

We have already mentioned that one can augment the Einstein-Hilbert action (\ref{eq:EHaction}) to include a cosmological constant
$\Lambda$ without introducing new degrees of freedom. In this case, the action is
\begin{equation}\label{eq:EHactionCC}
 S_{\rm EH}=\int \dd^4 x\sqrt{-g}\frac{\mpl^2}{2}\left(R-2\Lambda\right)+S_{\rm
m}[g_{\mu\nu};\psi_{\rm i}],
\end{equation}
which yields the modified equation 
\begin{equation}\label{eq:einsteinCC}
 G_{\mu\nu}=8\pi G T_{\mu\nu}-\Lambda g_{\mu\nu}.
\end{equation}
In the Newtonian limit, the effective Newtonian potential is
\begin{equation}
 \label{eq:newtonialimitcc}
\pn=\Phi_{\rm N}^{\Lambda=0}-\frac{1}{6}\Lambda r^2
\end{equation}
corresponding to a repulsive force. Current measurements \cite{Ade:2013ktc}
indicate that $\Lambda/\mpl^2\sim\mathcal{O}(10^{-120})$ and so one
generally has $|\Phi_{\rm N}^{\Lambda=0}|\gg|\Lambda r^2|$. The exception to
this is the largest scales since outside any matter distribution one generically
has $|\Phi_{\rm N}^{\Lambda=0}|\propto r^{-1}$. This term is then only relevant on cosmological scales. One can then see that the
addition of a cosmological constant to the Einstein-Hilbert action represents an infra-red modification of general relativity.

\section{Cosmological Solutions of General Relativity}

In this section we will briefly review some aspects of cosmology as predicted
by general relativity. This is a rich and vast subject and a full account would
be both lengthy and beyond the scope of this thesis. We will hence include only
those features that are directly related to the original work presented later. 

The standard cosmological model is predicated on the observation that on
large scales, the universe is spatially homogeneous and isotropic. This is
known as the \textit{cosmological principle}, and the most general space-time compatible
with its assumptions is the Friedmann-Lema\^{i}tre-Robertson-Walker (FLRW)
metric
\begin{equation}\label{eq:FRWmetrick}
 \dd s^2=-\dd t^2+a^2(t)\left[\frac{\dd x^2}{1-\kappa
x^2}+x^2\dd\Omega_2^2\right],
\end{equation}
where $\dd\Omega_2$ is the line element on a 2-sphere, $x$ is the co-moving
radial coordinate and $a(t)$ is the scale factor of the universe and is the only
dynamical degree of freedom. Its dynamics depend both on the matter content of the universe and the theory of gravity. When it is
non-constant, the
metric describes an expanding universe and physical distances are given by
\begin{equation}\label{eq:phycoorddef}
 \dd r=a(t)\dd x.
\end{equation}
We are free to choose the normalisation of $a$ and in this thesis we will
always choose a normalisation such that its value today $a_0=1$. $\kappa$ is the
spatial curvature of the universe; if it is identically zero then the universe
is spatially flat, if it is greater than zero the universe is spatially a
3-sphere and if it is negative the universe is a 3-hyperboloid. 

Before
specifying a theory of gravity there are some theory-independent quantities that
can be defined. The Hubble parameter is
\begin{equation}\label{eq:hubbledef}
H(t)=\frac{\dot{a}}{a},
\end{equation}
where a dot denotes a derivative with respect to coordinate time $t$. Unlike
$a(t)$ this is a physical observable and a useful quantity is its present day
value $H_0$.
This is often parametrised
using the dimensionless number $h$ as $H_0=100h$ km s$^{-1}$ Mpc$^{-1}$. Another
useful quantity is
the redshift
\begin{equation}\label{eq:zdef}
 1+z=\frac{\lambda_{\rm o}}{\lambda_{\rm e}},
\end{equation}
 where $\lambda_{\rm e}$ is the wavelength of
light emitted by some source and $\lambda_{\rm o}$ is the
wavelength observed at the present time. Using the FLRW metric
(\ref{eq:FRWmetrick}) one can show that
\begin{equation}
 a(t)=\frac{1}{1+z}.
\end{equation}
Finally, one needs to define the notion of distance in an expanding universe.
The instantaneous distance is not useful since it is unobservable and
may even refer to space-like separated points. Instead, we shall work with the
\textit{luminosity distance}. When we look out into the sky and observe we see
photons moving on radial null geodesics emitted at some redshift $z$ and
observed at $z=0$. These have been red-shifted and diluted by the expansion of
the universe and have also had their path length altered from the static case.
Using the FLRW metric (\ref{eq:FRWmetrick}), the co-moving distance
between us and the point at which the light was emitted is\footnote{We have set
$\kappa=0$ here because, as we shall see later on, observational evidence
indicates that the universe is very close to flat.}  
\begin{equation}\label{eq:lumdistcomov}
 x(z)=\int_0^z\frac{\dd z}{H(z)}.
\end{equation}
Now in Minkowski space, the flux F at radius $r$ is simply the luminosity per
unit area and so one has
\begin{equation}F=\frac{L}{4\pi r^2},\end{equation}
where $L$ is the bolometric luminosity of the source and $F$ is the observed flux. Generalising this to an expanding space-time we
define the luminosity distance via 
\begin{equation}\label{eq:lumdistdef}
{d_L}^2=\frac{L}{4\pi F }.
\end{equation}
The luminosity is not directly
observable and instead we must infer it using known principles such as standard candles. There is an added complication in an
expanding space-time that the observed flux integrated over a sphere of radius $4\pi r^2$ is not equal to the luminosity. This is
because the radiation we observe at some wavelength $\lambda_{\rm o}$ has been redshifted by the expansion of the universe from
its emitted value $\lambda_{\rm e}$, therefore diluting its energy content. Furthermore, the time-period over which we observe is
longer than the period over which the light was emitted owing to the increased wavelength. Ultimately, these two effects are due
to the non-conservation of the photon number and energy density in an expanding space-time.
Using equation (\ref{eq:zdef}) and the fact that the energy of a single photon is proportional
to the inverse of its wavelength, we have
\begin{equation}
 \label{eq:lumdisderv}
\frac{\Delta E_{\rm e}}{\Delta E_{\rm o}}=\frac{\lambda_{\rm
o}}{\lambda_{\rm e}}=1+z,
\end{equation}
where $E_{\rm o}$ and $E_{\rm e}$ are the observed and emitted photon energies respectively. Since the speed of light is
constant\footnote{Or rather, ignoring effects from regions of finite permittivity or permeability between us and the source.},
the time-period for one wave-cycle is proportional to the wavelength and so
\begin{equation}
 \frac{\Delta t_{\rm o}}{\Delta t_{\rm
e}}=1+z.
\end{equation}
The observed flux is then related to the luminosity of the source by
\begin{equation}\label{eq:lumobs}
F=\frac{L}{4\pi x^2(1+z)^2}
\end{equation}
\begin{equation}\label{eq:lumdistformula}
 d_L= \sqrt{\frac{L_{\rm emitted}}{4\pi F}}= (1+z)x(z)=(1+z)\int_0^z\frac{\dd
z}{H(z)}
\end{equation}
using equation (\ref{eq:lumdistcomov}). This result is independent of the
theory of gravity, however $H(z)$ is not and so any measured luminosity distance is sensitive to the theory of gravity for a
fixed matter content.

\subsection{The Standard Cosmological Model}

Using the FLRW metric (\ref{eq:FRWmetrick}) in the Einstein field equations
(\ref{eq:einsteinE}) will yield the equations governing the dynamics of $H(t)$ but first we
must specify an energy-momentum tensor. The standard cosmological model takes
the universe to consist of a non-interacting multi-component perfect fluid with energy-momentum
tensor
\begin{equation}\label{eq:emtensperfluid}
 T^{\mu\nu}=(\rho+P)u^\mu
u^\nu+Pg^{\mu\nu}=\textrm{diag}\left(\rho,P,P,P\right),
\end{equation}
where $u^\mu$ is again the 4-velocity, $\rho$ is the total energy density and
$P$ is the total pressure. Using this in the Einstein equation (with a
cosmological constant) (\ref{eq:einsteinCC}), one arrives at the Friedmann
equations
\begin{align}\label{eq:GRfriedmann}
 H^2&=\frac{8\pi G}{3}\rho-\frac{\kappa}{a^2}+\frac{\Lambda}{3}\\
\frac{\ddot{a}}{a}&=-\frac{4\pi
G}{3}\left(\rho+3P\right).\label{eq:GRacceleratione}
\end{align}
In general relativity, different types of matter do not exchange energy so that
the energy-momentum tensor for each species is covariantly
conserved independently ($\nabla_\mu T_{\rm i}^{\mu\nu}=0$). This leads to the continuity
equation for each species:
\begin{equation}\label{eq:GRcontinuity}
 \dot{\rho\iii}+3H\left(\rho\iii+P\iii\right)=0.
\end{equation}
The system of equations (\ref{eq:GRfriedmann})--(\ref{eq:GRcontinuity}) does not
close and one must specify an equation of state relating the pressure
to the density. This is usually taken to be of the form
\begin{equation}\label{eq:eos_cosmo}
 P_{\rm i} = w_{\rm i}\rho_{\rm i}
\end{equation}
where $w\iii=0 $ for non-relativistic matter, $1/3$ for radiation and $-1$ for
a cosmological constant and is referred to as the \textit{equation of state
parameter} or simply the \textit{equation of state}. Solving the continuity equation for an arbitrary equation of state
one finds
\begin{equation}\label{eq:contsol} \rho\iii= \left\{
  \begin{array}{l l}
   \frac{\rho_0}{a^{3(1+w)}} & \quad w\iii\ne -1\\
    \rho_0 & \quad w\iii=-1
  \end{array} \right. ,
\end{equation}
where $\rho_0$ is the present day density and is equal to $\mpl^2\Lambda$ for a
cosmological constant. One may then define the \textit{density
parameter} for species $\rm i$:
\begin{equation}
 \Omega\iii=\frac{\rho\iii}{3H_0^2\mpl^2}\quad w\iii\ne -1.
\end{equation}
The cosmological constant requires a special treatment so we define
$\Omega_\Lambda\equiv \Lambda/3H^2$ and we define
$\Omega_\kappa=-\kappa/a^2H^2$. The Friedmann equation (\ref{eq:GRfriedmann}) is
then 
\begin{equation}\label{eq:kappato1}
 \sum_{\rm i}\Omega\iii+\Omega_\Lambda+\Omega_\kappa=1.
\end{equation}
If we then define $\Omega=\sum_{\rm i}\Omega\iii+\Omega_\Lambda$ we can set
$\Omega_\kappa=0$ in (\ref{eq:kappato1}) to find 
\begin{equation}\label{eq:flatnesscond}
 \Omega=\sum_{\rm i}\Omega\iii+\Omega_\Lambda=1,
\end{equation}
which implies
\begin{equation}\label{eq:flatnessconddens}
 \sum_{\rm i}\rho\iii+\rho_\Lambda=3H^2\mpl^2
\end{equation}
This is an important equation, it tells us that if the universe is spatially
flat then the total density is constrained to be equal to the \textit{critical density} $\rho_{\rm c}=3\mpl^2H^2$. Moreover,
equations
(\ref{eq:flatnesscond}) and (\ref{eq:flatnessconddens}) hold at all times. In a
spatially flat universe $\Omega\iii(t)$ is then a measure of the fraction that
species i contributes to the total density. Current observational experiments
(see \cite{Ade:2013ktc} for example) indicate that $\Omega_\kappa$ is indeed
very close to zero and so from here on we will work with a spatially flat
universe\footnote{Technically this is only correct for general relativity
since the result is derived from a Bayesian fit to the CMB data and assuming
general relativity. The theories considered later all behave like
general relativity at early times and outside the horizon. Furthermore, null geodesics are unaffected by the theories we will
study and so we expect this observation to be largely independent of our theory of gravity. That being said, there are minor
second-order effects when the radiation fluid is tightly coupled to the baryons since the equations governing
non-relativistic perturbations are altered \cite{Brax:2011ta}.}. 

\subsection{The $\Lambda$CDM Model}

\subsubsection{Background Cosmology}

The $\Lambda$-Cold Dark Matter ($\Lambda$CDM) model is the Planck best fit
\cite{Ade:2013ktc} model to several different cosmological observables and is a
model of the late-time universe well after big bang nucleosynthesis (BBN). It
describes a spatially flat universe composed of baryons, cold dark matter (CDM),
radiation and a cosmological constant. Including these components and using the
solution to the continuity equation (\ref{eq:contsol}), we can write the
Friedmann equation (\ref{eq:GRfriedmann}) in the form
\begin{equation}\label{eq:friedmannlcdm}
 H^2=H_0^2\left(\frac{\Omega_{\rm c}^0+\Omega_{\rm
b}^0}{a^3}+\frac{\Omega_\gamma}{a^4}+\Omega_\Lambda\right),
\end{equation}
where c refers to cold dark matter, b to baryons and $\gamma$ to radiation.
Since $a(t)$ is a monotonically increasing function of time, the various
terms come to dominate at different times resulting in different epochs where
different species dominate. The details of the very early universe are
unimportant for this thesis and so we assume that the universe exits from
inflation (or some other early universe process) with a scale-invariant power
spectrum and leave the processes of reheating and preheating unspecified.
At this time, $a$ is very small and the radiation term dominates leading to a
radiation dominated era. As the scale factor grows, the baryons and dark matter
come to dominate leading to a matter dominated era that begins at a redshift
$z\sim1000$. Finally, as these dilute and the scale factor continues to grow,
only the cosmological constant is left and this dominates the subsequent
evolution of the universe. The solutions of equation (\ref{eq:friedmannlcdm})
deep in each of these eras so that the other terms can truly be neglected are
\begin{equation}\label{eq:friedsolns} a(t)= \left\{
  \begin{array}{l l}
   \left(\frac{t}{t_0}\right)^{\frac{2}{3}} & \quad \textrm{matter domination}\\
    \left(\frac{t}{t_0}\right)^{\frac{1}{2}} & \quad \textrm{radiation
domination}\\
e^{\frac{\Lambda}{3}t} & \quad \Lambda\textrm{ domination}
  \end{array} \right. ,
\end{equation}
where $t_0$ is the time today (i.e. the age of the universe). The \textit{Planck} best
fit parameters are shown in table \ref{tab:planckcosmo}.
\begin{table}
\centering
\heavyrulewidth=.08em
	\lightrulewidth=.05em
	\cmidrulewidth=.03em
	\belowrulesep=.65ex
	\belowbottomsep=0pt
	\aboverulesep=.4ex
	\abovetopsep=0pt
	\cmidrulesep=\doublerulesep
	\cmidrulekern=.5em
	\defaultaddspace=.5em
	\renewcommand{\arraystretch}{1.6}
\begin{tabu}{c|[2pt]c}
Parameter & Planck best-fit value\\
\tabucline[2pt]{-}
$\Omega^0_{\rm b}h^2$ & $0.022$   \\
$\Omega^0_{\rm c}h^2$ & $0.12$  \\
$\Omega^0_\gamma$ & $\mathcal{O}(10^{-5})$  \\
$\Omega^0_\Lambda$ & $0.68$  \\
$H_0$ & $68.14$ km s$^{-1}$ Mpc$^{-1}$\\
\end{tabu}
\caption{The Planck best fit cosmological parameters, taken from
\cite{Ade:2013zuv}.}
\label{tab:planckcosmo}
\end{table}
The values of $\Omega_\Lambda^0$ and $\Omega_{\rm c}^0$ are of the same order
of magnitude and so we are currently living in the transition period between
matter and cosmological constant domination. The mystery of why we observe
during this small transition time and not any other has been dubbed the
\textit{coincidence problem} and we will have nothing to say about it in this
thesis. 

\subsubsection{Linear Perturbations}

So far we have only looked at the background cosmology but many interesting
cosmological probes including the CMB arise due to departures from the
homogeneous and isotropic background in the form of linear perturbations.
Linear perturbation theory in general relativity and cosmology is a broad and
interesting subject and a full treatment is well beyond the scope of
this thesis. Here we will only focus on the evolution of perturbations to the
cold dark matter density deep inside the horizon since we will investigate this for chameleon-like models in chapter
\ref{chap:seven}. 

Throughout this thesis we will work in the \textit{conformal Newtonian gauge},
in which the perturbed form of the flat FLRW metric (\ref{eq:FRWmetrick}) is
fully specified by the two potentials $\Phi$ and $\Psi$\footnote{Note that these are different metric perturbations from those
defined in equation (\ref{eq:minkpertconfnewt}).}
\begin{equation}\label{eq:confnewtgauge}
 \dd s^2=-(1+2\Phi)\dd t^2+a(t)^2(1-2\Psi)\dd x^2.
\end{equation}
We also need to specify the perturbations to the cold dark matter
energy-momentum tensor. For our purposes, it will be enough to specify the
perturbation to the 00-component only, which is the density perturbation
$\delta\rho$ in this gauge. This choice completely fixes all the gauge degrees
of freedom and there is no residual gauge redundancy provided the potentials
decay to zero at spatial infinity. We define the linear density contrast $\Dc$
via
\begin{equation}\label{eq:CDMcontrast}
 \Delta_{\rm c}(x)\equiv \frac{\delta\rho_{\rm c}}{\bar{\rho}_{\rm c}},
\end{equation}
where barred quantities refer to unperturbed, background quantities. This is
defined in position space but in practice it is useful to Fourier-transform the
spatial part and so we define
\begin{equation}\label{eq:ftDC}
 \Dc=\int\frac{\dd^3\vec{k}}{\left(2\pi\right)^3}\tdc(\vec{k},t)e^{i\vec{k}
\cdot\vec{x}}.
\end{equation}
Since the theory is both translationally and rotationally invariant\footnote{We
will not deal with theories where this is not the case in this thesis.} we
generally have $\tdc=\tdc(t,k)$. Physically, a given $k$-mode corresponds to the
wave number of a perturbation of proper wavelength $\lambda$:
\begin{equation}\label{eq:klambda}
 k=\frac{2\pi a}{\lambda}.
\end{equation}
Upon solving the perturbed Einstein and energy
conservation equations, one finds that the linear density contrast evolves
according to
\begin{equation}
 \label{eq:cdmlinearGR}
\ddot{\tdc}+2H\dot{\tdc}-4\pi G\bar{\rho}_{\rm c}\tdc=0.
\end{equation}
Eliminating the factor of $G\bar{\rho_{\rm c}}$ using the Friedmann equation
(\ref{eq:GRfriedmann}) we have
\begin{equation}
 \label{eq:CDMlinear2}
\ddot{\tdc}+2H\dot{\tdc}-\frac{3}{2}H^2\frac{\bar{\rho}_{\rm c}}{\bar{\rho_{\rm
c}}+\bar{\rho}_{\gamma}}\tdc=0,
\end{equation}
where we have neglected the contribution from the baryons and the cosmological
constant. During matter domination when $\bar{\rho}_{\rm
c}\gg\bar{\rho}_{\gamma}$ the growing solution is
\begin{equation}\label{eq:CDlinearGRev}
 \tdc\propto t^n;\quad
n=-\frac{1}{6}+\frac{1}{2}\sqrt{\frac{1}{9}+\frac{8}{3}\Omega_{\rm c}^0}.
\end{equation}
Now $\Dc$ and $\tdc$ are not physical observables since $\dc$ is a gauge-dependent
quantity and so we need to relate it to something we can physically measure in
order to extract the model predictions. The relevant quantity is the two-point
correlation function or \textit{power spectrum} defined via
\begin{equation}\label{eq:powerspecdef}
 \langle\tdc(t,\vec{k})\tdc^*(t,\vec{k}')\rangle=\frac{2\pi^2}{k^3}P_{\dc{\dc}^*}
(k)\delta^{(3)}\left(\vec{k}-\vec{k}'\right).
\end{equation}
For brevity, we will often denote $P_{\dc{\dc}^*}(k)$ by $P(k)$.

\section{Dark Energy}

We have already alluded to the presence of dark energy in the universe in the
previous section where we included a cosmological constant in our model of the
universe and noted that $\Lambda$CDM is currently the best-fit to the
cosmological data. Equation (\ref{eq:friedsolns}) shows that the universe expands exponentially when the
cosmological constant dominates. Indeed, if
one takes the second Friedmann equation (\ref{eq:GRacceleratione}) and sets
every component except $\Lambda$ equal to zero we have $\ddot{a}>0$ so
that the universe is accelerating. This acceleration seems
counter-intuitive. Gravity dominates the evolution of the universe on large
scales and this is an attractive force between all matter and so surely the
universe should be decelerating? The acceleration is due to the fact that
$w_\Lambda=-1$ so that the pressure of the cosmological constant is negative. This behaviour is not
unique to the cosmological constant. Indeed, examination of equation
(\ref{eq:GRacceleratione}) shows that any fluid satisfying 
\begin{equation}\label{eq:wforacceleration}
 w<-\frac{1}{3}
\end{equation}
will accelerate if it dominates the universe's expansion. In 1998, two teams
\cite{Perlmutter:1998np,Riess:1998cb} published the luminosity distance
measured from supernovae distances estimates and found that a flat universe
composed only of matter and radiation could not fit the data well; a component
with $w\approx-1$ is needed. The universe is indeed accelerating. Since then there have been many independent experiments using
cosmological probes such as weak lensing, baryon acoustic oscillations and the integrated Sachs-Wolfe effect that all point to the
need for dark energy and a full review may be found in \cite{Copeland:2006wr}. The most compelling evidence by far comes from the
temperature-temperature power spectrum of the CMB including the effects of lensing \cite{Sherwin:2011gv}\footnote{The
power spectrum alone does not determine $\Omega\mmm$ and $\Omega_\Lambda$ uniquely; their values are degenerate with $H_0$.
Including the effects of lensing breaks this degeneracy.}. The Planck
limits on $w$ vary depending on the model assumed \cite{Ade:2013zuv} but there
is a large region around $w=-1$ that is not excluded and so the underlying mechanism driving this acceleration is far from
clear. The mysterious component causing this acceleration has been dubbed \textit{dark energy}. Its physical origin is one of the
biggest unsolved problems in modern physics.

%
\subsection{A Simple Example: Quintessence}

Quintessence models are an attempt to promote the cosmological constant to a dynamical field. The simplest models have a
canonically normalised scalar field slowly rolling down a potential $V(\phi)$ and are described by the action:
\begin{equation}
 \label{eq:quintessenceaction}
S=\int\dd^4x\sqrt{-g}\left[\frac{\mpl^2}{2}R-\frac{1}{2}
\nabla_\mu\phi\nabla^\mu\phi-V(\phi)\right]+S_{\rm m}[g_{\mu\nu};\psi\iii].
\end{equation}
Note that the field is \textit{minimally coupled} to gravity in the sense that there are no direct couplings to any curvature
tensors or to the matter degrees of freedom. The (conserved) energy-momentum tensor for the field is
\begin{equation}\label{eq:quintessenceemtens}
 T^{\mu\nu}=\nabla^\mu\phi\nabla^\nu\phi-g^{\mu\nu}\left[\frac{1}{2}
\nabla_\mu\phi\nabla^\mu\phi+V(\phi)\right],
\end{equation}
from which one can obtain the energy density
\begin{equation}\label{eq:densityquint}
 \rho=T^{00}= \frac{1}{2}\dot{\phi}^2+V(\phi)
\end{equation}
and the pressure
\begin{equation}\label{eq:pressquint}
 P=T^{|ii|}= \frac{1}{2}\dot{\phi}^2-V(\phi),
\end{equation}
where the use of $|ii|$ notation indicates that we are not summing over repeated indices. The equation of state is then
\begin{equation}\label{eq:quinteos}
 w_\phi=\frac{\dot{\phi}^2-2V(\phi)}{\dot{\phi}^2+2V(\phi)}.
\end{equation}
One can see that when $V(\phi)\gg\dot{\phi}^2$ we have $w_\phi\approx-1$ and so provided that the field is slowly-rolling, this
model mimics the effects of a cosmological constant. In practice, these models suffer from fine-tuning problems since one needs
to tune the initial conditions and the parameters in the potential in order to reproduce the measured values of the equation of
state and the energy density in dark energy. Furthermore, if one wishes to address the coincidence problem then one must
fine-tune further so that field starts to dominate the energy density of the universe around the current epoch. More complicated
models attempt to address these issues by looking for late-time attractors that give a dark energy dominated universe no
matter the initial conditions. Some also exhibit \textit{scaling solutions} where the dark energy tracks the evolution of the
dominant fluid component and remains sub-dominant until some mechanism causes it to deviate from this solution and quickly
dominate the universe. We will not be concerned with these models here and the interested reader is referred to
\cite{Copeland:2006wr} and references therein. 

\section{Stellar Structure in General Relativity}\label{sec:grstars}

Part of this thesis is concerned with the structure and evolution of stars in modified gravity and so in this section we will
change direction and briefly discuss the structure of stars in general relativity. We will see in chapter \ref{chap:two} that the
theories studied in this thesis only show
novel effects in non-relativistic stars and so we will limit our discussion to these.

\subsection{The Equations of Stellar Structure}

The equilibrium structure of non-relativistic stars are described by the stellar structure equations. In chapter
\ref{chap:three} we will derive them\footnote{In fact, we will derive them for modified theories of gravity but the general
relativity
equations can be obtained in a straightforward manner by setting the additional terms to zero.} formally from
modified
gravity hydrodynamics and so here we shall simply present them and discuss their solutions.

By equilibrium, we refer to hydrostatic equilibrium where a star maintains a spherical shape with constant radius and supports
itself against gravitational collapse by balancing the inward force with an outward pressure gradient. This is described by the
\textit{hydrostatic equilibrium equation}
\begin{equation}\label{eq:hsegr}
 \frac{\dd P}{\dd r}=-\frac{GM(r)\rho(r)}{r^2},
\end{equation}
where $r$ is the radial coordinate (with the centre of the star at $r=0$), $P(r)$ is the pressure and $\rho(r)$ is the density.
$M(r)$ is the mass enclosed within radius $r$. If the radius is $R$ then the total stellar mass is $M\equiv M(R)$. 
Since the start is spherically symmetric the mass is related to the density via the continuity equation,
\begin{equation}
\frac{\dd M}{\dd r} = 4\pi r^2\rho.\label{eq:continuity1}
\end{equation}
Photon propagation in the interior of the star is described by the radiative transfer equation,
\begin{equation}\label{eq:radiative1}
\frac{\dd T}{\dd r} = -\frac{3}{4 a} \frac{\kappa(r)}{T^3} \frac{\rho(r) L(r)}{4\pi r^2},
\end{equation}
where $L(r)$ and $T(r)$ are the luminosity and temperature respectively at coordinate $r$. The quantity $\kappa$ is known as the
\textit{opacity} and represents the cross section for radiation absorption per unit mass; it is generally a function of the
temperature and density. Finally, if energy is generated --- by nuclear (or possibly other)
processes --- at a rate $\epsilon(r)$ per unit volume then the luminosity gradient is
determined by the energy generation equation,
\begin{equation}
\frac{\dd L}{\dd r} = 4\pi r^2 \rho \epsilon(r).\label{eq:engen1}
\end{equation}

Taken by themselves, these equations do not close and one must specify the
equations of state relating $P, \rho, \kappa$ and $\epsilon$, which are
themselves determined by further equations involving energy transfer and nuclear
burning networks.

One can see that the stellar structure equations include aspects of nuclear, atomic, thermal and gravitational physics and are
hence perfect laboratories for testing fundamental physics.

\subsection{Lane-Emden Models}\label{subsec:legr}

The equations of stellar structure are incredibly complicated and in order to achieve realistic models one must couple them to a
complete set of atmosphere models, nuclear burning networks and opacity tables. It is often necessary to include effects such as
convection and mass-loss.
In practice, this requires complicated numerical codes if one wishes to find models that are realistic enough to compare to data.
If one is only interested in the gross physical features then there are some simple
approximations that one can make. In this thesis we will be interested in the effects
of changing the theory of gravity whilst leaving the other stellar physics unaltered\footnote{This is not to say that the other
physics is unaltered. Indeed, we will see in the coming chapters that changing the theory of gravity will result in changes to
the non-gravitational features such as the luminosity and the temperature. The important difference is that the physical theory
governing these features has not been altered and the new phenomena exhibited are the response to the change in the gravitational
theory.}. A particularly
useful class of simple
models for investigating the effects of gravity are \textit{Lane-Emden} models. Lane-Emden models make the simplifying
assumption that the star can be described as a barotropic fluid with a \textit{polytropic} equation of state 
\begin{equation}\label{eq:polytropic1} 
 P=K\rho^{\gamma},
\end{equation}
where $K$ is a constant and $\gamma$ is known as the \textit{adiabatic index}. Main-sequence stars are well described by
$\gamma=4/3$ \cite{prialnik2000introduction} whereas convective regions in post-main-sequence stars such as red giants are well
approximated by $\gamma=5/3$ \cite{kippenhahn1990stellar}. In practice, it is more convenient to work with the polytropic index
$n$ defined by
\begin{equation}\label{eq:polyindex1}
 \gamma=\frac{n+1}{n}
\end{equation}
so that $\gamma=4/3$ corresponds to $n=3$ and $\gamma=5/3$ corresponds to $n=1.5$. Using this assumption, the hydrostatic
equilibrium equation and the continuity equation do not depend on temperature and so are decoupled from the radiative transfer and
energy generation equations and this allows us to solve the first two for the structure of the star independent of the energy
source or opacity profile. This is why these models are good for investigating the effects of changing the theory of gravity.
These are two coupled first-order differential equations and so we require two boundary conditions at
the centre of the star. The central pressure $P_{\rm c}$ is arbitrary and we take this to be one boundary condition. The central
density $\rho_{\rm c}$ is then fixed by equation (\ref{eq:polytropic1}). The second boundary condition comes from the assumption
of spherical symmetry, which requires that the pressure (or equivalently the density) is a smooth function of $r$ and so its
derivative
must vanish at $r=0$. The problem is then reduced to that of solving for the pressure as a function of the radial position. 

The stellar structure equations are self-similar and so we
can work with both a dimensionless coordinate and a dimensionless pressure. We begin by defining the Lane-Emden
coordinate\footnote{Note that this is typically called $\xi$ in the literature. Later on we will use $\xi$ to refer to the
Fourier transform of the radial perturbation when studying stellar oscillations and so here we use $y$ instead.}
\begin{equation}\label{eq:LEcoorddef}
 y=r/r_{\rm c},
\end{equation}
where
\begin{equation}\label{eq:rc}
 r_{\rm c}^2 \equiv \frac{(n+1)P_{\rm c}}{4\pi G\rho_{\rm c}^2}.
\end{equation}
Next, we define the dimensionless pressure variable $\thn$ via 
\begin{equation}\label{eq:LEpressurevar}
 P(y)=P_{\rm c}\thn^{n+1}(y);\quad\rho(y)=\rho_{\rm c}\thn^n(y).
\end{equation}
Substituting the equation of state (\ref{eq:polytropic1}) into the hydrostatic equilibrium equation (\ref{eq:hsegr}) we can
differentiate
once and eliminate $\dd M(r)/\dd r$ using equation (\ref{eq:continuity1}) in order to find the Lane-Emden equation
\begin{equation}\label{eq:Lane-EmdenGR}
 \frac{1}{y^2}\frac{\dd}{\dd y}\left(y^2\frac{\dd \thn}{\dd y}\right)=-\thn^n
\end{equation}
with boundary conditions $\thn(0)=1$ ($P(0)=P_{\rm c}$) and $\dd\thn/\dd y(0)=0$ ($\dd P/\dd r(0)=0$). The Lane-Emden equation can
then be solved numerically for any value of $n$ for the function $\thn(y)$, which fully
specifies the structure of the star for a given $P_{\rm c}$ and $\rho_{\rm c}$ (or equivalently $K$). The definition of the
stellar radius is slightly ambiguous. In reality, one observes light emitted from the photosphere, which is defined as the
surface at which the optical depth falls to $2/3$. Stars that are simple enough to be well-described by Lane-Emden models have
radii that are well approximated by the condition that the pressure falls to zero at the stellar radius and so we
have $R=r_{\rm c}y_R$ where $y_R$ is defined by $\thn(y_R)=0$\footnote{One must be careful because the
solution of the Lane-Emden equation does not go to zero at finite $y$ when $n>4$ and so one needs to choose the
radius according to some prescription set by the problem. This will not be a problem for the work presented in this thesis since
we will only investigate the cases $n=3$ and $n=1.5$,
however this is a problem for isothermal core models that have $n=\infty$. }. Finally, one can integrate equation
(\ref{eq:continuity1}) to find the mass of the star:
\begin{equation}\label{eq:LEmass}
 M=\int_0^{y_R}4\pi r_{\rm c}^3\rho_{\rm c} y^2\thn^n\dd y= -4\pi r_{\rm c}^3\rho_{\rm c}y_R^2\left.\frac{\dd \thn}{\dd
y}\right\vert_{y=y_R}=4\pi r_{\rm c}^3\rho_{\rm c}\omr
\end{equation}
where we have used the Lane-Emden equation (\ref{eq:Lane-EmdenGR}) and have defined 
\begin{equation}\label{eq:omegaGR}
 \omr\equiv y_R^2\left.\frac{\dd \thn}{\dd
y}\right\vert_{y=y_R}
\end{equation}
for later convenience.

One can see that the solutions of the Lane-Emden equation are enough to fully determine the structure and mass of a polytropic
star given any choice of central pressure and density. We will return to the Lane-Emden equation in chapter \ref{chap:three}
where we will derive and investigate its generalisation in modified theories of gravity.

\section{Supersymmetry}

In chapters \ref{chap:six} and \ref{chap:seven} we will study supersymmetric models of modified gravity. A comprehensive
treatment of supersymmetry starting from first principles is far beyond the scope of this thesis and the interested reader is
referred to
review articles such as \cite{oai:arXiv.org:1011.1491} and references therein. Here we will only present the salient features ---
often without proof --- necessary for the understanding of the construction of the models presented later on. We will always work
within an $\mathcal{N}=1$ framework.

\subsection{Foundations of Supersymmetry}

Supersymmetry is a space-time symmetry that extends the Poincar\'{e} group $\mathrm{ISO(1,3)}$, which acts on the
space-time coordinates $x^\mu$, to the superpoincar\'{e} group, which acts on the coordinates
$\{x^\mu,\theta^\alpha,\,\bar{\theta}^{\dot{\alpha}}\}$\footnote{Extended supersymmetry introduces more spinorial dimensions.}
with $\alpha=\dot{\alpha}=1,\,2$. Here, $\theta^\alpha$ and $\bar{\theta}^{\dot{\alpha}}\equiv(\theta^\alpha)^\dagger$ are
anti-commuting two-component spinors\footnote{Our convention will be such that contractions of spinors are defined as
follows: $\psi\chi=\psi^\alpha\chi_\alpha$ and $\bar{\psi}\bar{\chi}=\bar{\psi}_{\dot{\alpha}}\bar{\chi}^{\dot{\alpha}}$, where
contractions are performed with the anti-symmetric epsilon symbol $\epsilon_{\alpha\beta}$ and
$\epsilon_{\dot{\alpha}\dot{\beta}}$ where $\epsilon_{12}=\epsilon_{\dot{1}\dot{2}}=-1$.}. In addition to the
usual generators of the Poincar\'{e} group, the Lie algebra of the superpoincar\'{e} group includes four new generators
$\{Q_\alpha,\,\bar{Q}_{\dot{\alpha}}\}$. The total momentum $C_1\equiv P_\mu P^\mu$
(where $P^\mu$ is the generator of translations) is still a Casimir of the superpoincar\'{e} group but the Pauli-Ljubanski
vector $W_\mu=1/2\epsilon_{\mu\nu\rho\sigma}P^\nu M^{\rho\sigma}$, where $M^{\mu\nu}$ is the generator of Lorentz transformations,
is not. If the representation is massless then the second Casimir operator is zero. In this case we can choose to label states in
each representation by their helicity given by the eigenvalue $\lambda$ of $J_3$ but this will vary between different states. The
Casimir operator for massive representations is a combination of $P^\mu,\, M^{\nu\sigma},\,
Q_\alpha$ and $\bar{Q}_{\dot{\alpha}}$ and is given by $C_2\equiv Y_\mu Y^\mu$ where $Y_\mu$ is known as the \textit{super spin}.
It satisfies the same commutation relations as the generators of $\mathrm{SU}(2)$ and hence states are labelled by the eigenvalue
$y$ of $Y^2$. Its precise definition is not important in what follows. Representations of the superpoincar\'{e} group are
therefore
classified by their mass $m$ such that $P_\mu P^\mu=-m^2$ and either their helicity if the representation is massless or the
eigenvalue of $Y^2$ if not. Unlike the Poincar\'e group, given a starting value for the helicity $\lambda$
($J_3|\lambda>=\lambda|\lambda>)$ or the superspin $y$ ($C_2|y>=y(y+1)|y>$), the rest of the states in the representation have
different helicities or spins and so any representation of the superpoincar\'{e} group has a fixed particle content. This is
because the total spin $J^2$ and helicity $J_3$ do not commute with the other generators of the superpoincar\'{e}
group. Each representation is referred to as a \textit{multiplet} and in table \ref{tab:multiplets} we list some multiplets that
we will encounter in chapters \ref{chap:six} and \ref{chap:seven}. Note that the number of fermionic degrees of freedom is equal
to the number of bosonic degrees of freedom. There is a general theorem proving that this is the case for any representation of
the supersymmetry algebra. Furthermore, when supersymmetry is unbroken the masses of each species are identical.
\begin{table}
\centering
\heavyrulewidth=.08em
	\lightrulewidth=.05em
	\cmidrulewidth=.03em
	\belowrulesep=.65ex
	\belowbottomsep=0pt
	\aboverulesep=.4ex
	\abovetopsep=0pt
	\cmidrulesep=\doublerulesep
	\cmidrulekern=.5em
	\defaultaddspace=.5em
	\renewcommand{\arraystretch}{1.6}
\begin{tabu}{l|[2pt] l}

Multiplet & Field Content \\
\tabucline[2pt]{-}
massless chiral & massless scalar $\phi$ \& massless Majorana fermion $\psi_\alpha$  \\
massive chiral & massive scalar $\phi$ \& massive Majorana fermion $\psi_\alpha$ \\
massless vector & massless vector $V_\mu$ \& massless Majorana fermion $\lambda_\alpha$ \\
gravity & massless graviton $g_{\mu\nu}$ \& gravitino $\psi_{\mu\alpha}$ \\
\end{tabu}
\caption{Representations of the superpoincar\'{e} group that will be useful later on. In each case $\mu=0,\,1,\,2,\,3,$ labels
space-time components and $\alpha=1,\,2$ labels spinorial components.}
\label{tab:multiplets}
\end{table}
There is a particularly simple method for constructing $\mathcal{N}=1$ Lagrangians by packaging the fields in different
multiplets into \textit{superfields} defined on superspace which are hence functions of
$\{x^\mu,\theta^\alpha,\bar{\theta}^{\dot{\alpha}}\}$. For the massless and massive chiral multiplet, the scalar and fermion are
packaged into \textit{chiral superfields}:
\begin{equation}\label{eq:chiralfield}
 \Phi(y)=\phi(y)+\sqrt{2}\theta\psi(y) +\theta\theta F(y),
\end{equation}
where $y^\mu=x^\mu+i\theta\sigma^\mu\bar{\theta}$ and $\sigma^\mu=(\mathbb{I}_2,\sigma^{\rm i})$, where $\sigma^{\rm i}$ are the
Pauli matrices that generate $\mathrm{SU}(2)$\footnote{In the context of supersymmetry $\sigma^\mu$ has the index structure
$\sigma^\mu_{\alpha\dot{\alpha}}$.}. The field $F$ is unusual in that it has mass dimension 2. At the level of the action, it is
an auxiliary field that is eliminated using the equations of motion in order to find the complete equations of motion. We will
not elaborate on the co-ordinate $y^\mu$ here since we will not present the technical derivation of any of the results below. It
is introduced here purely for notational convenience and the interested reader is referred to \cite{oai:arXiv.org:1011.1491}. The
massless vector multiplet can be described by a vector superfield $V=V^\dagger$
\begin{equation}\label{eq:vecfield}
 V(x)=\theta\sigma^\mu\bar{\theta}V_\mu(x)+\left(\theta\theta\right)\bar{\theta}\bar{\lambda}(x)+\left(\bar{\theta}
\bar{\theta}\right)\theta\lambda(x)+\frac{1}{2}\left(\theta\theta\right)\left(\bar{\theta}
\bar{\theta}\right)D(x),
\end{equation}
where $D(x)$ is another auxiliary field that is eliminated to produce the equations of motion. We do not write down the
superfield corresponding to the gravity multiplet since there are several, all of which correspond to geometric tensors on
superspace. A complete treatment of this can be found in \cite{wess1992supersymmetry}. These are all the tools that we require in
order to construct supersymmetric Lagrangians. We will briefly indicate how this is done below. 

\subsection{Global Supersymmetry}

Global supersymmetry refers to actions where the supersymmetry transformations are a symmetry of the action and not a gauge
redundancy i.e. the transformations act on each point in superspace in exactly the same manner. Globally supersymmetric actions
for a $\mathrm{U}(1)$ Abelian vector multiplet interacting with $i$ chiral multiplets are fully specified by two
functions\footnote{Technically one also requires a kinetic term for the vector superfield and this is accompanied by a third
gauge-kinetic function. In this thesis we will only consider vector fields whose kinetic terms are canonical and so this will not
be relevant.}: the K\"{a}hler potential $K(\Phi_i,\Phi_j^\dagger,V_j)$ and the superpotential $W(\Phi_i)$. Note that the
K\"{a}hler potential depends on both $\Phi$ and its complex adjoint whereas the superpotential is holomorphic. Rather than write
down the action in full, we will simply state how each of these functions act to specify the kinetic functions for the scalar, the
fermion masses and the scalar potential. 

The K\"{a}hler potential sets the kinetic term for the scalars. It is useful to define
the \textit{K\"{a}hler metric}
\begin{equation}\label{eq:kahlermetric}
K_{ij}=\left.\frac{\partial^2K}{\partial\Phi_i\partial\Phi_j}\right\vert_{\Phi=\phi, V=0}, 
\end{equation}
whose inverse is $K^{ij}$. The kinetic term for the scalars is then
\begin{equation}
 \frac{\mathcal{L}_{\rm kin}}{\sqrt{-g}}\supset K_{ij}\nabla_\mu\phi_i\nabla^\mu\phi_j^*.
\end{equation}
For a single chiral superfield $\Phi$, the field will only be canonically normalised if $K(\Phi,\Phi^\dagger)=\Phi\Phi^\dagger$.
The superpotential sets the mass of the fermions:
\begin{equation}\label{eq:fermionmasses}
 M_{ij}=\frac{\partial^2 W}{\partial\Phi^i\partial\Phi^j},
\end{equation}
where $M_{ij}$ is the fermion mass-matrix\footnote{The reader may worry that this can be complex according to
(\ref{eq:fermionmasses}). Since the fermions are described by Majorana spinors this is perfectly acceptable and the true masses
are given by the eigenvalues of $M^2=M^\dagger M$. }. A combination of the K\"{a}hler potential and the superpotential sets the
scalar
potential for $\phi_i$. There are two terms contributing to the scalar potential. The first is the \textit{F-term}
potential, so-called because it is found by eliminating the auxiliary field $F$ from the equations of motion. This is given by
\begin{equation}\label{eq:f-term}
 V_{\rm F}=K^{ij}\left.\frac{\partial W}{\partial\Phi_i}\frac{\partial W^\dagger}{\partial
\Phi^\dagger}\right\vert_{\Phi=\phi,\Phi^\dagger=\phi^*}.
\end{equation}
The second term comes from eliminating the auxiliary field $D$ from the equations of motion and is hence known as the
\textit{D-term} potential. In this thesis, we will only be interested in $\mathrm{U}(1)$ gauge theories, in which case the D-term
potential is
\begin{equation}\label{eq:D-termpot}
 V_{\rm D}=\frac{1}{8}\left(\xi+\sum_j q_j\left\vert\varphi_j\right\vert^2\right),
\end{equation}
where $\{\varphi_j\}$ are the subset of the scalar fields that transform under the $\mathrm{U}(1)$ symmetry with charge $q_j$.
Here we have assumed that there is a single vector multiplet coupled to an arbitrary number of chiral multiplets; the
generalisation to several multiplets is straightforward. We
have also included a Fayet-Illiopoulos term $\xi$. This term is only allowed if the gauge symmetry for the vector is
$\mathrm{U}(1)$. The extension to multiple vector fields and non-Abelian symmetry groups is straightforward. We will not
present the general formula here since we will have no need of it in this thesis.

The supersymmetry algebra is so constraining that only when $V=V_{\rm F}+V_{\rm D}=0$ is supersymmetry unbroken. Since these two
terms can be non-zero independently, there are two methods of breaking supersymmetry. The first is F-term supersymmetry breaking
where
\begin{equation}\label{eq:F-termbreaking}
 F_i=-\frac{\partial W}{\partial\Phi_i}\ne0
\end{equation}
and D-term breaking, where
\begin{equation}\label{eq:D-termbreaking}
 D=\xi+\sum_j q_j\left\vert\varphi_j\right\vert^2\ne0.
\end{equation}

\subsection{Supergravity}

The introduction of any Lorentz-invariant spin-2 field mandates that the theory is invariant under local diffeomorphisms in order
for the graviton to propagate the two degrees of freedom imposed by Lorentz symmetry. When one introduces the gravity multiplet,
one is then forced to
introduce diffeomorphism invariance on the entire superspace i.e. one must promote global supersymmetry to general superspace
co-ordinate transformations so that supersymmetry is now a gauge redundancy. This greatly
complicates the Lagrangian and, in particular, the F-term scalar potential is different from (\ref{eq:f-term}). A full
construction of
$\mathcal{N}=1$ supergravity can be found in \cite{wess1992supersymmetry}; here we will only write down the F-term scalar
potential and discuss the relevant new features that arise from gauging the theory. Defining
\begin{equation}\label{eq:covdervkahl}
D_i W\equiv \frac{\partial W}{\partial\Phi_i}+\frac{W}{\mpl^2}\frac{\partial K}{\partial\Phi_i}, 
\end{equation}
which has the physical interpretation as the covariant derivative on the manifold defined by the K\"{a}hler metric, the scalar
potential in supergravity is
\begin{equation}\label{eq:Vsugra}
 V_{\rm F}= e^{\frac{K}{\mpl^2}}\left(K^{ij}D_iW(D_jW)^\dagger-3\frac{|W|^2}{\mpl^2}\right).
\end{equation}
One can see that the global supersymmetry formula (\ref{eq:f-term}) is recovered in the limit
$\mpl\rightarrow\infty$\footnote{The correct limit is $m_{3/2}/\mpl\rightarrow0$ ($m_{3/2}$ is the gravitino mass) since
$\mpl$ is a dimensionless quantity. This distinction is relevant if supergravity is broken.}. The D-term
potential is left unaltered. Unlike global supersymmetry, the scalar potential can be non-zero and still preserve supersymmetry.
The strict condition for F-term supersymmetry breaking is
\begin{equation}\label{eq:F-termbreakingsugra}
 \langle F_i\rangle=-e^{\frac{K}{2\mpl^2}}D_i\Phi=0\;\forall i.
\end{equation}
In global supersymmetry the F-term is given by equation (\ref{eq:F-termbreaking}). In supergravity we have an extra term
$-3|W|^2/\mpl^2$ in the potential and so one can have a negative scalar potential and still preserve supersymmetry. A positive
scalar potential always breaks supersymmetry since equation (\ref{eq:Vsugra}) is only positive if at least one $F_i\ne 0$. Note
also that $V_{\rm F}=0$ does not necessarily imply that supersymmetry is preserved since this will only be true if $W=0$ at the
minimum. K\"{a}hler potentials where $V_{\rm F}=0$ at the minimum are known as \textit{no-scale} potentials. The simplest example
for a single chiral superfield is
\begin{equation}\label{eq:noscale}
 K(\Phi,\Phi^\dagger)=-n\mpl^2\ln\left(\frac{\Phi+\Phi^\dagger}{\mpl^2}\right).
\end{equation}
$n=3$ models are known as \textit{pure no-scale} models\footnote{This is because $n=3$ models give $V_{\rm F}=0$ provided that
one chooses the superpotential correctly.} and we will refer to $n\ne3$ as \textit{no-scale type} models. Finally, we note that
the gravitino mass is not arbitrary in this theory but is
given by\footnote{Note that the gauging of the supersymmetry algebra means that a non-zero gravitino mass does not imply that
supersymmetry is broken.}
\begin{equation}\label{eq:gravitinomass}
m_{3/2}^2=\frac{|W|^2}{\mpl^4}e^{\frac{K}{\mpl^2}}.
\end{equation}

\begin{savequote}[30pc]
The absence of evidence is not the evidence of absence.
\qauthor{Donald Rumsfeld}
\end{savequote}

\chapter{Modified Gravity}
\label{chap:two}
Before discussing any particular theory, the first question to address is
\textit{what exactly constitutes a modified theory of gravity?} We have already
mentioned in the introduction that general relativity is the unique Lorentz
invariant theory of a massless spin-2 particle so perhaps we should define
modified gravity as any theory containing more than one spin-2 degree of
freedom? Such a theory has already been presented in the form of Quintessence
(\ref{eq:quintessenceaction}), which includes another scalar degree of freedom
minimally coupled to gravity. The problem with this definition is that it
is ambiguous whether this is indeed a modification of gravity or whether we
should consider the scalar as a matter component that interacts with gravity according to
general relativity. One point of view is that gravity is modified on large scales because
including this scalar modifies the expansion history (and other cosmological
properties) when the same FLRW metric is used to describe cosmology. On the
other hand, this modification can equivalently be interpreted as adding another fluid
to
the matter content of the universe with a variable equation of state with the underlying cosmological dynamics being governed by
general relativity. This issue
is purely philosophical and has no absolute resolution. In this work we will not
consider this theory a modification for the following reason: suppose we were to
look at the non-relativistic limit of the geodesic equation (\ref{eq:geode1})
for matter moving in this theory. The energy-momentum tensor of the individual
matter species are still covariantly conserved and so we obtain the same
Newtonian force law that we found for general relativity. The difference here
is that the scalar profile could in principle modify the Newtonian potential of
a given source compared to what would have been obtained if it were absent.
Since there is no coupling of the field to the curvature the scalar field has a spatially homogeneous
source. In this case the field value on all scales is its cosmologically
rolling value\footnote{In this case it is the boundary conditions at infinity that ensure that the field is spatially homogeneous
but time-dependent.}. This can at most modify the Newtonian potential by a
time-dependent contribution and hence does not affect the dynamics in the Newtonian
limit. It is important to stress that this is simply an aesthetic choice for
what constitutes a modified theory of gravity and that one generally expects
deviations from the motion described by general relativity at the
post-Newtonian level and beyond\footnote{The absence of any coupling of the field to matter means that one can think of this
scalar as a new source of matter that back-reacts on the space-time given by the appropriate vacuum solution of general
relativity. This gives rise to different geodesics at any order beyond the Newtonian limit and hence alters the post-Newtonian
motion. Given this interpretation, we will not consider these theories as modifications of general relativity. }. We will see
presently that this is a well
motivated choice for this thesis since it deals with astrophysical effects of
theories of gravity that do modify the Newtonian force law. 

As an example of a theory that does alter the Newtonian limit consider the
non-minimally coupled action
\begin{equation}\label{eq:nmactionexample}
 S=\int\dd^4x\sqrt{\tilde{g}}\left[
\frac{\mpl^2}{2}\Omega^2(\phi)\tilde{R}-\frac{1}{2}k^2(\phi)\tilde{\nabla}
_\mu\phi\tilde {
\nabla}^\mu\phi-\Omega^4(\phi)V(\phi)\right]+S_{\rm
m}[\tilde{g}_{\mu\nu};\psi\iii]
\end{equation}
with
\begin{equation}
 k^2(\phi)=\Omega^2\left[1-\frac{3}{2}\left(\frac{\partial
\ln\Omega^2}{\partial\phi}\right)^2\right].
\end{equation}
At first glance, this looks like a trivial generalisation of the quintessence case
with the potential redefined and a non-canonical kinetic term for the field. The
field is not present in the matter action and so matter follows geodesics $\tilde{g}_{\mu\nu}$ but are these geodesics
necessarily the same as those found in general relativity? To answer this
question, we will perform the same non-relativistic analysis using the same
metric perturbations as (\ref{eq:minkpertconfnewt}). Since the geodesic
equation is the same we again find
\begin{equation}
\ddot{\vec{x}}=-\nabla\pn,
\end{equation}
However, this time the Poisson equation is sourced by both the density and the
scalar:
\begin{equation}\label{eq:nmexamplepoisson}
 \nabla^2\pn=\frac{1}{\Omega^2}\left[4\pi G
\rho+\frac{1}{4\mpl^2}k^2\nabla_i\phi\nabla^i\phi+\frac{1}{2\mpl^2}
\Omega^4V+\frac {1}{2\mpl^2}\nabla^2\Omega^2\right].
\end{equation}
Now in the case of quintessence we had $k=\Omega=1$ and so the only
modification is the potential term, which is $\ll G\rho$\footnote{If this is not the case then the model gives rise to an
unacceptable cosmology where the onset of dark energy domination happens at very early times.} and so we indeed find
that the Newtonian limit is unaltered. Of course, one must find the scalar
field's equation of motion in order to solve the system, which in our case is
\begin{equation}
 \label{eq:nmexamplescalareom}
\frac{1}{\Omega^2}\Box\phi+\frac{1}{\mpl\Omega^4}\frac{\dd\Omega^2}{\dd\phi}
\nabla_i\phi\nabla^i\phi=\frac{\dd
V}{\dd\phi}-\frac{\mpl}{\Omega^4}\frac{\dd\ln\Omega^2}{\dd\phi}4\pi G\rho.
\end{equation}
Unlike the case of quintessence, the non-minimal coupling term $\Omega^2R$ in
the action (\ref{eq:nmactionexample}) has resulted in an inhomogeneous source
term for the scalar field and hence we expect an inhomogeneous solution. In this
case, the new Poisson equation (\ref{eq:nmexamplepoisson}) contains an extra
source apart from the density and in general the solution for the Newtonian
potential sourced by the same density profile will be different if one uses
general relativity or this theory to describe the dynamics of gravity. For this
reason, it is sensible to consider theories such as these modified theories of
gravity and we will do so from here on. Before leaving this example we will
briefly show how theories such as these can be interpreted as including
additional scalars coupled to matter. 

So far, we have worked in the \textit{Jordan frame} where the scalar is non-minimally
coupled to gravity and matter is coupled only to the \textit{Jordan frame metric}
$\tilde{g}_{\mu\nu}$. It is possible to restore minimal coupling and
diagonalise the kinetic term for the graviton using the field redefinition
\begin{equation}\label{eq:Weyl1nmexample}
 g_{\mu\nu}=\Omega^2(\phi)\tilde{g}_{\mu\nu}
\end{equation}
known as a \textit{Weyl rescaling}. After this transformation (see appendix
\ref{app:weyl} for the transformation laws of various geometric
quantities, including the Ricci scalar, under this transformation) the
action takes the form
\begin{equation}\label{eq:nmexampleef}
 S=\int\dd^4x
\sqrt{g}\left[\frac{\mpl^2}{2}R-\frac{1}{2}
\nabla_\mu\phi\nabla^\mu\phi-V(\phi)\right]+S_{\rm
m}\left[\Omega^{-2}g_{\mu\nu};\psi\iii\right].
\end{equation}
This is known as the \textit{Einstein frame} and $g_{\mu\nu}$ the
\textit{Einstein frame metric}. In this frame the gravitational action looks
almost like general relativity with a minimally coupled scalar\footnote{note
the factor $k(\phi)$ was chosen such that $\phi$ is canonically normalised in
this frame.}, the difference being that $\phi$ is now coupled directly to matter through the
function $\Omega$. This direct coupling has the effect that the energy-momentum
tensor defined using this metric, $T_{\rm m}^{\mu\nu}=2/\sqrt{-g}\delta S_{\rm
m}/\delta g_{\mu\nu}$ for matter is not conserved in this frame:
\begin{equation}\label{eq:emtensnmexample}
 \nabla_\mu T_{\rm m}^{\mu\nu}= \frac{\dd\ln\Omega}{\dd
\phi}T_{\rm m}\nabla^\mu\phi,
\end{equation}
where $T_{\rm m}=g_{\mu\nu}T_{\rm m}^{\mu\nu}$ is the trace of the
energy-momentum tensor (see appendix \ref{app:weyl} for a derivation of this formula). According to equation (\ref{eq:geode1}),
this
non-conservation implies that matter does not move on geodesics of $g_{\mu\nu}$
and so if one were to take the non-relativistic limit of the modified equation
one would find extra terms in the Newtonian force law proportional to the
gradient of the scalar. In this frame, the Poisson equation is identical to
that found in general relativity\footnote{Provided one interprets the density as
coming from both the scalar and the matter, we will return to this point
shortly.} and it is the force law that is altered. This additional term thus
represents a new or \textit{fifth}-force. The two frames are
completely equivalent and so the only difference between the two lies in how
one interprets the resultant dynamics. Any debate as to which frame is more ``fundamental'' is purely philosophical. That being
said, the Einstein frame may be more convenient if one wishes to study the quantum properties of gravity since the kinetic term
for the graviton is canonical in this frame.

This thesis is primarily concerned with the non-relativistic and cosmological
dynamics of modified theories of gravity and not the strong field regime. For
this reason, we will consider any theory containing a graviton
and additional degrees of freedom non-minimally coupled to matter a modified
theory of gravity. In particular, this choice means that we do not consider
quintessence to be a modified theory of gravity. The example presented above in equation
(\ref{eq:nmexampleef}) is considered a modified theory of gravity on account of the function $\Omega(\phi)$ appearing in the
matter action. 

So far we have only presented a
discussion of modified theories of gravity in terms of theories which can be
written as general relativity with a non-minimal coupling of a single scalar to
matter. In terms of Weinberg's theorem \cite{Weinberg:1965rz}, this corresponds
to allowing more than one degree of freedom but there are other ways to violate
this. One method is to simply drop the requirement of Lorentz invariance, which
leads to interesting theories such as Einstein-Aether theory
\cite{Jacobson:2000xp} and Ho\v{r}ava-Lifshitz gravity \cite{Horava:2009uw}.
Another is to add multiple scalar fields or other particles such as vectors. A
third alternative is to look at other curvature invariants such as torsion, the
Weyl tensor or higher-derivative terms such as $R_{\mu\nu}R^{\mu\nu}$. These
modifications typically introduce Ostrogradsky ghost degrees of freedom and so their
stability must be considered very carefully. One can avoid these problems by
asking what is the most general theory describing a scalar and a massless
graviton such that the equations of motion are second order so that there are
no ghosts? The answer was written down by Horndeski in 1974
\cite{Horndeski:1974wa} and a derivation using modern methods was presented in
\cite{Deffayet:2009wt}. Another approach is to abandon the requirement that
gravity is described by a massless spin-2 particle. The simplest alternative is
to allow it to have a mass. The linear theory for Lorentz invariant massive
spin-2 particle, the Fierz-Pauli action, has been known since 1939
\cite{Fierz:1939ix} however any attempts to generalise it to non-linear orders
resulted in the introduction of an additional degree of freedom beyond the five
mandated by Lorentz invariance. This extra mode is always a ghost
\cite{Boulware:1973my} and so the theory seemed doomed. Very recently, de-Rham,
Gabadadze and Tolley \cite{deRham:2010kj} have constructed the most general
action where this ghost does not appear at any order in perturbation theory and
since then a substantial amount of work on the subject has appeared (see
\cite{Hinterbichler:2011tt} for a review). Massive gravity and its generalisations can provide a technically natural solution to
the cosmological constant problem and screen the modifications on small scales using the Vainshtein mechanism. This thesis is
primarily concerned with testing the chameleon mechanism and so we will not investigate it in detail here. The Vainshtein
mechanism is more efficient at screening than the chameleon mechanism and these nice properties certainly motivate a search for
new observational signatures. More exotic models of modified
gravity such as non-local modifications (see e.g. \cite{Woodard:2014iga}) and
higher-dimensional generalisations, for example, brane-world and Kaluza-Klein
models, have also been considered. 

The few models mentioned above are just a
small fraction of the models that have been studied and the literature is
overflowing with a plethora of models with more appearing daily (see
\cite{Clifton:2011jh} for a recent 312 page review). The challenge then is to
decide which theories are viable alternatives to general relativity. Clearly any viable
alternative must be compatible with all current data on all scales from compact
objects to the laboratory to the solar system to cluster scales and cosmology\footnote{See \cite{Clemson:2012im} for a
discussion on the cosmological distinguishability of interacting dark energy and modified gravity models.}.
Presently, there are many theories which can achieve this (for example, $f(R)$
theories can always have their parameters tuned such that the cosmological
expansion history mimics $\Lambda$CDM). In this thesis, we are interested in
infra-red modifications that may be able to resolve the dark energy problem and
many of these theories fail at the first hurdle: in order to produce
interesting effects on cosmological scales the additional degrees of freedom
need to have masses of order the Hubble parameter. This implies that they
mediate a long-range force of order $10^4$ Mpc and such a force range violates
laboratory bounds of a few microns \cite{Kapner:2006si}.

One may then think that we have failed before we have started but this is not
the case. As alluded to in chapter \ref{chap:one}, it is possible to construct theories with screening
mechanisms. These can circumvent this issue by changing the strength of the
fifth-force relative to the Newtonian one in dense environments. Exactly how
this is achieved will be the subject of the next section.

\section{Screening Mechanisms}
\label{sec:screenmechs}
From here on we will specialise to the case of a single scalar field $\phi$
coupled to matter. In particular, we will focus on the Einstein frame coupling
\begin{equation}\label{eq:EFgencoup}
 \frac{\mathcal{L}}{\sqrt{-g}}\supset C(\phi)T_{\rm m},
\end{equation}
where $T^{\mu\nu}_{\rm m}=2/\sqrt{-g}\delta S_{\rm m}/\delta g_{\mu\nu}$ is the
(non-conserved) energy-momentum tensor for
matter and $T_{\rm m}$ is its trace. Furthermore, we define the
\textit{coupling} $\beta(\phi)$ via
\begin{equation}\label{eq:betadef}
\beta(\phi)\equiv\mpl\frac{\dd C(\phi)}{\dd \phi}.
\end{equation}
This non-minimal coupling of the scalar to the energy-momentum tensor means
that it is not conserved in this frame and instead one has (see appendix \ref{app:weyl} for the derivation of this formula) 
\begin{equation}\label{eq:geode2}
\nabla_\mu T^{\mu\nu} = \frac{\beta(\phi)}{\mpl}T_{\rm m}\nabla^\nu\phi.
\end{equation}
Now we showed in equation (\ref{eq:geode1}) that a point particle with
$T\mmm^{\mu\nu}=\rho\mmm u^\mu u^\nu$ gives the geodesic equation for the left hand
side of this expression (multiplied by the density, which cancelled in
(\ref{eq:geode1}) since the entire expression was equal to zero). Furthermore,
in equation (\ref{eq:Newton1}) we showed that the non-relativistic limit of the
geodesic equation resulted in Newton's second law. This is unchanged in this
case and so all the effects of the non-conservation of the energy-momentum tensor manifest as a deviation from the geodesic
equation. Particles do not follow
geodesics of the metric in the Einstein frame. The non-relativistic limit of the left hand side of equation (\ref{eq:geode2}) was
calculated in equation (\ref{eq:Newton1}) and is equal to $\rho\mmm(\ddot{\vec{x}}+\nabla\pn)$. Therefore, the non-relativistic
limit of the right hand side gives the fifth-force. In the non-relativistic limit we have $T\mmm=-\rho\mmm$\footnote{Here we
interpret $\rho\mmm$ as the energy density in matter. The reader should be aware that since
the scalar is coupled to matter directly this interpretation is not always clear-cut. We will see below that the theories this
thesis is concerned with do not have this ambiguity.} and so cancelling the factors of $\rho\mmm$
we find the fifth-force per unit mass
\begin{equation}\label{eq:fifth-forcegeneral}
F_\phi=-\frac{\beta(\phi)}{\mpl}\nabla\phi.
\end{equation}

%
All known screening mechanisms can be classified by subsets
of the general Lagrangian expanded in the field perturbation
$\phi=\bar{\phi}+\delta \phi$ in the Einstein frame:
\begin{equation}\label{eq:efgenscreening}
 \frac{\mathcal{L}}{\sqrt{-g}}\supset
-\frac{1}{2}Z^2(\bar{\phi})\nabla_\mu\dpp\nabla^\mu\dpp-m_{\rm
eff}^2(\bp)\dpp^2+\beta(\bp)\frac{\dpp}{\mpl}T_{\rm
m}+\cdots,
\end{equation}
where the dots denote higher-order terms not relevant for this discussion. Here the wavefunction normalisation $Z(\bp)$, the
effective mass $m_{\rm eff}(\bp)$ and the coupling to the trace of the energy momentum
tensor $\beta(\bp)$ are all field-dependent and can vary as a function of
position. It is this positional dependence that allows for the existence of
screening mechanisms. If we can somehow arrange for the local value of the field to be such that the
fifth-force (\ref{eq:fifth-forcegeneral})
is rendered negligible compared with the Newtonian one then we will not be able to detect it and the force is screened. There are
several ways this
can be achieved: firstly, if the wavefunction normalisation $Z(\bp)$ is
different from unity then the canonically normalised field is $\dpp/Z(\bp)$. If
$Z(\bp)$ is large enough in dense environments such that the effective coupling
to matter, $\beta(\bp)/Z(\bp)\ll1$ then the fifth-force
(\ref{eq:fifth-forcegeneral}) will be negligible. This is the method employed
by the Vainshtein mechanism \cite{Vainshtein:1972sx}. If the effective mass of
the perturbation at solar system densities is large enough that the force range
$\lambda= m\eff^{-1}$ is smaller than a few microns then the theory will
satisfy all laboratory bounds and the force is again screened. In the language
of equation (\ref{eq:fifth-forcegeneral}), this corresponds to suppressing field
gradients and is employed by the chameleon mechanism
\cite{Khoury:2003aq,Khoury:2003rn}. Finally, if the coupling $\beta(\bp)$ is
small enough in dense environments so that the fifth-force is negligible compared with the Newtonian force then
the fifth-force is again screened. This method is utilised by the symmetron
\cite{Hinterbichler:2010es} mechanism and the environment-dependent
Damour-Polyakov effect \cite{Brax:2010gi}.

The Vainshtein mechanism screens in a very different manner to chameleon-like models. This thesis is primarily concerned with
the latter class of theories and so below we will describe their properties in detail. The chameleon mechanism, the symmetron
effect and the EDDP all arise from a general scalar-tensor theory where the scalar is conformally coupled to matter through the
metric and we will refer to these as \textit{conformal scalar tensor theories}. First, we present the general framework and
describe the salient new features that are not present in general relativity. We then give detailed examples of each screening
mechanism and present a model-independent description of the screening mechanism. In particular, we will derive the criteria for
an object to be self-screening and will show how any mechanism can be parametrised by two model-independent parameters:
one controlling how efficient an object is at screening itself and another that sets the strength of the fifth-force relative to
the Newtonian force in unscreened objects. Next, we discuss the current observational constraints on the model-independent
parameters and present a screening map of galaxies in the nearby universe that will be useful for the astrophysical tests
presented in chapters \ref{chap:three}, \ref{chap:four} and \ref{chap:five}. Finally, we discuss the cosmological dynamics of the
scalar, which will be useful for the discussion in chapters \ref{chap:six} and \ref{chap:seven}. At the end of the chapter we will
provide a short introduction to the Vainshtein mechanism using Galileon theories as an example. Many of the results in this thesis
apply equally to Vainshtein screened theories and so it will be enlightening to discuss the differences between the two screening
mechanisms. The Vainshtein mechanism is more efficient at screening than chameleon-like theories. For this reason, whereas it is
often the case that the formulae presented apply to both mechanisms, in practice only chameleon-like theories show novel effects
once we specialise to collapsed astrophysical objects. We will always begin by presenting the most general equations and will
unambiguously indicate where we have specialised to chameleon-like theories.

\section{Conformal Scalar-Tensor Theories}
\label{sec:suppresscharge}
Our starting point is the
Einstein frame action for a scalar field coupled to matter via the
\textit{coupling function} $A(\phi)$
\begin{equation}\label{eq:chamact1}
 S=\int\dd^4x
\sqrt{-g}\left[\frac{\mpl^2}{2}R-\frac{1}{2}k^2(\phi)
\nabla_\mu\phi\nabla^\mu\phi-V(\phi)\right]+S_{\rm
m}[A^2(\phi)g_{\mu\nu};\psi\iii],
\end{equation}
where $\psi\iii$ represent the various matter fields and we have allowed for the
fact that the field may not be canonically normalised by including the factor
$k(\phi)$. One may instead work in the Jordan frame (see appendix \ref{app:weyl} for the transformation laws of various geometric
quantities under
Weyl rescalings)
\begin{equation}\label{eq:chamactJF}
 S=\int
\dd^4x\sqrt{-\tg}\left[\frac{\mpl^2}{2A^2(\phi)}\tilde{R}-\frac{1}{2}
\left[\frac{k^2(\phi)}{A^2(\phi)}-6\left(\frac{A^\prime(\phi)}{A}\right)^2\right]\tilde{\nabla}_\mu\phi\tilde{\nabla}
^\mu\phi-\frac { V(\phi)}{
A^4(\phi)} \right] +S_{\rm m}[\tg_{\mu\nu};\psi\iii].
\end{equation}
Whereas the choice of frame is irrelevant, one should note that specifying the functional form of the (so
far) free functions in one frame will yield an inequivalent theory to the same choice made in a different frame. For example, the
Jordan frame potential with $V(\phi)$
defined in the Einstein frame is $\tilde{V}(\phi)=V(\phi)/A^4(\phi)$, however,
we could just as easily fix the Jordan frame potential to some specified
function $\tilde{V}(\phi)$, in which case the Einstein frame potential is
$V(\phi)=A^4(\phi)\tilde{V}(\phi)$. The same is true of the normalisation of
the field. We could arbitrarily decide to have a canonical kinetic term in
either the Jordan or Einstein frame, in which case the other frame will pick up
a non-canonical factor multiplying the kinetic term. The choice of
normalisation in one frame then represents a different theory to the same
choice in the other frame\footnote{Of course two theories are equivalent if a consistent normalisation and scalar potential is
used in both frames.}. This arbitrariness is simply
a matter of definition. In what follows we will work in the Einstein frame
wherever possible\footnote{The issue of frame definitions will be important
when we discuss $f(R)$ theories and the environment-dependent dilaton and we will have no choice other than to define the
theory in the Jordan frame and find the Einstein frame action using a Weyl rescaling.}, both because it is more transparent and
it is easier to discern the physics when applied to astrophysical systems.
A full account of screening in the Jordan frame can be found in
\cite{Hui:2009kc}, Appendix C. We will set the wavefunction normalisation in
this frame equal to unity. Given a choice of potential, coupling function and
wavefunction normalisation, one can always work with the canonically
normalised field $\dd\varphi=k(\phi)\dd\phi$ and re-write these functions in
terms of the new field. For this reason, we can make this choice without loss
of generality. The one exception to this is the case where the fundamental
theory is defined using a factor $k(\phi)$ such that $\varphi(\phi)$ cannot be
found analytically and one is forced to retain it in the analysis as is the
case with the environment-dependent dilation. In cases such as these we will
clearly indicate that $k(\phi)\ne1$ and discuss the modifications to the
standard results thoroughly.

Varying the action (\ref{eq:chamact1}) with respect to the field (note we have set $k(\phi=1)$ as discussed above) gives the
equation of motion
\begin{equation}
 \label{eq:chameom}
\Box\phi= -\frac{\dd V(\phi)}{\dd\phi}+\frac{A^\prime(\phi)}{A(\phi)}T_{\rm m}.
\end{equation}
This can be derived from the effective Lagrangian
\begin{equation}\label{eq:effL1}
 \frac{\mathcal{L}}{\sqrt{g}}\supset -V(\phi)+T\ln A
\end{equation}
and hence is exactly the type of theory described in section
(\ref{sec:screenmechs}) with $C(\phi)=\ln A$. The fifth-force in this theory
is then given by equation (\ref{eq:fifth-forcegeneral}) with
\begin{equation}\label{eq:betacham}
\beta(\phi)=\mpl\frac{\dd \ln A}{\dd\phi}.
\end{equation}
In order to have gravitational strength fifth-forces that are screened locally
one typically has $\beta(\phi)\sim\mathcal{O}(1)$ or greater when an object is
unscreened. Theories without screening mechanisms typically need to impose
$\beta(\phi)\ll1$ in order to be compatible with current observations.
Using the fact that $T\mmm=-\rho\mmm$ for non-relativistic matter, equation
(\ref{eq:effL1}) defines and effective potential
\begin{equation}
 V\eff(\phi)\equiv V(\phi)+\rho\mmm\ln A(\phi).
\end{equation}
In fact, this definition is not so useful because the density $\rho\mmm$ is not
conserved on account of equation (\ref{eq:geode2}). In any theory of gravity
such as this, the conserved density is the Jordan frame density found using
$\tilde{T}^{\mu\nu}\mmm=A^{-6}T^{\mu\nu}\mmm$. In this case one has
$\tilde{\rho}\mmm=-\tg_{\mu\nu}\tilde{T}^{\mu\nu}=A^4\rho\mmm$. It is this
density that results when one integrates over microphysical distribution
functions, the Einstein frame density contains interactions with the scalar
and the true properties of objects, such as their mass, can only be found once
the scalar interactions have been accounted for. In general, an interpretation
in terms of an effective potential would then require frame mixing, however, we
will see later that theories with screening mechanisms have the property that
\begin{equation}\label{eq:a1}
 A(\phi)=1+\mathcal{O}\left(\beta(\phi)\frac{\phi}{\mpl}\right).
\end{equation}
$A(\phi)$ never differs too significantly from $1$ since we require
$\phi\ll\mpl$ in order to have a sensible infra-red modification of gravity and
this allows us to define a conserved density in the Einstein frame. Consider
the 0-component of equation (\ref{eq:geode2}) with a non-relativistic source
$T^{\mu\nu}\mmm=\left(\rho\mmm,0,0,0\right)$. We have
\begin{equation}\label{eq:consdens1}
 \dot{\rho}\mmm+\Gamma^\mu_{\mu0}\rho\mmm+\Gamma^0_{00}\rho\mmm=\frac{
A^\prime(\phi)}{ A(\phi)}\frac{\dot{\phi}}{\mpl}\rho\mmm.
\end{equation}
Defining the conserved density $\rho\mmm=A(\phi)\rho$ this reduces to 
\begin{equation}
\dot{\rho}+\Gamma^\mu_{\mu0}\rho+\Gamma^0_{00}\rho=0,
\end{equation}
which is the continuity equation for a conserved non-relativistic density
$\rho$ ($=A^3\tilde{\rho}\mmm$ in the Jordan frame). From here on we will use
$\rho$ as the density of non-relativistic matter, both in cosmology and
astrophysics, however we stress that while it is conserved, it does not
correspond to the trace of any conserved energy-momentum tensor describing the
translational invariance of the theory in any frame\footnote{One may define the
tensor $\mathcal{T}^{\mu\nu}=A^5\tilde{T}^{\mu\nu}\mmm$, whose trace
corresponds to $\rho$, however it does not correspond to a physical
energy-momentum tensor and so we will not use it here.}. Later in the chapter we will see that $\beta(\phi)\phi/\mpl\le10^{-6}$
and so quantities such as mass defined by integrating
over this density do not differ significantly from those found using the Jordan
frame density by virtue
of equation (\ref{eq:a1}). For this reason, it does not matter which density we use to define the density of an object such as a
non-relativistic star; the choice of the conserved density is made for calculational convenience and is more intuitive than a
non-conserved quantity. Using this definition of the density, equation (\ref{eq:chameom})
can be written
\begin{equation}\label{eq:chameom2}
 \Box\phi= -\frac{\dd V(\phi)}{\dd\phi}-\rho\frac{\dd A(\phi)}{\dd \phi},
\end{equation}
which allows for a definition of the effective potential in terms of the
conserved density\footnote{Note that the effective potential is often defined as $V\eff=V+\rho A$ in the literature. Since
$V\eff$ is
not a fundamental quantity but is instead inferred from the equation of motion it is only defined up to an arbitrary
field-independent function. We have chosen to define it with the factor of $(A(\phi)-1)$ in order to keep track of
the energy density in the field and the matter separately. This definition does not impact the field dynamics but is important
when describing the cosmological dynamics, which we will investigate later, because the effective potential (and not its
derivative) appears in the Friedmann equation. For this reason, we have chosen to use this definition throughout this thesis in
the interest of consistency.}:
\begin{equation}\label{eq:veff}
 V\eff=V(\phi)+\rho (A(\phi)-1).
\end{equation}

It is the density-dependence in the effective potential that allows these theories to screen. In
particular, if the effective potential has a minimum whose position depends on
the density then the field will move to different positions in field space as a
function of the local density. If we then choose the functions $A(\phi)$ and
$V(\phi)$ such that the mass of the field in high density regions is large
enough to evade laboratory bounds or the coupling $\beta(\phi)\ll1$ then the
theory will screen. The first of these methods is employed by the chameleon
effect \cite{Khoury:2003aq,Khoury:2003rn} and the latter by the symmetron
mechanism \cite{Hinterbichler:2010es} and the
environment-dependent Damour-Polyakov effect (EDDP)
\cite{Brax:2010gi}\footnote{Including generalised symmetrons and dilatons
\cite{Brax:2012gr}.}. We will briefly review how all three mechanisms work
using simple examples before presenting a model-independent description of how they can
screen a spherical object.

\subsection{The Chameleon Effect}

\begin{figure}
\subfigure[Small{$\rho$}]{\label{fig:small_rho}\includegraphics[
width=0.5\textwidth]{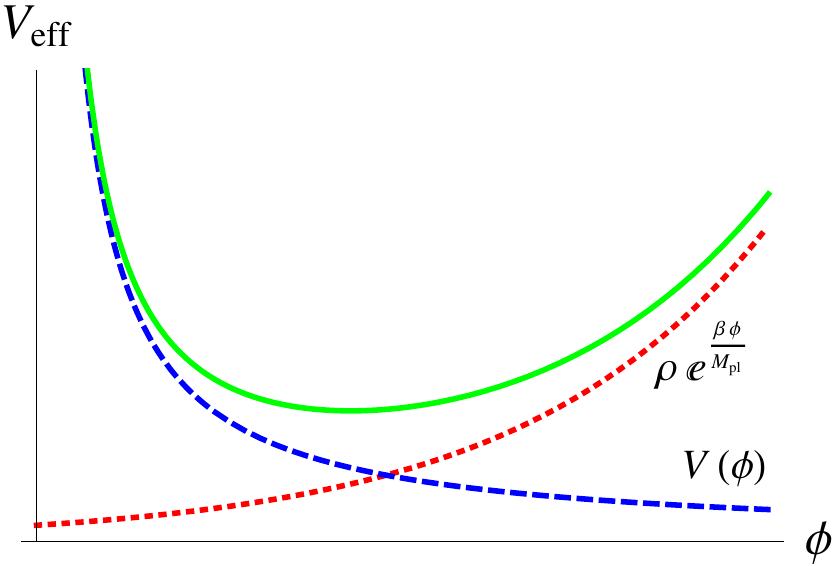}}
\subfigure[Large{$\rho$}]{\label{fig:large_rho}\includegraphics[
width=0.5\textwidth]{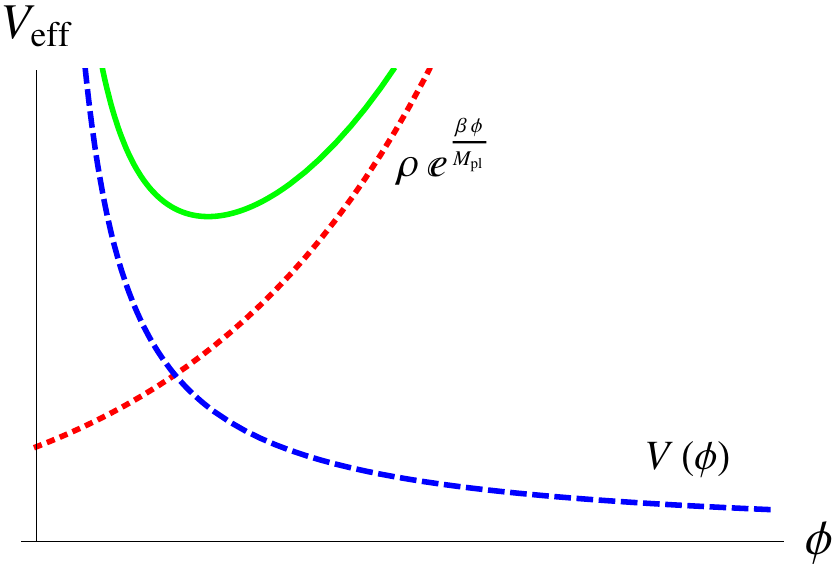}}
\caption{The chameleon effective potential (solid line) for small and
large densities. The blue dashed lines show the contribution from the potential
and the red dotted lines show the contribution from the coupling.}\label{fig:min_cham}
\end{figure}

First introduced by \cite{Khoury:2003rn}, the simplest example of models
exhibiting the chameleon effect are those with run-away potentials and
exponentially increasing coupling functions
\begin{equation}\label{eq:champotcoupex}
 V(\phi)=\frac{M^{4+n}}{\phi^n};\quad A(\phi)=e^{\beta(\phi)\frac{\phi}{\mpl}},
\end{equation}
the simplest case being that of constant $\beta(\phi)\equiv\beta$. We shall use
this case as an example. These models have a minimum at 
\begin{equation}\label{eq:phimincham1}
 \phi_{\rm
min}=\left(\frac{n\mpl}{\beta\rho}\right)^{\frac{1}{n+1}}M^{\frac{n+4}{n+1}}
\end{equation}
and so the field is pushed to smaller values in denser environments. The
effective mass of oscillations about the minimum is 
\begin{align}\label{eq:masscham}
 m\eff^2&=V^{\prime\prime}+\rho A^{\prime\prime}(\phi)\\
&\approx
\frac{n(n+1)}{M^{\frac{n+4}{n+2}}}\left(\frac{\beta\rho}{\mpl}\right)^{\frac{n+2
} { n+1 } }
\end{align}
and so one can see that this mass is larger in denser environments. This is
shown in figure \ref{fig:min_cham} where we plot the effective potential.
Figure \ref{fig:small_rho} shows this for low densities and figure
\ref{fig:large_rho} for high densities. One can see by eye that the potential
near the minimum in \ref{fig:small_rho} is far shallower than the potential
near the minimum in \ref{fig:large_rho}.

This is how the chameleon mechanism screens. In high density environments, the
field is pushed to smaller values where the potential is steeper and the
effective mass is larger. Provided the parameters are chosen such that the
Compton wavelength of the chameleon on Earth is less than a few microns, the
fifth-force will evade all current bounds.

The class of models (\ref{eq:champotcoupex}) are just the simplest and more
complicated variants have indeed been studied. Any model where the field is
pushed to smaller values where the mass is larger will screen and these include
power law potentials \cite{Mota:2006fz}, field-dependent couplings
\cite{Brax:2010kv} and even supersymmetric models
\cite{Hinterbichler:2010wu,Hinterbichler:2013we,Brax:2012mq,Brax:2013yja}.
Laboratory tests probe the parameter space of each model on a model by model
basis whereas astrophysical tests probe model-independent combinations of these
parameters that can be mapped to any specific model. Since this thesis is
focused on astrophysical tests, we will not present a complete list of models
here.

\subsection{$f(R)$ Theories}\label{subsec:f(R)}

One popular theory of modified gravity is the $f(R)$ class of models where the Einstein-Hilbert term in the action is
generalised to an arbitrary function of the Ricci scalar:
\begin{equation}\label{eq:f(R)def}
 S=\int\dd^4x\sqrt{-\tg}\frac{\mpl^2}{2}f(R) + S\mmm[\tg_{\mu\nu};\psi\iii],
\end{equation}
where the Ricci scalar is computed using $\tg_{\mu\nu}$. It is well-known that these theories are equivalent to scalar-tensor
theories and here we will follow the derivation of \cite{Brax:2008hh}\footnote{One could instead use a Lagrange multiplier to
find the equivalent Jordan frame action and proceed from there. The two approaches yield identical results and we have chosen to
follow this alternate derivation in order to make contact with the chameleon literature.}. We begin by defining the scalar field
$\phi$ via
\begin{equation}\label{eq:phifr}
 f^\prime(R)=e^{-\frac{2\phi}{\sqrt{6}\mpl}},
\end{equation}
where a prime denotes a derivative with respect to $R$. We can invert this relation to find
\begin{equation}
 f(R)=\int f^\prime(R)\dd R=\int e^{-\frac{2\phi}{\sqrt{6}\mpl}}\frac{\dd
R}{\dd\Phi}\dd\phi=-\frac{2\phi}{\sqrt{6}\mpl}R+\frac{2}{\sqrt{6}\mpl}\int e^{-\frac{2\phi}{\sqrt{6}\mpl}}R\dd R.
\end{equation}
Using the definition of $\phi$ (\ref{eq:phifr}), we have 
\begin{equation}
 \frac{\dd\Phi}{\dd R}=-\frac{\sqrt{6}\mpl}{2}\frac{f^{\prime\prime}(R)}{f^\prime(R)}
\end{equation}
and the last integral is
\begin{equation}
 \frac{2}{\sqrt{6}\mpl}\int e^{-\frac{2\phi}{\sqrt{6}\mpl}}R\dd R=-\int Rf^{\prime\prime}\dd R=-(Rf^\prime-f).
\end{equation}
The action can then be written as a scalar-tensor theory of the form 
\begin{equation}\label{eq:f(R)1}
 S=\int\dd^4x\sqrt{-\tg}\left[\frac{\mpl^2e^{-\frac{2\phi}{\sqrt{6}\mpl}}R}{2}-\tilde{V}(\phi)\right] +
S\mmm[\tg_{\mu\nu};\psi\iii],
\end{equation}
where using $\phi=\phi(R)$ the Jordan frame scalar potential is 
\begin{equation}
 \tilde{V}(\phi)=\frac{\mpl^2\left(Rf^\prime(R)-f(R)\right)}{2}.
\end{equation}
Equation (\ref{eq:f(R)1}) is precisely of the form (\ref{eq:chamactJF}) with no kinetic term for the scalar and so we can find
the equivalent Einstein frame formulation by performing a Weyl rescaling of the metric with
\begin{equation}\label{eq:afr}
 A(\phi)=e^{\frac{2\phi}{\sqrt{6}\mpl}}.
\end{equation}
The transformation of the Ricci scalar under this rescaling can be found in appendix \ref{app:weyl} equation
(\ref{eq:ricciscalarJR}) and the
square root of the determinant transforms as $\sqrt{-\tg}=A^4(\phi)\sqrt{-g}=f^{-2}(R)\sqrt{-g}$. Performing this rescaling,
we find the Einstein frame action
\begin{equation}\label{eq:f(R)ef}
 S=\int\dd^4x\sqrt{-g}\left[\frac{\mpl^2}{2}R-\frac{1}{2}\nabla_\mu\phi\nabla^\mu\phi-V(\phi)\right] +
S\mmm[A^2(\phi)g_{\mu\nu};\psi\iii],
\end{equation}
where the scalar potential
\begin{equation}\label{eq:Vf(R)}
 V(\phi)=\frac{\mpl^2(Rf^\prime(R)-f(R))}{2f^\prime(R)^2}.
\end{equation}
Note that the factor of $1/\sqrt{6}$ was chosen so that the field is canonically normalised in this frame. Equation
(\ref{eq:f(R)ef}) is a conformal scalar-tensor theory with a coupling function given by (\ref{eq:afr}). This is a chameleon
coupling with constant
$\beta(\phi)=1/\sqrt{6}$. The theory is not yet a chameleon because one must choose a potential that will give rise to the
chameleon mechanism such as a run-away potential. One well-studied example of this is the model of Hu and Sawicki
\cite{Hu:2007nk}
\begin{equation}\label{eq:husawicki}
 f(R)=-m^2\frac{c_1\left(R/m^2\right)^n}{1+c_2\left(R/m^2\right)^n}
\end{equation}
which is often studied in the context of N-body simulations (see, for example, \cite{Li:2011pj} and references
therein). One generally tunes the values of $c_1$ and $c_2$ so that $c_1/c_2=6\Omega\mmm/\Omega_\Lambda$ in order to yield an
identical expansion history to the $\Lambda$CDM model. $m^2\equiv8\pi G\tilde{\rho}/3$ is also fixed leaving $n$ and a choice of
either $c_1$ or $c_2$ as free parameters. In fact, it is not necessary to use the scalar field formulation of the theory in order
to understand the screening mechanism. Instead of interpreting the modified gravity effects as a fifth-force augmenting the
Newtonian force, the effective value of $G$ felt by non-relativistic matter is $4/3 G_{\rm N}$ where $G_{\rm N}$ is the measured
value of $G$ in the solar system. This would then immediately violate solar system bounds, except that the Poisson equation
(\ref{eq:poisson1}) contains another source proportional to the Ricci scalar in addition to the density. Consider an object of
mass $M$ and radius $R$. When the new term is negligible the solution is $\pn= GM/R$ outside the object and so the effective value
of $G$ is really $4/3$ times as large as predicted by general relativity. In this case the object is unscreened. When the new term
is comparable to the density term one finds that the effective mass found by integrating the Poisson equation is $M\eff\approx
3/4 M$ so that the force-law is identical to general relativity and the object is screened. We will shortly present the general
screening mechanism for a conformal scalar-tensor theory, which includes $f(R)$ theories but is more general and encapsulates any
screening mechanism using two model-independent parameters. For this reason we will always work with this more general framework
but will often refer to $f(R)$ theories owing to their ubiquity.

\subsection{The Symmetron Mechanism}
\label{subsec:symmetron}
\begin{figure}
\subfigure[{$\rho\ll M\mu$}]{\label{fig:nosym}\includegraphics[width=0.5\textwidth]{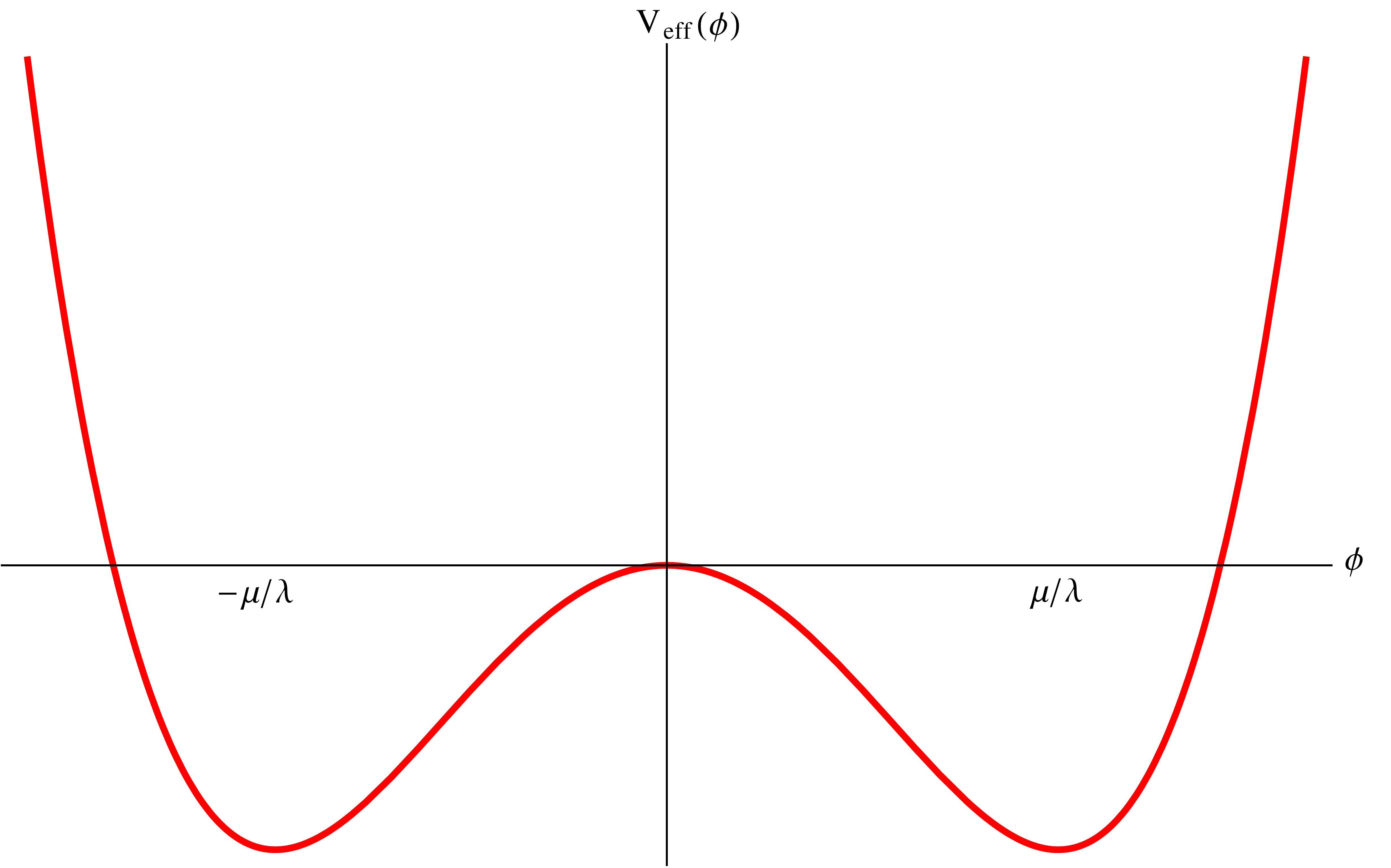}}
\subfigure[$\rho>M\mu$]{\label{fig:symm}\includegraphics[width=0.5\textwidth]{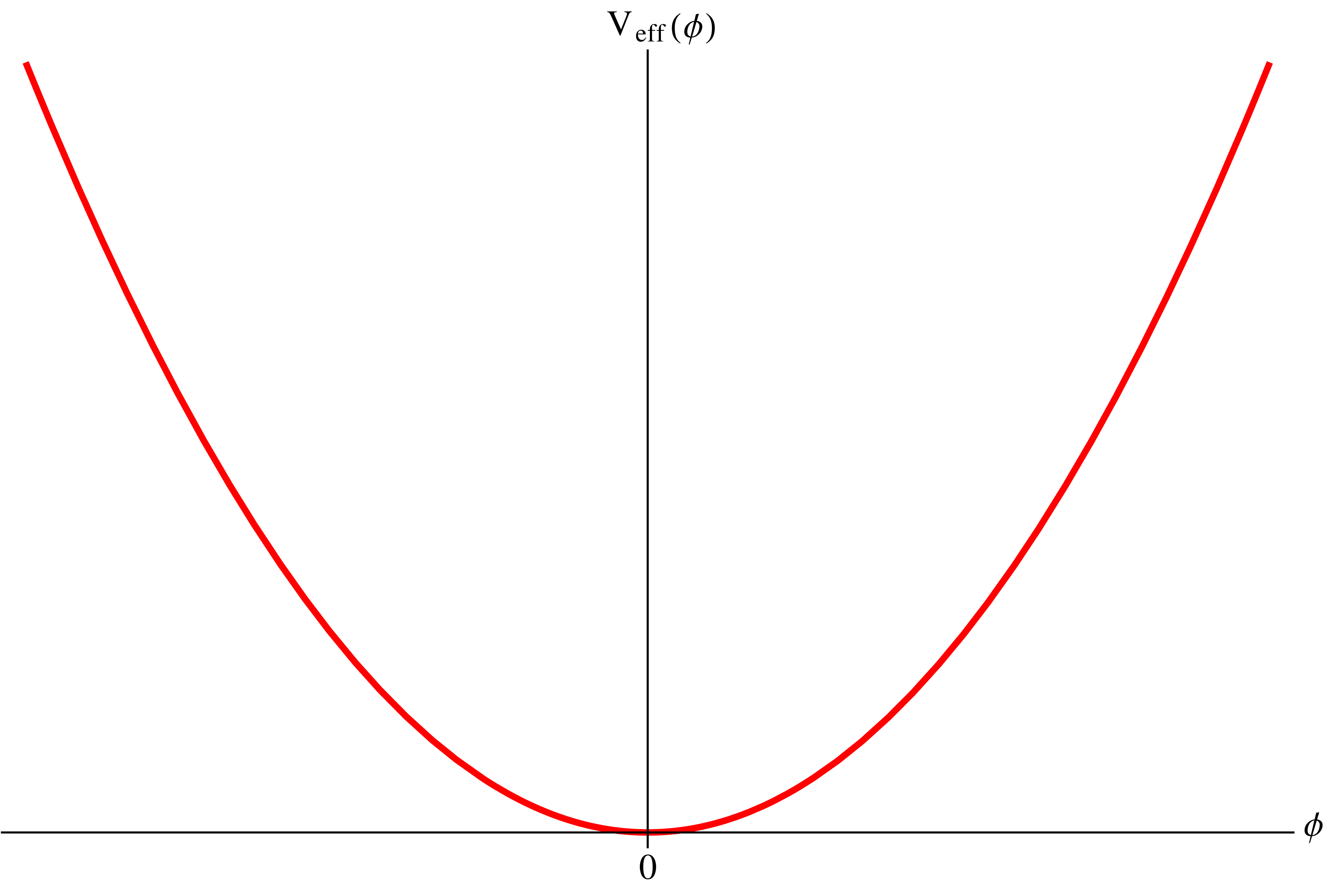}}
\caption{The symmetron effective potential for small and
large densities.}\label{fig:min_symmetron}
\end{figure}

In contrast to chameleon models, symmetrons have a light mass in all environments and instead work by moving $\beta(\phi)$ to
small values when the
density is large. They are defined by a $\mathbb{Z}_2$ symmetry breaking
potential and a quadratic coupling function:
\begin{equation}
 V(\phi)=-\frac{1}{2}\mu^2\phi^2+\frac{\lambda}{4}\phi^4;\quad
A(\phi)=1+\frac{\phi^2}{2M^2}
\end{equation}
so that the effective potential is
\begin{equation}
V\eff=\frac{\mu^2}{2}\left(\frac{\rho}{M^2\mu^2}-1\right)\phi^2+\frac{\lambda}{4}
\phi^4 ,
\end{equation}
which is invariant under the $\mathbb{Z}_2$ symmetry $\phi\rightarrow-\phi$. The
shape of this potential then depends on the local density. When $\rho>M\mu $
the coefficient of the quadratic term is positive and the minimum lies at
$\phi=0$ so that symmetry is unbroken. Conversely, when $\rho\ll M\mu$ the
minimum lies at
\begin{equation}
 \phi\approx\pm\frac{\mu}{\sqrt{\lambda}}
\end{equation}
and the symmetry is broken. This is shown in figure \ref{fig:min_symmetron}.
Calculating the coupling, we find
\begin{equation}\label{eq:betasymmgen}
 \beta(\phi)\approx\frac{\phi\mpl}{M^2}
\end{equation}
and so $\beta(\phi)\approx0$ in the symmetry unbroken phase and
\begin{equation}\label{eq:betasymbroken}
 \beta(\phi)\approx \frac{\mu\mpl}{\sqrt{\lambda}M^2}
\end{equation}
in the broken phase. This immediately reveals how the symmetron screens: in small density environments the
symmetry is broken and the field sits at one of the two minima leading to a non-negligible fifth-force. In
high density environments the symmetry is restored and the field moves rapidly to zero in order to minimise
the new effective potential, at which point $\beta(\phi)=0$ and the fifth-force is absent. Of course, one
must choose the parameters such that the symmetry is restored in densities corresponding to the solar system
if fifth-forces are to be absent locally. Cosmologically, we are interested in theories where the phase
transition happens in the recent past\footnote{More specifically, we are interested in theories where the phase transition
occurs around the present epoch so that the appearance of modified gravity coincides
with the onset of dark energy domination. One may then hope that it is a possible explanation for the cosmic
acceleration and the coincidence problem. It is now known that it is not \cite{Wang:2012kj}.} and so it is common to
set\footnote{One may
wonder what happens if we drop this requirement and push the phase transition to earlier times. In fact, we
require the theory to screen in the dark matter halo of the milky way, which corresponds to a density of
$10^6\rho_0$ and so the phase transition could not have occurred at redshifts greater than $10^2$. This leaves
a little leeway for moving the transition but not so much that there are any new features compared with the
standard case. Ultimately, the field rolls to its new minimum very quickly and remains there throughout the
subsequent cosmic evolution \cite{Hinterbichler:2011ca,Davis:2011pj}.}
\begin{equation}\label{eq:symmcosmo1}
 \mu^2M^2\sim\rho_0\sim H_0^2\mpl^2.
\end{equation}
We are also interested in modifications of gravity that are comparable with the Newtonian force when
unscreened\footnote{Again, this is an arbitrary choice and there is nothing precluding force enhancements
much larger than this.} and so we set $\beta(\phi)\sim1$ in equation (\ref{eq:betasymbroken}):
\begin{equation}
 \frac{H_0\mpl^2}{\sqrt{\lambda}{M^3}}\sim1
\end{equation}
where we have used equation (\ref{eq:symmcosmo1}). We will see later that current constraints impose $M<10^{-3}\mpl$, in which
case we find \cite{Hinterbichler:2011ca}
\begin{equation}\label{eq:symmlambda}
 \lambda\sim\frac{H_0^2\mpl^4}{M^6}\ge10^{-96},
\end{equation}
where the equality is satisfied when $M$ assumes its largest possible value. Assuming the bounds from local
tests of gravity are saturated, the mass of the field in the unbroken phase is
\begin{equation}\label{eq:symmunbrokenmass}
 m\eff\approx\sqrt{2}\mu\sim10^4H_0\sim\mathcal{O}(\textrm{Mpc}^{-1}).
\end{equation}
Hence, the symmetron mediates a force with a range (Compton wavelength) $\lambda_{\rm C}=m\eff^{-1}\le$ Mpc. Since the matter
coupling is irrelevant at late times --- when the cosmic density is well below the critical density for the phase transition ---
the symmetron behaves a quintessence field rolling down its scalar potential $V(\phi)$. Dark energy driven by quintessence-type
models requires the mass at the minimum be of order Hubble in order to achieve slow-roll \cite{Copeland:2006wr} so that the
Compton wavelength is $H_0^{-1}\sim\mathcal{O}(10^4\textrm{Mpc})$. Hence, the requirement (\ref{eq:symmunbrokenmass}) means that
the symmetron force is too short-ranged to account for the cosmic acceleration.

The key element behind this screening mechanism was the second order phase transition and so one may wonder
if more general models described by potentials and coupling functions of the form
\begin{equation}\label{eq:extendedsymmetrons}
 V(\phi)=-\left(\frac{\phi}{\phi_n}\right)^n+\left(\frac{\phi}{\phi_m}\right)^m;\quad
A(\phi)=1+\frac{\phi^{n}}{M^n}
\end{equation}
with $m>n>2$ can be constructed. Provided that both $m$ and $n$ are even, the effective potential will be even and will
indeed exhibit a second order phase transition when the density is below some threshold and can indeed
screen fifth-forces locally in the same manner as all of the other mechanisms. Such models go by the name
\textit{generalised symmetrons} and were discovered using a tomographic reverse-engineering of the potential
and coupling function from a generalised form for the cosmological evolution of the coupling for symmetrons
\cite{Brax:2012gr}. Since then, they have received little attention and their only other mention in the
literature is in the form of a no-go theorem precluding them from being realised within supersymmetric models
\cite{Brax:2012mq}.

\subsection{The Environment-Dependent Damour-Polyakov Effect}

Historically, this screening mechanism was discovered in the context of the environment-dependent dilaton
\cite{Brax:2010gi} and subsequently generalised to what have become known as \textit{generalised dilatons}
\cite{Brax:2012gr,Brax:2012nk}. This is, in some sense, a misnomer because the underlying mechanism has little
to do with whether the particle is a dilaton or not and is more transparent in the general framework. For this reason, here we
will construct the mechanism from the bottom-up and then specialise to the case of the dilaton.
Generalised dilatons are not easily written down in terms of potentials and coupling functions and are
instead reconstructed tomographically \cite{Brax:2012nk}. Since we will have no need for Tomography in this
thesis we will not discuss them here and simply remark that they are specific choices for the coupling
functions and potentials described in the general case below.

We have already noted that the force is screened in dense environments when 
\begin{equation}\label{eq:dilscreen}
 \beta(\phi)=\mpl\frac{\dd\ln A}{\dd\phi}=\frac{\mpl}{A(\phi)}\frac{\dd A(\phi)}{\dd\phi}=0.
\end{equation}
The symmetron mechanism described in subsection \ref{subsec:symmetron} has $\beta(\phi)\propto\phi$ and
used a phase transition to push $\phi$ to zero in dense environments. Another way of moving $\beta(\phi)$ to
zero is to utilise the last relation in equation (\ref{eq:dilscreen}) and somehow set $\dd A/\dd\phi$ to zero
in dense environments. This is tantamount to minimising the function $A(\phi)$. Consider then the effective
potential in the Einstein frame
\begin{equation}\label{eq:effpot2}
 V\eff=V(\phi)+(\rho A (\phi)-1)=A^n(\phi)\tilde{V}(\phi)+\rho (A(\phi)-1),
\end{equation}
where we demand that $A(\phi)$ has a minimum. Minimising this, one has 
\begin{equation}\label{eq:gendilmin}
 \beta(\phi_{\rm min})=-\frac{\mpl\tilde{V}^\prime(\phi_{\rm min})}{n\tilde{V}(\phi_{\rm min})+\rho},
\end{equation}
where, by virtue of (\ref{eq:a1}), we have set $A(\phi)\approx1$. In the limit $\rho\gg\tilde{V}(\phi_{\rm min})$, minimising
the effective potential is identical to
minimising the coupling function and screening the fifth-force. This screening by minimising the coupling function in dense
environments
is the environment-dependent Damour-Polyakov effect\footnote{This is named after a similar mechanism
introduced by Damour and Polyakov \cite{Damour:1994zq} where the cosmological evolution of a scalar
conformally coupled to matter minimises the coupling function and suppresses fifth-forces.}.

Now one may wonder why we have bothered to introduce the factor of $A^n$
in the potential. Surely if we leave it arbitrary then at high enough densities minimising the effective
potential is equivalent to minimising $A(\phi)$ and so the theory should screen. The problem with this
statement is what does one mean by the term \textit{high enough density}? Omitting the factor of $A^n$, the
relevant equation for the minimum is (again setting $A(\phi)\approx1$)
\begin{equation}\label{eq:dilgenmin2}
 \beta(\phi_{\rm min})=-\frac{\mpl V^{\prime}(\phi_{\rm min})}{\rho}
\end{equation}
and so the condition for screening is then $\rho\rightarrow\infty$. Without the $A^n$ term we need to go
to infinite densities in order to realise the screening mechanisms and this limit is ill-defined in conformal
scalar-tensor theories (see \cite{Koivisto:2012za,Zumalacarregui:2012us} for a discussion on this). Whilst
potentials of this form may look contrived in the Einstein frame, they have a very natural interpretation in
terms of fundamental theories where the potential is defined in the Jordan frame. We have already argued in
section \ref{sec:suppresscharge} that a potential $\tilde{V}(\phi)$ defined in the Jordan frame is described
by the potential $A^4\tilde{V}(\phi)$ in the Einstein frame and that the choice of which frame the potential
is defined in arbitrary. A natural class of models that screen using this mechanism then have $n=4$. Many
fundamental theories such as string theory and supergravity have low-energy effective actions that specify
the form of these functions in the Jordan frame and so these models are closely connected with fundamental
physics.

\subsubsection{The Environment-Dependent Dilaton}

The starting point for this mechanism \cite{Brax:2010gi} is the low-energy effective action for the string
dilaton coupled to gravity in the strong coupling limit \cite{Damour:1994zq,Gasperini:2001pc}:
\begin{equation}\label{eq:dilaton1}
 S=\int\dd^4x\sqrt{\tg}\left[\frac{e^{-2\psi(\phi)}}{2l_{\rm s}^2}\tilde{R}+\frac{Z(\phi)}{2l_{\rm
s}^2}\nabla_\mu\phi\nabla^\mu\phi-\tilde{V}(\phi)\right]+ S_{\rm
m}\left[\tg_{\mu\nu},g\iii(\phi);\psi_{\rm i}\right]
\end{equation}
where $\psi(\phi)$ is an unknown function that depends on the details of the string compactification, $l_{\rm s}$ is the string
length scale, and, unlike the previous models, the coupling constants
$g\iii$ are dilaton-dependent. Transforming to the Einstein frame by defining
\begin{equation}
 A(\phi)=\mpl l_{\rm s}e^{\psi(\phi)}
\end{equation}
we have\footnote{Note that this requires the inverse transformation to that used to find equation
(\ref{eq:chamactJF}), which can be found by setting $A\rightarrow A^{-1}$ and allowing for the fact that the
Planck mass was not included in the Einstein-Hilbert-like term in equation (\ref{eq:dilaton1}).}
\begin{equation}\label{eq:dilatonEF}
 S=\int\dd^4x\sqrt{g}\left[\frac{\mpl^2}{2}R-\frac{1}{2}
k^2(\phi)\nabla_\mu\phi\nabla^\mu\phi-A^4(\phi)\tilde{V}(\phi)\right]+S_{\rm
m}[A^2(\phi)g_{\mu\nu},g\iii(\phi);\psi\iii]
\end{equation}
where
\begin{equation}\label{eq:k^2dilaton}
 k^2(\phi)=6\beta^2(\phi)-\frac{A^2(\phi)Z(\phi)}{l_{\rm s}^2}. 
\end{equation}
In the strong coupling limit, which corresponds to $\phi\rightarrow\infty$ so that $e^{-\phi/\mpl}\ll1$, one can expand the
functions appearing in (\ref{eq:dilatonEF}) as \cite{Gasperini:2001pc}
\begin{align}\label{eq:expansionsdilaton}
 \tilde{V}(\phi)&=V_0e^{-\frac{\phi}{\mpl}}+\mathcal{O}\left(e^{-2\frac{\phi}{\mpl}}\right)\\
Z(\phi)&=-\frac{l_{\rm s}^2}{\lambda^2}+b_Ze^{-\frac{\phi}{\mpl}}+\mathcal{O}\left(e^{-2\frac{\phi}{\mpl}} \right)\\
g^{-2}\iii(\phi)&=\bar{g}^{-2}+b\iii e^{-\frac{\phi}{\mpl}}+\mathcal{O}\left(e^{-2\frac{\phi}{\mpl}}\right),
\end{align}
where the constants $\lambda, b_Z$ and $b\iii$ etc. are set by the details of the specific string compactification. We will treat
them as free parameters in what follows. One generally expects $b_Z\,,b\iii\sim\mathcal{O}(1)$ and
$\lambda\sim\mathcal{O}(1)$--$\mathcal{O}(l_{\rm s}^{-1}\mpl^{-1})$
($\gg\mathcal{O}(1)$) and so the kinetic factor is
\begin{equation}\label{eq:kindil}
 k(\phi)\approx\frac{1}{\lambda}\sqrt{1+6\beta^2(\phi)},
\end{equation}
where we have again set $A(\phi)\approx1$ in accordance with our earlier discussion. The equation of motion for the canonically
normalised field $\dd\varphi=k(\phi)\dd\phi$ is
then
\begin{equation}\label{eq:fieldeomdilaton}
 \Box\varphi=-\frac{1}{\mpl k(\phi)}\left[\beta(\phi)\left(4V(\phi)+\rho
A^\prime(\phi)\right)-V(\phi)\right]+S\iii\frac{g\iii^2(\phi)b\iii e^{-\frac{\phi}{\mpl}}}{2\mpl k(\phi)},
\end{equation}
where it is understood that $\phi=\phi(\varphi)$ and
\begin{equation}\label{eq:Si}
 S\iii\equiv \frac{\delta S\mmm}{\delta \ln g\iii}.
\end{equation}
Typically, $S\iii\sim\mathcal{O}(\rho)$ and $\mathrm{exp}(-\phi/\mpl)\ll1$ in the strong coupling limit so we can safely
ignore the final term in equation (\ref{eq:fieldeomdilaton}). In this case, we can integrate the equation of motion to find the
effective potential for $\varphi$:
\begin{equation}\label{eq:veffdilaton}
 V\eff(\varphi)=V_0A^4(\phi)e^{-\frac{\phi}{\mpl}}+ \rho (A(\phi)-1).
\end{equation}
The coupling can then be found using the chain rule and one finds
\begin{equation}\label{eq:betadilaton}
 \beta(\varphi)=\frac{\beta(\phi)}{k(\phi)}.
\end{equation}
Minimising the effective potential with respect to $\phi$ is equivalent to minimising it with respect to $\varphi$ since
$\dd\varphi/\dd\phi\ne0$ and so we find
\begin{equation}\label{eq:betamindilaton}
 \beta(\pmi)=\frac{V(\pmi)}{4V(\pmi)+\rho},
\end{equation}
at the minimum of the effective potential. When $\rho\gg 4V(\pmi)$ we have $\beta(\pmi)\approx 0$ and using equation
(\ref{eq:betadilaton}) we can then see that this
mechanism screens via the EDDP since $k(\pmi)\ne0$. The functional form of $A(\phi)$ is still unknown and there is no
natural choice in the strong coupling limit of string theory\footnote{Put another way, the functional form is presently unknown in
this limit.} and so the authors of \cite{Brax:2010gi} assume that the function has the requisite minimum for the screening
mechanism to be present to expand the coupling function as
\begin{equation}\label{eq:coupdil}
 A(\phi)=1+\frac{A_2}{2\mpl^2}\left(\phi-\pmi\right)^2+\cdots
\end{equation}
and use laboratory tests to constrain the parameter $A_2$. We will not examine the dilaton model explicitly in this thesis nor
any other theories that screen via the EDDP \footnote{They will however be constrained in chapter \ref{chap:four} using a
set of model-independent parameters presented in the next section, although we will not translate these constraints into $A_2$.}
and so for completeness we will simply state the constraint that they obtain is $A_2\gg1$. The reader is referred to
\cite{Brax:2010gi} for the technical derivation of this constraint\footnote{Note that their action and expressions for
quantities such as $\beta(\phi)$ and $k(\phi)$ differ from ours because they use a dimensionless field variable whereas ours has
the canonical mass dimension. Furthermore, their function $k(\phi)$ differs from ours by a factor of $2$ because they do not work
with a canonically normalised field. We have done this in order to provide a consistent link with the general framework presented
in this chapter. There are also several typographical errors in this paper that have been corrected here.}.

\subsection{The Screening Mechanism}\label{sec:screen_mech}

The previous sections have discussed the three mechanisms by which the fifth-force can be rendered negligible in dense
environments but there are still two important questions to address: How is this realised in reality? And exactly which objects
are screened? We will answer these questions in this section.

\begin{figure}
\subfigure[Unscreened]{\label{fig:unscreened}\includegraphics[
width=0.5\textwidth]{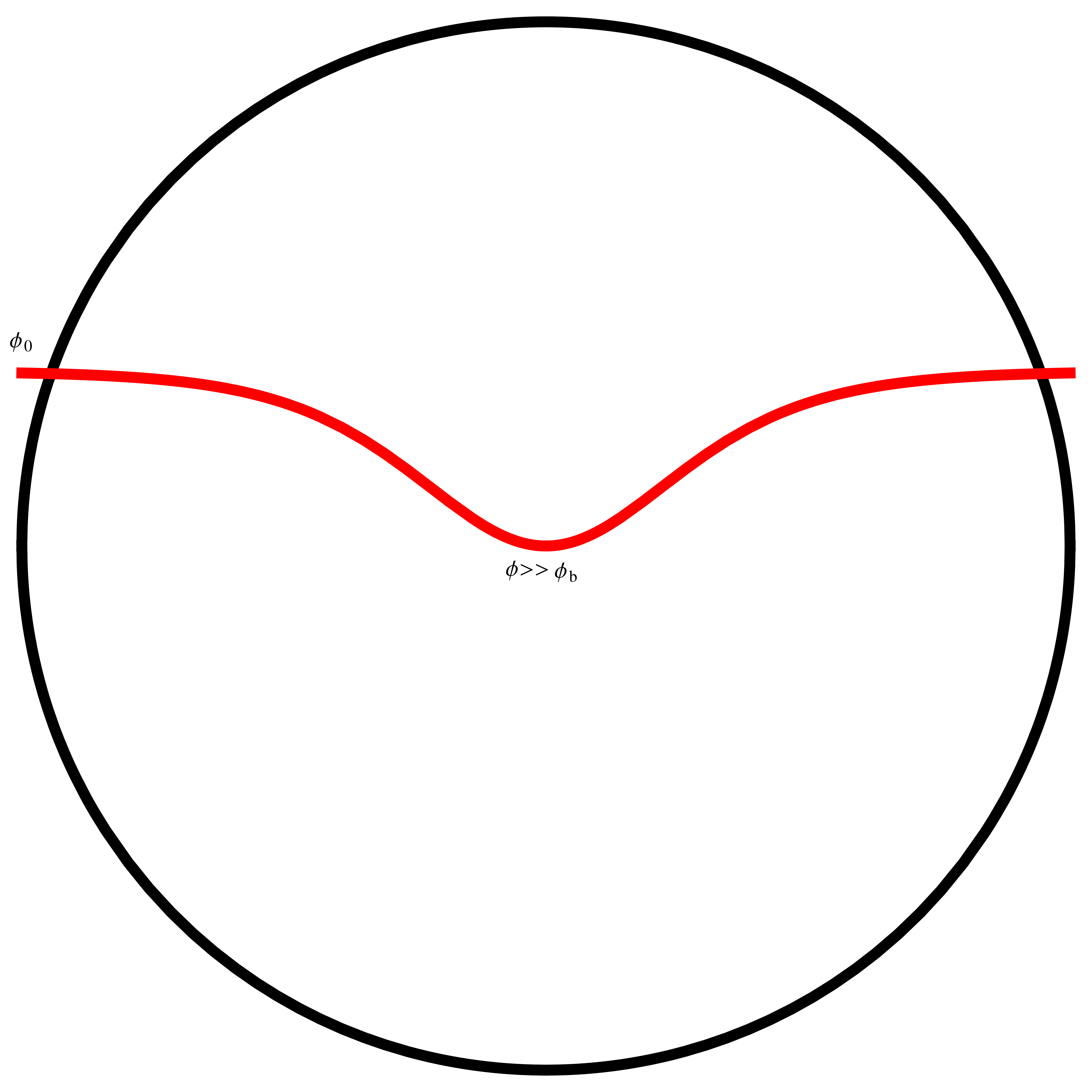}}
\subfigure[Screened]{\label{fig:screened}\includegraphics[
width=0.5\textwidth]{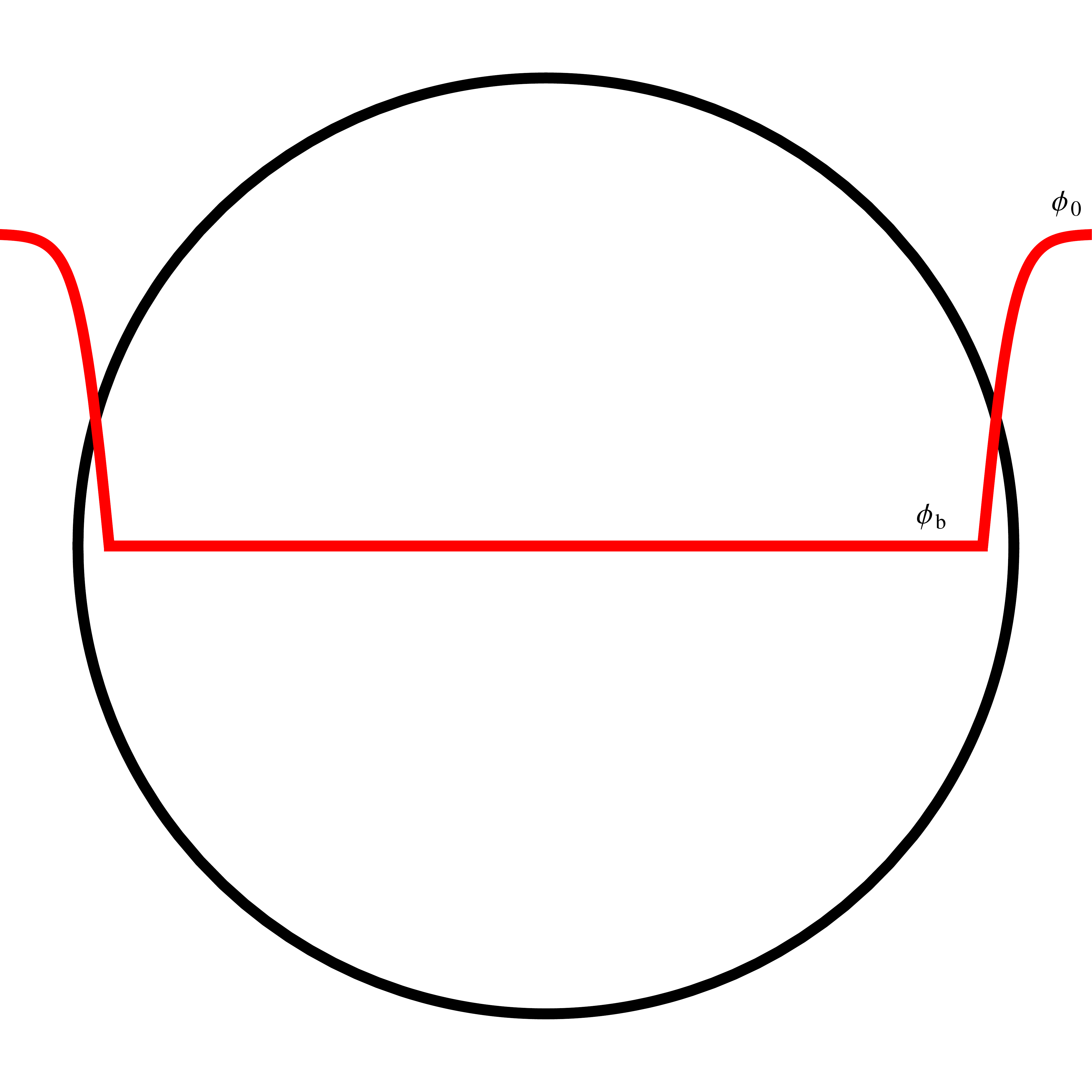}}
\caption{The field profile in the screened and unscreened scenarios.}\label{fig:screened/unscreened}
\end{figure}

Even if the fifth-force can in theory be screened in dense enthronements, this will only happen if the field can reach the minimum
of the effective potential where the screening occurs. Consider then a spherical over-density of radius $R$ with density profile 
$\rho_{\rm b}(r)$ placed into a much larger (by which we mean its characteristic length scale is $\gg R$) medium with a smaller
density $\rho_{\rm c}$. Far away from the object, the field can minimise its effective potential with density $\rho_{\rm c}$ and
assumes a field value $\phi_{\rm c}$. The field will want to minimise the effective potential with density $\rho_{\rm b}$ (with
corresponding field value $\phi_{\rm b}$) inside the object and so near the object (and inside of course) we expect a field
gradient, which according to equation (\ref{eq:fifth-forcegeneral}) implies the presence of a fifth-force. There are then two
possible scenarios. Suppose that the object is small enough (to be quantified presently) such that the field is unable to reach
this minimum at all. In this case, the field will be a small perturbation about the background value $\phi_{\rm c}$ and will hence
mediate a
fifth-force that is comparable with the Newtonian force since either $\beta(\phi)$ or $m\eff(\phi)$ correspond to those at the
minimum in low density environments. In this case, the object is \textit{unscreened} and we expect new and novel features compared
with those predicted by general relativity. This case is shown in figure \ref{fig:unscreened}. On the other hand, if the object
is large enough such that the field can minimise its effective potential at density $\phi_{\rm b}$ over most of the radius
of
the object then the field will only vary very close to the surface of the object and will quickly reach $\phi_{\rm c}$ leaving a
field gradient in a very thin shell near the surface. In this case, any perturbation in the field will be about $\phi_{\rm b}$ and
hence corresponds to either a negligible value of $\beta(\phi)$ or a very large effective mass. In this case the object is
\textit{screened} and any deviations from general relativity will be unobservable. This is shown in figure \ref{fig:screened}.
 
\begin{figure}
\centering
\includegraphics[width=0.5\textwidth]{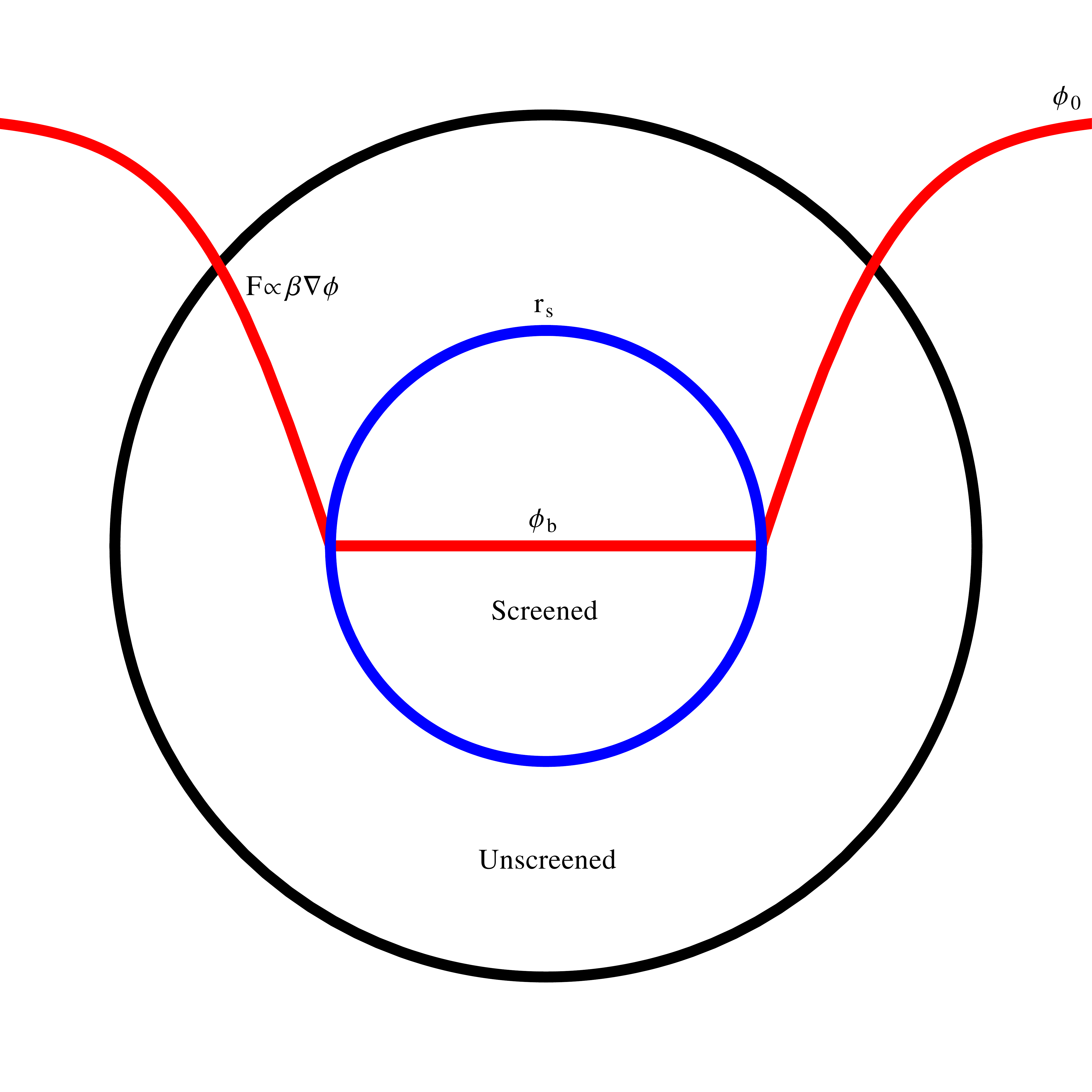}
\caption{The general field profile for a partially unscreened object.}\label{fig:genscreen}
\end{figure}

In general, one expects a situation somewhere between these two extremes so that the field can reach $\phi_{\rm b}$ at the
centre of the object and remains there until some radius $\rs$, which we shall refer to as the \textit{screening radius}, at
which point it begins to asymptote to $\phi_{\rm c}$. In this case, the region interior to the screening radius is screened and
there are no fifth-forces whereas in the region exterior to the screening radius there is a field gradient and this region is
unscreened. The screened case then corresponds to $\rs=R$ and the unscreened case to $\rs=0$; the general case is an intermediate
configuration between the two and we refer to this as the \textit{partially screened} case. This is shown schematically in figure
\ref{fig:genscreen}. We will now derive the field profile for this general case and use it to find the fifth-force.

We begin with the region $r\le\rs$. In the static, spherically symmetric limit equation (\ref{eq:chameom2}) reduces to
\begin{equation}\label{eq:staticeom}
 \nabla^2\phi=V^\prime(\phi)+\frac{\beta\rho}{\mpl}.
\end{equation}
Now suppose that the field minimises its effective potential at $r=0$ so that $\phi=\phi_{\rm b}$. This means that the effective
potential is minimised and so equation (\ref{eq:staticeom}) becomes
\begin{equation}
 \nabla^2\phi\approx0\label{eq:del^2=0}.
\end{equation}
Since there is no source for this equation, the field will remain at an approximately constant value until this approximation
breaks down. This defines the screening radius. In the unscreened region, the field will be a small perturbation $\delta\phi$
about the background value $\phi_{\rm c}$ so that $\phi(r)=\phi_{\rm c}+\delta\phi(r)$. Subtracting the equation of
motion (\ref{eq:staticeom}) for the over-dense region from that for the under-dense one and linearising we have
\begin{equation}\label{eq:screom}
 \nabla^2\delta\phi \approx m_{\rm c}^2\delta\phi +\frac{\beta_{\rm
{\rm c}}\delta\rho}{\mpl},
\end{equation}
where $\delta\rho\equiv\rho_{\rm b}-\rho_{\rm c}$, $m_{\rm c}^2\equiv
V_{,\phi\phi}(\phi_{c})$ is the mass of the field in the background (the mass of oscillations about the minimum of the effective
potential at density $\rho_{\rm c}$) and $\beta_{{\rm c}}\equiv\beta(\phi_{\rm {\rm c}})$. Theories that posses
screening mechanisms have the property that
\begin{equation}\label{eq:betaapprox0}
 \frac{\dd \beta}{\dd\phi}\delta\phi\ll\beta(\phi_{\rm c})
\end{equation}
in the unscreened region and so we have neglected all terms proportional to $\dd\beta/\dd\phi$ in equation (\ref{eq:screom}). In
all theories of interest, we have $m_cR\ll 1$ so that the fifth-force gives rise to novel features on large scales and so
we may ignore the first term in equation (\ref{eq:screom}), it is negligible compared with the Laplacian.
Now we know that $\delta\rho$ is related to the Newtonian potential $\Phi_{\rm
N}$ via the Poisson equation (\ref{eq:poisson1})  and so we 
may substitute this into (\ref{eq:screom}) and integrate twice to find the field
profile:
\begin{equation}
\delta \phi(r) \approx -\phi_{\rm c} + 2\beta_{\rm c} \mpl \left[\Phi_{\rm
N}(r)-\Phi_{\rm
N}(r_{\rm s}) + r_{\rm s}^2\Phi^{\prime}_{\rm N}(r_{\rm s})\left(
\frac{1}{r}-\frac{1}{r_{\rm s}}\right) \right]H(r-r_{\rm s}),
\label{eq:phiform1}
\end{equation}
where $H(x)$ is the Heaviside step-function. Using the definition of the Newtonian potential $\dd \Phi_{\rm N}/\dd r
= GM(r)/r^2$, where $M(r)$ is the mass enclosed within a sphere of
radius $r$ (the total mass of the over-density is $M=M(R)$), the fifth-force
(\ref{eq:fifth-forcegeneral}) in the unscreened region is given by the derivative of
(\ref{eq:phiform1}):
\begin{equation}\label{eq:gprof}
 \frac{\beta_{\rm c}}{\mpl}\frac{\dd\phi}{\dd r} =
\alpha_{\rm c}\frac{GM(r)}{r^2}\left[1-\frac{M(\rs)}{M(r)}\right],
\end{equation}
where $\alpha_{\rm c}\equiv 2\beta^2(\phi_{\rm c})$. We can see that equation (\ref{eq:gprof}) reproduces the fully screened and
unscreened cases. When $\rs = R$ there is no
fifth-force and the object is screened whereas when $\rs=0$ the object is full
unscreened and we have $F_\phi=\alpha_{\rm c} F_{\rm N}$, where $F_{\rm N}$ is
the Newtonian force. The parameter $\alpha_{\rm c}$ therefore determines the
strength of the fifth-force. 

All that remains is to find the screening radius $\rs$. Using the fact that
$\delta\phi\rightarrow0$ when $r/\rs\rightarrow\infty$ as does $\pn(r)$, equation (\ref{eq:phiform1}) gives us an
implicit expression for the
screening radius:
\begin{equation}
\frac{\phi_{\rm c}}{2\beta(\phi_{\rm c}) \mpl}\equiv \chi_{\rm c} = -\Phi_{\rm
N}(r_{\rm s}) -
r_{\rm s}\Phi_{\rm N}^{\prime}(r_{\rm s})\geq0 \label{eq:screenradius}.
\end{equation} 
It will be useful later to recast this as an integral equation:
\begin{equation}\label{eq:chiint}
 \chi_{\rm c} = 4\pi G\int_{\rs}^R r\rho_{\rm b}(r)\dd r,
\end{equation}
where $\chi_{\rm c}$ is to be viewed as a parameter of the theory to be input by hand and $\rs$ the screening radius to be found
using this equation. The parameter $\chi_{\rm c}$ is known as the \textit{self-screening parameter}
and is of paramount importance to the screening properties of these theories.
Since $\Phi_N <0$ and $\dd\Phi_N/\dd r>0$, there are no solutions when 
\begin{equation}
 |\Phi_{\rm N}(R)|= \frac{GM}{R} > \chi_{\rm c}.
\end{equation}
In this case, the object is fully screened and $\rs=R$. When $\Phi_{\rm
N}<\chi_{\rm c}$ the object will be at least partially unscreened. 

The screening and fifth-force properties in any region are fully specified
by $\alpha_{\rm c}$ and $\chi_{\rm c}$, however
these are environment-dependent and are non-linearly related to the field
values in different regions of the universe. The exception to this is unscreened
objects, where the field is only a small perturbation around the value in the cosmological background and so these values are
roughly constant. Whether or not an object is screened or not depends on its Newtonian
potential
relative to the cosmological value of $\chi_{\rm c}$; the same is true of denser
objects residing in larger unscreened objects. In each case, the strength of the fifth-force in
any unscreened region is proportional to the cosmological value of $\alpha_{\rm
c}$ and so any tests of these theories using unscreened objects probe the cosmological values of
$\alpha_{\rm c}$ and $\chi_{\rm c}$. The theory is then parametrised in a
model-independent manner by the cosmological values of these parameters, which
we denote by $\alpha$, which measures the strength of the fifth-force in
unscreened regions and
\begin{equation}\label{eq:chi0def}
\chi_0\equiv\frac{\phi_0}{2\mpl\beta(\phi_0)} ,
\end{equation}
which determines how small the Newtonian
potential must be in order for the object to be unscreened\footnote{One must
note however that this analysis applies to isolated objects only. In the
presence of other objects of Newtonian potential $\Phi_{\rm N}^{\rm ext}$ the
external potential can be large enough to screen an object that would otherwise
be self-unscreened \cite{Hui:2009kc}.}. Note that $f(R)$ theories have $\beta(\phi)=1/\sqrt{6}$ (see equation (\ref{eq:afr})) and
so $\alpha=1/3$ and there is only one free parameter. When working within an $f(R)$ framework the parameter that is often
constrained in the literature is $f_{R0}$, the present-day value of the first derivative of the function with respect to $R$.
This parameter is equivalent to $\chi_0$ and in this thesis we will only work with this parameter in order to avoid any ambiguity.

This thesis is concerned with both astrophysical probes of modified gravity and the cosmology of supersymmetric chameleon models.
$\alpha$ and $\chi_0$ are then perfect parameters to describe both of these scales. In general, cosmological tests probe the
cosmological mass and coupling (see \cite{Brax:2011aw,Brax:2012gr}), which can be mapped onto $\chi_0$ and $\alpha$ in a
straightforward manner. We will see presently that the Milky Way is screened and so laboratory and solar system tests probe the
mass as a function of the coupling.
These can be mapped onto $\chi_0$ and $\alpha$ in a model-dependent manner but this requires one to know the galactic evolution
of $\chi_{\rm c}$ and $\alpha_{\rm c}$ and how they relate to the cosmological values. This can be done using spherical collapse
models \cite{Li:2011qda} although to date no comprehensive analysis has been performed.

This parametrisation may seem esoteric at first, however it is nothing more than the mass and coupling of the field in
the cosmological vacuum. $\alpha=2\beta^2(\phi_0)$ and so $\alpha$ directly measures the coupling $\beta(\phi_0)$ in the
cosmological vacuum. Minimising the effective potential, one has
\begin{equation}\label{eq:mineffpot}
 \frac{\dd V(\phi)}{\dd \phi}= -\frac{\beta(\phi_0)\rho_0}{\mpl}.
\end{equation}
Setting $\rho_0\sim H_0^2\mpl^2$ and linearising the derivative of the potential about $\phi_0$ and using the fact that
$\delta\phi\approx-\phi_0$ one finds
\begin{equation}\label{eq:chi0mass}
 \chi_0\approx\left(\frac{H_0}{m_0}\right)^2
\end{equation}
where $m_0^2\equiv V^{\prime\prime}(\phi_0)$ is the free mass of the field in the cosmological vacuum. One can see that $\chi_0$
is a measure of the range of the force mediated on cosmological scales.

\subsection{Current Constraints}\label{sec:currentconstraints}

This subsection is somewhat historical in that it refers to the situation before 2012 when the original work
presented in this thesis was first reported in \cite{Jain:2012tn}. In chapter \ref{chap:four} we will present new and updated
constraints.
The material is presented in this fashion as a historical background and in order to highlight the relevance of this work to the
field.

We now have a method of classifying exactly which objects in the universe are screened and which are not. Any object whose
Newtonian potential is less than $\chi_0$ will be screened whereas objects where the converse is true will be at least partially
unscreened. We can therefore discern the ranges of $\chi_0$ that can be probed using different astrophysical objects by looking
at their Newtonian potentials. 

The first task is to ensure that our own solar-system is screened. The Newtonian potential of the Earth is
$\pn^\oplus\sim10^{-9}$ and so one may think that we need to impose $\chi_0\lsim\mathcal{O}(10^{-9})$ in order to satisfy
laboratory experiments. In fact, this is too stringent. The Newtonian potential of the Sun is
$\pn^\odot=2\times10^{-6}$ and so one would then need to impose $\chi_0\lsim\mathcal{O}(10^{-6})$ in order to satisfy
post-Newtonian tests of gravity using light bending around the Sun. This is still too n\"{a}ive a constraint because we have
yet to account for the fact that the solar system resides within the Milky Way. If the Milky Way were unscreened there would be
discernible effects coming from the motion of stars about the galactic centre and so we must demand that it is screened. Spiral
galaxies like the Milky Way have rotational velocities of $\mathcal{O}(100\textrm{ kms}^{-1})$ or more and so the virial
theorem ($GM/Rc^2\equiv\pn=v^2/c^2$) implies they have Newtonian potentials $\pn\gsim\mathcal{O}(10^{-6})$. The Milky Way has
$\pn\sim10^{-6}$ but it is situated in
the local group whose Newtonian potential is of order $10^{-4}$.  There are then two possibilities: we must either impose that the
local group screens the Milky Way, in which case we must impose $\chi_0<10^{-4}$ or we must demand that the Milky Way is
self-screening, in which case we need the stronger constraint $\chi_0\lsim 10^{-6}$. Either way, the fact that the Milky
Way is screened means that $\chi_{\rm c}^{\textrm{Milky Way}}\ll\chi_0$ and so the Sun and Earth are blanket screened and do not
impose
additional constraints. Whether or not the local group does screen the Milky Way has been debated in the literature. Later on, we
will present a new and independent constraint that $\chi_0<10^{-6}$ and so we shall adopt this constraint from here on. Note
that the symmetron has $\chi_0=M^2/2\mpl^2$ and so compatibility with this constraint demands that $M\le10^{-3}\mpl$ as we stated
without explanation in section \ref{subsec:symmetron}.

At the background level, these models do not alter the expansion history predicted by $\Lambda$CDM \cite{Wang:2012kj}, however
they give rise to modifications to the CDM power spectrum and other linear probes on scales inside the horizon. It is difficult
to define a Newtonian potential for cosmology because one generally requires a full relativistic treatment. It is tempting to use
the metric potential $\Phi$ defined in (\ref{eq:confnewtgauge}) in a similar manner to the potential defined in
(\ref{eq:minkpertconfnewt}), however saying that this assumes any value is a gauge-dependent statement and there is no guarantee
that one cannot find a gauge where a different metric potential assumes a vastly different value. Inside the horizon, the
equations of motion for perturbations are essentially non-relativistic and so should look Newtonian in any gauge. Therefore, we
can treat this potential as the equivalent of the Newtonian potential provided that one bears in mind the scales on which such an
interpretation is valid. In particular, this will not be a good approximation near the horizon. In this case, the potential on
sub-horizon scales is of order $10^{-6}$ and so we expect novel deviations on linear scales for $\chi_0\gsim 10^{-6}$. Now we
have argued above that $\chi_0\lsim10^{-6}$ and so according to equation (\ref{eq:chi0mass}) the mass of the field in the
cosmological background must satisfy $m_0\gsim 10^3H_0\sim 10^{-1}h$ Mpc$^{-1}$. Now it is well-known that linear perturbation
theory in general relativity is only valid on scales $k<10^{-1}$ Mpc$^{-1}$ and so the constraint coming from the necessity of
screening the Milky Way ensures that there are no observable effects on linear scales. On non-linear scales, one must use N-body
simulations to calculate the non-linear cosmological probes such as the halo mass function. A full account of N-body codes and
their predictions is beyond the scope of this thesis and so we will simply remark that all cosmological signatures of modified
gravity vanish when $\chi_0\le10^{-6}$. One can therefore use non-linear cosmological probes to constrain these theories when
$\chi_0\gsim 10^{-6}$. Currently, the strongest constraints come from cluster statistics and impose $\chi_0<10^{-4}$
\cite{Schmidt:2008tn}. Recently, a new method using the difference between the hydrostatic and lensing mass of the Coma cluster
has been used to impose an independent constraint $\chi_0\lsim 6\times10^{-5}$.

Given our constraint $\chi_0\le 10^{-6}$ one may wonder if there are more under-dense regions with lower Newtonian potentials
that can act as new probes of these theories. We have already seen than $GM_\odot/R_\odot=2\times10^{-6}$ but what about other
main-sequence stars? There is a well known mass-radius relationship for main-sequence stars
\begin{equation}\label{eq:M-Rms}
 \left(\frac{R}{R_\odot}\right)=\left(\frac{M}{M_\odot}\right)^\nu
\end{equation}
with $\nu=0.4$--$0.8$ depending on the dominant nuclear reaction network in the core. We then have
\begin{equation}\label{eq:pnstarsrelation}
 \pn=\frac{GM}{R}=\frac{GM_\odot}{R_\odot}\left(\frac{M}{M_\odot}\right)^{(1-\nu)}.
\end{equation}
The lowest mass stars have masses of order $0.1M_\odot$ and have $\nu=0.4$ corresponding to the PP1 chain giving Newtonian
potentials of order $10^{-6}$ and the highest mass stars have masses around $100M_\odot$ and have $\nu=0.8$ corresponding to the
CNO cycle, which gives Newtonian potentials of order $10^{-5}$. Main-sequence stars therefore all have Newtonian potentials
$\pn\ge10^{-6}$ and therefore cannot probe $\chi_0<10^{-6}$. Post-main-sequence stars such as red giant (RG) and asymptotic giant
branch (AGB) stars have masses equal to that of their progenitor main-sequence stars but have radii 10 to 100 times larger and
therefore have Newtonian potentials of order $10^{-7}$ ($1$--$2M_\odot$) to $10^{-8}$ ($10\!\!+\,M_\odot$). We can therefore use
post-main-sequence stars to probe the parameter range $10^{-8}\le\chi_0\le10^{-7}$. 

Now we have also argued that spiral galaxies are screened and so we cannot probe this parameter range unless we can find
unscreened galaxies in which to observe these stars. Elliptical galaxies have Newtonian potentials similar to spirals but dwarf
galaxies have lower rotational velocities of order $50\textrm{ km s}^{-1}$ and hence have Newtonian potentials
$\pn\sim\mathcal{O}(10^{-8})$. Dwarf galaxies are hence perfect laboratories for testing these theories. Dwarf galaxies in
clusters
will be blanket screened by their neighbours whereas those in cosmic voids will be unscreened. By comparing the properties of
these galaxies (or their constituents) in voids and clusters and looking for systematic offsets new constraints can be
placed. Tests using dwarf galaxies fall into two classes: first, the fact that the galaxies are unscreened means that the galactic
properties such as the rotation curves and morphology are altered relative to the general relativity prediction
\cite{Jain:2011ji} and second, other objects inside these galaxies such as post-main-sequence stars will also be unscreened and
show novel features compared with those situated inside dwarf galaxies in clusters. The first class has been investigated
elsewhere
\cite{Vikram:2013uba,Vikram:2014uza} and the data is not yet good enough to place new constraints, although this will change with the next
generation of galaxy surveys. The second class is the subject of a large proportion of this thesis and will be
the focus of chapters \ref{chap:three}, \ref{chap:four} and \ref{chap:seven}.

Below $\chi_0=10^{-8}$ dwarf galaxies are screened and so one must look for even more under-dense object to test these theories.
Unfortunately, there are no such objects; when $\chi_0<10^{-8}$ every object in the universe is screened\footnote{HI gas in the
outer disks of galaxies has $\pn\sim\mathcal{O}(10^{-11})$--$\mathcal{O}(10^{-12})$ \cite{Hui:2009kc} however this is blanket
screened in all galaxies when $\chi_0\lsim10^{-8}$.}. Whilst these theories can never be completely ruled out because one can
always
screen everything, when $\chi_0<10^{-8}$ the theory is rendered indistinguishable from general relativity on all
scales\footnote{One can construct models with arbitrarily weak matter couplings that cannot be probed astrophysically. In this
case, one then has a theory which shows no deviations from general relativity on any scale with the exception that one can look
for signatures in the laboratory. These cases are also physically uninteresting since they have no effect on any physical
phenomena.} and is hence physically uninteresting. In table \ref{tab:planckcosmo} we summarise the Newtonian potentials of
different objects in the universe and their degree of screening.

\begin{table}
\centering
\heavyrulewidth=.08em
	\lightrulewidth=.05em
	\cmidrulewidth=.03em
	\belowrulesep=.65ex
	\belowbottomsep=0pt
	\aboverulesep=.4ex
	\abovetopsep=0pt
	\cmidrulesep=\doublerulesep
	\cmidrulekern=.5em
	\defaultaddspace=.5em
	\renewcommand{\arraystretch}{1.6}
\begin{tabu}{l|[2pt]l|[2pt]l}
Object & $\pn$ & Screening Status   \\
\tabucline[2pt]{-}
Earth & $10^{-9}$ & Screened by the Milky Way  \\
The Sun & $2\times10^{-6}$  & Screened by the Milky Way  \\
Main-sequence stars & $10^{-6}$--$10^{-5}$ & Screened  \\
Local group & $10^{-4}$ & Screened\\
Milky Way & $\mathcal{O}(10^{-6})$ & Screened\\
Spiral and elliptical galaxies& $10^{-6}$--$10^{-5}$& Screened\\
Post-main-sequence stars& $10^{-7}$--$10^{-8}$& Unscreened in dwarf galaxies in cosmic voids\\
Dwarf galaxies & $\mathcal{O}(10^{-8})$& Screened in clusters, unscreened in cosmic voids\\
\end{tabu}\label{tab:screening}
\caption{The screening status of different collapsed objects in the universe given the constraint $\chi_0< 10^{-6}$. We do not
include linear and non-linear cosmological scales since there is no unequivocal equivalent of the Newtonian potential with which
to compare $\chi_0$ (see the discussion above). Post-main-sequence stars located in dwarf galaxies in clusters or in galaxies
with any other morphology are blanket screened by their host galaxy.}
\end{table}

\subsection{Identification of Unscreened Galaxies}\label{subsec:screeningmap}
 
So far, all of the work we have presented has been theoretical and we have not addressed the question of whether we can determine
whether or not an individual galaxy in the real universe is screened. We have already seen that this depends in part on the
self-Newtonian potential of the galaxy but this does not account for environmental screening due to nearby
galaxies. There has been a great effort to address this problem using N-body simulations and there is indeed a method of
determining the screening status of a galaxy. The criterion to determine screening exploits the fact that relativistic particles
that move on null geodesics do not distinguish between $g_{\mu\nu}$ and $\tg_{\mu\nu}$ and therefore do not feel the effects of
modified gravity. In particular, photons move on the same trajectories as they would in general relativity and hence the amount
of gravitational lensing due to a fixed mass is identical in both theories. This means that if one were to measure an object's
mass using
lensing measurements --- the so-called \textit{lensing mass} --- and compare it with the mass inferred from the dynamics of
non-relativistic particles about this mass --- the so-called \textit{dynamical mass} --- the two would only agree if the object is
screened. The necessary quantity for determining whether a dark matter halo of mass $M$ is screened or not is the ratio of
the distance to the nearest neighbour whose mass is at least as large as $M$ to the virial radius of the neighbouring halo
\cite{Haas:2011mt,Zhao:2011cu}. 

Observationally, this is difficult to obtain and so in order to obtain a screening map
\cite{Cabre:2012tq} have used a simpler set of criteria. The first is the condition we described at length above i.e. a galaxy is
classified as self-screened if
\begin{equation}\label{eq:selfscreenmap}
 \pn^{\rm self}\ge\chi_0,
\end{equation}
where
\begin{equation}\label{eq:phiself}
 \pn^{\rm self} \equiv \frac{GM}{r_{\rm vir}}
\end{equation}
where the virial radius $r_{\rm vir}$ is related to the critical linear density contrast $\delta_{\rm crit}=3M/4\pi r_{\rm
vir}^3$ and the authors take $\rho_{\rm crit}=200\rho_0$. The masses are measured using either X-ray and lensing observations,
or, where no such observations are available, using scaling relations such as the mass-luminosity or mass-velocity dispersion
relations. This condition is robust and there is little error in using it. A galaxy is classed as environmentally screened if
\begin{equation}\label{eq:envscreenmap}
 \pn^{\rm env}\ge \chi_0,
\end{equation}
where $\pn^{\rm env}$ is defined via
\begin{equation}
 \pn^{\rm env}=\sum_{\rm d\iii<\lambda_{\rm C}+r\iii} \frac{GM\iii}{d\iii}.
\end{equation}
Here, $d\iii$ is the distance to a neighbouring galaxy with mass $M\iii$ and virial radius $r\iii$ and $\lambda_{\rm C}$ is the
Compton wavelength, which is set by the specific modified gravity model and is a function of $\chi_0$ and $\alpha$. This condition
is an ansatz and as such needs careful consideration. It is motivated by the fact that the range of the fifth-force is
$\lambda_{\rm C}$ and hence the galaxy will only feel the fifth-force from neighbouring galaxies within a distance $\lambda_{\rm
C}+r\iii$. The authors of \cite{Cabre:2012tq} have tested this criterion rigorously against N-body simulations and unknown
systematics and the interested reader is referred there for the technical details (see figure 2 in particular). Here, we will only
remark
that this criterion is very successful at classifying the screening status of most of the galaxies and any errors are always such
that there are more unscreened galaxies than it would predict and so any new constraints found using this map are conservative.

We will use this screening map in chapter \ref{chap:four} in order to place new constraints on chameleon-like theories.

\subsection{Cosmological Behaviour}\label{subsec:cosmologyc2}

The majority of this thesis is devoted to astrophysical tests, however in chapter \ref{chap:seven} we will be interested in the
cosmology of supersymmetric chameleon models and so here we will briefly present the formalism for describing the cosmological
dynamics of these theories. In particular, the non-minimal coupling to matter has the consequence that the equation of state
parameter is not the
same as we found for simple quintessence models (\ref{eq:quinteos}) and so the main focus will be on deriving this. The
cosmological
behaviour can vary between models and so here we will state only the model-independent features without a lengthy derivation. We
refer the reader to \cite{Brax:2012gr} and references therein for further details. We will also derive the cold dark matter power
spectrum and so we also present the equations governing the evolution of the CDM linear density contrast in chameleon-like models.

\subsubsection{Background Evolution}

In order to be compatible with local constraints, the cosmological mass at the time-dependent minimum of the
effective potential of the field always satisfies
\begin{equation}\label{eq:masscos}
 \frac{m\eff(\phi)}{H}\gg1,
\end{equation}
where $H(t)$ is the Hubble parameter at any given time. This means that the time-scale ($m\eff^{-1}$) on which the field responds
to changes in the position of the minimum is far shorter than the time-scale on which the minimum moves ($H^{-1}$) and so given
any initial condition, the field tracks the minimum adiabatically throughout all subsequent cosmic evolution\footnote{Another way
of stating this is that the time-varying minimum is a global attractor for the dynamics.}. As time progresses, the matter density
redshifts as $a^{-3}$ and the position of the minimum (and hence field) moves to increasing field values. Now, the Einstein frame
Lagrangian (assuming a conserved matter density) is
\begin{equation}\label{eq:lagmass}
 \frac{\mathcal{L}}{\sqrt{-g}}\supset TA(\phi)
\end{equation}
and so the fermion masses ($T\supset m^0_{\rm f}\bar{\Psi}\Psi$ where $\Psi$ are Dirac spinors) are $m_{\rm f}=m^0_{\rm
f}A(\phi)$. This means that if the field starts off at some initial value $\phi_i$ and rolls over some distance in field space
$\Delta\phi$ before reaching the minimum then the fermion masses change by an amount
\begin{equation}\label{eq:fermassfield}
 \frac{\Delta m_{\rm f}}{m_{\rm f}}\approx\beta\frac{\Delta \phi}{\mpl},
\end{equation}
where we have used the fact that $\beta(\phi)$ does not vary too much in unscreened regions (recall the
discussion in section \ref{sec:screen_mech}). Now there are stringent constraints on the variation of fermion masses during big
bang nucleosynthesis (BBN) stating that they cannot vary by more than 10\% or so and so in order to comply with this constraint
the field must reach its minimum before the onset of BBN. 
In practice, this means that there is a small region near the minimum where the field can lie where it will reach the minimum
before the onset of
BBN. Exactly how this is achieved is irrelevant for this thesis and we will always assume the field has reached its minimum
before BBN. The interested reader can find details in \cite{Brax:2004qh}; see
\cite{Erickcek:2013oma,Erickcek:2013dea} for a discussion of quantum problems associated with this. 

An important question to ask is: can these models account for dark energy? This is tantamount to asking if the equation of state
can become $-1$ at late times. The field is coupled to matter and so finding the equation of state is not as simple as taking
$T_\phi^{00}/T_\phi^{ii}$ as in the case of quintessence. In order to find a physically meaningful expression for $w$ --- i.e.
one that has the correct interpretation for an observer whose cosmological measurements are predicated on minimally coupled
theories --- we will write down the Friedmann system and attempt to manipulate it into a form that looks like a
quintessence theory. The Friedmann equation is
\begin{equation}
 H^2=\frac{1}{3\mpl^2}\left[\rho_{\rm m} +\frac{1}{2}\dot{\phi}^2+V(\phi)\right],
\end{equation}
where we have temporarily reverted to using the non-conserved matter density $\rho\mmm$ in order to account for every factor of
$\phi$ in the problem. The conserved matter energy density satisfies
\begin{equation}\label{eq:rhoconscos}
 \dot{\rho}+3H\rho=0;\quad \rho\mmm\equiv A(\phi)\rho.
\end{equation}
If we try to match this onto a quintessence-like system where the total energy density in the Friedmann equation is $\rho_{\rm
T}=\rho+\rho_\phi$ then we can identify
\begin{equation}\label{eq:rhophicham}
 \rho_\phi=\frac{1}{2}\dot{\phi}^2+V(\phi)+\left(A(\phi)-1\right)\rho=\frac{1}{2}\dot{\phi}^2+V\eff(\phi).
\end{equation}
The matter is pressureless and so any pressure in the system is due to the field, in which case we have 
\begin{equation}\label{eq:pphi}
 P_\phi=T_\phi^{|ii|}=\frac{1}{2}\dot{\phi}^2-V(\phi).
\end{equation}
The equation of state is then
\begin{equation}\label{eq:wchams}
w_\phi=\frac{P_\phi}{\rho_\phi}= \frac{\dot{\phi}^2-2V(\phi)}{\dot{\phi}^2+V_{\rm eff}(\phi)},
\end{equation}
which differs from the quintessence expression by a factor of $V\eff(\phi)$. Of course, it is not enough to simply define
$w_\phi$ in this way. One must check that the continuity equation for $\rho_\phi$,
\begin{equation}
 \frac{\dd\rho_\phi}{\dd t}=-3H\left(1+w_\phi\right)\rho_\phi,
\end{equation}
and the acceleration equation 
\begin{equation}
 \frac{\ddot{a}}{a}=-\frac{1}{6\mpl^2}\left(\rho+(1+3w_\phi)\rho_\phi\right)
\end{equation}
are consistent. It is not difficult to verify that this is indeed the case \cite{Brax:2011qs}.

\subsubsection{Linear Perturbations}

The growth of linear
perturbations in screened modified gravity has been well studied \cite{Brax:2004qh,Brax:2005ew,Brax:2012gr} and the linear density
contrast $\delta_{\rm c}=\delta\rc/\rc$ in the conformal Newtonian Gauge evolves on sub-horizon scales according to
\begin{equation}\label{eq:pertfull}
 \ddot{\dc}+2H\dot{\dc}-\frac{3}{2}\Omega_{\rm c}(a)H^2\left(1+\frac{2\beta^2(\varphi)}{1+\frac{m_{\rm
eff}^2a^2}{k^2}}\right)\dc=0.
\end{equation}
This is the equivalent of equation (\ref{eq:CDMlinear2}) in general relativity. Comparing with (\ref{eq:cdmlinearGR}), the last
term in (\ref{eq:pertfull}) can be interpreted as a scale-dependent enhancement of Newton's constant:
\begin{equation}\label{eq:effG}
 \frac{G_{\rm eff}(k)}{G}=1+\frac{2\beta(\varphi)^2}{1+\frac{a^2m_{\rm eff}^2}{k^2}}.
\end{equation}
On large scales the screening is effective and $G_{\rm eff}\approx G$ whilst on smaller scales the full enhancement, $G_{\rm
eff}=G(1+2\beta(\varphi)^2)$ is felt. One would therefore expect that there is some wave number $\tilde{k}$ below which the modes
feel no significant fifth-forces and the general relativity power spectrum is
recovered. We will return to this equation in chapter \ref{chap:seven}.

\section{The Vainshtein Mechanism}

As alluded to in the introduction to this chapter, the Vainshtein mechanism screens in a very different manner to chameleon-like
models and here we present a brief discussion of its main features and how it differs from conformal scalar-tensor screening using
Galileon models as an example. 

\subsection{Galileon Theories of Gravity}

Unlike conformal scalar-tensor theories. The action for Galileon gravity is defined at the linearised level. A full
non-linear completion does exist \cite{Deffayet:2009wt} --- either in the form of Horndeski gravity \cite{Horndeski:1974wa} or
ghost-free massive gravity \cite{Babichev:2009us} --- but it will be sufficient to study the linearised theory here. First,
discovered as the 0-helicity mode of DGP gravity, the general model is that of a scalar field $\pi$ whose action is invariant (up
to a boundary term) under the Galileon shift symmetry $\partial_\mu\pi\rightarrow\partial_\mu\pi+c_\mu$ where $c_\mu$ is a
constant 4-vector. In particular, this does not forbid a coupling to matter of the form
\begin{equation}
 \mathcal{L}/\sqrt{-g}\supset\alpha_{\rm V}\frac{\pi}{\mpl} T,
\end{equation}
where $\alpha_{\rm V}$ is a dimensionless coupling constant\footnote{Note
that another coupling to matter of the form $\gamma T^{\mu\nu}\partial_\mu\pi\partial_\nu\pi$ is also allowed. Neglecting it does
not alter the main features of the screening mechanism and we will not consider it here.}. These theories then fall into the
class of modified gravity models defined by equation (\ref{eq:EFgencoup}) so the fifth force is then
\begin{equation}
 \vec{F}_\pi=-\frac{\alpha_{\rm V}}{\mpl}\nabla\pi.
\end{equation}
In four dimensions there are
five possible terms one can write down that satisfy this symmetry independently\footnote{One is a term linear in $\pi$ and this is
often neglected in the literature.}. Since we only wish to discuss the main features of the screening mechanism it will suffice to
study the \textit{cubic Galileon} only. The linearised action is
\begin{equation}\label{eq:galileon}
 S=\int\dd^4 x -\frac{1}{4}h^{\mu\nu}\left(\mathcal{E}h\right)_{\mu\nu} -
\frac{1}{2}\partial_\mu\phi\partial^\mu\phi-\frac{c_3}{2}\partial_\mu\phi\partial^\mu\phi\Box\phi + \alpha_{\rm
V}\frac{\phi}{\mpl}T,
\end{equation}
where $g_{\mu\nu}=\eta_{\mu\nu}+h_{\mu\nu}$, $\left(\mathcal{E}h\right)_{\mu\nu}$ is the Lichnerowicz operator and all
contractions are made with the Minkowski metric. The equation of motion for a static non-relativistic source of density $\rho(r)$
and mass $M$ is
\begin{equation}\label{eq:galeom}
 \frac{1}{r^2}\frac{\dd}{\dd
r}\left(r^3\left[\left(\frac{\pi^\prime}{r}\right)+2c_3\left(\frac{\pi^\prime}{r}\right)^2\right]\right)=\alpha_{\rm
V}\frac{\rho}{\mpl},
\end{equation}
where a prime denotes a radial derivative and we have assumed spherical symmetry. We are interested in the field-profile outside
the source and there are two limits to consider. At large enough distances from the source (exactly how large we will quantify
presently) the first term dominates and the solution is
\begin{equation}
 \frac{\dd\pi}{\dd r}\approx \alpha_{\rm V}\frac{M}{4\pi \mpl r^2}
\end{equation}
so that the fifth-force is 
\begin{equation}
 F_\pi=2\alpha_{\rm V} F_{\rm N}
\end{equation}
and so this limit corresponds to an unscreened fifth-force that is a factor of $2\alpha_{\rm V}$ larger than the Newtonian force.
The opposite limit, valid when the second term dominates yields the solution
\begin{equation}
 \frac{\dd \pi}{\dd r}= \left(\frac{\alpha_{\rm
V}}{c_3}\right)^{\frac{1}{2}}\left(\frac{M}{8\pi \mpl}\right)^{\frac{1}{2}}r^{-\frac{1}{2}}.
\end{equation}
In this case the fifth-force is
\begin{equation}\label{eq:forceratrltrv}
 \frac{F_\pi}{F_{\rm N}}= 2\alpha_{\rm V}^2\left(\frac{r}{r_{\rm V}}\right)^{\frac{3}{2}},
\end{equation}
where the \textit{Vainshtein radius} is
\begin{equation}\label{eq:Vainshteinrad}
 r_{\rm V}=\left(\frac{c_3\alpha_{\rm V}M}{2\pi\mpl}\right)^{\frac{1}{3}}.
\end{equation}
Equating the two terms in equation (\ref{eq:galeom}), we see that this is precisely the radius dividing the two regimes derived
above. One can then see that inside the Vainshtein radius where equation (\ref{eq:forceratrltrv}) is valid, the fifth-force is
screened relative to the Newtonian force by a factor of $(r/r_{\rm V})^{3/2}$. Writing the Vainshtein radius in terms of the
Schwarzchild radius $r_{\rm Sch}\equiv M/4\pi\mpl^2$ we have $r_{\rm V}=(\alpha_{\rm V}L^2r_{\rm Sch})^{1/3}$, with
$L^2=2\mpl c_3$. The strongest constraints on $L^2$ currently come from lunar ranging in the Earth-Moon system\footnote{See
\cite{Nordtvedt:2003pj} for a review of lunar ranging.}. \cite{Dvali:2007kt} and \cite{Afshordi:2008rd} have used data from
\cite{Murphy:2012rea} to constrain 
\begin{equation}\label{eq:lconst}
 L\gsim 150 \alpha_{\rm V}^{-\frac{3}{2}}\textrm{ Mpc}.
\end{equation}
We are interested in theories where $\alpha_{\rm V}\sim\mathcal{O}(1)$ and so taking this to be the case, the Vainshtein radius
for a solar mass object is\footnote{The reader may be confused that the Vainshtein radius appears to be universal for any object
of a given mass. This is because we have ignored the fact that any object has a finite extent and instead integrated equation
(\ref{eq:galeom}) to an arbitrary radius outside the object ignoring the sudden reduction in the density. Accounting for the
finite-extent of objects does induce changes in the Galileon field profile \cite{Hiramatsu:2012xj,Andrews:2013qva}, however these
are small and there is nothing to be gained here by including them.}
\begin{equation}\label{eq:rvsun}
 r_{\rm V}^{\odot}\gsim\mathcal{O}(10^3)\textrm{ pc}.
\end{equation}
The radius of the solar system is of order $10^{-4}$ pc and so the presence of the sun alone is enough to screen any smaller
objects within the entire solar system and beyond. Unlike the chameleon models described above, astrophysical objects necessarily
have Vainshtein radii that vastly exceed their extent and the orbital radii of their satellites. This makes potential
astrophysical tests scarce\footnote{Although see \cite{Hui:2009kc,Hui:2012jb,Hui:2012qt} for a discussion of potential
astrophysical
effects.} and it is for this reason that we describe this mechanism as more efficient than chameleon-like screening.
\thispagestyle{empty}
\clearpage

\newpage
\thispagestyle{empty}
\newpage
\clearpage

\newpage

\thispagestyle{empty}


\begin{flushright}
{\Huge{ \bf Part I:}\\
Astrophysical Tests of Modified Gravity}\\
\end{flushright}
\vspace{6cm} 
\begin{quote}
Because, sir, upon the strength of the strong nuclear interaction rests the
rate at which hydrogen fuses to helium in the core of the Sun. If the interaction strengthens even unnoticeably, the rate of
hydrogen fusion in the Sun
will increase markedly. The Sun maintains the balance between radiation and
gravitation with great delicacy and to upset that balance in favor of radiation,
as we are now doing---'' ``Yes?'' ``---will cause an enormous explosion. Under
our laws of nature, it is impossible for a star as small as the Sun to become a
supernova. Under the altered laws, it may not be.

\qauthor{Isaac Asimov, \textit{The Gods Themselves} }
\end{quote}
 \vspace*{\fill}

\newpage
\thispagestyle{empty}

\newpage

\begin{savequote}[30pc]
My candle burns at both ends;\newline
It will not last the night;\newline
But ah, my foes, and oh, my friends---\newline
It gives a lovely light. 
\qauthor{Edna St. Vincent Millay}
\end{savequote}

\chapter{Equilibrium Stellar Structure in Modified Gravity}
\label{chap:three}

In the previous chapter we examined the general properties of screened modified gravity and discussed the criteria that
determines whether an object is screened or not. Armed with this knowledge, the purpose of the next three chapters is to search
for novel astrophysical effects of these theories and to use them to place new constraints as well as discuss future tests
using upcoming experiments. In section \ref{sec:currentconstraints} we argued that objects with $\pn\lsim\mathcal{O}(10^{-6})$ may
be unscreened and suggested main- and post-main-sequence stars as potential candidates. In this chapter we will lay down the
theoretical groundwork for calculating how stars behave in modified theories of gravity. Specialising to chameleon-like models, we
will then discuss potential observational signatures and present a numerical tool for predicting the structure and evolution of
stars for arbitrary values of $\chi_0$ and $\alpha$ that is accurate enough to compare with observational data.

Before proceeding to perform any technical calculations, the new physics arising from stellar structure in modified gravity can
be discerned from simple physical considerations. Stars are spheres of gas that reach an equilibrium configuration with a constant
radius by burning nuclear fuel in their cores to create a pressure gradient that combats the inward gravitational force and
prevents collapse. Now consider a star in modified gravity compared with an identical star in general relativity\footnote{We will
see later that the notion of identical stars in different theories is somewhat subjective. It depends on which quantities one
wishes to keep fixed and in some cases cannot be defined at all. Nonetheless, the analogy presented here is still apt and
captures the main idea underpinning stellar structure tests of modified gravity.}. This star feels a stronger gravitational force
in
its outer layers and hence requires a larger pressure gradient to combat the extra inward force. It hence needs to burn more fuel
in its core per unit time in order to provide this gradient. One would therefore expect two new features: first, since a star of
fixed mass has a finite supply of fuel this will be depleted faster and the star will have a shorter life-time on any
evolutionary phase. Second, the increased burning rate will release more energy per unit time and so we expect that individual
stars will be brighter.

First, we will present the equations of motion governing hydrodynamics in
modified theories of gravity. This will apply equally to conformal scalar-tensor theories such as chameleon models and those that
screen using the Vainshtein mechanism like galileons. The static equations describe the equilibrium structure of stars and
perturbations about this configuration describe their oscillations. At the level of the equilibrium structure, theories that
screen via the Vainshtein mechanism are too efficient to show novel effects in any environment and so we will subsequently
specialise to chameleon-like theories and derive the modified equations of equilibrium stellar structure. At the level of linear
perturbations, it is possible that oscillating stars may
source scalar radiation in Vainshtein screened theories and so we will return to the general case when discussing the linear
perturbations in chapter \ref{chap:five}. 

Next, we derive the modified gravity analogue of the Lane-Emden stellar structure models presented in chapter
\ref{chap:one}. In this chapter we will only focus on main-sequence stars since their structure can be solved analytically
using Lane-Emden models. We investigate the new properties of these stars in modified gravity, predict the magnitude of the
luminosity enhancement and lifetime reduction and discuss possible observational tests of these predictions. We do this by using a
simple model of main-sequence stars, the Eddington Standard model, which is an $n=3$ Lane-Emden polytrope. Using this, we
calculate the
luminosity enhancement and lifetime as a function of stellar mass and $\chi_0$ for $f(R)$ gravity ($\alpha=1/3$, see section
\ref{subsec:f(R)}). Individual main-sequence stars cannot be resolved outside the Milky Way and we have already argued in
section \ref{sec:currentconstraints} that only dwarf galaxies in cosmic voids are unscreened. Hence, any test related
to
these predictions must rely on their contribution to the galactic properties such as the spectra, the luminosity and the colour.
In particular, since the constituent stars are brighter we expect that dwarf galaxies in voids will be brighter than those in
clusters and therefore looking for systematic offsets between a sample of galaxies in voids and clusters is a potential
observational probe of these predictions. This, in theory, should be possible using the screening maps of \cite{Cabre:2012tq}
although to date no such analysis has been performed and so here we present only the theory. We estimate the galactic luminosity
enhancement by integrating the stellar luminosity found using the Eddington standard model weighted with the initial mass
function (IMF). Both the
Eddington standard model and this simple calculation have several drawbacks and are not accurate enough to compare with real data
and so a more realistic numerical simulation is needed. In the final part of this chapter we present the details and results of
the implementation of the modified gravity equations into the publicly available stellar structure code MESA
\cite{Paxton:2010ji,Paxton:2013pj}. This is powerful enough to allow a comparison with data and although this modified gravity
version has not yet been used to predict the modified galactic properties\footnote{This is an ambitious project and is currently
in progress. Any results are still a long way off.} it will be used in subsequent chapters to provide models that have been
compared with data to place the strongest constraints to date.

\section{Modified Gravity Hydrodynamics}
\label{sec:MGhydro}

Hydrodynamics is the study of the motion and thermal properties of bulk fluids described by collective quantities such as the
pressure, density, temperature etc. under the motion of external forces. In stellar structure described by general relativity the
main contribution to the dynamics comes from the effect of the inward gravitational force, although other effects such as
convection, semi-convection, winds and thermohaline circulation may be important in certain post-main-sequence stars. When one
moves to a modified theory of gravity, the nature of the gravitational force is altered leaving all other properties such as
nuclear reaction rates, the opacity and convection unaltered\footnote{In fact, these can differ between identical screened and
unscreened stars but this is a second order effect. The radial profiles of these quantities depends on the solution of the
stellar structure equations which is different between the two stars but the equations governing their physics are identical in
both cases. It is also the case that changing the theory of gravity can result in changes due to quantum effects such as a shift
in the atomic energy levels. These changes are at most comparable with the changes due
to general relativity --- which are known to be negligible --- and hence do not affect the stellar properties.}. This is the
essence of stellar structure tests of modified gravity: Only the gravitational physics is altered, the
other sectors are unchanged. Therefore, the only equation to be altered is the fundamental force-law and no others. 

We have already seen that chameleon-like theories screen according to the Newtonian potential of the object and we have also
argued in section \ref{sec:currentconstraints} that since $\chi_0\lsim 10^{-6}$ we only expect to see novel effects in objects
with Newtonian potentials less than $\mathcal{O}(10^{-6})$. Neutron stars have potentials of order $10^{-1}$ and are hence
screened in these theories (although see \cite{Brax:2013uh} for a novel effect where the star may become slightly unscreened over
Hubble times). The status of these stars in Vainshtein screened theories is not certain, although given that they screen
incredibly efficiently in high densities it is likely that any effects are negligible. Hence, we will deal only with
non-relativistic stars where the general relativity corrections are of order $\pn\ll1$ and will solve the Newtonian force law for
the motion of the fluid elements and not the full Tolman-Oppenheimer-Volkov equation. 

We will describe bulk quantities such as the pressure and density in the
\textit{Eulerian picture}, where these quantities are to be considered as fields
which give the value of said quantities at any point in space as a function
of time. In contrast, we will describe the position of individual fluid
elements (and, when needed, the pressure perturbation) in the
\textit{Lagrangian picture}, where the motion of individual fluid elements are
followed as a function of time. In this case, the Lagrangian position $\vec{r}$
of a fluid element satisfies the (non-relativistic) momentum equation:
\begin{equation}\label{eq:momentumgen}
 \frac{\partial^2 \vec{r}}{\partial t^2} = -\frac{1}{\rho}\nabla P + \vec{F},
\end{equation}
where $P(\vec{r})$ is the pressure, $\rho(\ver)$ is the density and $\vec{F}$
is the external force density. In modified gravity, the fluid moves under its own Newtonian
gravity and the fifth-force
(\ref{eq:fifth-forcegeneral}) due to the scalar field so that
\begin{equation}\label{eq:momentumMG}
 \frac{\partial^2 \vec{r}}{\partial t^2} = -\frac{1}{\rho}\nabla P -
\frac{GM(r)}{r^3}\ver-\frac{\beta(\phi)}{\mpl}\nabla\phi,
\end{equation}
where $\beta(\phi)=\mpl\dd\ln A/\dd \phi$ for chameleon-like theories and $\beta(\phi)=\alpha_{\rm V}$ for Vainshtein screened
theories. This is the only hydrodynamical equation that is altered relative to general relativity; changing gravity only
changes the motion of the fluid elements and does not directly alter other processes such
as mass conservation, energy generation and radiative transfer. The quantity $M(r)$
($r\equiv|\ver|$) is the mass enclosed inside a radius $r$ from the centre, and
is given via the Poisson equation
\begin{equation}\label{eq:poisson}
 \nabla^2\Phi_{\rm N}=4\pi G\rho(r),
\end{equation}
which may be integrated once to give 
\begin{equation}\label{eq:poissint}
 \frac{\dd \pn}{\dd r}=\frac{GM(r)}{r^2}.
\end{equation} 
Since mass is a locally-conserved quantity we also have the continuity
equation:
\begin{equation}\label{eq:cont}
 \frac{\partial \rho}{\partial t}+\vec{\nabla}\cdot(\rho\vec{v})=0,
\end{equation}
where $\vec{v}\equiv\dd \ver/\dd t$ is the velocity of the fluid element. In
general, one must also consider the energy generation and radiative transfer
equations but these are only important if one wishes to study the effects
of perturbations coupled to stellar atmospheres, which is irrelevant in the
context of modified gravity. We will include their effects when describing the
equilibrium stellar configuration in order to produce the correct equilibrium stellar properties, however, we will not include
them in our perturbation
analysis. Since they do not depend on the theory of gravity, they are identical to equations (\ref{eq:radiative1}) and
(\ref{eq:engen1}) at the background level. At the level of perturbations, we will work in the so-called \textit{adiabatic}
approximation, where the density and pressure evolve according to
\begin{equation}\label{eq:dpdrho}
 \frac{\dd P}{\dd t} = \frac{\Gamma_1 P}{\rho}\frac{\dd \rho}{\dd t}.
\end{equation}
The quantity
\begin{equation}\label{eq:Gamma1def}
 \Gamma_1 \equiv \left(\frac{\dd \ln P}{\dd \ln \rho}\right)_{\rm adiabatic}
\end{equation}
is the \textit{first adiabatic index}. It is of
paramount importance to the study of stellar pulsation and stability and we
will return to discuss it later on in chapter \ref{chap:five}.

It is important to note that the first adiabatic index is not simply the equation of state relating $P_0(r)$ to $\rho_0(r)$ and in
particular is not equal to the adiabatic index described in equation (\ref{eq:polytropic1}) for polytropic gasses, although in
very simple cases the two may be equal. It describes the response of the pressure to \textit{adiabatic} compressions whereas the
adiabatic index is simply a relation between the density and the pressure and quite often relies on different assumptions. In
particular, whilst it is true that simple gasses have equations of state $P\propto\rho^\gamma$ with $\gamma=4/3$ ($5/3$) for
relativistic (non-relativistic) gasses it is not necessarily the case that $\Gamma_1=4/3$ indicates a relativistic system. In
fact, it means that the energy from adiabatic compressions are not raising the density like a simple non-relativistic gas. One
example of this is the progenitor of a type II supernova where the gas is non-relativistic but $\Gamma_1$ drops to $4/3$ because
the pressure does not increase upon compression, instead, the rate of photo-disintegration of iron is increased. Another example
is the He$^+$ ionisation region of Cepheid stars where pressure does not change upon a compression but the ionisation fraction is
increased.

\subsection{Equilibrium Structure}\label{sec:eqstruc}

The equilibrium stellar configuration
is both static and spherically symmetric and can be found by setting
time-derivatives to zero and $\ver= r$ in the hydrodynamic equations so that
this is now the Eulerian radial coordinate. We will denote all equilibrium
quantities with a subscript-zero except for $M(r)$, which is defined at the
background level only. It is important to note that $\chi_0$ is not a property
of the star but is the cosmological value of $\chi_{\rm c}$ found by evaluating
(\ref{eq:screenradius}) using the cosmological values of $\phi$ and $\beta$. In
what follows, $\phi_0(r)$ is the equilibrium field-profile throughout the star
and not the cosmological value. With no time-dependence, (\ref{eq:cont}) is
trivially satisfied and $\rho(r,t)=\rho(r)$. This simple form of the
density profile allows us to find the mass enclosed in any given radius:
\begin{equation}\label{eq:masscons}
 \frac{\dd M(r)}{\dd r}=4\pi r^2\rho_0(r).
\end{equation}
The momentum equation (\ref{eq:momentumMG}) then reduces to the \textit{modified
hydrostatic equilibrium equation} 
\begin{equation}\label{eq:MGHSE}
 \frac{\dd P_0(r)}{\dd r} =
-\frac{GM(r)\rho_0(r)}{r^2}-\frac{\beta(\phi_0)\rho_0(r)}{\mpl}\frac{\dd
\phi_0(r)}{\dd r}.
\end{equation}
Physically, this equation describes the pressure profile the star must assume
in order for the star to support itself against gravitational collapse. The
second term is the fifth-force due to the scalar field; stars in modified
gravity need to provide larger pressure gradients in order to combat this extra
inward component \cite{Chang:2010xh,Davis:2011qf,Jain:2012tn}. These equations
are then supplemented by the radiative transfer equation
\begin{equation}\label{eq:radtrans}
 \frac{\dd T_0(r)}{\dd r} = -\frac{3}{4 a} \frac{\kappa(r)}{T_0^3}
\frac{\rho_0(r) L_0(r)}{4\pi r^2},
\end{equation}
which describes how the temperature $T(r)$ varies due to the flux of energy with
luminosity $L(r)$ away from regions of energy generation governed by
\begin{equation}\label{eq:engen}
 \frac{\dd L_0}{\dd r} = 4\pi r^2 \rho_0(r) \epsilon(r),
\end{equation}
where $\epsilon(r)$ is the energy generation rate per unit mass. As mentioned above, these are identical to the
equations coming from general relativity and are repeated here for completeness.

\section{Lane-Emden Stars}\label{eq:LEStarsMG}

Lane-Emden stars are perfect tools for studying the effects of modified gravity without the complications of non-gravitational
physics. They are spheres of gas that have collapsed under their own gravity to reach a static equilibrium configuration, which is
set by the interplay of the pressure and gravitational physics alone. They are not realistic enough to compare with real stellar
data but all of the essential new physics is made transparent through their study and this serves as a direction for potential
observational probes using more complete numerical models. In this section we will present a general framework for calculating
their properties in screened modified gravity.

From here on we will specify to the case of chameleon-like theories and assume the fifth-force profile (\ref{eq:gprof}).
Since we are only concerned with the equilibrium structure in this chapter, we will drop the subscript zeros. We will restore them
in chapter \ref{chap:five} where we discuss perturbations about this equilibrium. In this
case, the modified hydrostatic equilibrium equation (\ref{eq:MGHSE}) becomes 
\begin{equation}\frac{\dd P}{\dd r} = -\frac{GM(r)\rho (r)}{r^2}  \left[1 + \alpha\left[1-\frac{M(r_{\rm
s})}{M(r)}\right]H(r-r_{\rm s})\right].  \label{eq:modhydrostatic}
\end{equation}
Since this is the only equation that is modified we can follow the same derivation in section \ref{subsec:legr} to
arrive at the equivalent of the Lane-Emden equation. Before doing this however, an examination of equation
(\ref{eq:modhydrostatic}) reveals that the equations of stellar structure are no longer self-similar in modified gravity; there
is a second length scale $\rs$ appearing and a second mass scale $M(\rs)$ and so continuing blindly will result in solutions that
cannot have every dimensionful quantity scaled out. This means the new solution cannot be directly compared with the general
relativity results. If we
wish to make a meaningful comparison we must make an approximation that preserves self-similarity and so we will make the
approximation that $G\rightarrow G(1+\alpha)$ in the region exterior to the screening radius, which is equivalent to ignoring the
factor of
$M(\rs)/M(r)$ in equation (\ref{eq:modhydrostatic}). This means that in any stellar model we compute over-estimates the effects
of modified gravity, however this is not as bad an approximation as it may first seem; the precise stellar structure is not
observable and most observable properties are those defined at the surface where this approximation is strongest.

Making this approximation and following the same steps as section \ref{subsec:legr} we arrive at the modified
Lane-Emden equation:
\begin{equation}\label{eq:MGLE}
\frac{1}{y^2} \frac{ \dd}{\dd y}\left[ y^2 \frac{\dd \thn}{\dd
y}\right]=-\left\{
\begin{array}{l l}
 (1+\alpha) \thn^{n},&    y>\ys,   \\
  \thn^{n},& y<y_{\rm s},  
 \end{array}\right. ,
\end{equation}
where $\ys$ is the Lane-Emden screening radius such that $\rs=r_{\rm c}\ys$. The boundary conditions are identical to the general
relativity case, namely $\theta(0)=1$ and $\theta^\prime(0)=0$ where a prime denotes a derivative with respect to the Lane-Emden
coordinate $y$, whose definition (as well as the definition (\ref{eq:rc}) of $r_{\rm c}$) is (\ref{eq:LEcoorddef}) unchanged in
modified gravity. For convenience, we introduce the quantities
\begin{align}
 \omr &\equiv -y_R^2\left.\frac{\dd \thn}{\dd y}\right\vert_{y=y_R}\quad
\textrm{and} \\
\oms & \equiv -y_{\rm s}^2\left.\frac{\dd \thn}{\dd y}\right\vert_{y=y_{\rm s}},
\end{align}
which are the modified gravity analogues of equation (\ref{eq:omegaGR}). We have already seen in subsection
(\ref{subsec:legr}) that in general relativity ($\alpha=\chi_0=0$, $\omr=\oms$) there is a unique solution for any given value of
$n$. We will denote the general relativity values of $y_R$ and $\omr$ by $\bar{y}_R$ and
$\bar{\omega}_R$ respectively. In modified gravity, there is a two-dimensional space of
solutions at fixed $n$ given by specific values of $\chi_0$ and $\alpha$, each
with different values of $\omr$, $\oms$, $y_{\rm s}$ and $y_R$. This is a consequence of the fact that, despite our
approximation, the modified Lane-Emden equation is still not self-similar. Self-similarity is weakly broken by the screening
radius, which still appears as a second length scale in the problem. Despite this, a meaningful comparison with general
relativity can still be made provided one is very careful about exactly which quantities are fixed in both cases and is clear
about what is actually being compared. We will return to this point when discussing specific solutions. Using these definitions,
we can find the mass of the star:
\begin{equation}\label{eq:LEMASS}
 M = \int_0^R4\pi r^2\rho(r)\dd r = 4\pi r_{\rm c}^3\rho_{\rm
c}\left[\frac{\omr+\alpha\oms}{1+\alpha}\right],
\end{equation}
where we have used the modified Lane-Emden equation. This differs from the general relativity expression (\ref{eq:LEmass}) in
that it does not simply depend on $\omr$ and the central density. Instead, it depends on the screening radius (and hence $\chi_0$)
and $\alpha$ as well. Hence, Lane-Emden models with fixed mass, $\chi_0$ and $\alpha$ correspond to stars with different central
densities
to those of the same mass in general relativity.

Equation (\ref{eq:chiint}) gives the screening radius once the density profile and $\chi_0$ is specified. By writing the density
in the Lane-Emden variable $\theta$ and using the modified Lane-Emden equation (\ref{eq:MGLE}), we can calculate the integral
exactly to find an implicit relation for $\oms$ (and hence the screening radius) in terms of $\chi_0$, $M$ and $R$:
\begin{equation}\label{eq:X}
 \frac{\chi_0}{GM/R}\equiv X = \left[\frac{y_R\thn(y_{\rm
s})+\omr-\frac{y_R}{y_{\rm s}}\oms}{\omr+\alpha\oms}\right],
\end{equation}
where we have used equation (\ref{eq:LEMASS}) to replace the factors of $r_{\rm c}$ and $\rho_{\rm c}$ with $M$. One can see how
it is the ratio of $\chi_0$ to $\pn=GM/R$ that determines the screening radius and not simply the density. As
$\rs\rightarrow0$\footnote{Strictly speaking, we should impose $\rs/R\rightarrow 0$ since $\rs$ is a dimensionful quantity.}, the
star becomes increasingly
unscreened, $\oms\rightarrow0$ and $\thn(y_{\rm s})\rightarrow1$. This gives the
maximum value of $X$ where the star is partially screened. For values greater
than this, equation (\ref{eq:X}) has no solutions and the star is always
unscreened. From (\ref{eq:X}), we have
\begin{equation}
 X_{\rm max} = \frac{y_R+\omr}{\omr}
\end{equation}
independent of $\alpha$. Later on, we will specify to the cases $n=1.5$ and $n=3$ and so for
future reference we note here that $X_{\rm max} \approx 2.346$ for $n=1.5$ and $X_{\rm max} \approx 4.417$ for $n=3$.

\section{Main-Sequence Stars}

In this section we will use $n=3$ polytropes to describe the structure of main-sequence stars in modified gravity. We have
already alluded to the fact that the structure itself is not observable. What is observable is the galactic luminosity and, as
discussed above, we expect that there will be systematic offset between the luminosity of dwarf galaxies in cosmic voids and
those in clusters. The first step towards calculating this is to find the luminosity enhancement of an individual unscreened star
compared with its screened counterpart as a function of the stellar mass and $\chi_0$. We will restrict to $f(R)$ theories by
setting $\alpha=1/3$ from here on, however the method can be applied to any other value without the need for any modifications.

\subsection{Scaling Relations}
\label{sec:scalaing}
Before proceeding to solve the stellar structure equations, we can gain a lot of physical intuition into the behaviour of
unscreened stars by using simple scaling relations\footnote{See \cite{Adams:2008ad} for a nice discussion on how scaling relations
can be used to provide a similar intuition for how stars behave when other physics is altered.}. We have already seen that the
stellar structure equations are
self-similar in general relativity and that a fully unscreened object can be described by $G\rightarrow G(1+\alpha)$ and hence the
same is true for any unscreened star. Self-similarity implies that we can replace each quantity e.g. pressure, density, mass etc.
by some characteristic quantity multiplying a dimensionless function and use the equations to derive relationships between these
quantities. For example, we can replace the pressure by setting $P=P_{\rm c}x_P(r)$ where $P_{\rm c}$ is the central pressure. As
an
example, making these replacements in the hydrostatic equilibrium equation (\ref{eq:hsegr}) leads to the relation
\begin{equation}\label{eq:scalingex}
 P_{\rm c}\frac{\dd x_P}{\dd x_r}=-\frac{GM^2x_Mx_\rho}{R^4 x_r^2},
\end{equation}
where we have set $\rho=M/R^3x_\rho$. Now since all the functions or derivatives of the $x_i$ functions are dimensionless this
gives us the scaling relation
\begin{equation}\label{eq:scalingex2}
 P_{\rm c}\propto \frac{M^2}{R^4}.
\end{equation}
Here we ignored the factor of $G$ because it is constant but what happens if we allow it to vary by a constant factor i.e. we
change our star from one described by general relativity to an unscreened star in a theory of modified gravity with coupling
$\alpha$? In this case we obtain a new scaling relation
\begin{equation}\label{eq:scalingex3}
 P_{\rm c}\propto \frac{GM^2}{R^4}.
\end{equation}
We know that the equations of stellar structure do not close and so one must specify an equation of state relating the pressure
to the density and temperature. Main-sequence stars are predominantly supported by two sources of pressure. Gas pressure is the
thermodynamic pressure associated with the motion of the individual atoms, which we take to be well-described by the ideal gas law
\begin{equation}\label{eq:gaspress}
 P_{\rm gas}=\frac{k_{\rm B}\rho T}{\mu m_{\rm H}},
\end{equation}
where $m_{\rm H}$ is the mass of a hydrogen atom and $\mu$ is the mean molecular weight, which represents the average number of
particles (nucleons and electrons) per unit hydrogen mass. Gas pressure is the dominant contribution to the pressure in low mass
stars. Radiation pressure
is the pressure due to the absorption of photons propagating through the gas and is given by integrating the Planck distribution
for radiation in thermal equilibrium:
\begin{equation}\label{eq:Prad}
 P_{\rm rad}=\frac{1}{3}aT^4.
\end{equation}
Radiation pressure is the dominant contribution to the pressure in high mass stars. Using these two expressions to find additional
scaling relations, one can use these in conjunction with the scaling relations found from the hydrostatic equilibrium and
radiative transfer equations to find the mass-luminosity scaling relation
\begin{equation}\label{eq:MLscaling}
 L\propto \left\{
  \begin{array}{l l}
   G^4M^3 & \quad \text{gas pressure}\\
    GM & \quad \text{radiation pressure}\\
  \end{array} \right. .
\end{equation}
We can then find the luminosity enhancement for an unscreened star of fixed mass relative to a screened one:
\begin{equation}\label{eq:Lehancescreen}
 \frac{L_{\rm unscreened}}{L_{\rm screened}} = \left\{
  \begin{array}{l l}
   (1+\alpha_0)^4 & \quad \text{low mass stars}\\
    1+\alpha_0 & \quad \text{high mass stars}\\
  \end{array} \right. .
\end{equation}
One can see that low mass stars receive a large luminosity enhancement when they are unscreened (recall $(1+\alpha)=4/3$) whereas
higher mass stars receive a not so large (but still $\mathcal{O}(1)$) enhancement. Whereas more
radiation is being released in high mass stars per unit time, more is being absorbed to provide the pressure and hence cannot
escape to the surface and so the stellar luminosity is not as affected. We therefore expect the luminosity enhancement as a
function of stellar mass will decrease with increasing
mass and, in particular, we expect a sharp turn-off when the radiation pressure comes to dominate. We will see precisely this
below.

\subsection{The Eddington Standard Model}\label{sec:eddsm}

The Eddington standard model is a simple set of assumptions about main-sequence stars that allows us to compute their properties
by solving one simple modified Lane-Emden equation rather than an entire system of coupled equations. It makes the simplifying
assumptions that the opacity is constant and due mainly to electron scattering $\kappa(T,\rho)\equiv\kappa_{\rm es}$\footnote{This
is a good approximation for most main-sequence stars except the lowest mass stars, which require Kramer's opacity law
$\kappa\propto\rho T^{-3.5}$.} and that the radiation entropy $S_{\rm rad} = 4a T^3/3\rho$ is constant throughout the star. This
allows us to simplify the equations considerably since the temperature is now a function of density only, and hence the pressure
is barotropic. We assume that the pressure is a combination of the gas and radiation pressure only and so we have 
\begin{equation}\label{eq:ESMpress}
 P=\frac{1}{3}aT^4+\frac{k_{\rm B}\rho T}{\mu m_{\rm H}}
\end{equation}
and we introduce the ratio of the gas pressure to the total pressure\footnote{This quantity is often denoted by $\beta$ in the
astrophysics literature. Since $\beta$ has already been used to denote the coupling we shall use $b$ here.}
\begin{equation}\label{eq:bdef}
 b\equiv\frac{P_{\rm gas}}{P}
\end{equation}
so that $P_{\rm rad}=(1-b)P$. Equating $bP_{\rm gas}$ with $(1-b)P_{\rm rad}$ we find
\begin{equation}\label{eq:trhoesm}
 T^3=\left(\frac{3k_{\rm B}}{a\mu m_{\rm H}}\right)\frac{1-b}{b}\rho.
\end{equation}
Now in theory $b$ could be a function of the radius, however the Eddington standard models' assumption implies that $T^3/\rho$ is
constant and hence so is $b$. In this case equation (\ref{eq:ESMpress}) can be written
\begin{equation}\label{eq:esmpoly}
 P=K(b)\rho^{4/3}
\end{equation}
where the constant
\begin{equation}\label{eq:esmkdef}
 K(b)=\left[ \frac{3}{a}\left(\frac{k_{\rm B}}{\mu m_{\rm H}}\right)^4 \frac{(1-b)}{b^{4}}\right]^{1/3}.
\end{equation}
We can hence see that the Eddington standard model is a $\gamma=4/3$ or $n=3$ polytrope and we can solve for its structure in
modified gravity by solving the modified Lane-Emden equation (\ref{eq:MGLE}) to find the structure of the star then using
the radiative transfer equation (\ref{eq:radtrans}) to calculate the luminosity.

We begin by calculating the surface luminosity. Differentiating equation (\ref{eq:Prad}) and substituting the radiative transfer
equation (\ref{eq:radtrans}) we have
\begin{equation}\label{eq:dpraddr}
 \frac{\dd P_{\rm rad}}{\dd r}=-\frac{\kappa_{\rm es}\rho L}{4\pi r^2}.
\end{equation}
Setting $P_{\rm rad}=(1-b)P$ in the modified hydrostatic equilibrium equation for $r>\rs$ and retaining the factor of
$M(\rs)/M(r)$ so that
we do not over-estimate the luminosity too much by making the approximation $G\rightarrow G(1+\alpha)$ we have
\begin{equation}\label{eq:intpraddr}
  \frac{\dd P_{\rm rad}}{\dd r}= -\frac{(1-b)\alpha_{\rm eff}GM\rho}{r^2};\quad \alpha_{\rm
eff}\equiv\alpha\left[1-\frac{M(\rs)}{M(r)}\right]
\end{equation}
from which we find the surface luminosity by equating the two expressions at the surface:
\begin{equation}\label{eq:Lesm}
 L=\frac{4\pi(1-b)\alpha_{\rm surf}GM}{\kappa_{\rm es}};\quad \alpha_{\rm surf}\equiv\alpha_{\rm
eff}(R)=\alpha\left[1-\frac{M(\rs)}{M}\right].
\end{equation}
We will see shortly that whereas $b$ is constant throughout a given star, its value depends on $\chi_0$ and $\alpha$ and hence
there are two sources of luminosity enhancement in main-sequence stars. First, the factor of $\alpha_{\rm surf}$ gives an
enhancement due to the increased rate of nuclear burning. Second, the ratio of gas to radiation pressure changes. Physically, the
factor of $1-b$ is present because the luminosity is due only to radiation and not the gas. Next, we must find the constant $b$,
which at this stage is still unknown. This can be found by using the polytropic relation (\ref{eq:esmpoly}) at $r=0$ in the
definition of $r_{\rm c}$ (\ref{eq:rc}) to find
\begin{equation}\label{eq:eddquarint1}
 r_{\rm c}^3=\left(\frac{1}{\pi G}\right)\frac{K(b)^{\frac{3}{2}}}{\rho_{\rm c}},
\end{equation}
which can be used in the formula for the mass (\ref{eq:LEMASS}) to find 
\begin{equation}\label{eq:esmmass}
 4\pi K(b)^{\frac{3}{2}} \left(\frac{1}{\pi G}\right)^{\frac{3}{2}} \left[ \frac{\omega_{R} + \alpha \omega_{\rm
s}}{1+\alpha_0}\right].
\end{equation}
Now we would like to make contact with the general relativity case in order to compare with modified gravity stars and so we note
that this corresponds to the fully screened case so that $\rs=R$. In this case we have $\omr=\bar{\omega}_R\approx
2.02$. In the other extreme case where the star is fully unscreened ($r_{\rm s}=0$) $\omega_{\rm s}=0$ and by rescaling
$\xi\rightarrow(1+\alpha)^{-1/2}$ to bring the modified Lane-Emden equation into the standard form (\ref{eq:Lane-EmdenGR}) we have
$\omega_{R} \approx 2.02 (1+\alpha)^{-1/2}$. It is convenient to define a function $\alpha_b$ that interpolates between these
two cases: 
\begin{equation}
1+\alpha_b = \left((1+\alpha)\frac{\bar{\omega}_{R}}{\omega_{R}+\alpha_0\omega_{\rm s}}\right)^{2/3}, \label{eq:alphab}
\end{equation}
such that when $r_{\rm s}=R$, $\omega_{R}=\omega_{\rm s} = \bar{\omega}_{R}$ we have $\alpha_{b}=0$ and when $r_{\rm
s}=0$ the
star is fully unscreened and $\alpha_{b}=\alpha$. We can then invert equation (\ref{eq:esmmass}) using equation
(\ref{eq:esmkdef}) to find
\begin{equation}\label{eq:eddquart}
 \frac{1-b}{b^4} = (1+\alpha_{b})^3 \left(\frac{M}{M_{\rm edd}}\right)^2,
\end{equation}
where the Eddington mass is
\begin{equation}
M_{\rm edd} = \frac{4 \pi^{1/2}}{G^{3/2}\bar{\omega}_R} \left(\frac{3}{a}\right)^{1/2} \left(\frac{k_{\rm B}}{\mu m_{\rm
H}}\right)^{2}, 
\end{equation}
which is defined in terms of fundamental constants and the general relativity Lane-Emden solution and is hence independent of
$\alpha$. Numerically, $M_{\rm edd} = 18.3 \mu^{-2} M_\odot$. In general relativity, equation (\ref{eq:eddquart}) is known as
\textit{Eddington's quartic equation} and here we will refer to it by the same name. 

We now have a system of equations that close. The only technical difficulty is that the screening radius is defined implicitly
via equation (\ref{eq:X}), which requires us to know the structure of the star. Now the structure of the star can only be
computed once we know the screening radius and so one must use an iterative procedure. This requires the mass-radius
relation (\ref{eq:M-Rms}) $R\propto M^{0.8}$\footnote{Technically, $n=3$ polytropes do not have a mass radius relation and the
relation we take here assumes an extra component not included in the Eddington standard model, namely nuclear burning. Without
the assumption of such a relation it is not possible to compare stars of fixed mass in general relativity and modified gravity or
even between stars with two different values of $\chi_0$ and $\alpha$. The choice of a mass-radius relation is tantamount
to deciding to
compare the properties of two stars of fixed mass and radius for a given value of $\chi_0$ and $\alpha$.}. Given a star of fixed
mass $M$ we can calculate $GM/R$. We then fix the value of $\chi_0$ and $\alpha$ and assume some test screening radius. This
allows us to solve the modified Lane-Emden equation numerically and extract the values of $\omr$ and $\oms$. We then use these in
equation (\ref{eq:X}) to calculate $X$. Next, we iterate through different screening radii until $X=\chi_0/(GM/R)$, in which case
we have
found the correct screening radius and so the solution of the modified Lane-Emden equation is the correct one. Using $\omr$ and
$\oms$, we can calculate $\alpha_b$ and $\alpha_{\rm surf}$ using
\begin{equation}
\frac{\alpha_{\rm surf}}{\alpha}  = 1- \frac{M(r_{\rm s})}{M} = \left[\frac{\omega_{R}-\omega_{\rm s}}{\omega_{R}+
\alpha}\omega_{\rm s}\right],
\end{equation}
which can be used in Eddington's quartic equation with the value $\mu=0.5$, which assumes only fully ionised hydrogen and is
appropriate for main-sequence stars, to calculate $b$. Once we have this, equation we can use equation (\ref{eq:Lesm}) to
calculate the luminosity of the star relative to the same star in general relativity
\begin{equation}\label{eq:lench}
 \frac{L}{L_{\rm GR}}=\frac{1-b(\alpha,\chi_0)}{1-b(\alpha=0,\chi_0=0)}\alpha_{\rm surf}.
\end{equation}
By calculating this for stars of different masses and different $\chi_0$ (recall we are fixing $\alpha=1/3$) we can numerically
fit the enhancement as a function of these two parameters. We have already seen that the star will be fully unscreened when
$X=4.417$ and so it is convenient to use the rescaled quantity $X_3 \equiv \chi_0/(4.417 GM/R)$ such that $0 \leq X_3 \leq
1$. The relation between $\alpha_b$ and $\alpha$ can then be recast as 
\begin{equation}\alpha_b(1+\alpha_b) =
\alpha(1+\alpha)f(X_3;\alpha)
\end{equation}
and 
\begin{equation}
\frac{\alpha_{\rm surf}}{\alpha} = g(X_3;\alpha),
\end{equation}
where the fitting functions $f$ and $g$ take values between $0$ and $1$. Numerically, we find
\begin{align} \label{eq:fits}
f(X_3;\alpha=1/3) &=  X_3^2(1.94+0.79X_3 -2.91X_3^2+1.18X_3^3), \nonumber \\
g(X_3;\alpha=1/3) &= \sqrt{-\frac{13}{14} +\sqrt{\frac{169}{196}+\frac{20 X_3}{7}}}. 
\end{align}

\begin{figure}[ht] 
   \centering
      \includegraphics[width=0.9\textwidth]{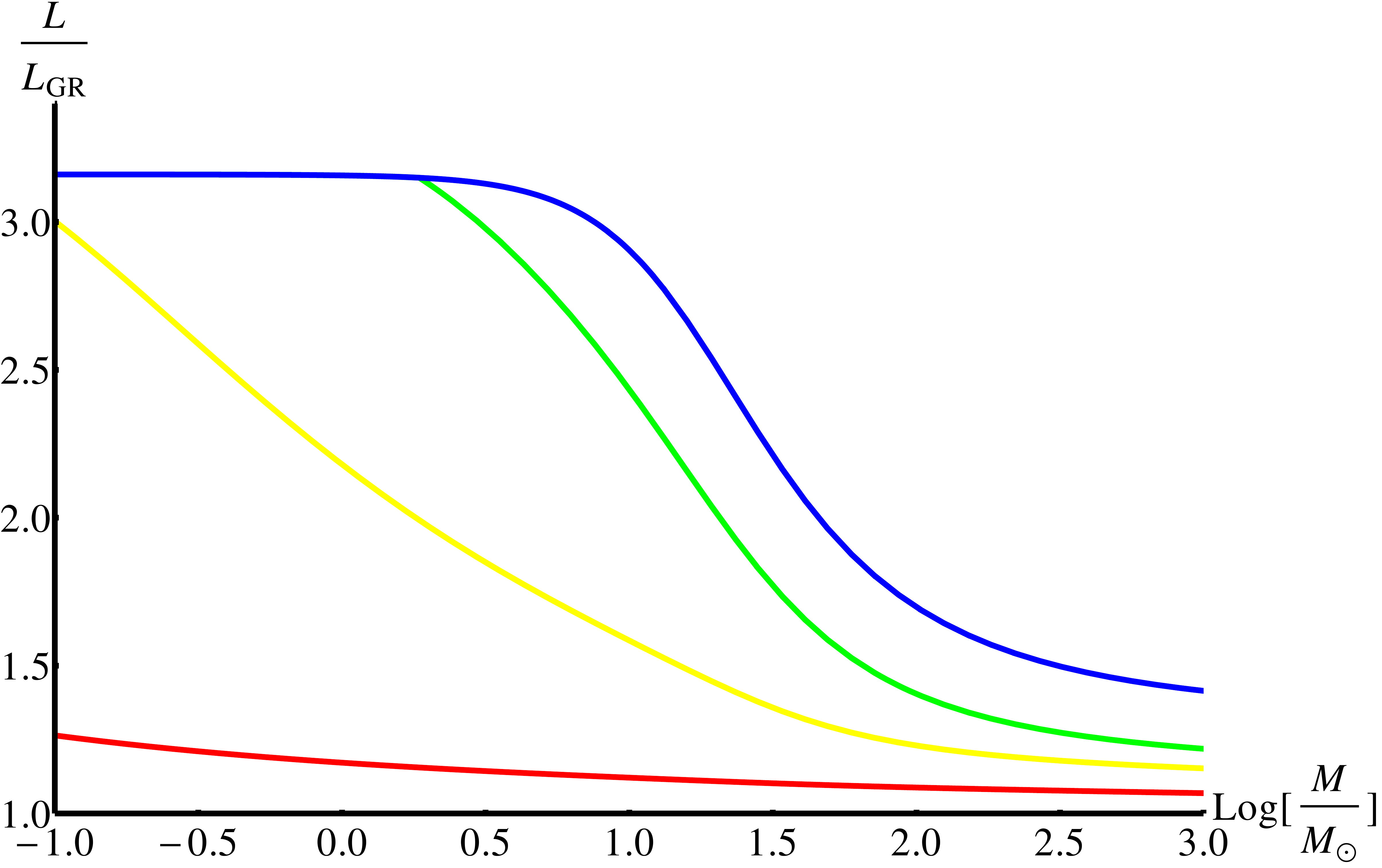}
         \caption{The ratio of the stellar luminosity of a partially screened star in $f(R)$ chameleon gravity compared with one
that is completely screened as a function of stellar mass $M$. $L_{\rm std}$ is the luminosity of a completely screened (general
relativity) star. From top to bottom: $\chi_0 = (10^{-4},10^{-5},5\times 10^{-6},10^{-6})$.}
          \label{fig:lumplot}
       \end{figure}

We plot the luminosity enhancement of a star for a given value of $\chi_0$ and $\alpha=1/3$ in figure \ref{fig:lumplot} for
$\chi_0=1\times10^{-6},\,5\times10^{-6},\,1\times10^{-5}$ and $1\times10^{-4}$. Several features that we have already discussed
are immediately obvious. Firstly, the ratio is always greater than unity; stars are indeed brighter in modified gravity than in
general relativity as we argued at the start of this chapter. We also remarked that the Sun has $\pn\sim2\times10^{-6}$ and that
according to equation (\ref{eq:pnstarsrelation}) lower mass stars have lower Newtonian potentials and so are more unscreened
whilst the converse is true for higher mass stars. This is evident in the figure where one can see that the curves with
$\chi_0\sim \mathcal{O}(10^{-6})$ are more unscreened (show higher luminosities) at lower masses. The other two curves with
$\chi_0\ge10^{-5}$ show a sharp turn-off of the luminosity enhancement at low masses. This is the effect we predicted in section
(\ref{sec:scalaing}) where we argued that high mass stars are dominated by radiation pressure, which shows a dramatically lower
luminosity enhancement. 
This is the reason for the sharp turn-off: these stars are
radiation dominated whereas their general relativity doppelg\"{a}ngers are gas dominated. Increasing $\chi_0$ at fixed $\alpha$
results in stronger pressure
gradients, larger temperatures and a higher surface luminosity and so its effects are degenerate with increasing the stellar
mass. For this reason, unscreened stars are more radiation supported than their screened counterparts and the reduction in the
luminosity enhancement begins at smaller stellar masses when $\chi_0$ is increased. At low stellar masses, these upper
curves flatten out and saturate at a value $(4/3)^4\approx3.16$ which is precisely the value we predicted in section
(\ref{sec:scalaing}) for a fully unscreened, gas pressure-supported star relative to the same star in general relativity. Using
the mass-radius relation, the Newtonian potential scales as $\pn\propto M^{0.2}$ and so decreasing the stellar mass also decreases
the Newtonian potential making the star more unscreened. When $X>4.417$, equation (\ref{eq:X}) has no solutions and the star is
fully unscreened. Since a fully unscreened star cannot become even more unscreened, lowering the mass (and
Newtonian potential) beyond the point where $X=4.417$ has no further effect and hence the curves flatten off. Finally, we note
that when $\chi_0\sim10^{-6}$ there
is little enhancement in the luminosity. This reflects the fact that these stars all have $\pn\sim\chi_0$ and are heavily
screened. Nevertheless, it is important to note that they are not completely screened as the simple aphorism $\pn=\chi_0$ implies
screening would suggest. This is because the self-screening condition was derived assuming a fixed density sourcing a field
profile whereas this formalism uses an iterative procedure to account for the back reaction of the field on the density. We can
see that when a star is treated as a dynamical object that adjusts its equilibrium configuration to match the effects of the
fifth-force the object is not as unscreened as one might have n\"{a}ively expected. This means that there are still astrophysical
signatures of modified gravity coming from main-sequence stars even at $\chi_0=10^{-6}$.

\subsection{Effects on the Galactic Properties}

So far, we have yet to discuss the reduced stellar lifetime we predicted qualitatively at the start of the chapter. Since the
stellar lifetime is related to the rate at which fuel is consumed in the core
we cannot predict it using the Eddington standard model alone. Indeed, this is an $n=3$ Lane-Emden star, which is a completely
static configuration. In order to include time-dependence, one must use rate equations to calculate the energy generation and use
equation (\ref{eq:engen}). Since we will present a complete numerical model below that includes both time-dependence
and does not use any of the other approximations we have made we will not attempt this here but will instead make a simple
estimate. The main-sequence lifetime can be well approximated by
\begin{equation}\label{eq:ms_life}
\tau_{\rm MS} \approx 10\left(\frac{M}{M_{\odot}}\right)\left(\frac{L_{\odot}}{L}\right)\textrm{ Gyr}
\end{equation}
and using equation (\ref{eq:MLscaling}) we have
\begin{equation}
 \frac{L}{L_\odot}=(1+\alpha)^4\left(\frac{M}{M_\odot}\right)^3
\end{equation}
for low-mass stars and so we can see that a fully unscreened solar-mass star in $f(R)$ gravity will go off
the main-sequence more than three times as quickly than the same star whose structure is governed by general relativity. This has
important observational consequences. Unscreened galaxies have hosted more generations of stars than
screened galaxies of the same mass and age and so we expect them to be more metal enriched and appear older. 


A quantitative prediction of the properties of unscreened dwarf galaxies is not possible using this simple model since it does not
keep track of the chemical evolution of the star. What is possible is a simple estimate of the enhancement of the galactic
luminosity at fixed age and mass which we present below in order to demonstrate that this enhancement can be significant and
merits further investigation.

The number of stars born with mass $M$ within a galaxy is given by the initial mass function $\Phi(M)=\dd N/\dd M$, i.e. the
number of stars $dN$ within the mass range $dM$. This relation is empirically found to be roughly universal \footnote{In
principle, since the physics of gravitational collapse is expected to be
encoded in the IMF, modified gravity may change its functional form. We do not attempt to investigate this here since the
IMF is empirical and there is no successful analytic or numerical derivation of this function from
known principles. Furthermore, it has been found to be universal in a number of very different
environments and so we expect it to be
largely robust.}. For simplicity we use the Salpeter IMF \cite{Salpeter:1955it}, $\Phi(M)\propto M^{-2.35}$ with
$0.08M_\odot\le M\le 100 M_\odot$. We can then
estimate the luminosity increase for an unscreened dwarf galaxy by integrating the luminosity using the fitting
functions (\ref{eq:fits}) over the IMF. 

Before doing so however, we must account for the stars that have gone off the main-sequence as the IMF only gives the number of
stars born. We do this by making use of equation (\ref{eq:ms_life}). If the age of the galaxy is $\tau_{\rm age}$ then we assume
that
all stars with $\tau_{\rm MS}>\tau_{\rm age}$ contribute in their entirety to the luminosity whereas stars whose main-sequence
lifetimes are less than the age of the galaxy contribute a fraction $\tau_{\rm MS}/\tau_{\rm age}$ of their luminosity. We note
that stars that have gone off the main-sequence still have a luminosity enhancement in the red giant phase (and beyond), however
we do not account for their contribution here due to the lack of an analytic model. In fact, these stars are brighter by several
orders of magnitude than their main-sequence counterparts and so this is a conservative estimate. This effect is accounted for in
by including a factor $f_0(M;\tau_{\rm age})$ where
\begin{equation}\label{eq:fffactor}
f_0(M;\tau_{\rm age}) = \left\{
  \begin{array}{l l}
   1 & \quad \tau_{\rm MS}>\tau_{\rm age}\\
   \frac{\tau_{\rm MS}}{\tau_{\rm age}} & \quad \tau_{\rm MS}<\tau_{\rm age}\\
  \end{array} \right.
\end{equation}
so that the galactic luminosity is 
\begin{align}\label{eq:galum}
L_{\mathrm{gal}}(\tau_{\mathrm{age}},&\chi_0) = \nonumber\\ &\int_{0.08{M_{\odot}}}^{100{M_{\odot}}} dM f_0(M,\tau_{\rm
{age}})L(M;\chi_0)\frac{\mathrm{d}N}{\mathrm{d}M}.
\end{align}
We can then immediately see that stars whose main-sequence lifetime are shorter than the age of the galaxy do not contribute to
the luminosity enhancement since the factor of $L(M)^{-1}$ in the main-sequence lifetime exactly cancels the factor in the
integral. Normalising the integrals so that the total luminous mass of screened and unscreened galaxies are identical (this is
required in order to account for the fact that more stars have gone off the main-sequence in the unscreened case), one can
perform the integral (\ref{eq:galum}) to calculate the ratio of the luminosity enhancement of an unscreened dwarf galaxy compared
to a screened one. As an example, in table \ref{tab:luminc} we show this ratio for a dwarf galaxy of mass $M_{\rm
gal}=$10$^{10}M_\odot$ and age
$\tau_{\rm age}=$13 Gyr for $\chi_0$ in the range 10$^{-6}$--10$^{-4}$ in $f(R)$ gravity.
\begin{table}[hb]
\centering
\heavyrulewidth=.08em
	\lightrulewidth=.05em
	\cmidrulewidth=.03em
	\belowrulesep=.65ex
	\belowbottomsep=0pt
	\aboverulesep=.4ex
	\abovetopsep=0pt
	\cmidrulesep=\doublerulesep
	\cmidrulekern=.5em
	\defaultaddspace=.5em
	\renewcommand{\arraystretch}{1.6}
\begin{tabu}{c|[2pt]c}
$\chi_0$ & Luminosity Enhancement \\
\tabucline[2pt]{-}
$1$x10$^{-4}$ & 42\% \\
$1$x10$^{-5}$ & 42\% \\
$5$x10$^{-6}$ & 29\% \\
$1$x10$^{-6}$ & 3\% \\
\end{tabu}
\caption{The luminosity enhancement in unscreened relative to screened dwarf galaxies as a function of $\chi_0$. All values were
computed using the fitting formulae (\ref{eq:fits}) and equation (\ref{eq:fffactor}) in (\ref{eq:galum}) taking $\alpha = 1/3$.}
\label{tab:luminc}
\end{table}
Table \ref{tab:luminc} reveals that the enhancement is significant for $\chi_0\gsim$10$^{-6}$, which, as argued above, we would
expect. The saturation around $\chi_0\sim$10$^{-5}$ is due to the effect of the decreased main-sequence lifetime. If $\chi_0$ is
around this value then, as seen in figure \ref{fig:lumplot}, the low mass stars are all completely unscreened and so increasing
its value further cannot make them any more luminous. Hence, the luminosity enhancement saturates around this value. As noted
in subsection \ref{sec:currentconstraints}, red giant stars are still unscreened if $\chi_0\lsim$10$^{-6}$ and so even at these
low values of $\chi_0$ it is possible that there will be a significant effect, for example, galaxies that have partially
unscreened
post-main-sequence stars and screened main-sequence stars will produce different spectra from those where all the stars are
partially screened. This is beyond the scope of our present model since we cannot calculate the enhancement in post-main-sequence
stars analytically. 

In practice, one cannot observe two identical galaxies --- one screened and the other not --- and
simply measure the difference in their luminosity. We would, however, expect there to be systematic differences between galaxies
located in clusters and those in voids. By looking for these systematic differences it is possible that independent constraints on
$\chi_0$ in the range 10$^{-4}$--10$^{-6}$ (or possibly lower) can be found.

\section{Numerical Modelling of Stars in Modified Gravity}\label{sec:mesa}

So far, we have presented a simple model of main-sequence stars in screened modified gravity and have used it to make simple
estimates of the luminosity enhancement of both individual stars and dwarf galaxies compared with their general relativity
counterparts. Along the way we were very careful to point out any shortcomings or approximations --- and indeed there were many
--- and
highlighted the fact that these models are not powerful enough to compare with observational data. For this, a more
accurate treatment is required. There are computer programs that include all of the non-gravitational stellar physics neglected
in our previous model that are powerful enough to be compared with current data and we have modified the publicly available
code MESA \cite{Paxton:2010ji,Paxton:2013pj} to include the effects of modified gravity. Below, we briefly describe the
implementation of modified gravity into this code and present some of the resultant predictions. A more detailed account of the
implementation is given in appendix \ref{app:nto}.

MESA \cite{Paxton:2010ji,Paxton:2013pj} is capable of solving the complete system of stellar structure equations
coupled to the radiative transfer system, stellar atmosphere models, nuclear
burning networks, convective motion and micro-physical processes such as
opacity and electron degeneracy. It also includes effects such as
overshooting, mass-loss and rotation in a fully consistent manner. Given some
initial mass, it generates a pre-main-sequence stellar model and dynamically
evolves it through the main-sequence and subsequent post-main-sequence to its
final state, be it a white dwarf, neutron star or core-collapse supernova. 

MESA is a one-dimensional code (in that it assumes spherical
symmetry) that divides the star into a series of variable-length cells, each
with a specific set of quantities such as temperature, density, mass fraction
etc. that may correspond to the values at either the cell centre or
boundary. Exactly which depends on the quantity in question, however it is
always possible to interpolate between the two. We implement the effects of
modified gravity by updating these assignments to include a cell-centred value
of $G$, which differs from the Newtonian value in the region exterior to the
screening radius. This implementation uses a quasi-static approximation where,
given some initial radial profile, the star is evolved to its new equilibrium
structure one time-step later. Using this profile, we integrate equation
(\ref{eq:chiint}) to successively deeper cells until it is satisfied. The
radius of this cell is then designated the screening radius and we then update
the value of $G$ in each cell according to equation (\ref{eq:gprof}) so that
\begin{equation}\label{eq:g(r)}
 G(r)=G\left[1+\alpha\left(1-\frac{M(\rs)}{M(r)}\right)\right]\quad r\ge \rs.
\end{equation}
We then let the model evolve one time-step further to find the modified
structure. This approximation is valid provided that the time-step between
successive models is smaller than the time-scale on which changes to $G(r)$ are
important and MESA provides a facility to ensure this is always
the case. Furthermore, \cite{Chang:2010xh} have modified MESA for
the same purpose of us using a scalar field ansatz and cell interpolation and
we have verified that our results are indistinguishable from theirs.

\begin{figure}
   \centering
      \includegraphics[width=0.9\textwidth]{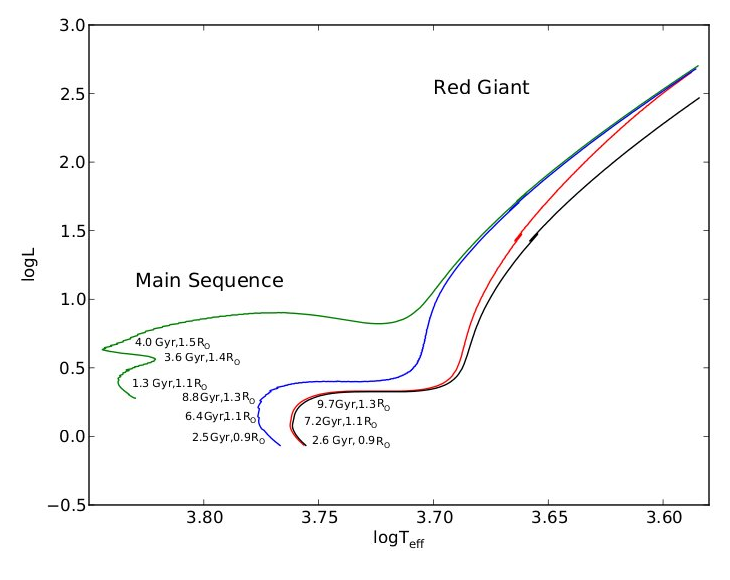}
         \caption{The Hertsprung-Russell diagram for stars of one solar mass with initial metallicity $Z=0.02$. The black line
shows the tracks for stars in general relativity while the red, blue and green tracks correspond to stars in modified gravity with
$\chi_0 = 10^{-7}, 10^{-6}$ and $5\times10^{-6}$ respectively. The radius and age at the point where the central hydrogen mass
fraction has fallen to 0.5, 0.1 and 10$^{-5}$ is shown for each star except the $\chi_0 = 10^{-7}$ case.}
          \label{fig:mesa}
       \end{figure}

In figure \ref{fig:mesa} we show the Hertsprung-Russell (HR) diagram for stars of one solar mass with initial metallicity
$Z=0.02$ (solar metallicity) evolving from the zero-age main-sequence (ZAMS) to the tip of the red giant branch for
$\chi_0=5\times10^{-6},10^{-6},10^{-7}$ and $\chi_0=0$ (general relativity). It is clear from the tracks that
stars that
are less screened are indeed hotter and more luminous. We can also see that for $\chi_0=10^{-7}$ the tracks are identical along
the main-sequence but separate in the red-giant phase corresponding to the Newtonian potential dropping due to the increased
radius and the star becoming less screened. It appears that the red giant tracks all converge to a similar track. This is due to
the fact that the Newtonian potential of these stars is so shallow that they are unscreened to a very high
degree, even at $\chi_0=10^{-7}$. In fact, if one examines the tracks in detail, small differences can be discerned. 
%
One important question to address is: \textit{How do we decide which two points on different tracks correspond to identical
stars?} One should be careful which properties to compare if a meaningful comparison is to
be made. This is often stars with identical temperatures and luminosities since these are the properties that are observed. In
this case, the stars will have different ages, radii and chemical compositions. If instead one wishes to make a more theoretical
comparison and ask which points along the tracks correspond to stars at identical evolutionary stages then a useful quantity is
the core hydrogen mass fraction\footnote{On the main-sequence. Quantities such as the helium mass fraction or core mass are more
appropriate for post-main-sequence stars.}, whose evolution defines the main-sequence. If one compares two stars with identical
core hydrogen mass fractions then the more unscreened star will be younger, brighter and hotter. To show this explicitly, we show
the star's age and radius at three points along the star's main-sequence with identical core hydrogen mass fractions. Notice that
it is clear that unscreened stars do indeed have shorter lives than their screened counterparts. It also shows that the radii at
the same evolutionary stage tend to be smaller as well. Physically, the extra pressure needed to support the star in modified
gravity is produced by increased densities and temperatures over the standard case, demanding a more compact star. We have not
shown the values for $\chi_0 = 10^{-7}$ since these stars are almost entirely screened on the main-sequence and hence have nearly
identical properties to the unmodified stars. Nor have we shown any information about the red giant phase. This is because
the red giant phase is far shorter than the main-sequence and so comparing quantities between stars that are screened to different
extents is misleading. Despite the assumptions of the simple model above, it is clear that the luminosity increase is still
present when more realistic models are used and the missing physics is accounted for. This is a good example of how Lane-Emden
models capture all of the new gravitational physics and that there are no major new features present when the non-gravitational
physics is re-introduced.

\begin{savequote}[30pc]
Astrophysics is fun.
\qauthor{Eugene Lim}
\end{savequote}

\chapter{Distance Indicator Constraints on Modified Gravity}\label{chap:four}

In this chapter we will change focus from main-sequence stars to post-main-sequence stars and how they can be used as probes of
modified gravity. We will again focus on chameleon-like models and will make use of some of the tools presented in chapter
\ref{chap:three}. 

\section{Distance Indicators as Probes of Modified Gravity}

To date, there are two main methods of probing screened modified gravity using dwarf galaxies. The first looks at their
kinematics and morphology and has been presented in \cite{Jain:2011ji,Vikram:2013uba}; currently these constraints are not
competitive with others in the literature. The second uses distance indicators. Suppose one were to set about measuring the
distance to a different galaxy from our own. Clearly we cannot measure this directly and so we must infer its value from other
measurements. This requires us to use some formula for converting these measurements into the distance
but where does this formula come from? There are two possibilities: either it is derived from our current theory or it is
calibrated on objects either in our own galaxy or neighbouring ones in the local group. Either way, there is an implicit
assumption that the physics that governs objects in our own and neighbouring galaxies is the same in every other galaxy in the
universe. Now chameleon-like theories of gravity do not posses this feature. The Milky Way and local group is screened but dwarf
galaxies may not be. In this case, the formula used to infer the distance to these galaxies would not be correct if it depends on
the theory of gravity. As an example, lets consider using the luminosity distance as a distance indicator and let us further
assume that the galaxy we are interested in is near enough that cosmological effects can be neglected. In this case, one has
\begin{equation}\label{eq:lumdistex}
 F=\frac{L}{4\pi d^2},
\end{equation}
where $F$ is the flux and $L$ is the luminosity of some object in the galaxy. Now the flux is a measured quantity, the number we
measure is not sensitive to the theory of gravity but we need some method of inferring the luminosity. One such method is to
use some standard candle where we can theoretically predict the value of $L$ and we find that it is universal for all objects of
the
same type and that we use general relativity to predict its value. Now suppose that the laws governing gravity are different in
this other galaxy to our own such that if we use general relativity we over predict the luminosity. The flux is a measured
quantity that is insensitive to our theory of gravity and so given any predicted value of the luminosity we must adjust the
inferred distance so that equation (\ref{eq:lumdistex}) is satisfied. Hence, we
over-predict the distance compared with its true value. This means that if we compare the distance inferred using this method to
the distance inferred using a different method that is independent of the theory of gravity, the two will only agree if we have
used the correct theory of gravity to infer the first distance. Note also that if we had not predicted the luminosity but
had instead measured it in nearby galaxies the conclusion would be the same and the two distances will only agree if the same law
of gravity holds in the galaxies used to make the calibration. 

The simple example above illustrates the premise behind distance indicator tests of modified gravity. We look for distance
indicators that are sensitive to the theory of gravity that we are interested in and make new predictions for the formula used to
infer the distance using their properties; we refer to these as \textit{unscreened distance indicators}. We then identify galaxies
where we expect this theory of gravity to show strong deviations from general relativity, in this case dwarf galaxies in voids,
and look for those with simultaneous distance measurements using \textit{screened distance indicators}, which do not depend on
the theory of gravity. Now the data generally gives distances to these galaxies and so what one must do is calculate the
difference between distance predicted using the general relativity formula and the new theory. If one then takes the distance
coming from the unscreened indicator and compares it with the measurement from the screened one there will be a discrepancy by a
known amount if the theory of gravity is not general relativity because the distance calculated using the unscreened indicator
has been found using an incorrect formula for that galaxy. The lack of any such deviation --- or more realistically the agreement
within errors --- places new constraints on modified gravity.

This is exactly what we will do in this chapter and we will use three different indicators as probes: water masers, tip of the
red giant branch (TRGB) distance indicators and Cepheid variable stars. We will describe each of these briefly below; a full
account of these distance indicators may be found in \cite{Freedman:2010xv}.

\subsection{Cepheid Variable Stars}

\begin{figure}[t,h]
\centering
\includegraphics[width=10cm]{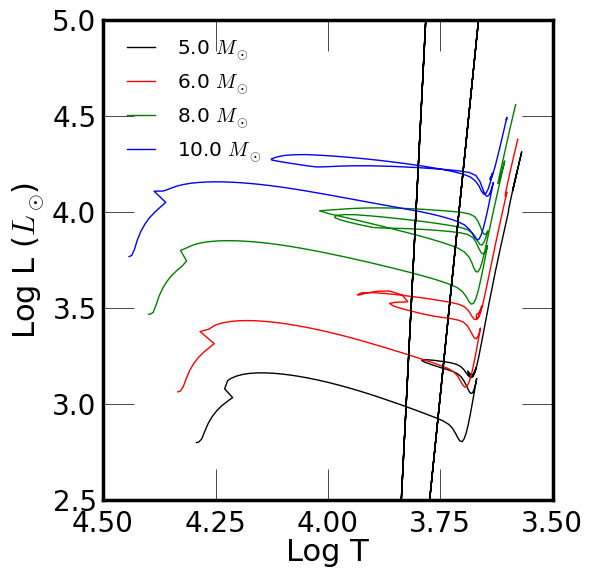}
\caption{The evolution of Cepheid stars in the HR diagram. The instability strip shown in the black lines is taken from
\cite{Alibert:1999an}. } 
\label{fig:T-L-mass}
\end{figure}

Cepheid variable stars are stars of 3-10$M_\odot$ that have gone off the main-sequence and evolved onto the red giant phase.
During this phase these star's tracks in the HR diagram undergo loops in the temperature direction at approximately constant
luminosity. This is shown in figure \ref{fig:T-L-mass} for stars in general relativity; the figure was produced using
MESA. Whilst traversing these loops they cross a region in temperature known as the \textit{instability strip},
which is shown in figure \ref{fig:T-L-mass} as two black diagonal lines. Inside the strip, the star is unstable to
\textit{Cepheid pulsations} and pulsate with a period of $\mathcal{O}(\textrm{days})$. This gives rise to a variation in the
star's
luminosity and there is a known period-luminosity relation calibrated on Cepheid stars in the local group:
 \begin{equation}
 M_V = a\log\tau + b(B-V) + c,
 \label{eq:PLC}
 \end{equation}
where $M_V$ is the V-band magnitude, which is related to the total luminosity, $\tau$ is the period of oscillation and
$(B-V)\propto\log T\eff$, where $T\eff$ is the effective temperature. This is the temperature inferred from the peak of the
Planck spectrum for the light that reaches us from the photosphere of the star. This is the radius at which the optical depth
becomes $2/3$ so that the stellar atmosphere is optically transparent and light can escape. $a,\,b$ and $c$ are constants and
$a\approx-3$ \cite{Freedman:2010xv}. A complete theoretical model of the physics driving period-luminosity relation is still to be
found, however equation (\ref{eq:PLC}) can be understood in a simple manner. Using simple scaling relations (we will show this
explicitly in chapter \ref{chap:five}) one can show that $\tau\propto (G\rho)^{-1/2}$ and so one can re-write equation
(\ref{eq:PLC}) schematically as
\begin{equation}
 \log L\propto \tilde{a}\log R+\tilde{b}\log T\eff+\tilde{c},
\end{equation}
which gives $L\propto R^2T\eff^4$ ($R$ is the stellar radius), which is nothing more than the Steffan-Boltzmann law. 

Physically, the oscillations are the result of acoustic waves resonating within the star. Hydrostatic equilibrium is not exact
and every star has a set of radial and non-radial eigenfrequencies for the propagation of internal waves. When these waves
propagate adiabatically they can neither grow nor decay and are standing waves within the star. When the motion is non-adiabatic,
these modes can be excited and the star can pulsate in one or a combination of these modes. The instability strip arises because
various mechanisms act to drive these pulsations and the motion within the star is non-adiabatic. In the case of Cepheid stars,
the pulsations are driven by the \textit{kappa mechanism}. Stars on the blue loop contain a layer where the temperature is around
$4\times10^{4}$ K. This corresponds to the ionisation potential of helium and so a small compression of the star will raise
the temperature and allow more photons to ionise helium in this layer. Usually, an increase in the temperature decreases the
opacity (recall Kramer's law give $\kappa\propto T^{-3.5}$) but in this case the opacity increases upon compression. This means
that heat is trapped in the layer upon a compression, damming up energy. Since the opacity has increased, more photons are
absorbed and the stellar luminosity decreases. Upon an expansion, the converse is true, ionised helium recombines with free
electrons releasing photons and the temperature decreases. These extra photons increase the luminosity of the star. This is the
driving mechanism behind Cepheid pulsations but we still need to worry about whether the motion is adiabatic or not. If the
motion is adiabatic then the photon number is conserved and the luminosity at the base of the ionisation zone must be equal to
that
at the top and hence the motion is damped. The blue edge of the instability strip corresponds to the motion becoming
non-adiabatic inside the ionisation zone. In theory, the pulsations should occur at all temperatures lower than this, however
convective effects act to damp the oscillations and the red edge corresponds to temperatures where this damping is
non-negligible. In what follows, we will use the instability strip given by \cite{Alibert:1999an}:
\begin{align}
 \log L &= 4.2 - 46 \left(\log T\eff - 3.8\right)\quad\textrm{blue edge},\label{eq:blueedge}\\
\log L& = 4.2 - 23 \left(\log T\eff - 3.7\right)\quad\textrm{red edge}.\label{eq:red edge}
\end{align}
We refer to the first time the star crosses the strip whilst moving on a blue loop as the first crossing and ignore the very
brief crossing during the transition from the main-sequence to red giant phase. This phase is so short lived that only a handful
of observations have been made here.

Cepheid distances have been calibrated using parallaxes for 10 Milky Way Cepheids in the distance range $\approx 0.3-0.6$ kpc,
with periods ranging from $\approx 3-30$ days. The error on the mean distance is $\pm 3\%$ or $0.06$ in magnitude. Outside the
Local Group, where modified gravity effects may be present, Cepheid distances have been measured to over 
50 galaxies. The final uncertainty in the distance modulus, which includes zero point calibration, metallicity, reddening and
other effects, is $\pm 0.09$ magnitude or $5\%$ in distance.

\subsection{Tip of The Red Giant Branch Stars}

Stars of 1.5--2 $M_\odot$ do not exhibit blue loops. Instead they continue to ascend the red giant branch with an ever increasing
luminosity and a decreasing temperature. After the star leaves the main-sequence its core is composed of neutral hydrogen
surrounded by a thin shell where hydrogen burning continues. The temperature in the outer layers is too low to ignite the
hydrogen there and so this is inert. This outer layer absorbs high-energy photons from the shell and expands the star to a
radius 10--100 times that of its main-sequence progenitor, causing a decrease in the effective temperature. At this point, the
entire
stellar luminosity is due entirely to the shell. As more and more hydrogen is converted to helium, the core grows more massive and
begins to contract, raising the temperature and further increasing the rate of hydrogen burning. Once the temperature in the core
reaches $T\sim10^{8}$ K helium may be ignited and will burn to produce carbon and oxygen via the triple alpha process. This moves
the star to the horizontal branch --- where it has a higher temperature and lower luminosity --- over a very small time-scale
leaving a
visible discontinuity. This is particularly pronounced in the I-band (the flux at 800 nm) and the discontinuity occurs at
\begin{equation}
 \label{eq:IbandTRGB}
I=-4.0\pm0.1
\end{equation}
independent of the stellar mass. The relation is robust over a large range of metallicities ($-2.2 < [\textrm{Fe/H}] < -0.7$
dex) and the small error is due only to a weak dependence. This comes about because the core mass is weakly dependent on the
metallicity. This makes the TRGB a standard candle: the luminosity can be inferred directly from the discontinuity since it is
related to the I-band magnitude there and hence a distance can be found by measuring the flux and using equation
(\ref{eq:lumdistex}).

TRGB distances have been measured to approximately $250$ galaxies. These
are applied to old, metal poor populations which enables distance estimates
out to about $20$ Mpc, slightly closer than Cepheid distances since TRGB stars are not as bright as Cepheids. However, since
single epoch photometry is enough to measure the luminosity at the tip, it is much easier to obtain the data required for a TRGB
distance estimate.

\subsection{Water Masers}\label{subsec:masers}

\begin{figure}[t,h]
\centering
\includegraphics[width=10cm]{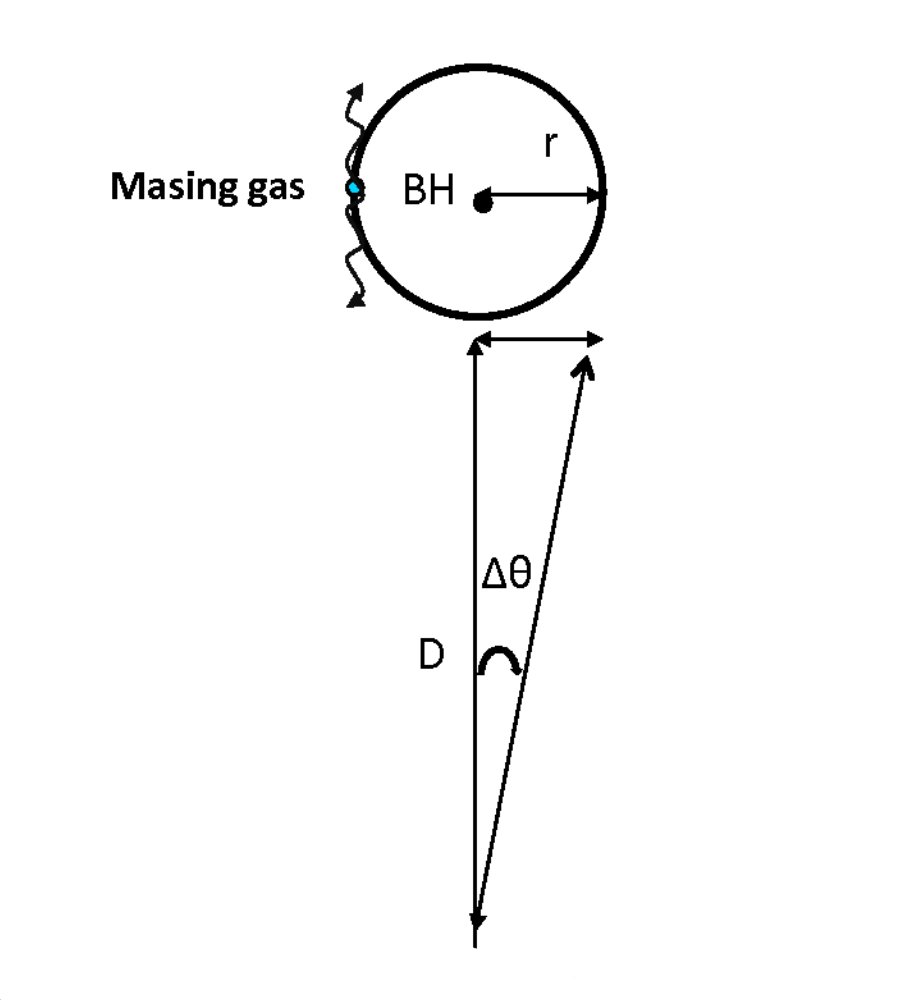}
\caption{Masing gas orbiting the central black hole (labelled BH) at a distance $D$ from an observer. Figure reproduced
from \cite{2011PhDT.......205K}. } 
\label{fig:masers1}
\end{figure}

H$_{\rm 2}$O vapour orbiting in Keplarian motion in the accretion disk at a distance of $\mathcal{O}(0.1\textrm{ pc})$ from the
central black hole of a galaxy has a far smaller number density of molecules than water vapour under terrestrial
conditions. This means that the gas cannot achieve thermal equilibrium and hence population inversion can occur. If the galaxy has
an active galactic nuclei then X-ray emission or shocks can cause a population inversion between the 6$_{\rm 56}$ and 5$_{\rm 23}$
rotational levels leading to microwave emission at $22.2$ GHz ($\lambda=1.35$ cm). Given the Keplarian orbits, a measurement of
the centripetal acceleration, rotational velocity, angle on the sky and the angle of inclination of the orbit can be used to
obtain a geometric distance estimate. Before proceeding to discuss the more technical aspects of the measurement, we will
elucidate the basic idea using a simple example. Figure \ref{fig:masers1} shows a simple illustration of one water maser in
circular orbit around the central black hole. Since the orbit is Keplarian we have
\begin{equation}\label{eq:centrip}
 a=\frac{v^2}{r},
\end{equation}
where $a$ is the centripetal acceleration and $v$ is the orbital velocity. Geometrically, we have
\begin{equation}\label{eq:maserD}
 D=\frac{r}{\Delta\theta}
\end{equation}
and so a measurement of $v$, $a$ and $\Delta\theta$ allows us to infer the geometric distance to the galaxy. In fact, since we
observe velocities and accelerations along the line of sight (LOS) we must correct for the inclination of the orbit. If the
orbit is inclined at an angle $i$ to the plane perpendicular to the line of sight we have $v_{\rm observed}=v\sin i$ and $a_{\rm
observed}=a\sin i$, in which case the distance can be found using
\begin{equation}\label{eq:maserD2}
 D=\frac{v^2}{a\Delta\theta}\sin i,
\end{equation}
where $v$ and $a$ are now the measured quantities. 

\begin{figure}[t,h]
\centering
\includegraphics[width=10cm]{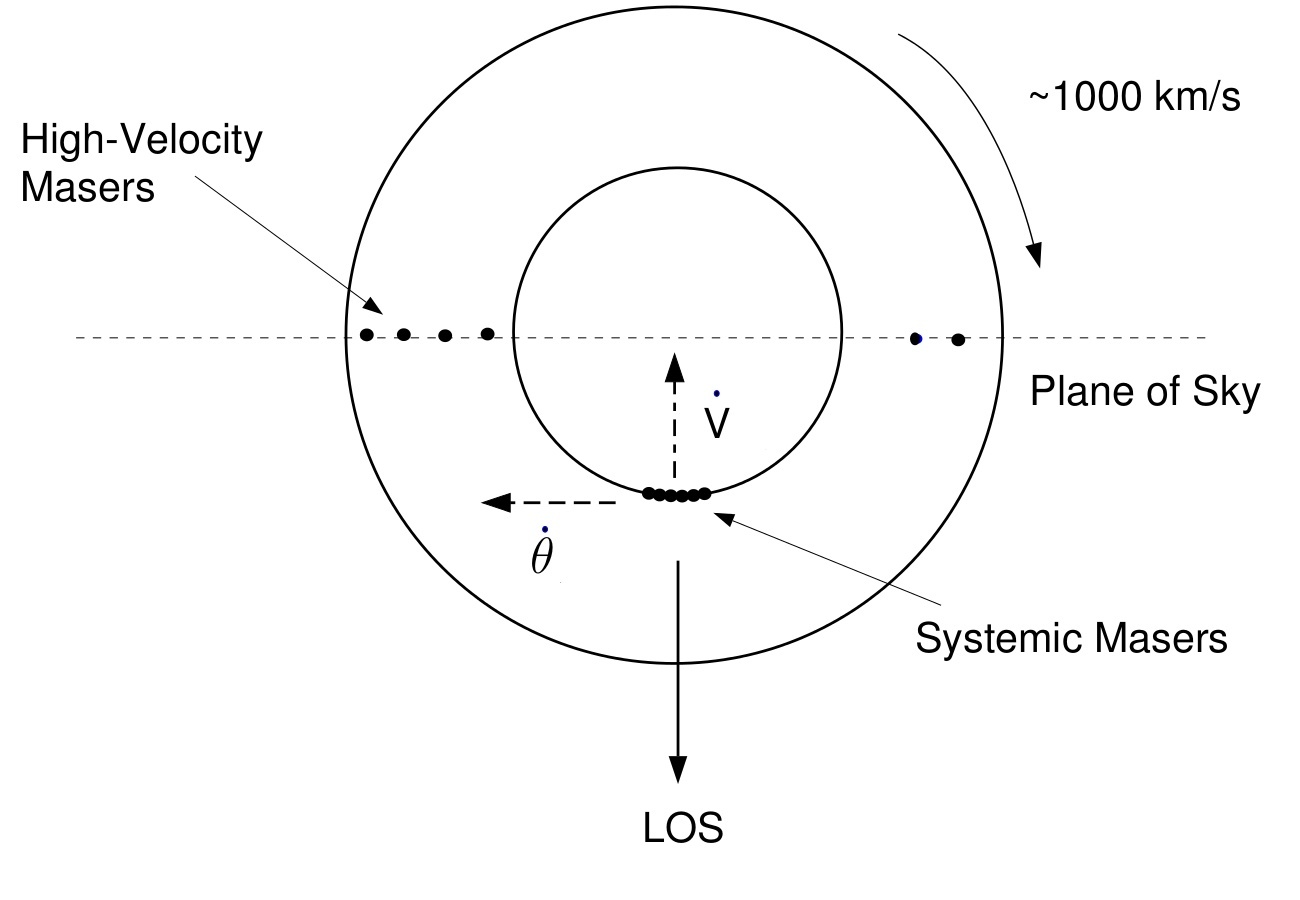}
\caption{An illustration of the accretion disc and water masers in NGC 4258; the black dots show the position of individual
masers. This figure has been adapted from \cite{Herrnstein:1998mh}.} 
\label{fig:maser2}
\end{figure}

\begin{figure}[t,h]
\centering
\includegraphics[width=10cm]{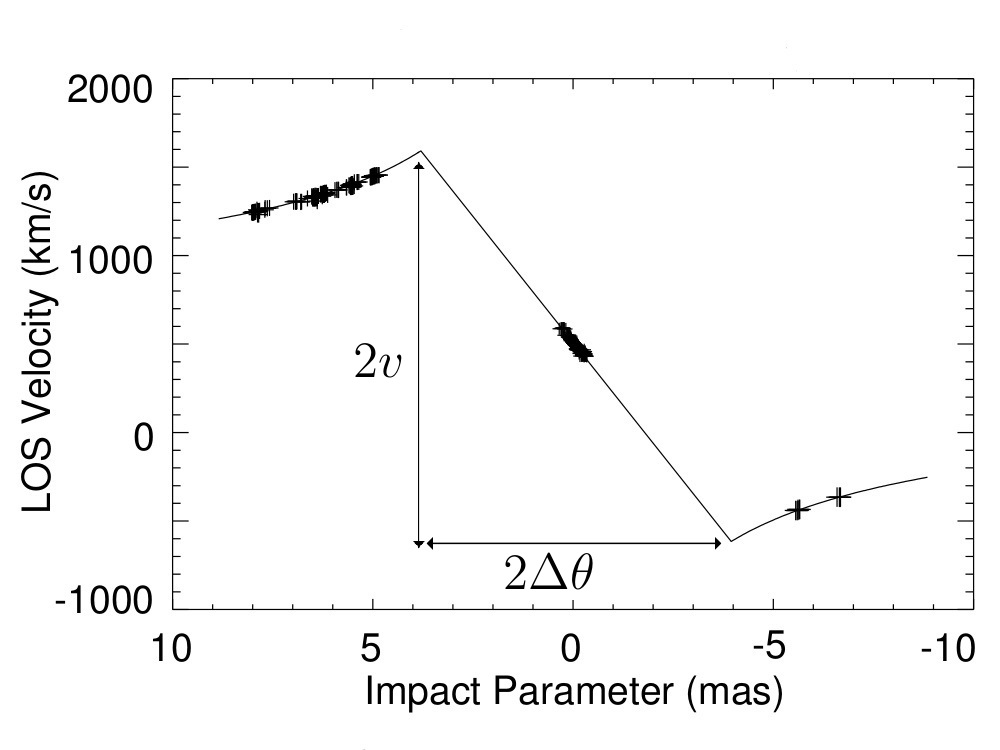}
\caption{The rotation curve for NGC 4258. This figure has been adapted from \cite{Herrnstein:1999cw}. } 
\label{fig:rotcurvemaser}
\end{figure}

In practice, one requires accretion disks that are edge-on since this allows for a complete measurement of the rotation curve,
which, as we will see shortly, is necessary to infer the distance. This means there are relatively few water maser galaxies
compared with those containing Cepheid and TRGB stars. In this work, we will only make use of the most studied galaxy, NGC 4258
\cite{Herrnstein:1999cw}, and so from here on we will concentrate on this galaxy only. The principle is similar for other
galaxies and a full account of water masers can be found in \cite{doi:10.1146/annurev.astro.41.011802.094927}. NGC
4258 has a thin, slightly warped accretion disk whose shape has been well-studied (see \cite{Moran:2007sm} and references
therein). This disk rotates at $\approx 1000$ kms$^{-1}$ and contains several masing clouds of water, which are classified into
two
types. Systemic masers are those on the near-edge of the disk and move with the systemic velocity of the galaxy ($\approx 470$
kms$^{-1}$). They drift in position and LOS velocity of $30$ $\mu$as yr$^{-1}$ and $9$ kms$^{-1}$ yr$^{-1}$ respectively with
respect to the motion of the disk. High-velocity masers are stationary with respect to the disk and hence show perfect Keplarian
rotation curves. This is illustrated in figure \ref{fig:maser2}.
The rotation curve for the water masers in NGC 4258 is shown in figure (\ref{fig:rotcurvemaser}).
At small angles on the sky, one is observing the systemic masers, whose velocities are constant. As one moves outward in angle
on the sky (impact parameter) the LOS velocity varies linearly since the component of the velocity along the line of sight is
$v_{\rm LOS}=v\sin\theta\approx v\theta$. At larger angles on the sky one is observing the high velocity masers, which trace
Keplarian orbits. Extrapolating the linear part of the rotation curve until it intersects the Keplarian curves then gives the
LOS velocity at the systemic radius and the angle on the sky subtended by the systemic orbit $\Delta\theta$. The LOS velocity is
measured using very long baseline interferometry and the acceleration is found using multi-epoch monitoring of the maser spectra
\footnote{This is a complicated exercise in data analysis and fitting and is well beyond the scope of the current section. The
interested reader is referred to \cite{2011PhDT.......205K}.}. Furthermore, since the structure of the accretion disk is
well-known, one can find the angle of inclination and make a geometric distance estimate using equation (\ref{eq:maserD2}).
Additionally, one can measure $\dot{\theta}$, the drift in position shown in figure \ref{fig:maser2}. Since $v\approx
D\dot{\theta}$ one can make a second, independent geometric estimate.

\section{Distance Indicators in Modified Gravity}

Water masers are generally found near the central black holes of galaxies and are mostly found in spiral galaxies. We have argued
in section \ref{sec:currentconstraints} that these are screened given our adopted constraint $\chi_0<10^{-6}$. Furthermore, the
region around the central black hole in any galaxy has $\pn\sim\mathcal{O}(10^{-5})$ and so water masers are screened distance
indicators. Cepheids and TRGB distances are sensitive to modified gravity in different regimes and so below we will derive the
effects on the distance estimates for each of these indicators.

\subsection{Tip of the Red Giant Branch Stars in Modified Gravity}\label{subsec:TRGBMG}

N\"{a}ively, one would expect these to be screened distance indicators since the location of the tip depends on nuclear and not
gravitational physics. In fact, modified gravity does effect the location of the tip if the hydrogen burning shell is unscreened,
in particular, the luminosity at the tip is lower the more unscreened the shell. This is because the increased gravity in the
shell leads to an enhanced rate of hydrogen burning, which causes the temperature to increase at a faster rate than the
equivalent star in general relativity and hence the triple alpha process can begin at a lower effective temperature
and surface luminosity. Ultimately, we will determine the magnitude of
the difference using MESA, but the physics driving the change can be understood using a simple model. 

It is a good approximation to treat the core as a solid isothermal sphere with temperature $T_{\rm c}$ and mass $M_{\rm
c}$. The hydrogen burning
shell is incredibly thin and can be treated as having constant mass equal to the core value and luminosity $L$. In this case, the
shell pressure and temperature is
given by the hydrostatic equilibrium and radiative transfer equations,
\begin{equation}
 \frac{\dd P}{\dd r}=-\frac{GM_{\rm c}\rho(r)}{r^2};\quad\frac{\dd T^4}{\dd r}=\frac{3}{4a}\frac{\kappa(r)\rho(r)L}{4\pi r^2},
\end{equation}
which can be used to find 
\begin{equation}
 P\propto \frac{G M_{\rm c}T^4}{L},
 \label{eq:P-T}
\end{equation}
where the opacity in the hydrogen shell is due mainly to electron scattering and so we have taken it to be constant. The pressure
in the shell is due mainly to the gas and so we ignore radiation pressure and take the equation of state to be that of an ideal
gas, $P\propto \rho T$. Using this and equation (\ref{eq:P-T}) in the radiative transfer equation we find
\begin{equation}\label{eq:TC}
 T(r)\propto \frac{GM_{\rm c}}{R_{\rm c}},
\end{equation}
where the integration constant is negligible near the base of the shell. Next, we can estimate the luminosity given an energy 
generation rate per unit mass $\epsilon\propto\rho(r)T(r)^\nu$
\begin{equation}
 L=\int 4\pi r^2\rho(r)\epsilon(r)\dd r.
 \label{eq:lumint}
\end{equation}
For temperatures above $10^7$ K, which is the case in the shell, hydrogen burning proceeds mainly via the CNO-cycle and so 
$\nu=15$. Using the equation of state and the results above in equation (\ref{eq:lumint}) one finds
\begin{equation}
 L\propto \frac{G^{\frac{8}{3}} M_{\rm c}^{7.7}} {R_{\rm c}^6} .
 \label{eq:TRGB}
\end{equation}
 
Now suppose that the core or shell becomes unscreened so that $G(r)\approx G(1+\alpha_{\rm e})$ where 
\begin{equation}
 \alpha_{\rm e}=\alpha\left[1-\frac{M(\rs)}{M(R_{\rm c})}\right]
\end{equation}
is the effective value of $\alpha$ in the shell (see equation (\ref{eq:g(r)})). The helium flash occurs at a fixed temperature,
independent of modified gravity, and so if we set $\xi=M_{\rm c}/R_{\rm c}$ at the point when $T_{\rm c}=10^8$ K then we have
$\xi_{\rm MG}/\xi_{\rm GR}=(1+\alpha_e)^{-1}<1$. The ratio of the core mass to the core radius at the helium flash in modified
gravity is then lower than that in general relativity. In general, this does not tell us anything about the
core mass and radius individually, however, in practice one finds that the core radius is the same in both cases (this is borne
out by MESA simulations) and so this is a relation between the core masses at fixed temperature. Substituting
equation (\ref{eq:TRGB}) into (\ref{eq:TC}) we can find the ratio of the shell luminosity in the unscreened case to the screened
case such that the core temperature is identical. One finds
\begin{equation}\label{eq:trgbLdrop}
 \frac{L_{\rm MG}}{L_{\rm GR}}=\frac{1}{(1+\alpha_{\rm e})^{5}}.
\end{equation}
Since $\alpha_{\rm e}\ge0$, the shell luminosity when the core is unscreened is lower than its value when the core is screened at
fixed temperature. Hence, the peak luminosity, which corresponds to the temperature required for the onset of helium burning, is
indeed lower in modified gravity.

\begin{figure}[t,h]
\centering
\includegraphics[width=10cm]{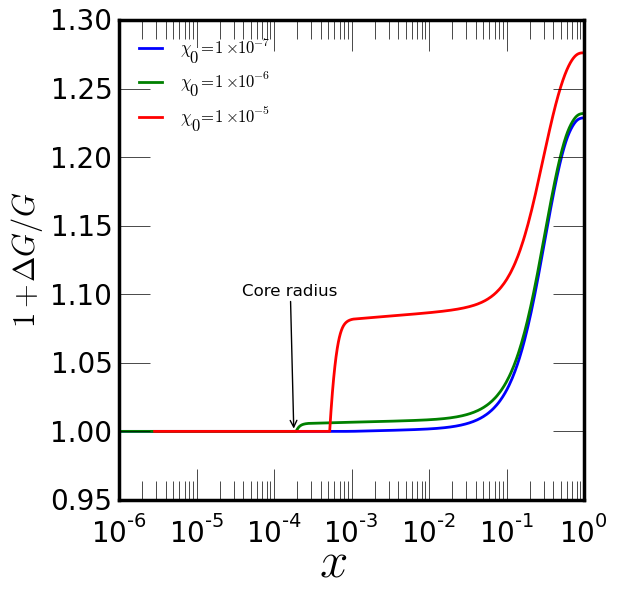}
\caption{The radial profile of $1+\Delta G/G$. We have indicated the core radius for the $\chi_0=10^{-5}$ model. Note that since
we are using the rescaled radius $x\equiv r/R$ the core radius does not necessarily occur at fixed $x$ for different $\chi_0$.
Numerically, we find that the core radius is largely insensitive to the value of $\chi_0$.} 
\label{fig:trgbmg}
\end{figure}

Using MESA we find numerically that when $\chi_0<10^{-6}$ the shell is screened and the change in the peak
luminosity is less than 1\%, however when $10^{-6}<\chi_0<10^{-5}$ the core becomes increasingly unscreened and the change in the
luminosity at the tip falls by 20\% to 50\% over this range\footnote{The dramatic change is due to the strong dependence on
gravity in equation (\ref{eq:trgbLdrop}).}. This is shown in figure \ref{fig:trgbmg} where we plot the radial profile of
$1+\Delta G/G$ where $\Delta G=G(r)-G$ (see equation (\ref{eq:g(r)})) as a function of $x\equiv r/R$ for a $1.5 M_\odot$ stellar
model with $\alpha=1/3$. It is evident that the hydrogen shell above the core feels the effects of modified gravity when
$\chi_0>10^{-6}$.

Hence, TRGB distance indicators are screened when $\chi_0<10^{-6}$ and unscreened
when $\chi_0>10^{-6}$. Using equation (\ref{eq:lumdistex}), failure to account for the reduced luminosity over-estimates the
distance when $\chi_0>10^{-6}$.

\subsection{Cepheid Distances in Modified Gravity}\label{subsec:cephMG}

Cepheid distance estimates are found using the period-luminosity relation (\ref{eq:PLC}) and so in theory, one should derive the
new period-luminosity relation in modified gravity. We shall take steps towards this in chapter \ref{chap:five}, however, a
full theoretical derivation of the relation requires sophisticated non-adiabatic pulsation models and are still not accurate
enough to fit the current experimental data correctly and so here we will estimate the error in the distance estimate if the
galaxy is unscreened by perturbing the empirically calibrated period-luminosity relation about the general relativity result. In
particular, this means
that we cannot address the question of whether the location of the instability strip changes by a significant amount in modified
gravity, nor can we predict the new amplitude of the period-luminosity relation.

Now the period-luminosity relation is an empirically calibrated formula based on Cepheids in the local group and the Milky Way
and so it is a formula derived using objects that behave according to general relativity. What happens if we then try to apply
this to a galaxy where gravity is described by a different theory? As a simple example, let us consider changing the value of
$G\rightarrow G+\Delta G$ for a constant $\Delta G$. In order to use the period-luminosity relation to
calculate the distance, one must measure the flux, the period and $B-V$. If one were to use the formula calibrated on systems
described by general relativity, one would end up with an incorrect value of $M_V$ because stars with a fixed period and fixed
effective temperature correspond to a different value of $M_V$ if the theory of gravity is changed. If $d_0$ is the value of
the distance that would be inferred had we used the correct formula (i.e. the true distance), using this incorrect formula will
result in a different distance $d_{\rm MG}$ being inferred. Cepheid measurements are most robust at the blue edge of the
instability strip and so when calculating $d_{\rm MG}$ we will always use MESA models at the first crossing of this
edge. In practice, this means that when comparing stars in different theories, we are changing the luminosity of the star at
fixed effective temperature. Using the definition of the V-band magnitude
\begin{equation}\label{eq:MV}
 M_V=-2.5\log\left(\frac{L}{d^2}\right)
\end{equation}
we have
\begin{equation}\label{eq:perturbedpl1}
 \frac{\Delta d}{d}=-0.3\frac{\Delta G}{G} - 0.025 \frac{\Delta L}{L}
\end{equation}
where we have used $L\propto R^2$ at constant $T\eff$ and $\Delta d= d_{\rm MG}-d_0$ with $\Delta d/d\equiv\Delta d/d_0$. Using
MESA, we find that
$\Delta L/L\ll 1$, in which case the main dependence comes from changing $G$. We then have 
\begin{equation}\label{eq:perturbedpl2}
 \frac{\Delta d}{d}\approx-0.3\frac{\Delta G}{G};
\end{equation}
if one tries to use the general relativity formula to infer the distance to a galaxy where gravity is stronger one will
under-estimate the distance. Physically, the stars in these galaxies pulsate faster than in general relativity at constant
temperature but they have an almost identical luminosity. Hence, using the general relativity formula is tantamount to
over-estimating the magnitude at a fixed luminosity and hence inferring a smaller distance.

So far, we have assumed that the effective value of $ G$ is a constant but we know from equation (\ref{eq:g(r)}) that this is not
the case, $G$
increases outward in the region exterior to the screening radius. This means that the magnitude of the deviation should be lower
than that predicted using a constant change in $G$ and so it is important to account for this. In chapter \ref{chap:five} we
will perturb the equations of modified gravity hydrodynamics to calculate the new period and so we can find $\Delta d/d$ in terms
of $\Delta \tau/\tau$ but here we will do something simpler and use an appropriately averaged value of $G$ so that $\Delta
G=\langle G\rangle$. We will see in chapter \ref{chap:five} that this is in fact an under-estimate and so any constraints we
obtain are conservative.

\begin{figure}[t,h]
\centering
\includegraphics[width=10cm]{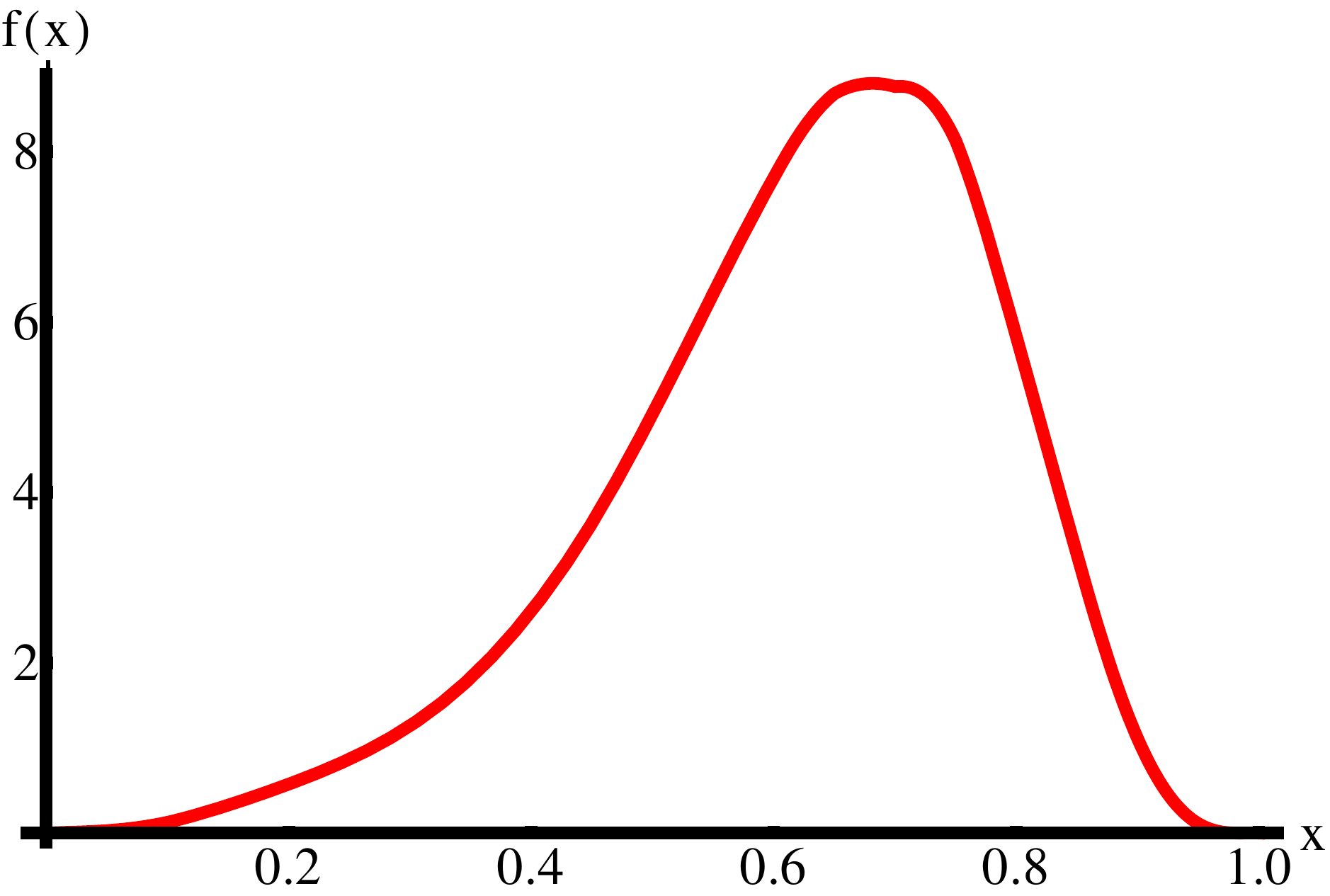}
\caption{The normalised Epstein function $f(x)$ with $x\equiv r/R$. The function was reconstructed by interpolating the data in
\cite{1950ApJ...112....6E}.} 
\label{fig:eps}
\end{figure}

In 1950, Epstein \cite{1950ApJ...112....6E} used numerical pulsation codes to find the relative importance of different regions
of the star for driving pulsations (see figures 1 and 2 in \cite{1950ApJ...112....6E}). Using the published values in
\cite{1950ApJ...112....6E}, we have recreated this function $f(r)$ and normalised it such that 
\begin{equation}
 \int_0^R f(r)\dd r=1.
\end{equation}
Figure \ref{fig:eps} shows this as a function of the dimensionless radial coordinate $x\equiv r/R$.
We then define the average value of $G$ via
\begin{equation}\label{eq:aveG}
 \langle G\rangle = \frac{1}{R} \int_0^R f(r)G(r)\dd r
\end{equation}
so that
\begin{equation}\label{eq:delta G}
 \frac{\Delta G}{G} \equiv \frac{\langle G \rangle - G}{G}.
\end{equation}
This procedure accounts for the fact that $G$ is not constant throughout the star but does not overly penalise the fact that
the core is heavily screened because this region does not contribute significantly to the pulsations. 

\begin{figure}[t,h]
\centering
\includegraphics[width=10cm]{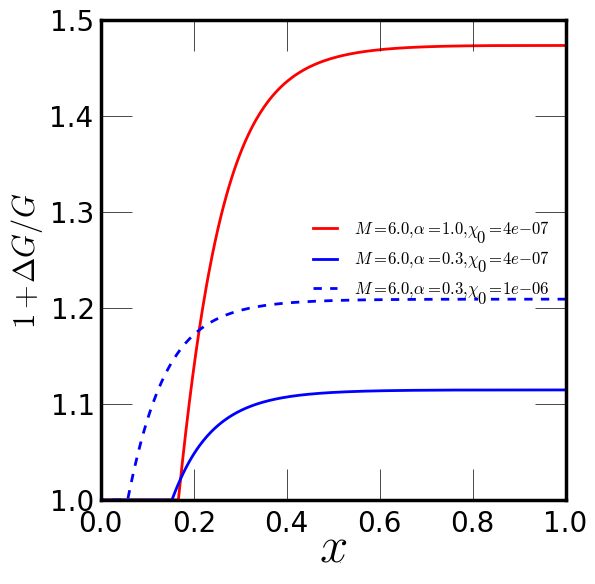}
\caption{The radial profile of $1+\Delta G/G$ as a function of $x\equiv r/R$ for a $6M_\odot$ star at the blue edge of the
instability strip. Different values of $\chi_0$ are indicated and $\alpha=1/3$ in each case.} 
\label{fig:cephmg}
\end{figure}

Using MESA, we find that (for $\alpha\sim\mathcal{O}(1)$) $\Delta G/G\sim\mathcal{O}(1)$ when $\chi_0\gsim
10^{-8}$ depending on the stellar mass (higher mass Cepheids are more unscreened) and so we can use these stars to probe this
parameter range. This is shown in figure \ref{fig:cephmg}, where we plot the radial
profile of $1+\Delta G/G$ (again $\Delta G= G(r)-G$) for a $6 M_\odot$ stellar model at the first crossing of the blue edge of the
instability strip for $\alpha=1/3$ and different values of $\chi_0$. The gravitational enhancement drops rapidly when
$\chi_0<4\times 10^{-7}$, however larger Cepheids ($M\sim10M_\odot$) show enhancements when $\chi_0\gsim 10^{-8}$.

\section{Distance Indicator Constraints on Modified Gravity}

We have argued above that water masers are screened distance indicators, TRGB stars are screened when $\chi_0\lsim 10^{-6}$ and
Cepheid stars are screened when $\chi_0\lsim10^{-8}$. This means that any deviation between water maser and TRGB distances can
probe the range $\chi_0>10^{-6}$ and any deviation between TRGB and Cepheid stars can probe the range $10^{-8}<\chi_0<10^{-6}$.
In theory, comparing water maser distances with Cepheid distances could also probe smaller values of $\chi_0$. In practice there
are no water maser distances to dwarf galaxies so this comparison is not currently possible. We will perform both of these
tests below and will derive new constraints in each case.

\subsection{Comparison of Water Maser and TRGB distances}

Using MESA, we have evolved $1.5 M_\odot$ stellar models to the onset of the helium flash for a
variety of different values of $\chi_0$ and $\alpha$ and compared the luminosity at the tip with that of the corresponding
general relativity
model. Using this, we can calculate $\Delta d/d= d_{\rm MG}-d_0$ using equation (\ref{eq:lumdistex}). In this case, $d_{\rm
MG}=d_{\rm TRGB}$ is the incorrect distance found by assuming the general relativity luminosity at the tip and $d_0=d_{\rm
maser}$ is the correct, geometric distance found using water maser measurements. As an example of the size of the discrepancy, we
tabulate
the change in the luminosity at the tip and the value of $\Delta d/d$ for various different
values
of $\chi_0$ and $\alpha=1/3$ (corresponding to $f(R)$ models) in table \ref{tab:TRGB}. We have also varied $\alpha$ as well but
these values are not given for brevity.

\begin{table*}[th!]
\centering
\heavyrulewidth=.08em
	\lightrulewidth=.05em
	\cmidrulewidth=.03em
	\belowrulesep=.65ex
	\belowbottomsep=0pt
	\aboverulesep=.4ex
	\abovetopsep=0pt
	\cmidrulesep=\doublerulesep
	\cmidrulekern=.5em
	\defaultaddspace=.5em
	\renewcommand{\arraystretch}{1.6}
\label{tab:TRGB}
\begin{tabu}{c|[2pt]  c|[2pt] c|[2pt]   c } 
$\alpha$ & $\chi_0$ &  $\log L/L_\odot$ & $\Delta d/d$ \\
\tabucline[2pt]{-}
$0 $&$ 0$ &$ 3.34$ &$ 0$ \\
$1/3$ & $1\times 10^{-6}$  &$ 3.32$ &$ 0.02$\\ 
$1/3$ & $2\times 10^{-6}$  & $3.30 $& $0.04$\\
$1/3$& $4\times 10^{-6}$  &  $3.25$ & $0.12$\\
$1/3$ & $8\times 10^{-6}$  & $<3$ & $>0.20$\\
\end{tabu}\caption{Change in the inferred distance using the TRGB indicator for $f(R)$ chameleon models for a $1.5 M_\odot$
stellar model.}
\end{table*}

As discussed in section \ref{subsec:masers}, there is only one galaxy for which we have simultaneous distance estimates from
water masers and other methods: the spiral galaxy NGC 4258. The two distance estimates using the positional drift and the
Keplarian motion methods agree within errors (see \cite{Herrnstein:1999cw,Humphreys:2007ir}) and a comparison with the
TRGB and Cepheid distance estimates (taken from \cite{Freedman:2010xv}) is shown below:
\begin{eqnarray}
{\rm NGC 4258\ Maser}: d &=& 7.2 \pm 0.2\ {\rm Mpc} \\
{\rm NGC 4258\ Maser}: d &=& 7.1 \pm 0.2\ {\rm Mpc} \\
{\rm NGC 4258\ Cepheid}: d &=& 7.18 \pm 0.07 ({
\rm statistical})\ {\rm Mpc} \\
{\rm NGC 4258\ TRGB}: d &=& 7.18 \pm 0.13 \pm 0.40\ {\rm Mpc}.
\end{eqnarray}
The distances agree within estimated errors that are at the
few percent level for the maser distances and (allowing for systematics, which are not included in the Cepheid measurement) at the
5-10\% level for Cepheid and TRGB distances. The agreement of TRGB and water maser distances probes $\chi_0>10^{-6}$. The precise 
range probed depends on the value of $\alpha$ and the stellar mass. For $\alpha=1/3$ the range probed is $\chi_0 > 4\times10^{-6}$
and a typical star of mass $1.5 M_\odot$, the TRGB luminosity is smaller by over 20\%, corresponding to an inferred distance that
is larger by over 10\%. Thus, given the measurements above, $f(R)$ models with this parameter range are excluded. Shortly after
this new constraint was presented in \cite{Jain:2012tn}, \cite{Humphreys:2013eja} reported a new maser distance to NGC
4258 using the Keplarian
method accounting for the warping of the accretion disk, elliptical orbits of the masers and orbital precession. This yielded an
updated distance $7.60\pm 0.17 \textrm{ (random) } \pm 0.15 \textrm{ (systematic) Mpc}$. Although not as good, this still agrees
with the TRGB and Cepheid distances within errors and the conclusions are unchanged.

The agreement of the Cepheid distance with the water maser distances would probe $\chi_0<10^{-6}$ if the galaxy were not a spiral
and the Cepheid was not screened. The megamaser cosmology project \cite{Henkel:2012by} will provide new maser galaxies and
estimate their distance. Since the goal of this survey is to determine $H_0$ the morphology of the galaxies surveyed is
not relevant and it is unclear whether this data will be suitable for testing modified gravity or indeed if masers can be found
in galaxies with shallower Newtonian potentials. 

Since TRGB distances vary from the general relativity value by larger amounts with increasing $\chi_0$ the agreement of the maser
and TRGB distances rules out the entire range $\chi_0\gsim 10^{-6}$ for $\alpha\ge\mathcal{O}(1)$. Previously, this constraint
had been assumed by demanding that the Sun or the Milky Way is unscreened but no comparison of data with theoretical models was
ever attempted and these objects could have been blanket screened by the local group if $\chi_0<10^{-4}$. This is a new and
independent constraint that resolved the debate as to how the Milky Way is screened: it is self-screening.

\subsection{Comparison of Cepheid and TRGB distances}

We now turn our attention to the parameter range $\chi_0<10^{-6}$, which can be probed by comparing TRGB distances, which are
screened in this range, with Cepheid distances, which are not. We require simultaneous Cepheid and TRGB distances to the same
galaxy, which we classify as screened or unscreened (this is a function of $\chi_0$) using the screening map of
\cite{Cabre:2012tq} discussed in section \ref{subsec:screeningmap}. The TRGB and Cepheid data used here is taken from the
literature; a full discussion of the various telescopes and experimental methods is given in \cite{Jain:2012tn} and references
therein. In appendix \ref{app:data} we list the various galaxies used in the comparison, their TRGB and Cepheid distance
measurements as well as their literary references. In order to calculate theoretical predictions of $\Delta d/d$ for various
$\chi_0$ and
$\alpha$ we evolve $6M_\odot$ stellar models of initial metallicity $Z=0.004$ using MESA to the first crossing of
the blue edge of the instability strip and apply the procedure detailed in section \ref{subsec:cephMG}. As an example, we show
both $\Delta G/G$ and $\Delta d/d$ for a selection of models used in the analysis in table \ref{tab:period}. In this case $\Delta
d=d_{\rm Cepheid}-d_{\rm TRGB}$, $d_{\rm MG}=d_{\rm Cepheid}$ and $d_0=d_{\rm TRGB}$.

\begin{table*}[th!]
\centering
\heavyrulewidth=.08em
	\lightrulewidth=.05em
	\cmidrulewidth=.03em
	\belowrulesep=.65ex
	\belowbottomsep=0pt
	\aboverulesep=.4ex
	\abovetopsep=0pt
	\cmidrulesep=\doublerulesep
	\cmidrulekern=.5em
	\defaultaddspace=.5em
	\renewcommand{\arraystretch}{1.6}
\label{tab:period}
\begin{tabu}{c|[2pt]  c|[2pt]  c |[2pt] c }

$\alpha_0$ & $\chi_0$ & $\Delta G/G$ & $\Delta d/d$ \\
\tabucline[2pt]{-}
$1/3 $& $4\times 10^{-7}$ &$ 0.11 $&$ -0.03$\\
$1/3$ & $1\times 10^{-6}$ &$ 0.21 $& $-0.06$\\
$1/2 $& $4\times 10^{-7}$ &$ 0.17 $& $-0.05$\\
$1/2$ & $1\times 10^{-6}$ &$ 0.34$ & $-0.09$\\
$1 $& $2\times 10^{-7}$ & $0.21 $&  $-0.06$\\
$1$ & $4\times 10^{-7}$ & $0.45$ & $-0.12$  \\
\end{tabu}\caption{Change in inferred distance due to the change in the Cepheid periods for
different modified gravity parameters. In each case the quantities were computed using $6M_\odot$ stellar models with initial
metallicity $Z=0.004$.  }
\end{table*}

\begin{figure}[t]
\centering
\includegraphics[width=12cm]{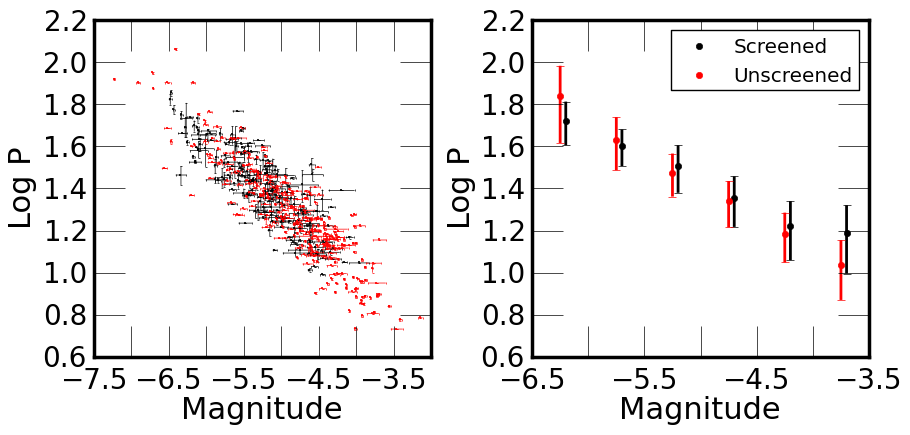}
\caption{The period-luminosity relation for the galaxies in our sample; the black and red points denote screened and unscreened
galaxies respectively. \textit{Left panel}: we show all the Cepheids observed along with the reported error bars. \textit{Right
panel}: the mean period and dispersion within bins in absolute magnitude of size 0.5.}
\label{fig:cephPL1}
\end{figure}

As mentioned in the introduction to this chapter, the analysis of this data was performed by the other authors of
\cite{Jain:2012tn}
and so here we will only briefly describe the method, the interested reader is referred to \cite{Jain:2012tn} for the full
details. We began with a
sample of 27 galaxies with both TRGB and Cepheid distances taken from the literature. In general, we only use galaxies with
several Cepheid distances, which are combined to give a lower error on the measurement. After removing one galaxy with only two
confirmed Cepheids and one with a TRGB distance greater than $10$ Mpc we are left with 25 galaxies. We perform a likelihood
analysis on the data to estimate the best fit value of $\Delta d/d$. The best values and 1-$\sigma$ errors are given in table
(\ref{tab:ceph1}) along with the reduced $\chi^2$ and the number of galaxies. We included empirically estimated systematic errors
in the estimate of the distance to each galaxy from multiple measurements, as well as in the average deviation $\Delta d/d$ for
each sub-sample of galaxies. For the latter, we made the ansatz that each galaxy has an additional unknown systematic error that
can
be added in quadrature to the reported error. By further assuming that the systematic error was the same for each galaxy, we could
estimate $\sigma_{\rm sys}$ iteratively such that the reduced $\chi^2$ was unity. We found that the systematic error thus
estimated is sub-dominant for the majority of galaxies. 
\begin{table*}[t!]
\centering
\caption{Best fit values for $\Delta d/d$ and uncertainty $\sigma$ in the fractional difference between
Cepheid and TRGB distances for the screened and unscreened sub-samples. Our estimated $\sigma$ includes systematic errors. The
number of galaxies $N$ in each sub-sample is given as is the reduced $\chi^2$.
}
\label{tab:ceph1}
\heavyrulewidth=.08em
	\lightrulewidth=.05em
	\cmidrulewidth=.03em
	\belowrulesep=.65ex
	\belowbottomsep=0pt
	\aboverulesep=.4ex
	\abovetopsep=0pt
	\cmidrulesep=\doublerulesep
	\cmidrulekern=.5em
	\defaultaddspace=.5em
	\renewcommand{\arraystretch}{1.6}
\centering
\begin{tabu}{c |[2pt] c|[2pt]  c |[2pt] c|[2pt] c }
Sample & $N$ & $\Delta d/d$ & $\sigma$ & Reduced $\chi^2$\\
\tabucline[2pt]{-}
Unscreened & 13 & $0.003$ & $0.017 $& $1.0$\\
Screened & 12 &$ -0.005$ &$ 0.022$ &$ 1.3$\\
\end{tabu}
\end{table*}
Figure \ref{fig:cephPL1} shows the observed period-luminosity relation for both the screened and unscreened galaxies in our
sample. There are no major differences evident and any disparity must be searched for using statistical
methods.

\begin{figure}[t,h]
\centering
\includegraphics[width=16cm]{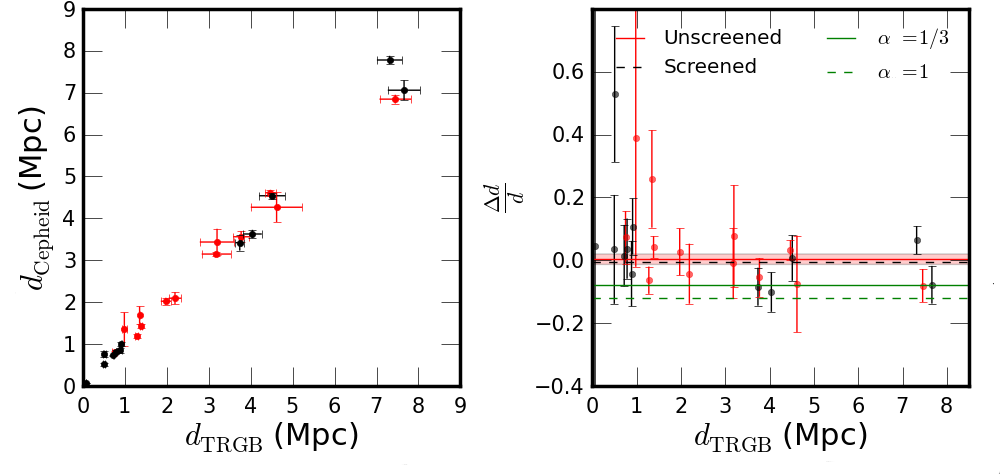}
\caption{{\it Left panel:} A comparison of distances measured using the Cepheid period-luminosity relation and TRGB
luminosities. The black and red points show galaxies from the screened and unscreened sub-samples respectively. {\it Right panel:}
$\Delta d/d$, the fractional difference between Cepheid and TRGB distances, as a function of TRGB distance. The shaded region in
the right panel shows the 68\% confidence region around our best fit to the unscreened sample (red line). The best fit to the
screened sample is shown by the dashed black line. Both sub-samples are consistent with the general relativity expectation of zero
deviation. The dotted and dashed green lines show two possible predictions of chameleon theories with 
$\alpha = 1$, $\chi_0 = 4 \times 10^{-7}$ and $\alpha = 1/3$, $\chi_0 = 1 \times 10^{-6}$, which corresponds to $f(R)$ gravity.}
\label{fig:comparison}
\end{figure}

Figure \ref{fig:comparison} shows the Cepheid distance compared with the TRGB distance for both sub-samples; both are
clearly consistent and again one must use statistical methods to quantify any small discrepancies. As an example, the two green
lines show the predictions for chameleon theories with coupling strength
$\alpha=1/3$ and $\alpha= 1$ and values of $\chi_0$ as indicated in the caption. These two models are ruled out at over 95\%
confidence. 

We have checked several sources of systematic errors in our analysis including metallicity corrections to the period-luminosity
relation, different screening criteria for classifying galaxies into specific sub-samples, including only the best distance
measurements to a specific galaxy and including galaxies we previously rejected (technical details can be found
in \cite{Jain:2012tn}). In all cases, the best-fit line for $\Delta d/d$ moves in the opposite direction to that predicted by
modified gravity i.e. more positive values and so we conclude that our analysis is robust to these sources of systematics.

\subsubsection{Constraints}

\begin{figure}[t]
\centering
\includegraphics[width=10cm]{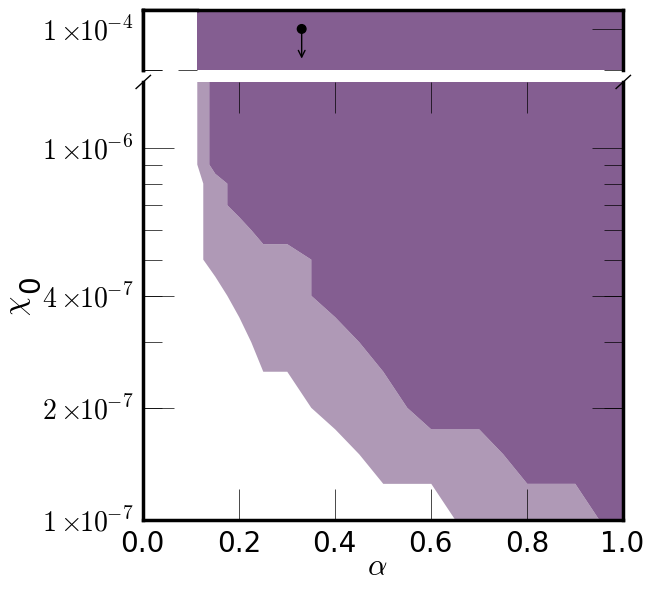}
\caption{The excluded region in the $\chi_0$--$\alpha$ plane for chameleon models. The boundaries of the shaded regions show the
upper limits at
68\% and 95\% confidence level. The black arrow shows the previous constraint, $\chi_0\le10^{-4}$, coming from cosmological
and cluster constraints, which was obtained for $f(R)$ theories with $\alpha=1/3$.}
\label{fig:confidence-regions}
\end{figure}

Whereas the modified stellar structure can be calculated in a model-independent way using $\chi_0$ and $\alpha$, the screening
of dark
matter haloes is model-dependent and this needs to be accounted for when constructing the screening map. So far, the map has been
calibrated using chameleon N-body codes \cite{Cabre:2012tq} and so this is where the most robust constraints can be found. In
figure \ref{fig:confidence-regions} we plot the regions in the
$\chi_0$--$\alpha$ plane that are excluded with 68\% (light region) and 95\% (dark region) confidence for chameleon models. The
jaggedness of the
contours is due to the small sample size; decreasing $\chi_0$ will cause galaxies to move from the unscreened to screened
sub-sample and so the quality of the data can change rapidly with a small change in $\chi_0$. Also plotted is the previous
constraint coming from cosmological probes \cite{Lombriser:2010mp,Schmidt:2010jr} $\chi_0\lsim10^{-4}$. These new constraints are
three orders of magnitude stronger than those coming from cosmological probes. After these results were published,
\cite{Terukina:2013eqa} published an independent constraint coming from the equivalence of the hydrostatic and lensing mass of the
Coma cluster. They find $\chi_0<6\times10^{-5}$ for $f(R)$ theories; our results are stronger than theirs by two orders of
magnitude.

Two theories of particular interest are $f(R)$ theories with $\alpha=1/3$ and $\mathcal{O}(1)$ chameleons. Our analysis reveals
the new upper limit on $\chi_0$ (at the 95\% confidence level) for these theories:
\begin{align}
 \alpha=1/3&\textrm{:}\quad\chi_0\le4\times10^{-7}\\
\alpha=1&\textrm{:}\quad\chi_0\le1\times10^{-7}.
\end{align}
These limits correspond to a cosmological Compton wavelength of the field of order $1$ Mpc. As discussed in section
(\ref{subsec:screeningmap}), the Compton wavelength is important for determining whether or not a galaxy is screened. Symmetron
models with $\alpha=2$ have similar Compton wavelengths to these models and so one can place the
constraint
\begin{equation}\label{eq:symmetroncons}
 \chi_0\le3\times10^{-7}
\end{equation}
for this model, however other models require a re-calibration of the screening map. We have not attempted this here, nor have we
attempted to place a constraint on theories which screen using the EDDP effect. In principle, a
new screening map calibrated on symmetron and dilaton N-body simulations (which have been performed \cite{Brax:2012nk}) could be
used in conjunction with the same data to place new constraints although no attempt has been made to date.

\section{Summary of Main Results}

Simultaneous measurements of the distance to unscreened dwarf galaxies will not agree if one of the methods used to infer the
distance is sensitive to the theory of gravity. In this chapter, we have used three distance indicators to place new constraints
on chameleon and chameleon-like theories of gravity. Water masers are insensitive to the theory of gravity considered here. TRGB
distances are screened when $\chi_0\gsim10^{-6}$ and unscreened otherwise, making them versatile tools for probing these theories.
Cepheid variable stars are unscreened when $\chi_0\gsim10^{-8}$ and hence have the potential to probe the entire parameter range
of interest. The only obstacle to this is a lack of unscreened galaxies for $\chi_0\lsim 10^{-7}$. Future surveys will improve
the number of Cepheid measurements to distant galaxies and we will discuss these in chapter \ref{chap:eight}.

Only the galaxy NGC 4258 has simultaneous maser, TRGB and Cepheid measurements. It is a spiral galaxy and the agreement of these
measurements places the new independent constraint $\chi_0<10^{-6}$. By comparing simultaneous measurements of Cepheid and TRGB
distances to a sample of screened and unscreened galaxies we have been able to probe into the previously unexplored range
$\chi_0<10^{-6}$. Our analysis is data limited due to the small sample size but nonetheless we have been able to place new
constraints summarised in figure \ref{fig:confidence-regions}. In particular, we can exclude $\chi_0>4\times10^{-7}$ for $f(R)$
($\alpha=1/3$) theories of gravity. These constraints are currently the strongest in the literature. 

In order to obtain these results, we had to use an approximation to find the change in the pulsation period of Cepheid stars. In
particular, we were unable to predict the new pulsation period as a function of $\chi_0$ and $\alpha$ due to an incomplete
theoretical framework and instead had to perturb the relation found using general relativity. In the next chapter, we will
construct such a framework and use it to assess the validity of this approximation and the robustness of our results. Ultimately,
we will find that the constraints presented here are conservative and, it is possible to improve them using the same data and
statistical analysis. 

\thispagestyle{empty}
\newpage

\thispagestyle{empty}
\begin{savequote}[30pc]
Self-education is, I firmly believe, the only kind of education there is.
\qauthor{Isaac Asimov}
\end{savequote}

\chapter{Stellar Oscillations in Modified Gravity}\label{chap:five}

In the previous chapter we obtained new constraints on chameleon-like theories of modified gravity by comparing Cepheid
and TRGB distance indicators to unscreened dwarf galaxies. In theory, it is possible to improve these constraints using the same
or upcoming data, however there were three main uncertainties which prevented this: systematic errors in the data, approximations
used to determine whether a galaxy is screened or not and an incomplete theoretical model of Cepheid pulsations. Recall that we
did not calculate the new period-luminosity relation but instead perturbed the general relativity calibrated formula. Furthermore,
we did not account for the radial profile of $G$ completely and used an averaging procedure. The systematics were investigated and
we found that the constraints are largely robust to their effects. A better understanding of which galaxies are screened requires
better data with lensing masses and dynamical masses \cite{Li:2011pj} and further N-body simulations and it is not currently
possible to address this issue. 

In this chapter, we will address the final issue and take the first steps towards a model of Cepheid pulsations in modified
gravity. As discussed in the previous chapter, this has applications beyond the current methods of placing constraints. A full
theoretical model would allow a prediction of both the amplitude and slope of the period-luminosity relation as well as the
location of the blue edge of the instability strip\footnote{In theory, including convection could predict the location of the red
edge as well, however there are already theoretical uncertainties in this in general relativity.}. All of these are potential new
and independent probes of modified gravity. So far, we have been treating the stars at the level of perturbations (Cepheid
pulsations are oscillations about the equilibrium profile) but we have not addressed the issue of perturbations of the scalar
field. At the background level, we assumed the field was static, however, since we have a time-varying density profile we expect
time-dependent perturbations in the scalar field. This corresponds to scalar radiation, which can back-react and modify
the oscillation period. 

To address these issues, in this chapter we will perturb the equations of modified gravity hydrodynamics coupled to the scalar
field equation of
motion to first-order and derive a system of equations governing the evolution of radial perturbations. The system is a coupled
Sturm-Liouville eigenvalue problem whose eigenfrequencies give the period of
oscillation. We will specialise to radial perturbations since we are interested in stars in other galaxies for which non-radial
modes are unobservable. This allows for a calculation of the new oscillation period in modified gravity that does not rely on
perturbing the general relativity result. Furthermore, using equilibrium profiles from MESA we can fully account
for the radial dependence of $G$. Since the equations of modified gravity hydrodynamics apply equally to all theories of modified
gravity that can be written in the form (\ref{eq:EFgencoup}), the resulting equations are valid equally for chameleon-like
theories and Vainshtein-screened theories provided one perturbs the relevant equation of motion for the scalar. There are three
new effects in modified gravity:
\begin{enumerate}
 \item When the star is unscreened, the period of oscillation can be reduced by an $\mathcal{O}(1)$ amount and is larger than one
would find by simply changing the value of $G$. For Cepheid stars in chameleon-like theories we find that the change in the period
can result in differences from the general relativity inferred distance measurement of up to three times what we found in the
previous chapter by perturbing the general relativity formula.
\item When the star is unscreened, the star is more stable to perturbations. There is a well-known result in stellar
astrophysics that when the first adiabatic index falls below $4/3$ the squared-frequency of the fundamental mode is negative and
so the mode is unstable. We will show that in modified gravity the first adiabatic index can fall below $4/3$ without any
instability. In
modified gravity, there is no universal bound on the critical index for the appearance of the instability; its precise value
depends on the structure and composition of the star as well as how unscreened it is. 
\item The perturbations of the star can source scalar radiation and vice versa.
\end{enumerate}
The first two effects require the star to be unscreened whilst the third does not. Vainshtein-screened theories are too efficient
to leave any star unscreened and so it is only the third effect that may be used to probe this theory. Here, we will
only investigate the first two effects with the aim of improving the Cepheid constraints of the previous chapters and so once the
general result has been presented we will specialise to chameleon-like theories. In fact, scalar radiation in chameleon theories
has been previously studied. \cite{Silvestri:2011ch} approximated pulsating stars as solid spheres whose radius and density are
oscillating and calculated the resultant scalar radiation using linear perturbations of the scalar equation of motion. This
reduced the problem to a forced harmonic oscillator being driven at the frequency of the oscillation of the star.
\cite{Upadhye:2013nfa} modelled the system as a thin-shell solution that adjusts itself adiabatically to match the
changing radius of the star. In both cases, they concluded that scalar-radiation is negligible. These approaches both ignored the
fact that the star is not an isolated system unaffected by modified gravity. Its internal structure is coupled to the scalar
field profile and hence the correct method of calculating the scalar radiation and the
oscillation period is to couple perturbations in the stellar structure equations to the scalar field perturbations. The formalism
presented here will capture this fully and it is for this reason that we obtain an eigenvalue problem and not a forced oscillator.
That being said, the two approaches above are enough to estimate the amount of scalar radiation to the correct order of
magnitude and so
we will decouple the equations by ignoring scalar perturbations. Scalar radiation is not necessarily negligible in other
theories of modified gravity and so the formalism presented here may be useful in the future to provide a consistent framework for
calculating new constraints on other theories --- galileons being one example --- using scalar radiation. 

When deriving the equations governing the stellar oscillations we will make an adiabatic approximation for simplicity and to
discern the new physics. This means that the formalism is not powerful enough to predict the amplitude of the period-luminosity
relation, its slope or the location of the blue edge of the instability strip. Non-adiabatic extensions are required for this.
Since the
non-adiabatic driving processes in Cepheids do not depend on gravitational physics, the adiabatic result can be extended in the
same manner as the equivalent equation in general relativity without introducing new features. All of the new physics driving the
change in the period-luminosity relation and the location of the instability strip is included in the modified equilibrium
structure of the star and the perturbation analysis presented below. 

The extension to non-adiabatic systems has yet to be investigated.
Non-adiabatic processes tend to drive Cepheids in the fundamental mode of oscillation and so the periods
calculated using the adiabatic equations do not differ significantly from those that would result had we solved the full
non-adiabatic system. 

We will solve the new equation governing radial oscillations of the star numerically using both Lane-Emden and MESA
models to calculate $\Delta \tau/\tau$ (which is found from the eigenfrequency of the Sturm-Liouville problem) and investigate
the modified stellar stability properties. The former has implications for the tests presented in the previous chapter and
upcoming experimental tests of modified gravity and we discuss these in light of these new predictions. The latter may lead to
interesting observational consequences such as an enhanced type II supernova rate in unscreened dwarf galaxies or the presence of
super-massive blue stars. It is currently unclear whether these effects can be used as the basis for any new observational tests
but a further investigation is certainly merited.

\section{Modified Gravity Hydrodynamics: Linear Perturbation Theory}

The dynamics of stellar oscillations are governed by the hydrodynamics of small
perturbations about the equilibrium configuration and so we shall linearise the
equations (\ref{eq:momentumMG}), (\ref{eq:poisson}), (\ref{eq:cont}) about some
assumed background profile $\{P_0(r), \rho_0(r), \pno(r), \gao(r), \phi_0(r)\}$
ignoring second-order terms in the perturbations. 

The fundamental object of interest is the linearised perturbation to the
Lagrangian position of each fluid element $\vec{\delta r}(\vec{r})$, which
describes the oscillation of the fluid from equilibrium at each radius. The equations of modified gravity hydrodynamics presented
in section \ref{sec:MGhydro} are completely general and describe the full
three-dimensional hydrodynamic problem. Continuing with the full three dimensional framework would result in a
complete treatment of both radial and non-radial modes of oscillation. The aim
of this section is to provide a consistent framework with which to predict the
oscillation properties of partially unscreened stars residing in unscreened
dwarf galaxies. The extra-galactic nature of these stars ensures that only their
fundamental radial mode (and possibly the first-overtone) are
observable and so non-radial modes are independent for observational tests of modified gravity. We therefore specify to the case
of radial oscillations so that $\vec{\delta r}(r)$ is a purely radial vector. A more convenient quantity to work with is the
relative
displacement, given by
\begin{equation}\label{eq:zetadef}
 \zeta(r) \equiv \frac{\delta r(r)}{r},
\end{equation}
and the aim of the present section is to derive its equation of motion.

We shall work with Eulerian perturbations of the background profile, which we
will distinguish from Lagrangian perturbations by the use of a tilde so that
\begin{align}\label{eq:backgroundperts}
 P(r,t)&=P_0(r)+\tilde{P}(r,t)\\
\rho(r,t)&=\rho_0(r)+\tilde{\rho}(r,t)\\
\Phi_{\rm N}(r)&=\Phi_{{\rm N}, 0}(r)+\tilde{\Phi}_{\rm N}(r,t)\\
\phi(r,t)&=\phi_0(r)+\tilde{\phi}(r,t).
\end{align}
As remarked above, the Lagrangian perturbations may provide more physical
insight on occasion and the two are related, for example, via
\begin{align}\label{eq:perteg}
 \delta P(r,t) &= P(\vec{r}+\vec{\delta r},t)-P_0(r) \\&=
\tilde{P}(r,t)+\delta r\frac{\dd
P_0(r)}{\dd r}.
\end{align}
It will be useful later to have the Lagrangian
pressure perturbation in terms of our system-variables. This is given by
\begin{equation}\label{eq:lagpresspert}
 \frac{\delta P}{P_0}= \gao\frac{\delta\rho}{\rho_0} =
-\gao\left(3\zeta+r\frac{\dd \zeta}{\dd r}\right),
\end{equation}
which only holds in the adiabatic approximation.

We begin by perturbing equation (\ref{eq:cont}) to obtain the Eularian density perturbation
\begin{equation}\label{eq:contpert}
 \tilde{\rho}=-\frac{1}{r^2}\frac{\partial }{\partial r}(r^3\rho_0\zeta).
\end{equation}
This may be used in the perturbed form of (\ref{eq:momentumMG}) to
find\footnote{We have neglected terms proportional to $\dd\beta(\phi)/\dd\phi$ in accordance with the discussion in section
\ref{sec:screen_mech}. Chameleon-like theories have $\delta\phi\dd\beta/\dd\phi\ll1$ in the unscreened region and
theories that utilise the Vainshtein mechanism have $\dd\beta/\dd\phi=0$.}
\begin{equation}\label{eq:accpert}
 \rho_0r\frac{\partial^2\zeta}{\partial t^2}=-\frac{\dd \tilde{P}}{\dd
r}-\rho_0\frac{\dd\tilde{\Phi}_{\rm N}}{\dd r}+\frac{\beta_0}{\mpl
r^2}\frac{\partial }{\partial r}(r^3\rho_0\zeta)\frac{\dd \phi_0}{\dd
r}-\frac{\beta_0\rho_0}{\mpl}\frac{\dd \tilde{\phi}}{\dd r}.
\end{equation}
The perturbed Poisson equation is
\begin{equation}\label{eq:pertpoisson}
 \nabla^2\tilde{\Phi}_{\rm N}=4\pi G\tilde{\rho},
\end{equation}
which can be integrated once using (\ref{eq:contpert}) yielding
\begin{equation}
 \frac{\dd \tilde{\Phi}_{\rm N}}{\dd r}=-4\pi Gr\rho_0\zeta.
\end{equation}
Now 
\begin{equation}
 \frac{\dd P}{\dd t}=\frac{\partial P}{\partial t}+\vec{v}\cdot\nabla
P=\frac{\partial \tilde{P}}{\partial t}+r\frac{\partial \zeta}{\partial
t}\frac{\partial P_0}{\partial r}
\end{equation}
and using (\ref{eq:dpdrho}) this is 
\begin{equation}\label{eq:pertprho}
 \tilde{P}+r\zeta\frac{\partial P_0}{\partial
r}=\frac{\Gamma_{1,0}P_0}{\rho_0}\left(\tilde{\rho}+r\zeta\frac{\partial
\rho_0}{\partial r}\right).
\end{equation}
We wish to eliminate $\tilde{P}$ and $\tilde{\rho}$ and so we substitute
(\ref{eq:contpert}) into (\ref{eq:pertprho}) to give
\begin{align}
 \tilde{P}&=-\frac{\Gamma_{1,0}P_0}{\rho_0}\left(\frac{1}{r^2}\frac{\partial}{
\partial r}(r^3\rho_0\zeta)-r\zeta\frac{\dd P_0}{\dd
r}+\frac{\rho_0}{\Gamma_{1,0}P_0}\frac{\dd P_0}{\dd r}r\zeta\right)\nonumber\\
&=-\Gamma_{1,0}P_0r\left(\frac{\partial \zeta}{\partial
r}+\frac{3}{r}\zeta+\frac{1}{\Gamma_{1,0}P_0}\frac{\dd P_0}{\dd r}\zeta\right).
\end{align}
We then have 
\begin{align}
 \frac{\partial \tilde{P}}{\dd r}&=-\frac{\partial}{\partial
r}(\Gamma_{1,0}rP_0\frac{\partial\zeta}{\partial
r})\nonumber\\&-\frac{\partial}{\partial
r}(3\Gamma_{1,0}P_0\zeta)-\frac{\partial}{\partial r}(r\frac{\dd P_0}{\dd
r}\zeta).
\end{align}
\newline\newline 
This may then be used with (\ref{eq:pertpoisson}) and
(\ref{eq:accpert}) to
find:
\begin{align}\label{eq:mlawepre}
 r^4\rho_0\frac{\partial ^2 \zeta}{\partial t^2}&=\frac{\partial}{\partial
r}\left(r^4\Gamma_{1,0}P_0\frac{\partial \zeta}{\partial
r}\right)+r^3\frac{\partial}{\partial
r}\left[\left(3\Gamma_{1,0}-4\right)P_0\right]\zeta\nonumber\\&-\frac{
r^3\beta_0\rho_0}{ \mpl } \frac{\partial \tilde{\phi}}{\partial
r}-\frac{\beta_0\rho_0}{\mpl}r^4\frac{\dd^2\phi_0}{\dd
r^2}\zeta-2\frac{r^3\beta_0\rho_0}{\mpl}\frac{\dd \phi_0}{\dd r}\zeta.
\end{align}
This is the equation of motion governing the evolution of $\zeta$. In general relativity,
$\partial \tilde{\phi}/\partial r=0$ and we obtain a single equation. In modified gravity,
however, we need a separate equation for $\tilde{\phi}$. This is where we will specify to the case of chameleon-like theories and
perturb the equation of motion (\ref{eq:chameom})\footnote{Equation (\ref{eq:mlawepre}) is
equally valid for any scalar-tensor theory of modified gravity where $\dd\beta(\phi)/\dd\phi$ can be neglected provided that one
perturbs the equation of motion for the scalar field coming from that theory.}:
\begin{equation}
 -\frac{\partial^2\tilde{\phi}}{\partial
t^2}+\nabla^2\tilde{\phi}=m_0^2\tilde{\phi}-3\frac{\beta}{\mpl}
\rho_0\zeta-\frac{\beta}{\mpl}r\frac{\partial }{\partial r}(\rho_0\zeta),
\end{equation}
where $m_0^2\equiv V_{\phi\phi}(\phi_0)$ is the mass of the unperturbed
field at zero density and equation (\ref{eq:contpert}) has been used. We are
interested in stationary-wave solutions and so we expand
\begin{align}
 \zeta(r,t)&=\xi(r)e^{i\omega t}\\
\tilde{\phi}(r,t)&=\varphi(r)e^{i \omega t}
\end{align}
to yield two coupled equations
\begin{align}
(\nabla^2+\omega^2)\varphi = m_0^2\varphi-3\frac{\beta}{\mpl}\rho_0\xi-&\frac{
\beta}{\mpl}r\frac{\partial }{\partial r}(\rho_0\xi),\label{eq:varphi}\\
\frac{\dd}{\dd r}\left(r^4\Gamma_{1,0}P_0\frac{\dd \xi}{\dd
r}\right)+r^3\frac{\dd}{\dd
r}\left[\left(3\Gamma_{1,0}-4\right)P_0\right]\xi&-\frac{r^4\beta_0\rho_0}{\mpl}
\frac{\dd^2\phi_0}{\dd r^2}\xi\nonumber\\-2\frac{r^3\beta_0\rho_0}{\mpl}\frac{\dd
\phi_0}{\dd r}\xi+r^4\rho_0\omega^2\xi=\frac{r^3\beta_0\rho_0}{\mpl}
\frac { \dd \varphi}{\dd r}.&\label{eq:MLAWE}
\end{align}
Equations (\ref{eq:varphi}) and (\ref{eq:MLAWE}) constitute the main result of
this section. One could combine them into a single equation, however, it is more
instructive to treat the system as two coupled equations. In general relativity, we have
$\varphi=\beta_0=\dd\phi_0/\dd r = 0$ and (\ref{eq:MLAWE})
reduces to
\begin{equation}\label{eq:LAWE}
\frac{\dd}{\dd r}\left(r^4\Gamma_{1,0}P_0\frac{\dd \xi}{\dd
r}\right)\nonumber+r^3\frac{\dd}{\dd
r}\left[\left(3\Gamma_{1,0}-4\right)P_0\right]\xi+r^4\rho_0\omega^2\xi=0,
\end{equation}
which describes linear, adiabatic, radial waves moving in the stellar interior.
It is known at the \textit{linear adiabatic wave equation} (LAWE) and its
eigenfrequencies $\omega^2$ give the frequency of stellar oscillations about
equilibrium. We will hence refer to (\ref{eq:MLAWE}) as the \textit{modified
linear adiabatic wave equation} (MLAWE). It's properties will be the subject of
the next section. Note that the effects of modified gravity alter the oscillation properties of the star via the MLAWE in two
different ways. Firstly, the MLAWE contains extra terms not present in the general relativity problem. These represent the effects
of perturbing the modified gravity hydrodynamics. In particular, the new term proportional to $\dd\phi_0/\dd r\sim G(r)$ encodes
the effects of the radial-dependence of Newton's constant not present in general relativity. The term
proportional to $\dd\phi_0/\dd r\sim \dd G(r)/\dd r$ encodes the effects of the rate of change of this constant.
Steeper gradients make it more difficult for acoustic waves to propagate and this will have important consequences for the
stability of unscreened stars. Secondly, the equilibrium profiles represented by $P_0$ etc. are computed using
the modified equations of stellar structure. We will discern how each of these contributes to the change in the oscillation period
below.

Using the profile (\ref{eq:gprof}), we have
\begin{align}\label{eq:mgcham}
\frac{\beta_0}{\mpl}\left[2\frac{\dd\phi_0}{\dd r}+r\frac{\dd^2\phi_0}{\dd
r^2}\right]=4\pi\alpha Gr\rho_0(r)\quad r>\rs,
\end{align}
which we shall use in all analytic and numerical computations from here on.

\section{Properties of the Modified Linear Adiabatic Wave
Equation}\label{sec:MLAWE}

The MLAWE describes the behaviour of stellar oscillations in modified
gravity. There are two major differences with respect to the general relativity equation.
Firstly, there are two additional terms in the homogeneous part, proportional to
the first and second derivatives of the background field. When $r<\rs$ these
are negligible and the homogeneous part behaves as it would in general relativity,
however, these are comparable to the other terms when $r>\rs$ and
encode the effect of modified gravity on wave propagation in the region
exterior to the screening radius. Physically, the term proportional to $\dd
\phi_0/\dd r$ acts as a varying enhancement of Newton's constant $G(r)$ given
by equation (\ref{eq:gprof}) and the term proportional to $\dd^2
\phi_0/\dd r^2$ can schematically be viewed as $\dd G(r)/\dd r$ and so it
encodes the effect of a radially varying Newtonian force in the outer regions.

The second effect is a driving term proportional to $\beta_0\dd \varphi/\dd r$.
This is clearly the effect of the fifth-force due to perturbations in the
field. This was modelled by \cite{Silvestri:2011ch} as an inhomogeneous forcing
term at a single frequency. Here it appears as coupling between the field and
stellar perturbations: the stellar perturbations source the scalar field
perturbations and vice versa. Physically, $\tilde{\phi}(r,t)$ corresponds to
scalar radiation (or rather the flux $T_{\phi\,0i}$ at infinity). As mentioned in the introduction to this chapter, there is
evidence from previous studies \cite{Silvestri:2011ch,Upadhye:2013nfa} that
this is negligible in chameleon-like systems and so from here on we will neglect
the dynamics of the field perturbations and treat only the homogeneous part of
the MLAWE (\ref{eq:MLAWE}). 

\subsection{Boundary Conditions}\label{sec:BCS}

The MLAWE requires two boundary conditions in order to fully specify the
solution given a specific value of $\omega$\footnote{Of course, we also need to
derive the value of $\omega^2$ from the solution of the equation. This can be
done by looking for values such that the solutions satisfy both boundary
conditions and is discussed in section \ref{sec:evalue}.}. Firstly, our system
is spherically symmetric and so we must impose $\delta r=0$ at $r=0$. The MLAWE
then requires
\begin{equation}\label{eq:BCC}
 \left.\frac{\dd \xi}{\dd r}\right\vert_{r=0}=0.
\end{equation}
The surface boundary condition depends on the stellar atmosphere model (see
\cite{cox1980theory} for a discussion) but the lowest modes, where the
period of oscillation is longer than the inertial response time of the
atmosphere can be described by solutions with vanishing surface pressure so that $\delta P(R)=0$.
This gives the surface condition \cite{cox1980theory}
\begin{equation}\label{eq:BCS}
 \left.\frac{\delta P}{P_0}\right\vert_{r = 0}=\left(\frac{\omega^2R^3}{GM}+4\right)\zeta(R),
\end{equation}
where the Lagrangian pressure perturbation is given by (\ref{eq:lagpresspert}).
Note that the additional terms in the MLAWE vanish at the stellar centre and
radius if we take $\rho_0(R)=0$ so that these conditions are identical to those required by general relativity.

\subsection{Sturm-Liouville Nature of the Equation}\label{sec:evalue}

The MLAWE can be written in Sturm-Liouville form
\begin{equation}
 \hat{\mathcal{L}}\xi + w(r)\omega^2\xi,=0
\end{equation}
where the weight function $w(r)=r^4\rho_0(r)$ and the operator can be written
\begin{align}
 \hat{\mathcal{L}}^{\rm GR}&=\frac{\dd}{\dd
r}\left(r^4\Gamma_{1,0}P_0\frac{\dd}{\dd r}\right)+r^3\frac{\dd}{\dd
r}\left[\left(3\Gamma_{1,0}-4\right)P_0\right],\label{eq:lhatGR}\\
\hat{\mathcal{L}}^{\rm MG}&=
\hat{\mathcal{L}}^{\rm GR}-\frac{\beta\rho_0}{\mpl}r\frac
{\dd^2\phi_0}{\dd r^2}-2\frac{\beta\rho_0}{\mpl}\frac{\dd \phi_0}{\dd r}.\label{eq:lhatmg}
\end{align}
The problem of finding the pulsation frequencies is then one of finding the
eigenvalues of these equations that correspond to eigenfunctions satisfying the
boundary conditions (\ref{eq:BCC}) and (\ref{eq:BCS}). In practice, it is not
possible to solve these equations analytically for physically realistic stars
and numerical methods must be used. We will do just this in section
\ref{sec:numerics}. Despite the need for numerics, a lot of the new modified
gravity features can be discerned and elucidated using well-known
Sturm-Liouville techniques and so we shall investigate these first. 

\subsection{Scaling Behaviour of the Eigenfrequencies}\label{sec:scaling}

Using the dimensionless quantities:
\begin{align}
 \bar{P}_0(r)&\equiv\frac{R^4}{GM^2}P_0(r),\label{eq:dimlessP}\\ 
\bar{\rho}_0(r)&\equiv\frac{R^3}{M}\rho_0(r)\quad\textrm{and}
\label{eq:dimlessrho} \\ x&\equiv
\frac{r}{R},\label{eq:dimlessr}
\end{align}
the MLAWE (\ref{eq:MLAWE}) can be cast into dimensionless form:
\begin{align}\label{eq:dimlessLAWE}
 &\frac{\dd}{\dd x}\left(x^4\gao\bar{P}_0\frac{\dd \xi}{\dd
x}\right)\\&+x^3\frac{\dd}{\dd
x}\left[\left(3\gao-4\right)\bar{P}_0\right]\xi+x^4\bar{\rho}_0\left[
\Omega^2-4\pi\alpha\bar{\rho}_0\right]=0,
\end{align}
where
\begin{equation}\label{eq:dimOmega}
 \Omega^2\equiv\frac{\omega^2R^3}{GM}
\end{equation}
is the dimensionless eigenfrequency and the term proportional to $\alpha$ is
only present when $r>\rs$. In general relativity, $\alpha=0$ and one can solve this given some
equilibrium stellar model to find $\Omega^2$. Since this must be a
dimensionless number one has $\omega^2\propto GM/R^3$. In modified gravity, there are two independent effects that act to
change this value at fixed mass: the change due to the different
equilibrium structure and the change due to the
additional term in the MLAWE. At the level of the background,
we expect that stars of fixed mass have smaller radii and larger values of
$\langle G\rangle$ (where by $\langle\rangle$ we mean some appropriate average
over the entire star) so that the frequencies are higher in modified gravity. At the
level of perturbations, one can replace $\Omega^2$ in the general relativity equation by the
effective frequency $\Omega^2-4\pi\alpha\langle\bar{\rho}_0\rangle$ so that
$\Omega^2_{\rm MG}\approx \Omega^2_{\rm
GR}+4\pi\alpha\langle\bar{\rho}_0\rangle$ and we therefore expect the modified gravity
eigenfrequency to be larger still.

One can gain some insight by considering
scaling relations in a similar manner to the methods of section \ref{sec:scalaing} when the
star is fully unscreened so that $G\rightarrow G(1+\alpha)$. Let
us assume a polytropic equation of state of the form 
\begin{equation}\label{eq:eosgam}
P=K\rho^\gamma,
\end{equation}
where $K$ is a
constant and $\gamma$ differs from $\gao$ since the system need not be
adiabatic. In this case, equations (\ref{eq:masscons}) and (\ref{eq:MGHSE})
give
\begin{align}
 \rho_{\rm c}&\propto \frac{M}{R^3}\\
\rho_{\rm c}^{\gamma-1}&\propto \frac{GM^2}{R},
\end{align}
which can be combined to find the scaling of the radius for a fully-unscreened
star in modified gravity:
\begin{equation}\label{eq:Rscaling}
 \frac{R_{\rm MG}}{R_{\rm GR}}=(1+\alpha)^{-\frac{1}{3\gamma-4}}
\end{equation}
at fixed mass. Ignoring the modified gravity perturbations, one would
then expect
\begin{equation}\label{eq:backgroundomegascaling}
 \frac{\omega^2_{\rm MG}}{\omega^2_{\rm
GR}}=(1+\alpha)^{\frac{3\gamma-1}{3\gamma-4}}.
\end{equation}
We shall confirm this limit numerically for some simple models later in section
\ref{sec:LEmodels}. In the fully unscreened limit, we would then expect the
eigenfrequencies to scale approximately like
\begin{equation}
 \frac{\omega^2_{\rm MG}}{\omega^2_{\rm
GR}}\sim(1+\alpha)^{\frac{3\gamma-1}{3\gamma-4}}\left(1+
\frac{4\pi\alpha}{\Omega_{\rm GR}^2}\langle\bar
{ \rho }_0\rangle\right)
\end{equation}
so that they are always larger than the general relativity prediction (assuming $\gamma>4/3$), at least when
$\omega^2_{\rm GR}>0$.

\section{Stellar Stability}\label{sec:stellarstability}

Given the Sturm-Liouville nature of the problem, we can find an upper bound on
the fundamental frequency $\omega_0$ using the variational principle. Given an
arbitrary trial function $\Psi(r)$, one can construct the functional
\begin{equation}
 F[\omega]\equiv-\frac{\int_0^R\dd r \,\Psi^*(r)\hat{\mathcal{L}}\Psi(r)
}{\int_0^R\dd r\,\Psi^*(r)\Psi(r)\rho_0r^4},
\end{equation}
which has the property that $\omega_0^2\le F[\omega]$. Ignoring modified gravity for now and
taking the simplest case where $\chi$ is constant, the fundamental
eigenfrequencies of the LAWE (\ref{eq:LAWE}) satisfy
\begin{equation}\label{eq:GRstab}
\omega_0^2\le \frac{\int_0^R\dd r\,3r^2(3\Gamma_{1,0}-4)P_0}{\int_0^R\dd
r\,\rho_0r^4},
\end{equation}
where we have used $\hat{\mathcal{L}}= \hat{\mathcal{L}}^{\rm GR}$ defined in (\ref{eq:lhatGR}).
When the right hand side is negative we have $\omega_0^2<0$ and the
eigenfunctions have growing modes. This is the well-known result in stellar
astrophysics that stars where the first adiabatic index falls below $4/3$ are unstable
to linear perturbations and cannot exist\footnote{Corrections from general
relativity increase this critical value to $4/3+\mathcal{O}(1) GM/R$
\cite{Chandrasekhar:1964zza}, where the $\mathcal{O}(1)$ factor
depends on the specific composition of the star. We are interested in the
properties of main-sequence stars with $GM/R\sim 10^{-6}$ and Cepheid stars with
$GM/R\sim 10^{-7}-10^{-8}$ and so this correction is always negligible compared
with the effects of modified gravity, which are of the same order as the non-relativistic
contribution when the star is unscreened.}.

In modified gravity, we have $\hat{\mathcal{L}}= \hat{\mathcal{L}}^{\rm MG}$ and so using equation (\ref{eq:lhatmg}) we have
\begin{equation}\label{eq:modstab}
 \omega_0^2\le\frac{\int_0^R\dd
r\,3r^2\left[(3\Gamma_{1,0}-4)P_0\right]+\int_{\rs}^R\frac{\beta\rho_0}{\mpl}\left(r^4\frac{
\dd^2\phi_0}{\dd r^2}+2r^3\frac{\dd\phi_0}{\dd r}\right)}{\int_0^R\dd r\,\rho_0 r^4}
\end{equation}
and so this stability condition is altered in stars which are at least
partially unscreened. This is not surprising given the form of equation
(\ref{eq:MLAWE}). The term proportional to the derivative of $[(3\gao-4)P_0]$
behaves like a position-dependent mass for $\xi$, which is negative when
$\gao<4/3$ so that one would expect growing modes. The additional terms in
(\ref{eq:modstab}) are due to the two new terms in (\ref{eq:MLAWE}), which are
of precisely the same varying mass form with the opposite sign. A negative mass,
which would signify an instability, coming from the general relativity term can then be
compensated by the new terms in modified gravity, restoring stability\footnote{The reader may
wonder why the star is more stable in modified gravity when the stability criterion in general relativity
does not change if one changes the value of $G$. When the star is unscreened we
have $G\rightarrow G(1+\alpha)$ and so one may expect any modified gravity
effects to vanish in this limit. In general relativity, there is an exact cancellation coming
from the perturbations to the momentum equation (\ref{eq:momentumgen}) and the
Newtonian potential. In modified gravity, the additional gravitational force is not derived
from the Newtonian potential but from the field profile and so any cancellation
must come from the perturbation to the field equation. A priori, there is no
reason why the field perturbations should cancel this new contribution and,
indeed, we see here that they do not.}.

Physically, the first adiabatic index is a measure of how the pressure responds to a
compression of the star. Given a compression from one radius $R_1$ to a smaller
radius $R_2$, a larger adiabatic index will result in more outward pressure.
If this increase in pressure is faster than the increase in the gravitational
force, the star can resist the compression and is hence stable. Below the
critical value of $4/3$, the converse is true and the star is unstable. We have
already seen above that the new terms contributing to the stability correspond
to a varying value of $G$ and its derivative in the outer layers. Since
modified gravity enhances the gravity, one may n\"{a}ively expect that its
effect is to destabilise stars, however, we will argue below that this is not
the case. The MLAWE describes acoustic waves propagating in the star. If the force of gravity and its gradient is larger in the
outer layers then it is more
difficult for these waves to propagate and hence modes which would usually have
been unstable are stabilised.

Once again, we must disentangle the effects of the modified equilibrium
structure and the perturbations on the critical value of $\gao$. Consider first
the modified equilibrium structure only. In this case, the stability condition
is given by the general relativity expression, equation (\ref{eq:GRstab}), however the pressure
and density profiles will be different. Clearly, the critical value for the
instability is still $4/3$ since this is the only value which makes the
integral vanish but this does not necessarily mean the stability is altered
away from this value. Scaling the pressure and density using the dimensionless
quantities defined in (\ref{eq:dimlessP}), (\ref{eq:dimlessrho}) and
(\ref{eq:dimlessr}), we have
\begin{equation}\label{eq:equildestab}
 \omega_0^2\le(3\gao-4)\frac{GM}{R^3}f(\chi_0,\alpha),
\end{equation}
 where $f$ is a dimensionless function which depends on the composition of
the star\footnote{The reader should note that this is a varying function of
$\chi_0$ and $\alpha$ and so is not universal.}. In modified gravity, the radius of the star will be
smaller than its general relativity counterpart and the effective value of $G$ is larger.
Hence, when $\gao>4/3$ the maximum possible frequency is greater than in general relativity
whereas when $\gao<4/3$ the maximum frequency is more negative. If a star is
unstable in general relativity then modified gravity enhances the instability, moving the frequency further
away from zero. This can also be seen from the scaling relation
(\ref{eq:backgroundomegascaling}). If $\omega_{\rm GR}^2<0$ then $\omega_{\rm
MG}^2$ is even more negative. At the background level, the effects of modified gravity are to
destabilise stars that are already unstable, without altering the stability
condition.

Let us now turn our attention to the effects of perturbations. Using equation
(\ref{eq:mgcham}), we have
\begin{equation}\label{eq:mgstab2}
 \omega_0^2\le\frac{\int_0^R\dd
r\,3r^2\left[(3\Gamma_{1,0}-4)P_0\right]+\int_{\rs}^R4\pi\alpha Gr^4\rho_0^2}{\int_0^R\dd
r\,\rho_0r^4}.
\end{equation}
The additional term is clearly positive and so one may lower the value of
$\gao$ below $4/3$ and still find positive eigenfrequencies, confirming our
earlier intuition that the effect of modified gravity is to stabilise stars compared with general relativity. We will denote the
critical index by $\Gamma_{1,0}^{\rm crit}$. When $\Gamma_{1,0}<\Gamma_{1,0}^{\rm crit}$ the star is unstable to linear
perturbations. Unlike general relativity, there is no universal critical index in modified gravity. The precise value depends on
$\chi_0$ and
$\alpha$ and is composition-dependent. We will investigate the stability of some simple semi-analytic models in
section \ref{sec:stab_LE} but before doing so, one can gain some insight
into the full effects of modified gravity on the stability by using the same scaling relations
as section \ref{sec:scaling}. In particular, we can set the numerator in
(\ref{eq:mgstab2}) to zero to find the modified critical index:
\begin{equation}\label{eq:newgammacrit}
 \gao^{\rm critical}=\frac{4}{3}-g(\chi_0,\alpha)\alpha,
\end{equation}
where 
\begin{equation}\label{eq:gstabdef}
g(\chi_0,\alpha) =\frac{4\pi G}{3}\frac{\int_{\rs}^Rr^4\rho_0^2(r)}{\int_0^R r^2P_0(r)}      
\end{equation}
is a dimensionless function that encodes the effects of the
structure of the star. In $f(R)$ theories, $\alpha = 1/3$
\cite{Brax:2008hh} and so we expect the critical value of $\gao$ to
change by $\mathcal{O}(10^{-1})$ assuming that $g\sim\mathcal{O}(1)$. We will
verify numerically that this is indeed the case for a simple model in
section \ref{sec:stab_LE}. 

\section{Numerical Results}\label{sec:numerics}

We now proceed to solve the MLAWE for various different stars. We will do
this for two different stellar models: Lane-Emden and MESA models. The first models are simple compared with
the second but they have the advantage that the non-gravitational physics (e.g.
nuclear burning) is absent, which will allow us to gain a lot of physical
intuition about the new modified gravity features. They are simple
semi-analytic models and this allows us to first investigate the MLAWE
using a controlled system with known scaling properties and limits without
the complications arising from processes such as radiative transfer. This also allows
us to test that the code is working correctly since we can compare our results
with both the general relativity case, which has been calculated previously, and the fully
unscreened case, which can be predicted analytically given the general relativity one. Their
perturbations can also be described using an arbitrary value of $\gao$,
independent of their composition and so we will use them to study the
modifications to stellar stability. These models are not realistic enough to
compare with observational data and the power of MESA lies in that
it can produce realistic models of stars such as main-sequence and Cepheid
stars, which will allow us to predict the effects of modified gravity on realistic stars in
unscreened galaxies. In the previous chapter, we used MESA predictions to
obtain the strongest constraints on chameleon-like models to date. Combining these models with the modified gravity oscillation
theory has the
potential to provide even tighter constraints.

Details of the numerical procedure used to
solve the MLAWE are given in appendix \ref{app:nto}. The shooting method has been used to
solve the MLAWE in all instances.

\subsection{Perturbations of Lane-Emden Models}\label{sec:LEmodels}

In this section we numerically solve the MLAWE for stars whose equilibrium configurations are given by the Lane-Emden models
described in section \ref{eq:LEStarsMG}.

\subsubsection{Oscillation Periods of Lane-Emden Models}\label{sec:LEperts}

We solve the MLAWE by first tabulating solutions of the modified
Lane-Emden equation and using these to numerically solve the MLAWE. The dimensionless eigenvalues (see section \ref{sec:lawele}
in appendix \ref{app:nto} for a derivation and discussion of this expression) 
\begin{equation}\label{eq:omtild}
 \tilde{\omega}^2\equiv \frac{(n+1)\omega^2}{4\pi G \rho_{\rm c}}
\end{equation}
for Lane-Emden models in general relativity were numerically calculated in 1966 by
\cite{Hurley:1966} and so as a code comparison, we have compared our fundamental
frequencies and first overtones with theirs for different values of $n$ and
$\gao$. Their values are given to five decimal places and in each case our
results agreed with theirs to this accuracy.

Self-similarity is not completely preserved in modified gravity Lane-Emden models. This means that given values of
$\chi_0$ and $\alpha$ we are not free to choose the mass and radius of the modified gravity star to be identical to those of the
general relativity star because the ratio $GM/R$ is constrained via equation (\ref{eq:X}). Therefore, given a star in general
relativity of mass $M$ and radius $R$, one must decide upon the correct comparison in modified gravity. In what follows, we will
fix the mass and composition (this implies that $K$ is fixed)
of the star and allow the radius to vary so that the stars we compare are stars
of the same mass whose radii (and pressure and density profiles) have adjusted
to provide an equilibrium configuration given a specific value of $\chi_0$ (and
hence $\rs$)\footnote{This is not possible in the case $n=3$ since there is
no mass-radius relation. The absence of such a relation has the result
that the constant $K$ must vary as a function of stellar mass, $\chi_0$ and
$\alpha$. Indeed, this is what we found in chapter \ref{chap:three} where we studied the Eddington standard model. For this
reason, there is no meaningful way to compare perturbations
of stars in modified gravity since one is comparing stars with different equations of state.
For this reason, we will not consider perturbations of these models.}. In particular, this means that it is not possible (or
rather not meaningful) to compare stars of fixed $GM/R$ since these stars have different masses and radii to their general
relativity ($\alpha=0$) counterparts. This means that the stars we compare have fixed masses but different Newtonian
potentials\footnote{This is another illustration of how including the effects of the fifth-force on the structure of an object
can violate the condition that $\pn\sim\chi_0$ implies screening. The field-profile is only set once one knows the structure of
the object and, as we have seen here, even when observable properties (such as the mass) are fixed the Newtonian potential may
vary if the star is unscreened and a n\"{a}ive estimate using the general relativity value may not give the correct screening
radius.}. In order to fix the mass, one must fix the central density and so in modified gravity we have
$\rho_{\rm c}=\rho_{\rm c}(\chi_0,\alpha)$, highlighting the consequences of
breaking self-similarity. For concreteness, we will work with $n=1.5$. This is a
good approximation to stellar regions which are fully convective (see
\cite{kippenhahn1990stellar}, sections 7 and 13 for more details) and hence have
physical applications to red giant and Cepheid stars\footnote{The $n=1.5$ polytrope model is a far better approximation to red
giant stars than Cepheids.}. In terms of an equation of
state of the form (\ref{eq:eosgam}), this model corresponds to $\gamma=5/3$. In
what follows, we will assume that $\gamma$ is identical to the first adiabatic index
and set $\gao=5/3$, however, we will relax this assumption when considering
stellar stability and allow for more general models.

Using equation (\ref{eq:LEMASS}), we have
\begin{equation}\label{eq:rhocmg}
 \rho_{\rm c}=\left(\frac{M}{4\pi}\right)^2\left[\frac{8\pi
G}{2K}\right]^{3}\left(\frac{1+\alpha}{\omr+\alpha\oms}\right)^2,
\end{equation}
which may be used to find the modified Mass-Radius-$\chi_0$ relation (as
opposed to the Mass-Radius relation found in general relativity).
\begin{equation}\label{eq:MGR-M}
 R=\frac{1}{\left(4\pi\right)^{\frac{3}{2}}}\left(\frac{5K}{2G}
\right)\left(\frac{\omr+\alpha\oms}{1+\alpha}\right)^{\frac{1}{3}}.
\end{equation}
In the cases of red giant stars and low-mass Cepheids, we have
$GM/R\sim10^{-7}$ and so we can pick $M=M_{\rm GR}$ and $R=R_{\rm GR}$ in the general relativity
case ($\alpha=0$) such that $GM_{\rm GR}/R_{\rm GR}=10^{-7}$ and
\begin{equation}
 \frac{GM}{R}=10^{-7}\left(\frac{M_{\rm GR}}{M}\right)\left(\frac{R}{R_{\rm
GR}}\right)
\end{equation}
and equation (\ref{eq:X}) is (fixing $M=M_{\rm GR}$ and recalling that $\bar{y}_R$ is the value of $y_R$ found using general
relativity)
\begin{equation}\label{eq:scrfind}
\frac{\chi_0}{10^{-7}}=
\frac{\bar{y}_R}{y_R}\left[\frac{y_R\thn(y_{\rm
s})+\omr-\frac{y_R}{y_{\rm
s}}\oms}{\left(\bar{\omega}_R(1+\alpha)\right)^{\frac{1}{3}}
\left(\omr+\alpha\oms\right)^{ \frac{2}{3} }} \right ].
\end{equation}
The procedure is then as follows: Given specific values of $\chi_0$ and $\alpha$, we use
a trial value of $y_{\rm s}$ to solve the Modified-Lane Emden equation until
equation (\ref{eq:scrfind}) is satisfied. We then use the Lane-Emden solution
in the MLAWE to numerically calculate the value of $\tilde{\omega}^2$ given a
value of $\gao$. Using equation (\ref{eq:rhocmg}), we can find the ratio
of the period in modified gravity to that predicted by general relativity:
\begin{equation}\label{eq:taurat}
\frac{\tau_{\rm MG}}{\tau_{\rm GR}}=\sqrt{\frac{\tilde{\omega}^2_{\rm
GR}}{\tilde{\omega}^2_{\rm
MG}}}\frac{\omr+\alpha\oms}{\bar{\omega}_{R}(1+\alpha)}.
\end{equation}
We can then calculate this ratio for any value of $\chi_0$ and $\alpha$. 

Before presenting the numerical results, it is worth noting that the
fully-unscreened behaviour of the star, at least in the case where only the
effects of the modified equilibrium structure are considered, can be calculated
in terms of the general relativity properties of the star. In the fully-unscreened case, one
has $\oms=y_{\rm s}=0$. One can then set $y\rightarrow
(1+\alpha)^{-\frac{1}{2}}y$ to bring equation (\ref{eq:MGLE}) into the same form
as in general relativity. This then gives $y^{\rm unscreened}_R =
(1+\alpha)^{-\frac{1}{2}}\bar{y}$ and $\omr^{\rm
unscreened}=(1+\alpha)^{-\frac{1}{2}}\bar{\omega}_R$. For $n=1.5$, one has
$\bar{y}_R\approx 3.654$ and $\bar{\omega}_{\rm R}\approx2.72$. One then has, by
rescaling the MLAWE (see appendix \ref{app:nto} for the equation in these coordinates),
$\tilde{\omega}^2_{\rm{ unscreened}}/\tilde{\omega}_{\rm GR}^2=(1+\alpha)$ and
so, using (\ref{eq:rhocmg}), $\omega^2_{\rm{ unscreened}}/\omega_{\rm
GR}^2=(1+\alpha)^4$, which exactly matches our prediction in
(\ref{eq:backgroundomegascaling}) for $\gamma=5/3$. From equation
(\ref{eq:taurat}), we have $\tau^{\rm unscreened}_{\rm MG}/\tau_{\rm
GR}=(1+\alpha)^{-2}=0.5625$ for $\alpha=1/3$. These unscreened results can be
used to check that the numerical results are behaving as expected\footnote{One should note that whereas (\ref{eq:taurat}) is an
exact analytic expression it is not possible to evaluate its numerical value analytically. One must still solve the modified
Lane-Emden equation and the MLAWE for quantities such as $\omr$ and $\tilde{\omega}^2_{\rm GR}$.}. 

We would like to investigate the effect of the different modifications coming
from the altered equilibrium structure and the modified perturbation equation
separately. They appear at the same order in the MLAWE and so we expect them to
contribute equally and it is important to disentangle their effects,
especially since we have already argued in section \ref{sec:stellarstability}
that they may contribute differently to the stellar stability and that the
equilibrium structure acts to make negative general relativity frequencies more negative.
Lane-Emden models are perfectly suited for this study since we do not have to
worry about altered evolution histories and so we will consider two cases:
\begin{itemize}
\item Case 1: We solve the LAWE (\ref{eq:LAWE}) using modified Lane-Emden
profiles. This case only includes the modified equilibrium structure.
\item Case 2: We solve the full MLAWE using the modified background
structure. This is the physically
realistic case.
\end{itemize}
The case where we ignore the modified equilibrium structure and solve the
MLAWE is highly unphysical. There is no screening radius and so we are
introducing the perturbations about an arbitrary radius. Furthermore, the size
of the effect depends on whether we take the radius of the star as
corresponding to the general relativity solution (in which case we are ignoring the effects on
the period coming from the change in the central density (\ref{eq:rhocmg})) or
the equivalent modified gravity solution (in which case our profiles do not satisfy the
boundary conditions and we are including some of the modified equilibrium
properties in our analysis). For these reasons, we do not investigate this
scenario.  In each case, we assume the profile (\ref{eq:gprof}). Since we have
solved the modified Lane-Emden equation with constant $\alpha$ in the region
$r>\rs$ (this is required for a physically meaningful comparison with general relativity) this
profile is not technically correct since it does not satisfy the stellar
structure equations. In fact, it is an under-estimate \footnote{Compared to the
result that would be obtained if we had used a fully-unscreened profile. The
equilibrium structure is still a small over-estimate of the effects of modified gravity. For
our purposes, this is not an issue since we seek only to investigate the new
effects of modified gravity oscillations and do not compare any of these results with real
stars. When we analyse MESA predictions we will use a fully
consistent approach.}. In each case we will fix $\alpha=1/3$, corresponding to
$f(R)$ gravity and vary $\chi_0$.

\begin{figure}
\centering
\includegraphics[width=15cm]{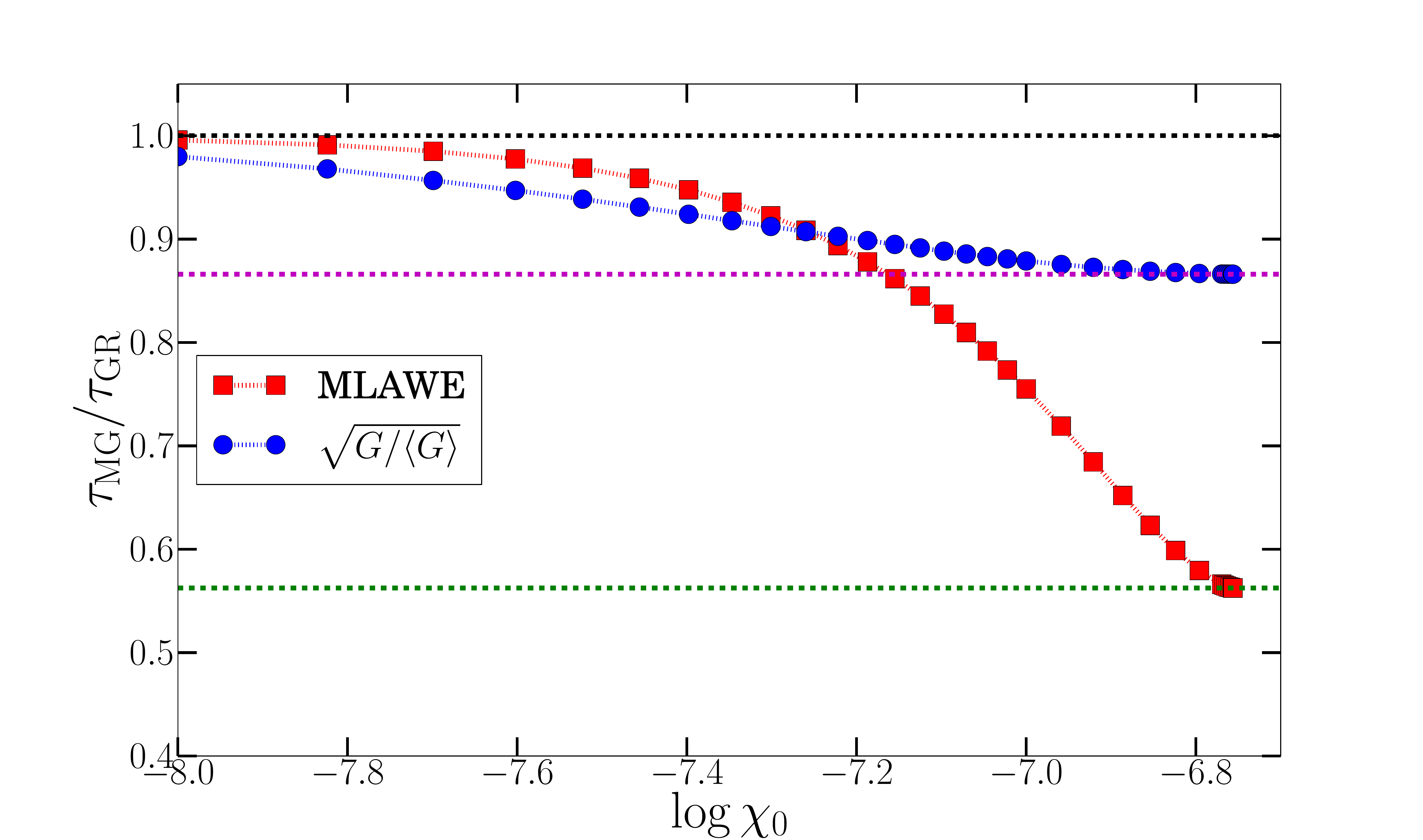}
\caption{The fractional change in the stellar pulsation period as a function of
$\log_{\rm 10}\chi_0$ when only the change to the equilibrium structure is
considered (case
1). The red squares correspond to eigenfrequencies of the LAWE whereas the blue
circles show the approximation (\ref{eq:periodapprox}). The green dashed line
shows the ratio for a fully unscreened star and the black dashed line shows a
ratio of 1, corresponding to a general relativity star. The magenta line shows the
fully-unscreened value of $\sqrt{\frac{G}{\langle
G\rangle}}=(1+\alpha)^{-\frac{1}{2}}$.}\label{fig:noperts}
\end{figure}

In figure \ref{fig:noperts} we plot the ratio of the modified gravity period to the general relativity
one as a function of $\log\chi_0$ for case 1. In chapter \ref{chap:four}, the approximation
\begin{equation}\label{eq:periodapprox}
\frac{\tau_{\rm MG}}{\tau_{\rm GR}}=\sqrt{\frac{G}{\langle G\rangle}},
\end{equation}
where $\langle G\rangle$ is the average value of the effective Newtonian
constant using the Epstein weighting function was used in order to obtain
new constraints on the model parameters. This was found by perturbing the general relativity prediction that $\tau\propto
G^{-1/2}$ at fixed radius and describing the effects of the radial variation of $G$ using $\langle G\rangle$. This approximation
is based on the
change in the equilibrium structure only and so it is important to test not only
how it compares with the predictions from the full numerical prediction at the
background level but also how well it can be used to approximate the frequency
once modified gravity perturbations are taken into account. Hence, we also plot the prediction coming from this approximation. In
each case we have calculated $\langle
G\rangle$ using the modified Lane-Emden solution at given $\chi_0$. One should
emphasise that in chapter \ref{chap:four} we used this approximation for MESA models whereas this comparison is purely for the
hypothetical case of
Lane-Emden models. We will investigate how well this holds for MESA models in section \ref{sec:MESAperts}. The figure reveals that
the
approximation (\ref{eq:periodapprox}) is an over-estimate for very screened
stars whereas it is a large under-estimate for stars that are significantly
unscreened. The Epstein function favours the regions of Cepheid stars that are
most important for pulsations. This tends to be the outer layers and so it is
no surprise that it over-estimates the effects in the screened case: it places
a large emphasis on the small region where the gravity is enhanced even though
this region has little to no effect on the structure of the star. The
approximation (\ref{eq:periodapprox}) assumes that the stellar radius is fixed
but in Lane-Emden models this is clearly a decreasing function of
$\chi_0$ and $\alpha$ according to (\ref{eq:MGR-M}). According
to (\ref{eq:dimOmega}), the period scales as $R^{\frac{3}{2}}$, which explains
why the approximation is an underestimate when the star is very unscreened and
the change in the radius is significant. This is also the reason for the slow divergence of the two curves: since we are solving
the LAWE, the deviation of the two curves is driven entirely by the change in the stellar radius, which is not significant when
the star is very screened. As soon as the effects of the fifth-force are significant the curves begin to diverge. We can also see
that the ratio asymptotes to the value predicted in (\ref{eq:taurat}) when the star is fully-unscreened confirming that the
numerics are behaving as expected.

\begin{figure}
\centering
\includegraphics[width=15cm]{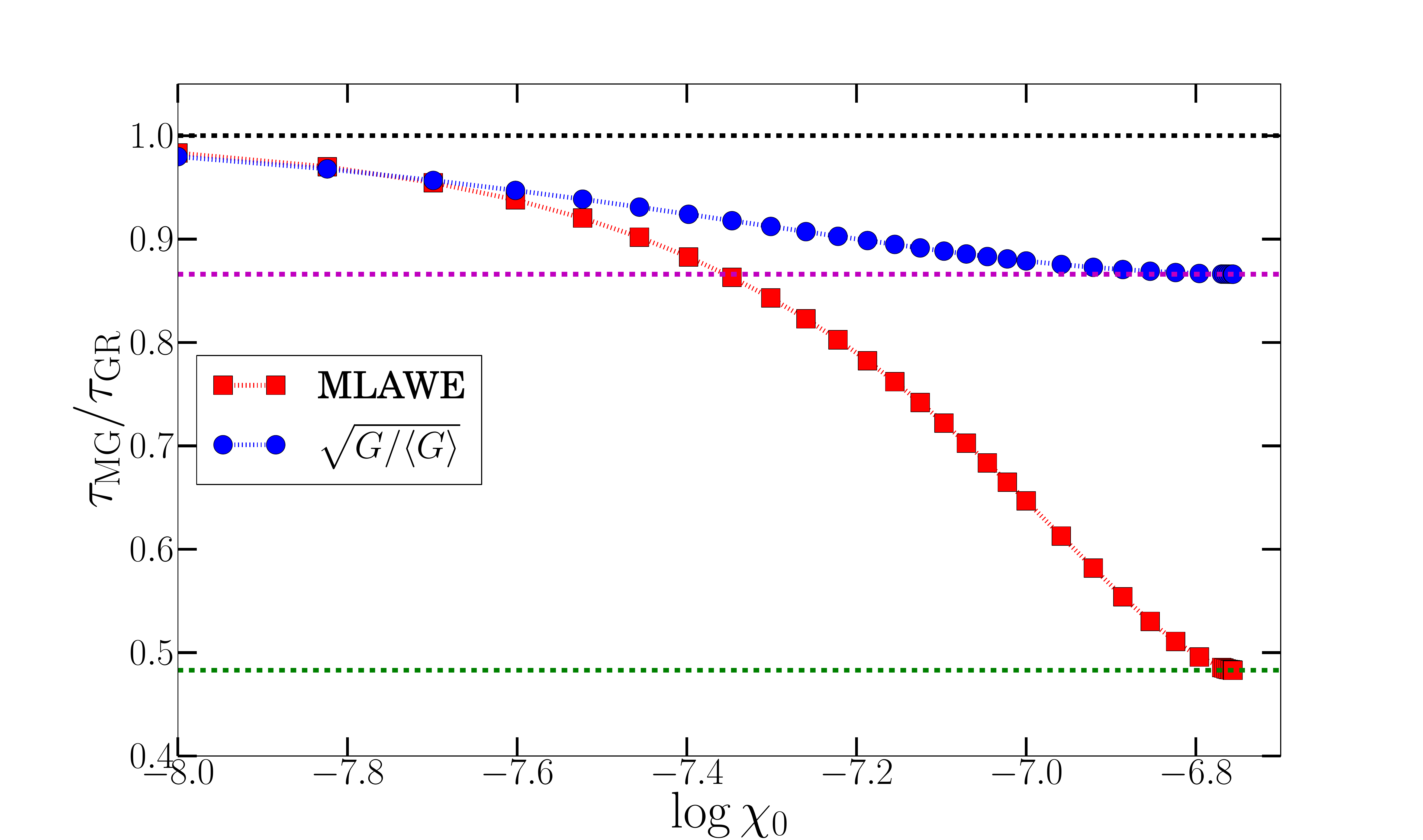}
\caption{The fractional change in the stellar pulsation period as a function of
$\log_{\rm 10}\chi_0$ when both the effects of the modified equilibrium
structure and
modified perturbation equation are considered (case 2). The red squares
correspond to eigenfrequencies of the MLAWE whereas the blue circles show the
approximation (\ref{eq:periodapprox}). The green dashed line shows the ratio for
a fully unscreened star and the black dashed line shows a ratio of 1,
corresponding to a general relativity star. The magenta line shows the fully-unscreened value
of $\sqrt{\frac{G}{\langle
G\rangle}}=(1+\alpha)^{-\frac{1}{2}}$.}\label{fig:case3}
\end{figure}

In figure \ref{fig:case3} we plot the ratio of the modified gravity period to the general relativity one for
case 2. We can see that the approximation (\ref{eq:periodapprox}) fails very
rapidly and that the change in the period is significant and can be as large
as 50\% for significantly unscreened stars. Unlike the previous case, we can see that the red and blue curves diverge very
rapidly, even when the star is very screened. This is because the blue curve is based entirely on the LAWE and does not capture
the additional $\mathcal{O}(1)$ effects included in the MLAWE. When the star is very screened any deviations from general
relativity are small so the curves agree well but as soon as the star becomes even slightly unscreened and the additional terms
in the MLAWE are very important, hence the rapid divergence. 

\begin{figure}\centering
\includegraphics[width=15cm]{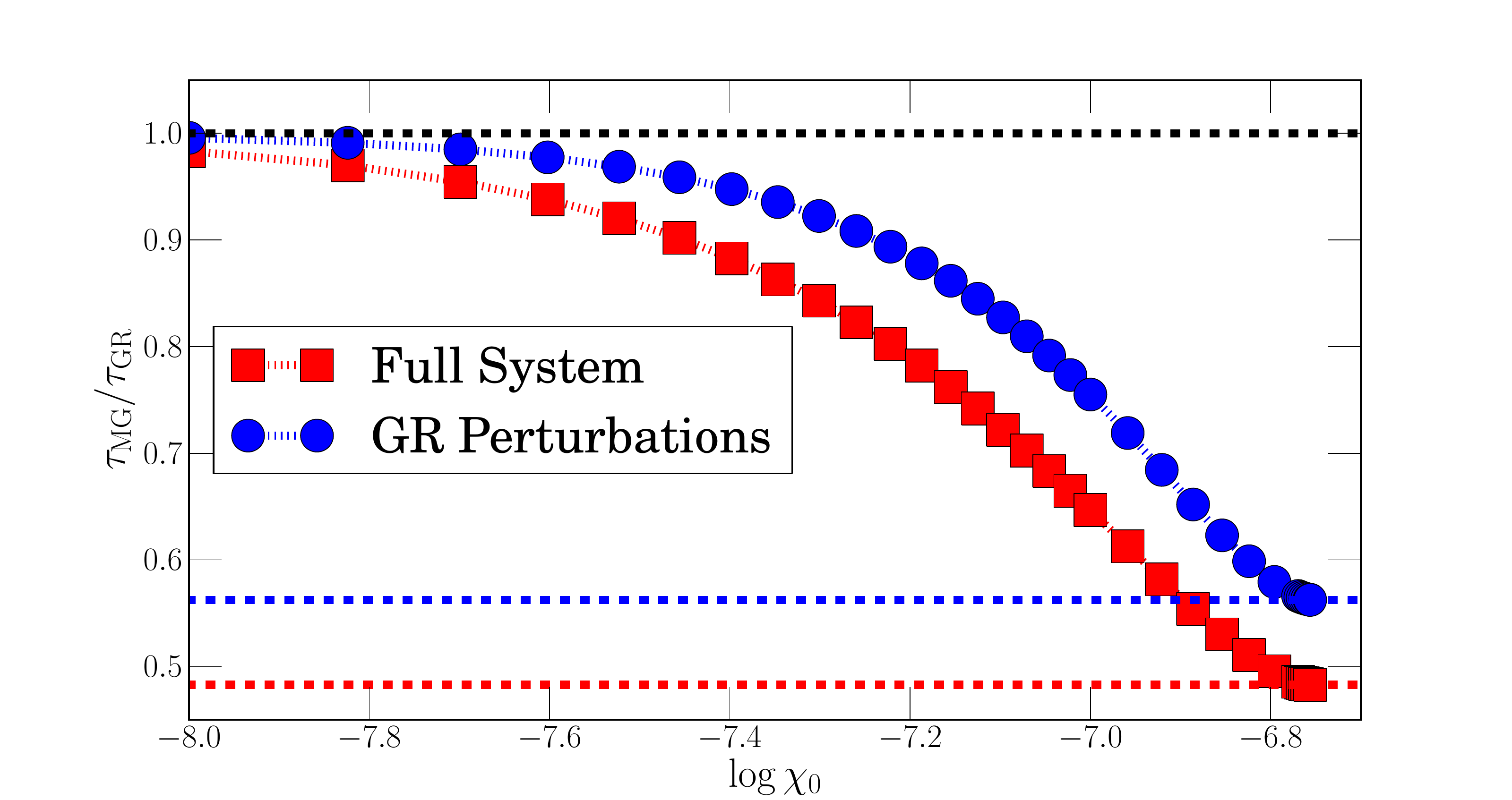}
\caption{The fractional change in the stellar pulsation period as a function of
$\log_{\rm 10}\chi_0$ for case 1 (blue circles) and case 2 (red
squares). The black dashed line shows the general relativity ratio of $1$ and the red and blue
lines show the fully-unscreened ratio for the
full simulation and the one including only the modified
equilibrium structure respectively.}\label{fig:all3}
\end{figure}

We plot the two cases together in figure \ref{fig:all3}. One can see that
the effect of the new terms coming from the modified structure of hydrodynamics
has a significant effect on the period and that if one were to consider only
the equilibrium structure, the change in the period would be a large
under-estimate. That being said, the change in the period from the general relativity value is
$\mathcal{O}(1)$ as soon as one calculates using the modified equilibrium
structure and the effect of the perturbation is to increase this by an amount
not as large as this initial change. These results seem to suggest that
convective stars such as Cepheids and red giants may show very large changes in
the oscillation periods due to their modified background structure and that the
approximation will tend to under-estimate this change. Furthermore, the effects
of the hydrodynamic perturbations will make these changes more drastic but not
as large as those coming from the modified background. In fact, we will see
below that this is not the case for Cepheid models. We will see that the
approximation holds very well when only the background structure is considered
but when the perturbations are included the resulting change in the period can be three times as large as that due to the modified
equilibrium structure alone.
One assumption we have made here is that $K$ is constant. This is tantamount to having a uniform
composition throughout the entire star. This is a good approximation for red
giant stars, which are fairly homogeneous, but Cepheids have shells of varying
composition and several ionised layers and so this is not an accurate
approximation. In particular, we will see below that the radius of Cepheid stars
does not change significantly in modified gravity, contrary to what this model would predict
and this is why we find the approximation holds well when the modified gravity perturbations are ignored, despite this model's
predictions. 

\subsubsection{Stability of Lane-Emden Models}\label{sec:stab_LE}

Before moving on to look at realistic models from MESA, we will
first use Lane-Emden models to investigate the modification to the stellar
stability criterion. In section \ref{sec:stellarstability} we derived the new
properties relating to stellar stability in modified gravity and argued that there are two new features:
first, that when there are unstable modes present in general relativity such that
$\omega_0^2<0$, then, when only the change equilibrium structure is taken into
account, the instability is worse i.e. $\omega_0^2$ is more negative;
second, that the new term appearing in the MLAWE makes stars more stable, 
the critical value of $\gao$ required for $\omega_0^2<0$ is less
than the general relativity value of $4/3$ and the correction is of order
$g(\chi_0,\alpha)\alpha$ given in (\ref{eq:newgammacrit}). $g(\chi_0,\alpha)$ encodes the
competing effects of the new term in the MLAWE and the modified structure and
composition coming from the new equilibrium structure. Here, we will verify
these predictions numerically.

\begin{figure}\centering
\includegraphics[width=15cm]{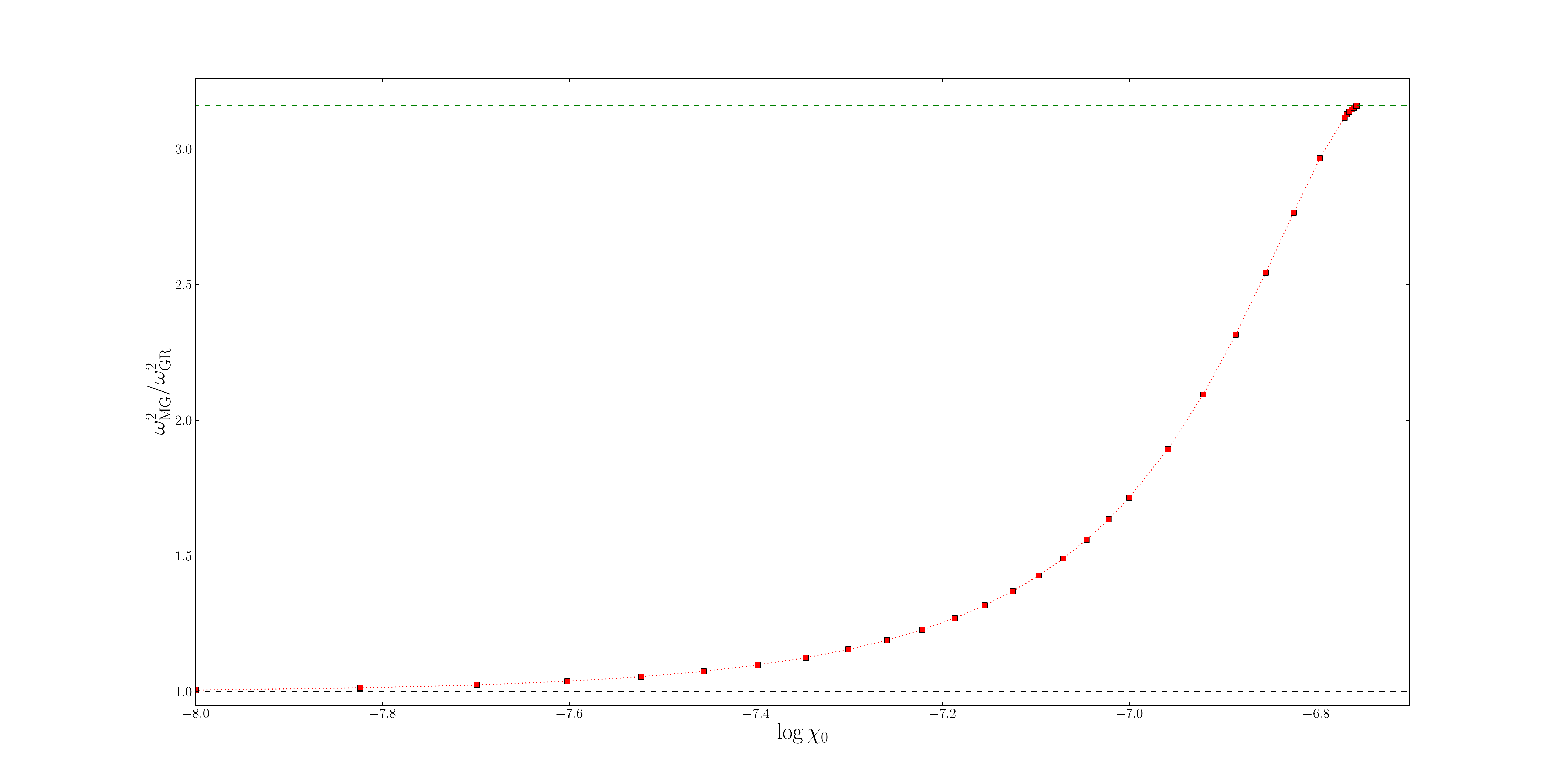}
\caption{The ratio of the modified gravity to general relativity eigenfrequencies as a function of $\log_{\rm
10}\chi_0$ when $\gao\approx 1.2333$. The black dotted line shows the general relativity ratio of 1 and the green
dotted line shows the ratio for a fully unscreened star.}\label{fig:gl0}
\end{figure}

In order to investigate the first, we have solved for the modified
eigenfrequencies of the same $n=1.5$ modified Lane-Emden model investigated in
the previous section using the LAWE for various values
of $\gao<4/3$. This corresponds to a star whose adiabatic perturbations are
governed by a different index to that appearing in the equation of state that
fixes the equilibrium structure. In each case, the modified eigenfrequencies are
indeed more negative the more unscreened the stars are and, as an example, we
plot the ratio $\omega^2_{\rm MG}/\omega^2_{GR}$ in the case
$\gao=37/30\approx1.23333$, which is close to being stable. In this case one has
$\omega^2_{\rm GR}=-0.314$ and so the larger this ratio, the more negative the
modified gravity value. This is plotted in figure \ref{fig:gl0} and it is evident that the
instability is indeed worse in stars which are more unscreened. We have checked that this is the case for other values of
$\gao<4/3$.

\begin{figure}
\centering
\includegraphics[width=15cm]{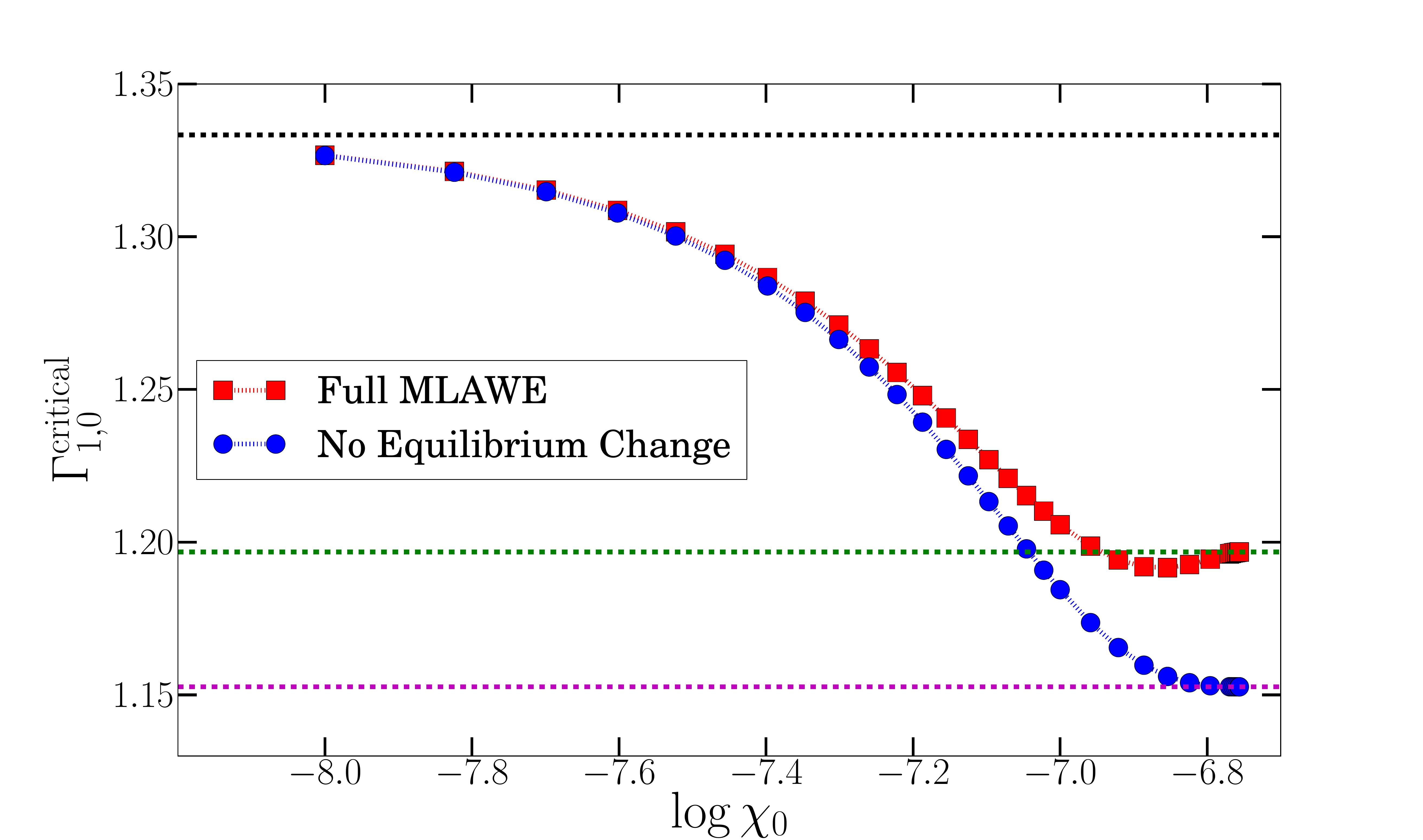}
\caption{The critical value of $\gao$ as a function of $\log_{\rm 10}\chi_0$.
The blue circles show the critical values when the modifications to the
equilibrium structure are ignored and the red squares show the critical values
when the full MLAWE is solved. The black dashed line shows the general relativity value of $4/3$
and the magenta line shows the fully unscreened value when we ignore the modified equilibrium structure. The green line shows
the fully unscreened value found by solving the full MLAWE.}\label{fig:critgam}
\end{figure}

Next, we turn our attention to the modification of the critical value of
$\gao$. In order to investigate this, we again use the $n=1.5$ model above and
vary $\gao$ as a function of $\chi_0$. We scan through different values of
$\gao$ at fixed $\chi_0$ in order to find the value where $\omega^2\approx0$
(to 8 decimal places), which is the new critical value in modified gravity. We solve for the
zero-eigenfrequencies in two cases: the case where we ignore the modified
equilibrium structure\footnote{We have already argued in the previous subsection
that this case is highly unphysical and ambiguous. This is true if we
wish to discern how the modified gravity perturbations affect the numerical value of the
oscillation periods but here we seek only to qualitatively investigate how the
modified equilibrium structure influences the critical index. Hence,
for this purpose it is a reasonable case to investigate.} and include only the
modified gravity perturbations and the full MLAWE. The values of the critical value of $\gao$
vs $\log\chi_0$ are plotted in
figure \ref{fig:critgam}.

It is evident from the figure that the critical value is indeed lower than
$4/3$ showing that these models are indeed more stable in modified gravity. One can also see
that the red curve lies above the blue one so that the effect of the modified
equilibrium structure is to stabilise the star compared with how stable it
would have been had only the modified perturbation structure been present. This can be seen from equation (\ref{eq:gstabdef}),
where the effects of the modified equilibrium structure appear in the
denominator of $g(\chi_0,\alpha)$. When the star is more unscreened the integrated pressure is larger, which increases
the denominator and therefore reduces the size of the correction to the general relativity value. The critical value when the
modified background structure is ignored decreases monotonically,
however, the full MLAWE predicts an increase in stars that are significantly
unscreened, showing that the effects of this altered equilibrium structure
become more important when the star is more unscreened. Finally, we note that the change in the
critical index is indeed of order $10^{-1}$, which we predicted using analytic
arguments in section \ref{sec:stellarstability}.

\subsection{Perturbations of MESA Models}\label{sec:MESAperts}

Having studied some simple stellar models and gained some intuition about the
MLAWE, we now turn our attention to realistic stellar models from MESA. We will limit the discussion to Cepheid
stars, firstly because we have already studied them in detail in chapter \ref{chap:four} and secondly because they are the only
stars whose oscillations can be
observed in distant galaxies\footnote{Some other pulsating objects such as RR
Lyrae stars can be resolved in the local group but this is necessarily
screened and so these objects cannot be used to probe modified gravity.}.

\subsubsection{Perturbations of Cepheid Models}

Equation (\ref{eq:perturbedpl2}) was derived by assuming that the effect of modified gravity is to rescale Newton's constant by a
constant factor. We would now like to account for the fact that this is not the case and that the effective value of $G$ is
radially varying and so it is no longer appropriate to work with $\Delta G$ since this requires some sort of averaging. Instead
we will leave the perturbed period-luminosity relation in terms of $\Delta \tau/\tau$ so that one has
\begin{equation}\label{eq:plmg}
 \frac{\Delta d}{d}\approx-0.6\frac{\Delta \tau}{\tau}.
\end{equation}
Note that we still need to perturb the empirically calibrated period-luminosity relation since this formalism is not powerful
enough to predict it. We are now in a position to calculate
$\Delta\tau/\tau$ for the same models we used in table \ref{tab:period} and check how well the
approximation we used in the previous chapter holds. We first calculate the modified periods by ignoring the effects of the
modified gravity perturbations. This is equivalent to solving the general relativity problem with the modified gravity equilibrium
structure and hence corresponds to finding the same period that we attempted to find in the previous chapter using the averaging
procedure, this time including the full radial dependence of the
effective value of $G$. Recall that in the previous chapter we assumed that the formula for the period in modified gravity is
identical to that in general relativity by replacing $G$ by an appropriate average. Implicit in this is the assumption that the
dynamics of the pulsations are governed by the same underlying equations. We have already seen that this is not the case
and hence our procedure was oblivious to the extra terms in the MLAWE. Next, we solve the full MLAWE and calculate the new
oscillation periods, which allows us to discern how well the approximation of perturbing the general relativity formula works. The
models with $\alpha=1/3,\,\chi_0=10^{-6}$, $\alpha=1/2,\,\chi_0=10^{-6}$ and $\alpha=1,\,\chi_0=4\times10^{-7}$ all execute one
loop that does not cross the instability strip and cross only on the second loop. This is because these stars are so unscreened
that their tracks in the HR diagram are significantly different from the general relativity tracks\footnote{Recall that we are
using the observed instability strip to define the first crossing, which inherently assumes general relativity since it is
found empirically using Cepheid stars in our own galaxy and the local group. Its location may change once one solves the full
non-adiabatic problem. As discussed above, this is beyond the scope of this thesis.}. There is hence no model to compare to the
general relativity first-crossing and so we do not analyse these models\footnote{This is not to say that the approximation used in
the previous chapter is not applicable, rather that numerically calculating the period will give spurious results due to the lack
of a suitable general relativity model to compute the unmodified period.}. Comparing table \ref{tab:period} with figure
(\ref{fig:confidence-regions}), we see that the remaining models correspond to those with the lowest values of $\chi_0$ at fixed
$\alpha$ that were ruled out by the analysis in the previous chapter and hence represent the tightest constraints placed using the
previous approximation.

In table \ref{tab:cepheids} we show $\Delta \tau/\tau$ calculated for each model using the approximation, the case where
modified gravity perturbations are neglected and the full MLAWE. 
\begin{table*}[ht]
\centering
\heavyrulewidth=.08em
	\lightrulewidth=.05em
	\cmidrulewidth=.03em
	\belowrulesep=.65ex
	\belowbottomsep=0pt
	\aboverulesep=.4ex
	\abovetopsep=0pt
	\cmidrulesep=\doublerulesep
	\cmidrulekern=.5em
	\defaultaddspace=.5em
	\renewcommand{\arraystretch}{1.6}
\begin{tabu}{c|[2pt]c|[2pt]c|[2pt]c|[2pt]c}
$\alpha$ & $\chi_0$ & $\Delta \tau/\tau$ (approximation) & $\Delta \tau/\tau$
(no perturbations) & $\Delta \tau/\tau$ (full MLAWE)   \\
\tabucline[2pt]{-}
$1/3$ & $4\times10^{-7}$ & $0.086$ & $ 0.092$ & $0.266$ \\
$1/2$ & $4\times10^{-7}$ & $0.054$ & $ 0.064$ & $0.207$ \\
$1$ & $2\times10^{-7}$ & $0.102$ & $ 0.122$ & $0.314$ \\
\end{tabu}
\caption{The change in the period of Cepheid pulsations due to modified gravity effects.}
\label{tab:cepheids}
\end{table*}
In table
\ref{tab:cepheids2} we show how these changes propagate to $\Delta d/d$, which is
the astrophysical quantity used to place constraints. 
\begin{table*}
\centering
\heavyrulewidth=.08em
	\lightrulewidth=.05em
	\cmidrulewidth=.03em
	\belowrulesep=.65ex
	\belowbottomsep=0pt
	\aboverulesep=.4ex
	\abovetopsep=0pt
	\cmidrulesep=\doublerulesep
	\cmidrulekern=.5em
	\defaultaddspace=.5em
	\renewcommand{\arraystretch}{1.6}
\begin{tabu}{c|[2pt]c|[2pt]c|[2pt]c|[2pt]c}
$\alpha$ & $\chi_0$ & $\Delta d/d$
(approximation) & $\Delta d/d$ (no perturbations)& $\Delta d/d$
(full MLAWE)   \\
\tabucline[2pt]{-}
$1/3$ & $4\times10^{-7}$ & $-0.03$ & $ -0.04$ & $-0.12$  \\
$1/2$ & $4\times10^{-7}$ & $-0.05$ & $ -0.06$ & $-0.16$  \\
$1$ & $2\times10^{-7}$ & $-0.06$ & $ -0.07$ & $-0.19$  \\
\end{tabu}
\caption{The change in the inferred Cepheid distance due to modified gravity. In each case
$\Delta d/d$ was found using the perturbed period-luminosity relation \ref{eq:plmg}.}
\label{tab:cepheids2}
\end{table*}
MESA gives the change in the radius between modified gravity
and general relativity as $\Delta R/R\sim\mathcal{O}(10^{-2})$ and so most of the change in the period is due to the enhanced
gravity and not the size of the star. This is very different from the
$n=1.5$ Lane-Emden model in section \ref{sec:LEperts}, which predicts a large
reduction in the radius when the star is unscreened. One can see that when we neglect the modified gravity perturbations the
approximation we used in the previous chapter is very close to the true value, indicating that the Epstein function is very
successful at describing which regions of the star determine the period of oscillation. This agreement is hardly surprising. The
radius is almost constant between the two theories and the approximation was found by perturbing the LAWE prediction $\tau\propto
G^{-\frac{1}{2}}$ at fixed radius. When the modified gravity perturbations are included, we see that this
approximation breaks down and the relative difference in the period is $\mathcal{O}(10^{-1})$, which is approximately a day. We
can see that $\Delta d/d$ can be up to three times as large as we predicted in the previous chapter using the approximation. We
therefore conclude that the constraints we placed there (and that were reported in \cite{Jain:2012tn}) are conservative, and it is
possible that they could be improved using the same data and analysis.
\begin{table*}
\centering
\heavyrulewidth=.08em
	\lightrulewidth=.05em
	\cmidrulewidth=.03em
	\belowrulesep=.65ex
	\belowbottomsep=0pt
	\aboverulesep=.4ex
	\abovetopsep=0pt
	\cmidrulesep=\doublerulesep
	\cmidrulekern=.5em
	\defaultaddspace=.5em
	\renewcommand{\arraystretch}{1.6}
\begin{tabu}{c|[2pt]c}
$\alpha$ & $\chi_0$  \\
\tabucline[2pt]{-}
$1/3$ & $9\times10^{-8}$  \\
$1/2$ & $7\times10^{-8}$   \\
$1$ & $3\times10^{-8}$  \\
\end{tabu}
\caption{The lower bounds on $\chi_0$ and $\alpha$ that could potentially be
placed if one were to use the same procedure and data-sets as
\cite{Jain:2012tn} using the full MLAWE instead of the approximation.}
\label{tab:constraints}
\end{table*}
With this in mind, we estimate the values of $\chi_0$ and $\alpha$ that can be probed using
the full MLAWE rather than the approximation used in chapter \ref{chap:four}. We accomplish this by taking the same initial
stellar
conditions and running a series of new simulations using MESA for
successively decreasing values of $\chi_0$ at fixed $\alpha$. Using the same procedure as chapter \ref{chap:four} to
identify the Cepheid models at the blue edge, we calculate
$\Delta d/d$ using the MLAWE until it is equal to the value predicted by the
approximation i.e. the value that gave the tightest constraint before experimental errors prevented any further analysis. In
table \ref{tab:constraints} we show the values of $\chi_0$
and $\alpha$ such that the MLAWE gives the same result as the approximation.

These then represent an estimate of the range of parameters that one could hope to
constrain using the same data and the MLAWE. Of course, this is just a
simple estimate and a more rigorous method would be to repeat the data analysis.
Nevertheless, this simple
estimate serves to show that we expect the new constraints to be significantly stronger. In particular, the MLAWE predictions
suggest that the
constraints could be pushed into the $\mathcal{O}(10^{-8})$ regime\footnote{One must also worry about the fact that at lower
values of $\chi_0$, galaxies will move from the unscreened to the screened sub-sample, reducing the quality of the statistics. We
do not attempt to estimate the effect this will have on the lowest values of $\chi_0$ that could potentially be constrained but
we note that it will raise them above the values predicted in table \ref{tab:constraints}.}.

\section{Summary of Main Results}

In this chapter we have perturbed the equations of modified gravity hydrodynamics to first-order and have found the new
equations governing the oscillations of stars about their equilibrium configurations. We have specialised to radial modes since
only these are observable in distant galaxies where chameleon-like theories can be tested. The MLAWE (\ref{eq:MLAWE}) was
the resultant equation and the rest of the chapter was dedicated to its study. We have identified two new effects: firstly, the
new oscillation periods of stars in these theories are always smaller than the general relativity prediction and can be up to
three times smaller than one would predict if one only accounted for the modified equilibrium structure. Secondly, stars are more
stable in chameleon-like theories, by which we mean that the critical first adiabatic index for the onset of unstable radial
modes is lower than the value of $4/3$ predicted by general relativity. The nev value is not universal but is instead model- and
composition-dependent.

Using a pulsation calculator that calculates the new eigenfrequencies by solving the MLAWE, we investigated these effects
quantitatively using Lane-Emden models. These are useful to discern the new physics but are not realistic enough to compare with
observational data and so we then used the calculator to find the new frequencies of MESA models. In the previous chapter, we
used these MESA models to place the strongest constraints on the model parameters $\chi_0$ and $\alpha$ to date using an
approximation based on the LAWE. By using the same models and solving the MLAWE exactly we have found that this approximation is
very conservative and that the true modified periods can be up to three times shorter than it predicts. This means that it is
possible to improve the constraints shown in figure \ref{fig:confidence-regions} using the same data and statistical techniques.
In order to estimate the amount by which the constraints can be improved we chose three fiducial models and fixed $\alpha$ whilst
lowering the value of $\chi_0$ until the MLAWE prediction for $\Delta d/d$ matches the value that gave the strongest constraint
in the previous chapter. This is an optimistic estimate since lowering $\chi_0$ reduces the number of galaxies in the unscreened
sample and therefore increases the statistical error. The results are shown in table \ref{tab:constraints}, which shows that
it may be possible to probe into the regime $\chi_0\sim\mathcal{O}(10^{-8})$ provided there are a sufficient number of galaxies
in the unscreened sample at these low values.

\clearpage

\newpage
\thispagestyle{empty}
\newpage
\clearpage

\newpage

\thispagestyle{empty}


\begin{flushright}
{\Huge{ \bf Part II:}\\
Supersymmetric Models of Modified Gravity}\\
\end{flushright}
\vspace{6cm} 
\begin{quote}
There should be a science of discontent. People need hard times to develop psychic muscles.
\qauthor{Frank Herbert, \textit{Dune} }
\end{quote}
 \vspace*{\fill}

 \newpage
 \thispagestyle{empty}


\begin{savequote}[30pc]
You can prove anything you want by coldly logical reason --- if you pick the proper postulates.
\qauthor{Isaac Asimov, \textit{Reason}}
\end{savequote}

\chapter{Supersymmetric Models of Modified Gravity}\label{chap:six}

In this chapter, we will change directions completely and focus on more theoretical aspects of chameleon-like models. In
particular, we are interested in studying supersymmetric models. There are many reasons why one might wish to look for
supersymmetric completions of these theories. One might be interested in unifying these models with particle physics
within a more fundamental framework. Many beyond the standard model theories such as string theory are supersymmetric and so any
supersymmetric model is a step towards this goal. Another reason is the powerful non-renormalisation theorems that supersymmetric
theories enjoy. Recently, the quantum stability of these models has come into question. Symmetron models do not have a shift
symmetry yet we have fine-tuned the mass to allow for screening in our own galaxy and novel features on cosmological scales.
Without some symmetry protecting this value, we expect scalar loops in the matter sector to raise it considerably and so inherent
in this tuning is the assumption that such a symmetry exists at some level. Chameleon models screen by increasing the mass by
several orders of magnitude in dense environments. Now the Coleman-Weinberg one-loop quantum correction to the scalar potential is
\begin{equation}\label{eq:cw1loop}
 \Delta V_{\rm 1-loop}=-\frac{1}{64\pi^2}m_\phi^4\ln\left(\frac{m^2_\phi}{\mu^2}\right),
\end{equation}
where $\mu$ is the renormalisation group scale. Clearly the more efficient the mechanism is the more we expect that quantum
corrections are important and may act to negate the mechanism completely. This has been studied by \cite{Upadhye:2012vh} who have
found there are large regions of parameter space where chameleon models are dominated by quantum corrections. Furthermore,
\cite{Erickcek:2013oma,Erickcek:2013dea} have studied the quantum effects during the radiation era and have again found that
quantum corrections can act to destroy the classical dynamics. One may hope to circumvent all of these problems by including some
sort of symmetry that either removes or greatly reduces the quantum corrections. This would then imply that the field is a
Goldstone boson with a technically natural small mass, however this is incompatible with the screening mechanisms. Chameleon
models require a very large mass in order to screen and hence this symmetry would be completely broken, thus making its
introduction obsolete. Symmetron models have the field screen by moving over large regions in field space to zero and so even if
the symmetry is present at one field value, it will be broken as soon as the field moves and quantum corrections will reappear.
This
is not the case if the symmetry imposed is supersymmetry. Supersymmetric theories have a Goldstone fermion instead and the field
can therefore have a large mass whilst still enjoying small quantum corrections.

This has motivated several previous attempts at finding supersymmetric completions of chameleon-like models. The first was that
of Brax \& Martin \cite{Brax:2006kg,Brax:2006dc,Brax:2006np} who constructed a general framework within $\mathcal{N}=1$
supergravity where supersymmetry breaking in a hidden sector can generate non-supersymmetric corrections to the scalar potential
for an arbitrary number of dark sector fields and induce a coupling between these fields to the minimum supersymmetric standard
model (MSSM). This is precisely what we need for chameleon-like models; the corrections to the scalar potential may allow us to
find run-away models, which is difficult in supergravity due to the power-law nature of the K\"{a}hler potential and the
superpotential. Furthermore, the coupling
to matter provides the scalar-tensor structure and an extra density-dependent term in the effective potential. Unfortunately,
the final result was a no-go theorem precluding the existence of viable chameleon models \cite{Brax:2006np}. When supersymmetry
is broken in the hidden sector it is generally at a scale $M_{\rm s}$ and
particle physics typically requires this to be around $1$--$10^3\textrm{TeV}$. The
corrections to the scalar potential are then of $\mathcal{O}(\mpl^2M_{\rm s}^2)$, leading to an effective potential for the
chameleon of order
\begin{equation}\label{eq:SUGRAeffpot} 
 V\eff(\phi)\sim \mpl^2M_{\rm s}^2f(\phi)+\rho (A(\phi)-1)+\cdots,
\end{equation}
where $f$ is some $\mathcal{O}(1)$ dimensionless function. In this case the density-dependent part of the effective potential is
negligible on all scales of interest for modified gravity and the mechanism cannot operate.

Another attempt has been made by \cite{Hinterbichler:2010wu} who have used a type IIB string theory approach where the low-energy
effective action is no-scale\footnote{The original model had $n=3$, however this is ruled out in this set-up by accelerator
searches for extra dimensions and so an updated model uses $n=2$ \cite{Hinterbichler:2013we}.} $\mathcal{N}=1$ supergravity with
corrections coming from the KKLT mechanism \cite{Kachru:2003aw} to provide a run-away chameleon potential. In this case the volume
modulus for the six-dimensional internal manifold plays the role of the chameleon in the four-dimensional theory. The coupling
to matter is
provided by the warp factor on the six-dimensional internal manifold and arises from the
ansatz that once the compactification has been performed, the four-dimensional matter fields couple to the 10-dimensional Jordan
frame metric. This is motivated by the fact that the 10-dimensional matter fields are coupled to the 10-dimensional Jordan frame
metric, although a direct demonstration that this remains the case has yet to be seen. The combination of the compactification to
$\mathcal{N}=1$
with a no-scale K\"{a}hler potential, corrections from the KKLT mechanism and gaugino condensation leads to an effective
potential of the form
\begin{equation}\label{eq:stringchampot}
 V\eff(\phi)=V(\phi)+\rho e^{\beta\frac{\phi}{\mpl}}
\end{equation}
where $\beta$ is a constant and $V(\phi)$ is found using the supergravity formula for the scalar potential (\ref{eq:Vsugra}) and
adding
the KKLT correction. The exact form is not relevant here and the reader is referred to \cite{Hinterbichler:2010wu} for the
technical details. This model can indeed produce a chameleon theory with constant $\beta$ but only if one changes the sign of one
of the constants in the gaugino condensation superpotential. Typically, $W\sim
\mathrm{exp}(ia\sigma)$ where
$\sigma$ is the volume modulus and $a>0$. This does not give rise to a run-away potential and so one is forced to take $a<0$,
which
is not realised within the standard KKLT set-up, however the authors argue that situations where this is the case are possible.
When the standard case of $a>0$ is assumed the chameleon mechanism operates in reverse and the evolution of the universe acts to
decompactify the extra dimensions \cite{Conlon:2010jq}. This model has subsequently been generalised to racetrack like models
\cite{Nastase:2013ik}, however this does not improve the situation.

In this chapter we will take a more bottom-up approach. Chameleon-like theories are low-energy IR modifications of gravity and so
a bottom-up approach will capture all of the new supersymmetric features without the
technical complications of more fundamental theories. Of course, one would ultimately like to realise these models in more
fundamental theories and using results from a bottom-up approach such as this will allow us to determine which models
are viable once they are supersymmetrised and therefore exactly where to concentrate the search efforts in more fundamental
theories. In particular, we will see that under the most general assumptions (only global supersymmetry and corrections coming
from supergravity) only $n=3$ no-scale K\"{a}hler potentials can give rise to models that show any deviation from
general relativity in the laboratory and astrophysical objects. This then gives us a clear indication of the types of models we
should examine
in supergravity and string theory.

Since supergravity is broken at higher energy scales than those where the screening mechanism operates we will concentrate on
global supersymmetry coupled to gravity. This avoids the problem of large breaking scales appearing in the effective
potential since supersymmetry is not broken in a hidden sector at very high scales\footnote{Or rather, it is but we assume that
the hidden sector fields do not couple to the sector containing the modified gravity field and so there are no corrections and the
Lagrangian
retains global supersymmetry.}. Instead, we will see that supersymmetry is broken at finite density where the density-dependence
in the effective potential moves the field away from the zero-density minimum, which preserves supersymmetry. In this case the
scale of supersymmetry breaking is set by the ambient density and not by particle physics. This is an interesting new feature
that allows one to decouple the supersymmetry breaking scales in the dark and observable sector, which may have important
consequences for supersymmetric cosmology. We will only consider a theory where the field is coupled to dark matter fermions and
not the observable sector. This will allow us to discern all of the new supersymmetric features without dealing with the
technical complications of the entire MSSM. A generalisation of this model that includes couplings to the standard model is
straight forward, although we will not examine it here.

\section{The Framework}\label{sec:susyframework}

We will take a bottom-up approach and attempt to find the most general $\mathcal{N}=1$ globally supersymmetric model that gives
rise to an effective potential of the form (\ref{eq:effpot2}). The gravitational scalar is
taken to be the lowest component of a chiral superfield $\Phi=\phi+\ldots$. This is coupled to two other fields
$\Phi_\pm=\phi_\pm+\sqrt{2}\theta\psi_\pm+\ldots$, whose fermions act as dark matter. The K\"{a}hler potential is
\begin{equation}
 K(\Phi,\Phi^\dagger,\Phi_\pm,\Phi_\pm^\dagger)= \Phi_+\Phi_+^\dagger+\Phi_-\Phi_-^\dagger+\hat{K}(\Phi,\Phi^\dagger),
\end{equation}
where $\hat{K}(\Phi,\Phi^\dagger)$ is left arbitrary for now. Its specific form must be chosen carefully to obtain an effective
potential that realises one of the screening mechanisms described in chapter \ref{chap:two}. When $\hat{K}(\Phi\Phi^\dagger)\ne
\Phi^\dagger\Phi$ the field is not canonically normalised, indeed
\begin{equation}
 \mathcal{L}_{\rm kin}\supset K_{\Phi\Phi^\dagger}\nabla_\mu\Phi\nabla^\mu\Phi^\dagger
\end{equation}
so that the mass of the field is
\begin{equation}\label{eq:Phimass}
 m^2_\Phi=\frac{1}{K_{\Phi\Phi^\dagger}}\frac{\partial^2V(\Phi)}{\partial\Phi\partial\Phi^\dagger}.
\end{equation} The superpotential is
\begin{equation}\label{eq:suppotgen}
 W(\Phi,\Phi_\pm)=\hat{W}(\Phi)+mA(\Phi)\Phi_+\Phi_-,
\end{equation}
where $A(\Phi)$ is an arbitrary holomorphic function of the gravitational chiral scalar only. Ultimately, it will play the
role of the coupling function and so one must choose its functional form depending on which screening mechanism one wishes to
realise. Again, we leave $\hat{W}$ unspecified; a specific choice of its form leads to different models. With this arrangement,
$\langle\phi_+\rangle=\langle\phi_-\rangle=0$ and so the potential is
\begin{equation}\label{eq:VFpot}
 V(\Phi)=V_{\rm F}(\Phi)=\hat{K}^{\Phi\Phi^\dagger}\left\vert\frac{\dd \hat{W}}{\dd \Phi}\right\vert^2
\end{equation}
There is a $\Phi$-dependent contribution to the dark matter fermion mass
\begin{equation}
 \mathcal{L}_{\rm f}=\frac{\partial ^2 W}{\partial\Phi_+\partial\Phi_-}\psi_+\psi_-=mA(\Phi)\psi_+\psi_-.
\end{equation}
When these fermions condense to finite density such that $\langle\psi_+\psi_-\rangle=\rc/m$ this term provides an additional
contribution to the potential resulting in an effective potential
\begin{equation}\label{eq:veffPhi}
 V_{\rm eff}(\Phi)=V_{\rm F}(\Phi)+\rc (A(\Phi)-1),
\end{equation}
where $\rc$ is the conserved matter density. In practice, it will be necessary
to decompose $\phi$ as $\phi=|\phi|e^{i\theta}$ and stabilise the angular field at the minimum, however several general results
can be derived before specialising to specific models and so we shall continue to work with $\Phi$ for the time being. When this
decomposition is used we shall set $\phi\equiv |\phi|$ and use $\varphi$ to denote the field found by bringing the kinetic term
for $\phi$ into canonical form.

\section{Supersymmetric Features}\label{sec:features}

In this section we will discuss some of the new model-independent features that accompany the embedding of these theories into a
supersymmetric framework.

\subsection{Environment-Dependent Supersymmetry Breaking}

Minimising (\ref{eq:veffPhi}) with respect to $\Phi$, one has
\begin{equation}\label{eq:qvoid}
 \left(\frac{K_{\Phi\Phi^\dagger\Phi^\dagger}}{K^2_{\Phi\Phi^\dagger}}-\frac{1}{K_{\Phi\Phi^\dagger}}\frac{\dd^2\hat{W}}{\dd\Phi^2
}\right)\frac{\dd \hat{W}}{\dd \Phi}=\rc\frac{\dd A(\Phi)}{\dd \Phi}.
\end{equation}
The vacuum expectation value (VEV) of the dark matter scalars is $\langle\phi_\pm\rangle=0$ and so $F_\Phi=-\dd W/\dd \Phi=-\dd
\hat{W}/\dd\Phi$. Any
coupling to dark matter necessarily breaks supersymmetry at finite density. This is one of the new features of supersymmetric
screened modified gravity; by secluding the dark sector from the observable one (up to supergravity breaking effects described
below) the scale of supersymmetry breaking is not set by particle physics effects but rather by the ambient density and so our
model is not plagued with issues such as the cosmological constant being associated with TeV scale breaking effects or detailed
fine-tunings. That being said, this is far from a solution to the cosmological constant problem since we do not attempt to explain
why the vacuum energy in the observable sector associated with QCD and electroweak symmetry breaking does not contribute to the
cosmological dynamics. We also offer no explanation for the cancelling of the cosmological constant in the hidden sector which is
of order $\mpl^2m_{3/2}^2$.

\subsection{Absence of Observational Signatures}

When working in the low-energy framework of global supersymmetry it is important to ensure that any corrections coming from
supergravity breaking in the hidden sector are negligible. We will see presently that once these corrections are accounted for the
vast majority of supersymmetric screened theories of modified gravity necessarily have $\chi_0\ll10^{-8}$ and hence there are no
unscreened objects in the universe. Whereas this does not rule these theories out, it renders them observationally
indistinguishable from general relativity\footnote{Technically, this is not quite correct, we will see in chapter \ref{chap:seven}
that there is a small region in parameter space where there are still potential deviation in the cold dark matter power spectrum
on linear scales where the non-linear screening mechanism does not operate and $\chi_0$ is not an indicator of how much modified
gravity effects are suppressed.}.

\subsubsection{Supergravity Corrections}\label{sec:sugracorr}

The most important supergravity correction for these models are those coming from
$|D_\Phi W|^2$ of the form
\begin{equation}\label{eq:sugracorr}
 \Delta V_{\cancel{\rm
SUGRA}}=\frac{K^{\Phi\Phi^\dagger}|K_\Phi|^2|W|^2}{\mpl^4}e^{\frac{K}{\mpl^2}}=m_{3/2}^2K^{\Phi\Phi^\dagger} |K_\Phi|^2.
\end{equation}
This correction must be negligible compared to $V_{\rm F}$ and $\rc A(\Phi)$ if they alone
are to be responsible for the screening mechanism \footnote{If this is not the case then one is really working within the
framework of supergravity and can therefore not realise any screening mechanisms due to the no-go result of \cite{Brax:2006np}.}.
This correction introduces an important new feature into these models: the mass of the field is always at least as large as the
gravitino mass. To see this, one can take derivatives of (\ref{eq:sugracorr}) and focus on certain terms only to find
\begin{equation}\label{eq:m32massres}
 \frac{\partial^2V(\Phi)}{\partial\Phi\partial\Phi^\dagger}\supset m_{3/2}^2K_{\Phi\Phi^\dagger}.
\end{equation}
Recalling that the field may not be canonically normalised and applying (\ref{eq:Phimass}) one finds there is a contribution
to the field's mass of exactly $m_{3/2}$. This can be anywhere from $1$ eV as predicted by gauge mediated supersymmetry breaking
scenarios to $\mathcal{O}(\textrm{TeV})$ corresponding to gravity mediated breaking \cite{Nilles:1983ge}. Consequently, the
Compton wavelength of the field is $\lambda_{\rm C}\sim m_{3/2}^{-1}$ and so the range of the fifth-force in such models is always
less than $10^{-6}$ m depending on the gravitino mass. It should be noted that this result is completely independent of the form
of the matter coupling or the potential, it is not even sensitive to their origins or whether the field is coupled to dark matter
or the standard model. When one has scalars coupled to matter and the theory has an underlying $\mathcal{N}=1$ supergravity at
some high energy scale then the range of the fifth-force will always be less than $m_{3/2}^{-1}$. In supergravity breaking
scenarios with a large gravitino mass this precludes the need for screening mechanisms altogether.

Another immediate consequence of this is that canonical symmetrons \cite{Hinterbichler:2010es} cannot be accommodated within a
supersymmetric framework. The supersymmetric symmetron is found by imposing a $\mathbb{Z}_2$ symmetry upon the effective
potential. This is achieved by including only odd powers of $\Phi$ in $\hat{W}(\Phi)$ and only even powers in the coupling
$A(\Phi)$. The K\"{a}hler potential is $\hat{K}(\Phi\Phi^\dagger)=\Phi^\dagger\Phi$ so that the fields are canonically normalised
and, at lowest relevant order, the superpotential is
\begin{equation}\label{eq:suppotss}
 W(\Phi)=M^2\Phi+\frac{1}{3}g\Phi^3+m\left(1-h\frac{\Phi^2}{2m\Lambda_3}+f\frac{\Phi^4}{4m\Lambda_3^3}\right)\Phi_+\Phi_-,
\end{equation}
where the explicit introduction of the $-$ sign in the coupling will become clear momentarily. The F-term potential is then
\begin{equation}\label{eq:sspot1}
 V_{\rm F}(\phi,\theta)=M^4+g^2\phi^4+2gM^2\phi^2\cos(2\theta),
\end{equation}
which is minimised when $\cos(2\theta)=-1$ so that the model is a symmetron:
\begin{align}\label{eq:supersymmv}
 V(\phi)&=M^4
-2gM^2\phi^2+g^2\phi^4=(g\phi^2-M^2)^2,\nonumber\\A(\phi)&=1+h\frac{\phi^2}{2m\Lambda_3}+f\frac{\phi^4}{4m\Lambda_3^3}.
\end{align}
Note that this has a supersymmetric minimum ($V=0$) at $\phi_0=M/\sqrt{g}$; at finite density the field moves to smaller value
thereby breaking this supersymmetry. Now the symmetron mechanism requires that the bare mass be negative, however there is a
contribution from supergravity corrections of $\mathcal{O}(m_{3/2}^2)$ and so either we must demand that there is a fine-tuned
cancellation or we must take $M>m_{3/2}$ (recall that the canonical symmetron model requires $M\le10^{-29} eV$ whereas
$m_{3/2}\ge 1$ eV). If this is not the case the symmetron mechanism is lost. Suppose then that $M>m_{3/2}$. We have
\begin{equation}
 \beta(\phi)=\mpl\frac{\dd \ln A(\phi)}{\dd \phi}\sim \frac{\mpl\phi_0}{m\Lambda_3}
\end{equation}
in the cosmological background when the $\mathbb{Z}_2$ symmetry is broken.
So if the force is to be of comparable strength to gravity in free space we need $M\mpl\sim m\Lambda_3$. Now the symmetry is
broken (or restored) at a density
\begin{equation}
 \rho_\star\sim M^2m\Lambda_3\sim M^3\mpl>m^3_{3/2}\mpl>10^{27}\textrm{ eV}^{4}=10^{39}\rho_0,
\end{equation}
where we have taken the best case scenario of an eV mass gravitino. This means that in the late-time universe only objects whose
densities exceed $10^{27}$ eV$^4$ can restore the $\mathbb{Z}_2$ symmetry locally and screen
the fifth-force. This immediately precludes screening in all dark matter haloes (whose density is typically $10^6\rho_0$) and
Earth
based laboratories (with density $10^{29}\rho_0$). This problem is not ameliorated if we instead allow the force in free space to
be stronger than gravity since this increases the lower bound on $\rho_\star$. Either the symmetron mechanism does not exist or
$\mathcal{O}(1)$ fifth-forces are present in our solar system. One should note that this does not rule out the symmetron as a
viable model of modified gravity but it is the case that if the universe is supersymmetric (including an underlying supergravity)
then the symmetron mechanism cannot be realised.

One may then also wonder whether the same is true of generalised symmetrons? In this case, a correction to the mass of
$\mathcal{O}(m_{3/2})$ does not affect the choices for the parameters in the theory since a mass term is not present in the
effective
potential before supergravity corrections are included (see equation (\ref{eq:extendedsymmetrons})), however, it does add a term
proportional to $m_{3/2}^2\phi^2$ to the effective potential and hence changes the transition from second to first order. In this
case, the mechanism is lost and $\beta(\phi)$ does not approach zero smoothly in increasingly dense environments.
Therefore, it is not possible to realise generalised symmetrons within a supersymmetric framework either.

\subsubsection{No-Scale-Type Models}

Given that the mere presence of an underlying supergravity imposes such stringent restrictions on the mass of the field one might
naturally wonder how general these restrictions really are and whether they can be circumvented. There are indeed a class of
models where the supergravity correction, i.e. the mass (\ref{eq:m32massres}) is not present. Clearly if
$K^{\Phi\Phi^\dagger}|K_\Phi|^2$ is constant then (\ref{eq:m32massres}) is spurious since the second derivative of the corrections
are zero and there are no corrections to the field's mass. These are the no-scale type models, a particularly common example of
which is the logarithmic K\"{a}hler potential that arises in type IIB superstring theory $K=-n\mpl^2\ln[(\Phi+\Phi^\dagger)/\mpl]$
($n=1$ for the dilaton and $n=3$ for T-moduli, which corresponds to the pure no-scale case where $V_{\rm F}(\phi)=0$ at the
minimum). In more complicated scenarios one
typically has many chiral scalars which parametrise a no-scale type manifold given by
$K^{\Phi_i\Phi_j^\dagger}K_{\Phi_i}K_{\Phi_j^\dagger}=c$ with $c=3$ in the pure no-scale case.

These models evade the corrections so one may wonder if they are re-introduced by loop corrections. The
one-loop effective potential is
\begin{equation}\label{eq:1loopV}
 \Delta V_{\rm 1-loop}=-\frac{1}{64\pi^2}\textrm{STr}\left[M^4\ln\frac{M^2}{\mu^2}\right],
\end{equation}
where $M$ is the mass matrix and $\mu$ is the renormalisation group scale. At tree level we have, for the scalar, $M^2\sim
|\hat{W}_{\Phi\Phi}|^2$ and so if $\hat{W}\sim \mathcal{O}(\mathcal{M}^3)$ we expect $\mathcal{M}\ll\mpl$ since $\hat{W}$ is
associated with low-energy behaviour well below the supergravity breaking scale. In this case, the quantum corrections are set
entirely by the tree-level parameters\footnote{Note that since the theory is supersymmetric we do not expect corrections to the
scalar mass coming from fermion loops that would usually be present in non-supersymmetric theories and be of a similar order to
the largest fermion mass. We have already seen above that the underlying supersymmetry is broken at finite density and that the
scale of breaking is set by the local density and hence we expect corrections to be present at this scale, which we will see in
the next chapter is generally a lot lower than any standard model (and hence also supersymmetric partner) fermion masses.}, which
a priori are independent of the gravitino mass.

The equation (\ref{eq:1loopV}) encompasses only supersymmetric corrections and so we must also account for the soft masses induced
by hidden sector supersymmetry breaking. These have been studied extensively by
\cite{Brignole:1993dj,Brignole:1997dp,Farquet:2012cs} (and references therein) who find that whenever the manifold is not pure
no-scale i.e. $K^{\Phi_i\Phi_j^\dagger}K_{\Phi_i}K_{\Phi_j^\dagger}\ne3$ the soft masses are always of order $m_{3/2}$ and so one
can conclude that these models do not evade the supergravity breaking constraints. Furthermore, in the pure no-scale case the
same analyses have shown that only no-scale models where the isometry group of the scalar manifold is
\begin{equation}\label{eq:manifod}
 \mathcal{M}=\frac{\mathrm{SU}(1,n)}{\mathrm{U}(1)\times\mathrm{SU}(n-1)},
\end{equation}
do not acquire soft masses. Any no-scale model whose isometry group differs from this must necessarily include gravitino-mass
scalars in its low-energy effective theory.

One should note however that it is very difficult to find screening mechanisms with this class of K\"{a}hler potentials. For
example, in the simplest case where $K=-3\mpl^2\ln(T+T^\dagger)/\mpl$, the canonically normalised field is $\Phi={\rm
exp}(\sqrt{2/3}\phi/\mpl)$. Now any term in the superpotential is
of the form $W(\Phi)\propto \Phi^n$ and so at best one has an exponentially decreasing scalar potential and an exponentially
increasing coupling function. It is very difficult
to obtain a thin shell solution for an Earth-like density profile in such a model \cite{Brax:2010gi}. One must then rely on
non-perturbative effects to generate a viable potential and no satisfactory mechanism has been found to date. 

We can then discern the conditions under which globally supersymmetric theories are not bound by constraints from supergravity
breaking; they must be no-scale models with the isometry group (\ref{eq:manifod}). At the level of string theory, models such as
these receive corrections to their K\"{a}hler potentials in string perturbation theory, which are then used in the tree-level
supergravity formula to find the scalar potential. Hence, only string theory models which preserve this no-scale property to all
orders in perturbation theory and under non-perturbative corrections can evade the supergravity corrections. At the
level of field theory, any no-scale model with this isometry group will always lead to a low-energy model which is not bound
by these constraints.

\subsubsection{Efficient Screening}

When the models do not evade the supergravity corrections the presence of a contribution of order $m_{3/2}$ to the field's mass is
enough to ensure that the screening in these models is so efficient that no object in the universe is unscreened. This can be
deduced as follows. Working with the canonically normalised
field $\varphi$ and assuming that $\theta$ is stabilised at its minimum, the new effective potential is\footnote{Note that we are
using $\rho$ and not $\rc$ since this argument holds even when the field is coupled to baryons}
\begin{equation}
 V_{\rm eff}(\varphi)=V_{\rm F}(\varphi)+\rho (A(\phi)-1)+\frac{1}{2}m_{3/2}^2\varphi^2.
\end{equation}
At the minimum, we have
\begin{equation}
 -\frac{\beta(\varphi_0)\rho_0}{\mpl\varphi_0}= m_{3/2}^2+\frac{1}{\varphi_0}\frac{\dd V_{\rm F}(\varphi)}{\dd \varphi}
\end{equation}
where we have used the fact that $A(\varphi_0)\approx1$. If this equation has no solutions then there is no minimum and the theory
is not one of screened modified gravity. We are interested in situations where this is not the case and so we will assume that a
minimum exists. Now the left hand side of this equation --- barring any fine-tuned cancellation --- must be as large as
$m_{3/2}^2$, in
which case
\begin{equation}\label{eq:betcond}
 \frac{\varphi_0}{\bv(\varphi_0)}\le\frac{\rho_0}{m_{3/2}^2\mpl}.
\end{equation}
The quantity on the left hand side of the inequality is precisely $\chi_0$ (see equation (\ref{eq:chi0def})).
Taking equation (\ref{eq:betcond}) and inserting it into (\ref{eq:chi0def}) we have
\begin{equation}\label{eq:chicond}
\chi_0\le \left(\frac{H_0}{m_{3/2}}\right)^2,
\end{equation}
where $\rho_0\sim 3\Omega_{\rm c}^0H_0^2\mpl^2$. In the best case scenario we have $m_{3/2}\sim\mathcal{O}(\textrm{eV})$ and so we
have $\chi_0\le10^{-66}$. In section \ref{sec:currentconstraints} we showed that most unscreened objects in the universe are
dwarf galaxies with $\Phi_{\rm N}\sim 10^{-8}$ and so this condition ensures that no object in the
universe is unscreened; hence, there are no observational signatures of fifth-forces in any astrophysical object. Since
$\chi_{\rm MW}\ll \chi_0$ ($\chi_{\rm MW}$ is the value of $\chi_c$ in the Milky Way), the mass of the field in the Milky Way is
$m_{\rm MW}\gg m_0\gg 10^{33}H_0$ (see the discussion in section \ref{sec:screen_mech}) whereby $\gg$ we mean several orders of
magnitude. The Compton wavelength corresponding to a mass this large is well below the length scales that can be probed using
current laboratory probes of fifth-forces \cite{Kapner:2006si}. Therefore, this bound also precludes any possibility of a
laboratory detection and
will continue to do so for the foreseeable future.

This behaviour can alternatively be seen by considering the derivation of the field profile
presented in section \ref{sec:screen_mech} in the unscreened region of a spherical over-density. Consider a spherical object of
constant dark matter density $\rho_b$ embedded in a much larger medium of density $\rho_c$ so that
\begin{equation}
\rho(r)=\left\{
  \begin{array}{l l}
   \rho_b,& r<R\\
    \rho_c,& r>R\\
  \end{array}\right. .
\end{equation}
When the object is static, equation (\ref{eq:chameom2}) becomes $\nabla^2\varphi=V_{\rm eff}(\varphi)_{\,,\varphi}$ and when the
object is unscreened the field only differs from the exterior value by a small perturbation $\delta\varphi$ and we can expand
this to first order to find
\begin{equation}
 \nabla^2\delta\varphi\approx m_c^2\delta\varphi+\frac{\bv(\varphi_0)\delta\rho}{\mpl},
\end{equation}
where $\delta\rho=\rho_b-\rho_c$ and $m_c^2$ is the mass of the field in the outer medium. The next step was to use the fact
that models with screening mechanisms have the property that the Compton wavelength of the field $\lambda_{\rm C}m_c^{-1}$ is much
greater than the size of the object and so we neglected the mass term for the perturbation. In these models however, we have
$m_c\ge m_{3/2}\ge1$ eV$\sim10^{28}$Mpc$^{-1}$ and therefore the Compton wavelength is incredibly small compared with typical
galactic, stellar and planetary scales. Therefore, the unscreened solution does not exist in any object in the universe.
\begin{savequote}[30pc]
If there is such a thing as reincarnation, knowing my luck I'll come back as me. 
\qauthor{John Sullivan, \textit{Rodney Trotter in: Only Fools and Horses}}
\end{savequote}

\chapter{Phenomenology of Supersymmetric Chameleons}\label{chap:seven}

In the previous chapter we presented a general framework for embedding screened modified gravity into global supersymmetry and
proved several model-independent results. We showed that supersymmetric symmetrons do not exist and that supersymmetric
chameleons (and theories that screen via the environment-dependent Damour-Polyakov effect) have $\chi_0<10^{-66}$. This implies
that there are no unscreened objects in the universe and hence astrophysical or laboratory tests of modified gravity cannot probe
these models. No-scale models are an exception, however we showed that the simplest models do not screen and, to date,
more complicated models have not been found. Whereas $\chi_0$ sets the screening properties of collapsed objects, there are two
regimes where it is not enough to specify whether modified gravity effects are present: the cosmological background and linear
scales in cosmological perturbation theory. In this chapter, we will construct a class of supersymmetric chameleon models using
the framework of the previous chapter and use them as prototypes to investigate both of these regimes.

On cosmological scales, the homogeneous time-evolution of the scalar field is a
source of pressure and density in the Friedmann equation in the Einstein frame, which may lead to different cosmological dynamics
from $\Lambda$CDM\footnote{\cite{Wang:2012kj} have proved a general no-go result stating that chameleon-like models are
indistinguishable from $\Lambda$CDM. One of their key assumptions is that the coupling to matter
$\beta(\phi)\sim\mathcal{O}(1)$. Supersymmetric models do not satisfy this property and since $\beta\gg\mathcal{O}(1)$ and
so this theorem does not apply to these models.}. The first part of this chapter examines the background cosmology of the
supersymmetric chameleon model we will construct using the tools presented in section \ref{subsec:cosmologyc2}. We will find
that the dynamics are indistinguishable from $\Lambda$CDM and so we require a cosmological constant to act as dark
energy\footnote{Technically, any
other dynamical degrees of freedom whose phenomenology can reproduce the current data could be included in the model, however this
is more complicated and we will consider dark energy to be a cosmological constant for simplicity.}. This
presents a new problem not present in more phenomenological chameleon models. Supersymmetry is broken if the vacuum energy
density is positive and so we cannot include a cosmological constant at the level of the action, one must be induced via the
dynamics. One could appeal to supergravity breaking, which contributes an amount of order $\mpl^2m_{3/2}^2$ to the vacuum energy
density. This is the same order as the supergravity breaking scale and is far too large
compared with the measured value and so this contribution needs to be fine-tuned away in most supersymmetric models and
chameleons are no exception. Instead, we will present a new and novel mechanism where a cosmological constant in the form of a
Fayet-Illiopoulos term appears at late times due to the coupling of the chameleon to two charged scalars. The scalar potential
for these scalars is $\mathbb{Z}_2$ symmetric with a mass term that depends on the chameleon VEV. At early times, the symmetry is
broken and the charged scalars have non-zero VEVs but as time progresses, the chameleon evolves towards larger field values and
the coefficient of the mass term becomes increasing larger, eventually becoming positive and restoring the symmetry. At this
time, the scalar's VEV moves towards zero, leaving only an FI term in the potential which acts as a cosmological constant.
FI terms have no natural value within global supersymmetry and run logarithmically at most with the energy scale
\cite{Jack:1999zs} and so tuning this term to match the current observed energy density is technically natural and an appealing
model, which merits further study in terms of the cosmological constant problem\footnote{Of course, this is far from a solution to
the cosmological constant problem. We are fine-tuning all the contributions to the vacuum energy density coming from supergravity
breaking in the hidden sector and any symmetry breaking in the matter sector including the electroweak and QCD phase transitions
to zero. Furthermore, we have no natural mechanism that sets the magnitude of the FI term and appeal to supergravity (or
possibly more
fundamental physics) to provide this mechanism}. We will investigate the range of parameters that allow this cosmological constant
to be generated before last scattering so that its effects are imprinted on the CMB and will use this to rule out certain regions
in
the model parameter space.

We next turn our attention to linear perturbations of CDM. The screening mechanism is inherently non-linear, and so we do not
expect to see any deviation in the CDM power spectrum on scales greater than $0.1h$ Mpc$^{-1}$\footnote{We assume that the scale
on which perturbations become non-linear is approximately the same as the non-linear scale in general relativity. This is
motivated by the efficient screening theorem of the previous chapter.}, however, on scales less than this the mechanism cannot
operate and $\chi_0$ is a meaningless parameter for describing the extent of the modifications of gravity. We therefore expect
that
deviations from general relativity may be present. In order to investigate this, we solve the modified equations governing CDM
perturbations and
calculate the modified CDM power spectrum. This indeed shows an enhancement on large scales due to the enhanced gravitational
force
arising from the unscreened fifth-force. This enhancement can be made arbitrarily large
or small by tuning the model parameters due to the lack of other constraints. We rule out some parameter ranges by demanding that
these deviations are not larger than
the error of 10\% reported by the WiggleZ survey \cite{Contreras:2013bol} and indicate the regions in parameter
space that could potentially be probed using Euclid.

\section{Supersymmetric Chameleon Models}\label{sec:susycham}

So far, we have not presented any specific chameleon models. In this section we will use the framework of section
(\ref{sec:susyframework}) to construct one possible class. In theory, it is possible to construct other classes of models
although to date none have been studied. We will argue at the end of this chapter than the phenomenology of these models will not
differ too drastically from the ones we will construct here and are no more appealing.

The K\"{a}hler potential for $\Phi$ is non-canonical --- this is a requirement for it to give rise to a run-away potential ---
whilst the dark matter fields are canonically normalised:
\begin{equation}
 K(\Phi,\Phi^\dagger)=\frac{\Lambda_1^2}{2}\left(\frac{\pdp}{\Lambda_1^2}
\right)^\gamma+\Phi_+^\dagger\Phi_++\Phi_-^\dagger\Phi_-,
\end{equation}
where $\gamma$ is an arbitrary integer exponent whose value determines how steep the run-away potential is. The self-interacting
part of the superpotential is
\begin{equation}
 \hat{W}=\frac{\gamma}{\sqrt{2}\alpha}\left(\frac{\Phi^\alpha}{\Lambda_0^{\alpha-3}}
\right)+\frac{1}{\sqrt{2}}\left(\frac{\Phi^\gamma}{\Lambda_2^{\gamma-3}}\right),
\end{equation}
where $\Phi=\phi+\sqrt{2}\theta\psi+\ldots$ contains a scalar $\phi$ whose modulus ultimately plays the role of the
superchameleon and $\Phi_\pm=\phi_\pm+\sqrt{2}\theta\psi_\pm+\ldots$ are chiral superfields containing dark matter fermions
$\psi_\pm$. Splitting the superchameleon field as $\phi(x)=|\phi|e^{i\theta}$ and identifying $\phi\equiv|\phi|$ from here on,
one
can minimise the angular field (this is done explicitly in appendix \ref{app:mgp} where a coupling to two $\mathrm{U}(1)$ charged
scalars, which we will introduce later, is also examined) and define the new quantities
\begin{equation}\label{eq:scales}
 \Lambda^4\equiv\left(\frac{\Lambda_1}{\Lambda_2}\right)^{2\gamma-2}{\Lambda_2}
^4, \quad
M^{n+4}=\left(\frac{\Lambda_1}{\Lambda_0}\right)^{2\gamma-2}{\Lambda_0}^{n+4},
\quad \pmi=\left(\frac{M}{\Lambda}\right)^{\frac{4}{n}}M,\quad
n=2(\alpha-\gamma)
\end{equation}
to find the F-term potential
\begin{equation}\label{eq:Fpot}
 V_{\rm F}(\phi)=K^{\Phi\Phi^{\dagger}}\left\vert\frac{\dd
W}{\dd\Phi}\right\vert^2=\left(\Lambda^2-\frac{M^{2+\frac{n}{2}}}{\phi^{\frac{n}{2}}}\right)^2=\Lambda^4\left[1-\left(\frac{\phi_{
\rm min}}{\phi}\right)^\frac{n}{2}\right]^2.
\end{equation}
The parameters $\Lambda_i$ appearing in the K\"{a}hler potential and superpotential are scales associated with non-renormalisable
operators and one would expect them to be large, however, we can see that the scales governing the low-energy dynamics are $M$ and
$\Lambda$. We will explore their values in detail when discussing the parameter space in section \ref{sec:cosgen}. In practice, it
is easier to work with other low-energy parameters, which will be introduced later, and their relation to these parameters is
given in appendix \ref{app:dcor}. The index $n$ should be even and one would expect $\gamma$ and $\alpha$ to be small (but not
$1$) given their origin as indices in the superpotential and so we will often consider the case $n=2$ when the need to elucidate
specific calculations arises. 

When $\phi\ll\pmi$ equation (\ref{eq:Fpot}) reduces to the Ratra-Peebles potential
\begin{equation}
V_{\rm F}(\phi)\approx \Lambda^4 \left(\frac{\pmi}{\phi}\right)^n,
\end{equation}
which has been well studied in the context of dark energy \cite{PhysRevD.37.3406} (although one should be aware that we have not
yet canonically normalised our field). At larger field values the potential has a minimum at $\phi=\pmi$ where $V(\pmi)=0$ and
$\dd W/\dd \Phi=0$. Supersymmetry is therefore broken whenever $\phi\ne\pmi$.

\begin{figure}[ht]\centering
\includegraphics[width=0.9\textwidth]{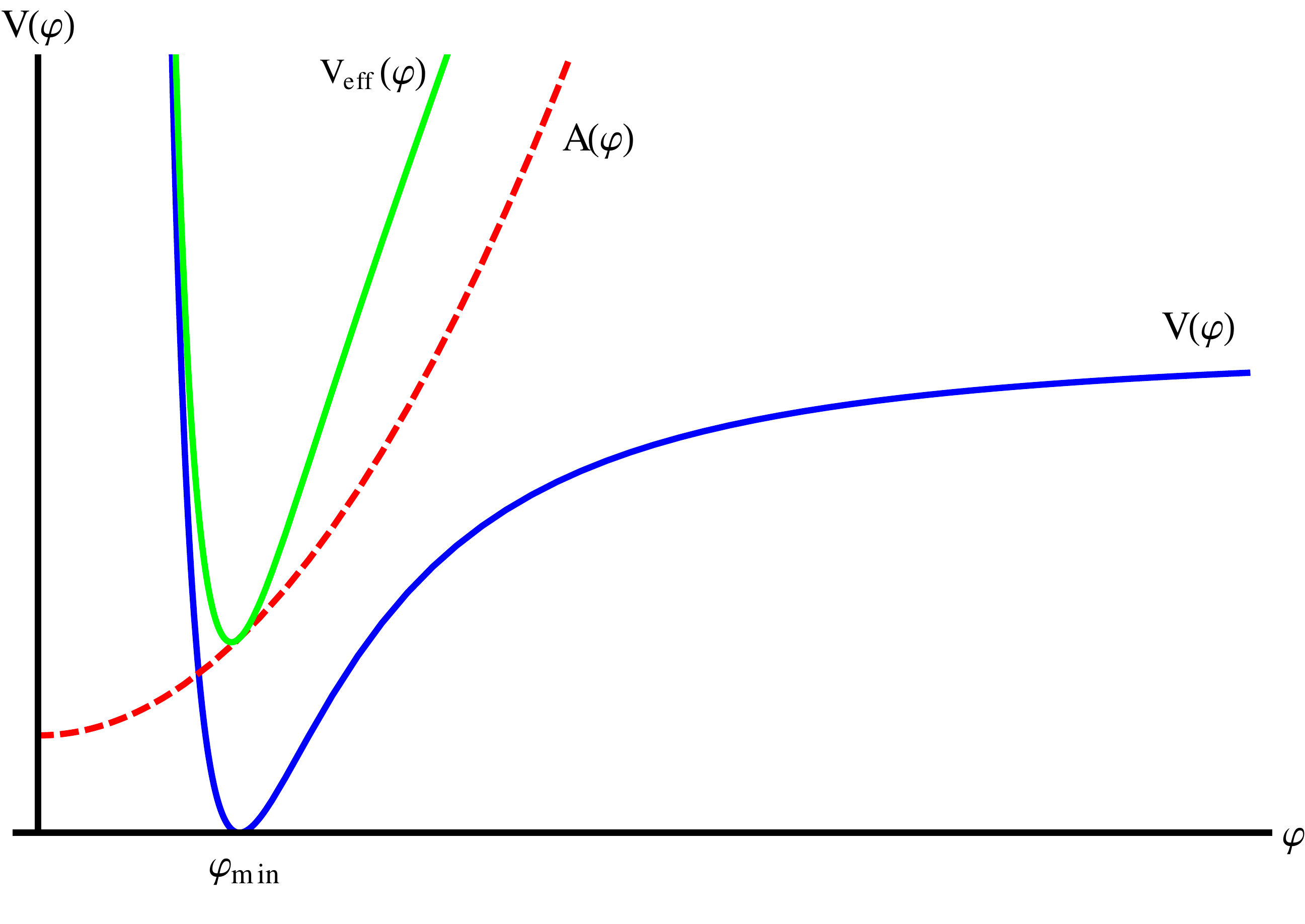}
\caption{The effective potential.}\label{fig:veff}
\end{figure}

The coupling function is found by considering the part of the superpotential containing the interactions of $\Phi$ and $\Phi_\pm$
\begin{equation}
 W_{\rm int}=m\left[1+\frac{g}{m}\frac{\Phi^\delta}{\Lambda_3^{\delta-1}}\right]\Phi_+\Phi_-,
\end{equation}
which gives a superchameleon-dependent mass to the dark matter fermions
\begin{equation}
 \mathcal{L}\supset \frac{\partial^2 W}{\partial\Phi_+\partial\Phi_-}\psi_+\psi_-.
\end{equation}
When the dark matter condenses to a finite density $\rc=m\langle\psi_+\psi_-\rangle$ this term provides a density-dependent
contribution to the scalar potential resulting in the scalar-tensor effective potential $V_{\rm eff}=V+\rc (A-1)$. With the above
choice of superpotential, the coupling function is
\begin{equation}\label{eq:AmodB}
 A(\phi)=1+\frac{g}{m{\Lambda_3}^{\delta-1}}\phi^{\delta}=1+\left(\frac{\phi}{\mu}\right)^{\delta};\quad \mu^\delta\equiv
\frac{m\Lambda_3^{\delta-1}}{g}.
\end{equation}
The scale $\Lambda_3$ is not present when $\delta=1$, which may make fine-tuning of the dark matter mass necessary. For
this reason, we will always consider $\delta\ge2$. As it stands, the field $\phi$ is not canonically normalised since the kinetic
term in the Lagrangian reads
\begin{equation}
 \mathcal{L}_{\rm
kin}=-K_{\phi\phi^\dagger}\partial_\mu\phi\partial_\mu\phi^\dagger=-\frac{1}{2}
\gamma^2\left(\frac{|\phi|}{\Lambda_1}\right)^{2\gamma-2}
\partial_\mu\phi\partial_\mu\phi^\dagger.
\end{equation}
The normalised field is
\begin{equation}\label{eq:vphi}
\varphi=\Lambda_1\left(\frac{\phi}{\Lambda_1}\right)^\gamma
 \end{equation}
so that the coupling function (\ref{eq:AmodB}) becomes
\begin{equation}\label{eq:x}
A(\varphi)=1+x\left(\frac{\varphi}{\varphi_{\rm
min}}\right)^{\frac{\delta}{\gamma}};\quad x\equiv\frac{g\phi_{\rm
min}^{\delta}}{m{\Lambda_3}^{\delta-1}}
\end{equation}
and the effective potential is
\begin{equation}\label{eq:bveff}
V_{\rm eff}(\varphi)=\Lambda^4\left[1-\left(\frac{\varphi_{\rm
min}}{\varphi}\right)^{\frac{n}{2\gamma}}\right]^2+x\rc\left(\frac{\varphi}{
\varphi_{\rm min}}\right)^{\frac{\delta}{\gamma}},
\end{equation}
which is shown in figure \ref{fig:veff}.
We may then find the coupling $\beta(\varphi)$:
\begin{equation}\label{eq:betaphi}
 \beta(\varphi)=\frac{x\delta\mpl}{\gamma\varphi_{\rm
min}}\left[1+\left(\frac{\varphi}{\varphi_{\rm
min}}\right)^{\frac{\delta}{\gamma}}\right]^{-1}\left(\frac{\varphi}{\varphi_{
\rm min}}\right)^{\frac{\delta}{\gamma}-1}.
\end{equation}

Equation (\ref{eq:bveff}) is the effective low-energy potential for a scalar-tensor theory described in section
\ref{sec:suppresscharge} with the scalar coupled to dark matter via the coupling function $A(\varphi)$. Minimising the effective
potential we have\footnote{The careful reader may notice that taking the limit $\delta=1$ gives a different equation from that
found in \cite{Brax:2011bh}, which contains a typographical error.}
\begin{equation}\label{eq:phimineq}
 \left(\frac{\varphi_{\rm
min}}{\varphi}\right)^{\frac{n+\delta}{\gamma}}-\left(\frac{\varphi_{\rm
min}}{\varphi}\right)^{\frac{n+2\delta}{2\gamma}}=\frac{\rc}{\rho_\infty},
\end{equation}
where
\begin{equation}\label{eq:rhoinf}
 \rho_\infty\equiv\frac{n\Lambda^4}{\delta x}\equiv\rc^0(1+z_\infty)^3=3\Omega_{\rm c}^0\mpl^2H_0^2(1+z_\infty)^3,
\end{equation}
which defines the quantity $z_\infty$. This is an important model parameter; we will see later that it is the redshift at which
the field settles into its supersymmetric minimum. It turns out that this redshift controls the large scale behaviour of the
modified linear CDM power spectrum, whose features change very rapidly for modes which enter the horizon after this redshift.

At zero density the field sits at the supersymmetric minimum $\varphi=\varphi_{\rm min}$ where its mass is
\begin{equation}\label{eq:vphieffpot}
 m_\infty^2\equiv\frac{n\delta x \rho_\infty}{2\gamma^2\varphi_{\rm min}^2}=
\frac{3n\delta x}{2\gamma^2}\Omega_{\rm
c}^0(1+z_\infty)^3\left(\frac{\mpl}{\varphi_{\rm min}}\right)^2H_0^2.
\end{equation}
When $\rc\gsim\rho_\infty$ the field minimum is moved to smaller values by the matter coupling term and supersymmetry is
broken. Supersymmetry is therefore broken locally in our model
depending on the ambient density and $\rho_\infty$. The scale of this breaking is set by the cold dark matter density and the
model parameters, however this is generally far lower than the TeV scale associated with particle physics in the observable
sector. This is one advantage of decoupling the dark and observable sectors, the dark sector does not suffer from an unnatural
hierarchy of scales set by the decoupling of standard model particles. Away from the
supersymmetric minimum, the
field's mass is \begin{equation}\label{eq:mphi}
 m_{\rm \varphi}^2=m_\infty^2\left[\frac{2(n+\gamma)}{n}\left(\frac{\varphi_{
\rm
min}}{\varphi}\right)^{\frac{n}{\gamma}+2}-\frac{n+2\gamma}{n}\left(\frac{
\varphi_ {\rm
min}}{\varphi}\right)^{\frac{n}{2\gamma}+2}+\frac{2(\delta-\gamma)}{n}\frac{\rc}
{ \rho_\infty}\left(\frac{\varphi_{\rm
min}}{\varphi}\right)^{2-\frac{\delta}{\gamma}}\right].
\end{equation}
Clearly $m_{\rm eff}(\varphi)>m_\infty$ when $\varphi<\varphi_{\rm min}$ and so these models are indeed chameleons with a mass at
the minimum of the effective potential which depends on the matter density.

\subsection{Supergravity Corrections and Screening}

We saw in section \ref{sec:sugracorr} that supergravity corrections place extremely stringent constraints on combinations of
the model parameters and ensure that the screening is too efficient to allow laboratory or astrophysical tests. Here we will
briefly elucidate how this works in practice by calculating the corrections to this specific class of models.

The correction to the potential coming from supergravity breaking is
\begin{equation}
 \Delta
V_{\cancel{\mathrm{SG}}}=\frac{m_{3/2}^2\left|K_{\Phi}\right|^2}{K_{
\Phi\Phi^\dagger}}\sim\frac{m_{3/2}^2\phi^{2\gamma}}{\Lambda_1^{2\gamma-2}},
\end{equation}
which competes with the density-dependent term in the effective potential (\ref{eq:bveff}) to set the position of the minimum.
Since we focus on the branch of the potential where $\phi\le\pmi$, this term can always be neglected provided that it is far less
than the density-dependent term when $\phi$ has converged to its supersymmetric minimum. This requires that the supergravity
corrections are negligible at densities around $\rho_\infty$ so that
\begin{equation}\label{eq:sugraphi}
 \left(\frac{\varphi_{\rm min}}{\mpl}\right)^2\ll \frac{x\rho_\infty}{\mpl^2m_{3/2}^2}.
\end{equation}
The denominator is proportional to the supergravity contribution to the vacuum energy, which is typically very large and is
usually fine-tuned away whereas the numerator is proportional to the vacuum energy when the supersymmetric minimum is reached (see
below), which we expect to be well below this. This condition tells us that the supersymmetric minimum must be well below the
Planck scale and is simply the statement that the matter coupling and fifth-force is a low-energy, IR phenomena. It will be useful
to express this condition in the alternative form
\begin{equation}\label{eq:SGB}
 \left(\frac{\varphi_{\rm min}}{\mpl}\right)^2\ll 3x\Omega_{\rm c}^0(1+z_\infty)^3\left(\frac{H_0}{m_{3/2}}\right)^2,
\end{equation}
from which it is immediately evident that $\varphi_{\rm min}\ll \mpl$ even when the gravitino mass is as low as the gauge mediated
supersymmetry breaking value of $1$ eV. Using (\ref{eq:rhoinf}) equation (\ref{eq:sugraphi}) can be recast as a condition on the
normalised field's mass at the supersymmetric minimum
\begin{equation}\label{eq:minfbound}
 m_\infty^2\gg \frac{3n\delta}{2\gamma^2}m_{3/2}^2.
\end{equation}
The field's mass is at least as large as the gravitino mass. The condition (\ref{eq:sugraphi}) also has implications for the
coupling $\beta(\varphi)$. Using it in conjunction with equation (\ref{eq:betaphi}), the coupling when the field reaches the
supersymmetric minimum (we will see in the next section that this happens at late times) satisfies
\begin{equation}\label{eq:bigbeta}
 \beta(\varphi_{\rm min})\gg \frac{x^{\frac{1}{2}}\delta}{\sqrt{3\Omega_{\rm c}^0}\gamma}\left(\frac{m_{3/2}}{H_0}\right)\gg
10^{33}x^{\frac{1}{2}}
\end{equation}
and so we see that $\beta(\varphi)\gg\mathcal{O}(1)$ unlike more conventional chameleon models that assume order-one couplings.

\section{Cosmology}\label{sec:cos}

In this section we will examine the cosmology of these models with the aim of accounting for dark energy. As alluded to in the
introduction to this chapter, we will ultimately find that a cosmological constant is required in order to match both the present
day equation of state $w$ and the energy density in dark energy.

\subsection{Background Cosmology}\label{sec:backcos}

Solving (\ref{eq:phimineq}) for the minimum in the limit of both large and small dark matter density we have
\begin{equation}
 \frac{\varphi}{\varphi_{\rm min}}\approx\left\{
 \begin{array}{l l}
   \left(\frac{\rho_\infty}{\rc}\right)^{\frac{\gamma}{n+\delta}},&
\rc\gg\rho_\infty \\
    1,&\rc\ll\rho_{\infty}\\
  \end{array}\right.. \label{eq:vphimin}
\end{equation}
We can now find the contribution to the vacuum energy density
\begin{equation}
 V_{\rm eff}(\varphi)\approx\left\{
  \begin{array}{l l}
    \frac{x(\delta+n)}{n}\rc\left(\frac{\rho_\infty}{\rc}\right)^{\frac{\delta}{n+\delta}},& \rc\gg\rho_\infty \\
    x\rc,&\rc\ll\rho_{\infty}\\
  \end{array}\right. \label{eq:Veffphimin}
\end{equation}
and the mass of the field using (\ref{eq:mphi})
\begin{equation}
 \left(\frac{m_\varphi}{m_\infty}\right)^2\approx\left\{
  \begin{array}{l l}
\frac{2(\delta+n)}{n}\left(\frac{\rc}{\rho_\infty}\right)^{\frac{n+2\gamma}{
n+\delta}},& \rc\gg\rho_\infty \\
    1,&\rc\ll\rho_{\infty}\\
  \end{array}\right. .\label{eq:mphimin}
\end{equation}
Finally, one can find the equation of state for the field $w_\varphi$ using equation (\ref{eq:wchams}).
so that
\begin{equation}
 w_\varphi=\left\{
  \begin{array}{l l}
    -\frac{\delta}{n+\delta},& \rc\gg\rho_\infty \\
    0,&\rc\ll\rho_{\infty}\\
  \end{array}\right. .\label{eq:wmin}
\end{equation}

In order to match this with current observations we would like $w_\varphi\approx-1$, which can be achieved by
taking the
limits $\rc\gg\rho_\infty$, $\delta\gg n$ and imposing the condition
\begin{equation}\label{eq:viacond}
 x\delta(1+z_\infty)^3\approx3n\Omega_{\rm c}^0.
\end{equation}
This corresponds to the case where $z_\infty<0$ and the supersymmetric minimum has not been reached by the current epoch. Both $n$
and $\delta$ appear as indices (or a combination of indices) in the superpotential and so we would expect them to be of similar
order; taking $\delta\gg n$ is then tantamount to neglecting many lower order operators in the superpotential, making the model
appear somewhat contrived. When these conditions are not met --- we will see shortly that this scenario gives rise to
unacceptably large deviations in the CDM power spectrum --- a cosmological constant is required in order to account for the
present-day dark energy observations. Unlike most models however, it is not so simple to add a cosmological constant by hand
within a supersymmetric framework. Global supersymmetry is broken if $\langle V\rangle\ne0$ and so the addition of a cosmological
constant to the system is non-trivial. One method is to appeal to supergravity breaking, which adds a cosmological constant of the
order $\mpl^2m_{3/2}^2\gg\rho_0$ and so one must somehow fine-tune to great extent in order to arrive at the small value observed
today. Here, we will take a different approach. If we assume that the cosmological constant problem in the matter and
observable sectors is solved then we can dynamically generate a cosmological constant at late times in the form of a
Fayet-Illiopoulos term provided that there exists a coupling between the chameleon and two $\mathrm{U}(1)$ charged scalars.
Unfortunately, this does not remove the need for some degree of fine-tuning, since the value of the Fayet-Illiopoulos constant
must be set by hand in this framework, however, this method has the advantage that this constant receives no quantum corrections
from decoupling particles and so if one can find a more UV complete theory where its value is set in terms of other constants then
this value would be preserved at low energy scales. The study of globally supersymmetric chameleons is aimed as a first step
towards realising them within a more UV complete theory and a lot of insight can be gained by studying this mechanism.

\subsection{A Late-time Cosmological Constant}\label{sec:cosgen}

An effective cosmological constant can be implemented by introducing two new scalars\footnote{One new scalar is not sufficient
since there is no $\mathrm{U}(1)$-invariant superpotential that can be written down for a single charged scalar and so it is
not possible to couple the gravitational scalar to the charged sector.} $\Pi_\pm=\pi_\pm+\ldots$ with charges $\pm q$
under a local $\mathrm{U}(1)$ gauge symmetry. These have the canonical K\"{a}hler potential
\begin{equation}
  K=\Pi_+^\dagger e^{2qX}\Pi_+ + \Pi_-^\dagger e^{-2qX}\Pi_-,
\end{equation}
where $X$ is the $\mathrm{U}(1)$ vector multiplet containing the gauge field and couple to the superchameleon via the
superpotential
\begin{equation}
 W_\pi=g^\prime\Phi\Pi_+\Pi_-.
\end{equation}
This construction gives rise to a new structure for the F-term potential as well as a D-term potential for the fields $\pi_\pm$:
\begin{equation}
V_{\rm D} =\frac{1}{2}\left(q\pi_+^2-q\pi_-^2-\xi^2\right)^2,
\end{equation}
where we have included a Fayet-Illiopoulos term $2\xi^2$ which will later play the role of the cosmological constant. The new
scalar potential is far more complicated with the addition of these new fields but when $\langle\pi_-\rangle=0$ it reduces to the
original effective potential for the superchameleon (\ref{eq:bveff}) and an effective potential for $\pi_+$:
\begin{equation}\label{eq:D-term}
 V(\pi_+)=\frac{1}{2}\left(q\pi_+^2-\xi^2\right)^2+{g^{\prime}}^2\phi^2\pi_+^2;\quad \langle\pi_-\rangle=0,
\end{equation}
where in this expression we have set $\pi_+=|\pi_+|$ and will continue to do so from here on. In appendix \ref{app:mgp} we
minimise
the entire global F- and D-term potentials with respect to the angular fields coming from $\pi_\pm$ and show that
$\langle\pi_-\rangle=0$ is indeed a stable minimum of the system.

The mass of the charged scalar $\pi_+$ is
$m_{\pi_+}^2=\gp^2\phi^2-q^2\xi^2$. At early times the superchameleon is small ($\ll\phi_{\rm min}$) and this mass is negative
and the $U(1)$ symmetry is therefore broken ($\langle\pi_+\rangle\ne0$). However, as the cosmological field evolves towards its
minimum this mass slowly increases until it reaches zero, restoring the symmetry so that $\langle\pi_+\rangle=0$. We would
therefore expect $\pi_+=0$ in the late-time universe leaving us with the FI term, which plays the role of a cosmological constant.
Indeed, minimising (\ref{eq:D-term}) with respect to $\pi_+$ one finds
\begin{equation}\label{eq:pi+}
q^2\pi_+^2=\left\{
  \begin{array}{l l}
    0 & \quad \phi\ge\Delta\\
    q\xi^2-{g^\prime}^2\phi^2 & \quad \phi < \Delta\\
  \end{array}\right. ,
\end{equation}
where
\begin{equation}
\Delta\equiv \sqrt{\frac{q}{{g'}^2}}\xi
\end{equation}
and $\phi=\Delta$ is equivalent to the statement $m_{\pi_+}=0$. When $\langle\pi_+\rangle=0$ equation (\ref{eq:D-term}) reduces to
$V(\pi_+)=\xi^2/2$ and so we shall set $\xi\sim10^{-3}$ eV in order to match the present-day energy density in dark energy. There
is no natural choice for this parameter within our globally supersymmetric framework and so this value is completely arbitrary.
Once set, this value is largely robust to quantum corrections; when supersymmetry is unbroken they do not
run and when this is not the case they run logarithmically at most \cite{Jack:1999zs}. Therefore, if one could find a natural
mechanism by which a small FI term is present in a more UV complete theory, for example a suitable combination of two or more
large mass scales, then its value at lower energy scales will remain at the same magnitude\footnote{Here we are assuming that the
cosmological constant problem in the hidden and observable sectors is resolved.}; the same is not true of scalar VEVs, which
receive large corrections from heavy particle loops.

\subsection{The Model Parameter Space}

Given the above mechanism, it is prudent to examine the model parameter space to determine the viable regions where a cosmological
constant can appear. Firstly, when $\langle\pi_+\rangle\ne0$ (i.e. at early times) there are corrections to the superchameleon
potential which can act to alter its cosmological dynamics. Secondly, we must ensure that the cosmological constant has the
correct properties to reproduce current observations. We require the symmetry to be restored at field values lower than the
supersymmetric minimum, which requires
\begin{equation}\label{eq:mincond}
 \phi_{\rm min}>\Delta\quad\textrm{or equivalently}\quad
\left(\frac{M}{\Lambda}\right)^{\frac{4}{n}}>\left(\frac{q}{2{g^\prime}^2}\right)^{\frac{1}{2}}\frac{\xi}{M}.
\end{equation}
Furthermore, if our model is to produce the correct
imprint on the CMB then the cosmological constant must be generated before last scattering. We shall do this by imposing that the
cosmological density $\rho_\Delta$ (given in (\ref{eq:rhoi})) at which the $\mathrm{U}(1)$ symmetry is restored is greater than
$1$ eV$^4$.

\subsubsection{Corrections to the Scalar Potential}
At late times (defined by the time at which $\phi=\Delta$) we have a FI cosmological constant, but at earlier times the non-zero
VEV of $\pi_+$ induces corrections to the effective potential for $\phi$\footnote{At first glance, one may be concerned that the
correction
proportional to $-\phi^4$ results in a potential that is unbounded from below, but this form of the potential is deceptive. If one
were to consider allowing the field to run away down this potential then at some point we would be in a situation where
$\phi>\Delta$ and these corrections are no longer present; what looks like an unbounded potential is in fact a hill in the global
potential.}:
\begin{equation}\label{eq:dcorr}
 V_{\rm corr}=\frac{{g^\prime}^2\xi^2}{q}\phi^2-\frac{{g^\prime}^4}{2q^2}\phi^4.
\end{equation}
These corrections compete with the density-dependent term coming from the dark matter coupling and therefore act to negate the
chameleon mechanism. When they are important, they lead to a new, density-independent minimum. Since the magnitude of the
density-dependent term decreases as the dark matter redshifts away it is possible to have a scenario where the field gets stuck at
the new minimum and the cosmological constant is never generated. 

There are several possible scenarios involving these corrections, which either allow or preclude the generation of a cosmological
constant depending on the model parameters. If the corrections are negligible compared to the density-dependent term throughout
the entire time that $\phi<\Delta$ then they are never important to the model dynamics and vanish once $\phi>\Delta$. If, on the
other hand, the corrections are important before they vanish then their dynamics must be included. However, if $\phi$ can still
pass $\Delta$ then a cosmological constant can still be generated since the corrections vanish after $\Delta$ is passed. If the
only important correction is the quadratic one then a minimum always develops and therefore the cosmological constant will only be
generated if the field value at this minimum is larger than $\Delta$. If either the quartic correction or both corrections
simultaneously are important then the potential may or may not develop a minimum. If no minimum develops then the field will
eventually pass $\Delta$ since the potential takes on a (locally) run-away form. If a minimum does develop then we again require
the field value at this minimum be larger than $\Delta$ in order to generate a viable cosmology. The exact details of how one can
determine which scenario is applicable to a certain choice of parameters and whether or not the dynamics are affected to the
extent that the model is not viable are given in full detail in appendix \ref{app:dcor}. Below, we shall only present the
resulting parameter space once every possible scenario is taken into account.

\subsubsection{Low-Energy Parameters}

In order to classify the parameter space into viable regions we will need to derive certain conditions on combinations of the
model parameters and so it is important to know which parameters are fixed in terms of certain combinations of the others. It will
be sufficient to examine the position of the minima and the values of $\phi$ relative to $\Delta$ and at no point will we need to
use the dynamics of $\varphi$. For this reason, we will work exclusively with the field $\phi$ and not its canonically normalised
counterpart since this avoids unnecessary powers of $\gamma$. We have already seen in section \ref{sec:susycham} that three of the
underlying parameters $\Lambda_i$ ($i=0,1,2$) combine to form two derived parameters $M$ and $\Lambda$. What are observable (or
rather, what governs the low-energy dynamics) however are the low-energy parameters ${n,\delta,\gamma,x,\mu,z_\infty,\gp}$, which
are either combinations of $M$ and $\Lambda$ or indices that appear in the low-energy effective potentials (\ref{eq:AmodB}) and
(\ref{eq:D-term}); $\mu$ is a combination of the underlying parameter $\Lambda_3$ and the dark matter mass $m$. It will prove
useful to introduce the parameter 
\begin{equation}\label{eq:Gdef}
 \mathcal{G}\equiv \gp/\sqrt{q}.
\end{equation}
A static analysis therefore probes the six dimensional parameter space ${n,\delta,x,\mu,z_\infty,\mathcal{G}}$ and leaves
$\gamma$ unspecified.

In what follows, we will be interested in regions of parameter space where the background cosmology is viable and the parameters
themselves assume \textit{sensible} values. In order to decide exactly what is meant by ``sensible'' it is instructive to pause
and think about their physical significance. $n$ and $\delta$ are indices (or are combinations of indices) that appear in a
superpotential and so these should naturally have values close to $1$ as argued in section \ref{sec:backcos}.
$g\pr=\sqrt{q}\mathcal{G}$ is
a $\mathrm{U}(1)$ coupling constant that appears in the coupling of the charged fields to $\phi$ and so we would expect values of
$\mathcal{O}(10^{-2}-10^{-3})$ so that the theory is not strongly coupled. Values much smaller than this would be tantamount to
fine-tuning. Similarly, $x$ parametrises the ratio of the vacuum energy density to the matter density when the field has converged
to its supersymmetric minimum. The energy density in the field today is (see equation (\ref{eq:Veffphimin})) $V_{\rm
eff}(\pmi)=x\rc$, which must be less than $\xi^4$ so that the dominant contribution to dark energy comes from the cosmological
constant and so we require $x\lsim\mathcal{O}(1)$. Na\"{i}vely, one might argue that $x$ should be small since it also
parametrises the coupling to matter (see equations (\ref{eq:x}) and (\ref{eq:betaphi})) and so directly controls the enhanced
gravitational force. We have already seen that supergravity corrections ensure that all astrophysical fifth-force effects
are screened and so this argument does not apply.

Finally, we are left $\mu=m^{\frac{1}{\delta}}\Lambda_3^{\frac{\delta-1}{\delta}}$. When $\delta=1$ this is simply the dark matter
mass and thus varying $\mu$ is tantamount to fine-tuning the dark matter mass to produce an acceptable cosmology. When
$\delta\ne 1$ however we are free to fix the dark matter mass and what we are really varying is $\Lambda_3$. In this sense we are
not fine-tuning when we vary $\mu$ but are in fact scanning the space of viable cosmologies as a function of $\Lambda_3$. For this
reason, we shall always fix $\delta\ne1$. Now the dark matter mass can be any where from $\mathcal{O}(\textrm{eV})$ to
$\mathcal{O}(\textrm{TeV})$ depending on the model and $\Lambda_3$ appears as a mass scale in the underlying supersymmetric theory
and so we would naturally expect it to be large (at least compared to the scales involved in the low-energy dynamics). Hence, in
what follows we will treat anywhere in the region $\mathcal{O}(\textrm{eV})\lsim\mu\lsim\mathcal{O}(\mpl)$ as sensible.

\subsection{Constraints on the Parameter Space}\label{subsec:susycons}

Given the above considerations and the procedure for dealing with corrections to the effective potential in appendix
\ref{app:dcor} we are now in a position to explore the parameter space at the background level.

We have performed a thorough investigation into the exact effects of varying each of the six parameters on the cosmology and can
find a large region where the parameters are indeed sensible and the background cosmology is viable. It is difficult to gain any
insight from the equations since they are all heavily interdependent in a complicated fashion and a large number of plots can be
misleading since they can change very abruptly when a single parameter is varied by a small amount. For these reasons, here we
shall simply describe the qualitative effects of varying some of the more constrained and less interesting parameters and present
only a few
two-dimensional cross-sections once these have been fixed at sensible values. We will also indicate the regions where the
linear CDM power spectrum deviates from the general relativity prediction by up to 10\% and can therefore be probed with upcoming
experiments such as Euclid \cite{Amendola:2012ys} whilst the other regions will show a negligible deviation. This
region is found using the power spectrum derived in section \ref{sec:perts} and parameters that give
deviations already ruled out by WiggleZ \cite{Contreras:2013bol} are not shown in the plots.

\begin{figure}[ht]\centering
 \includegraphics[width=0.75\textwidth]{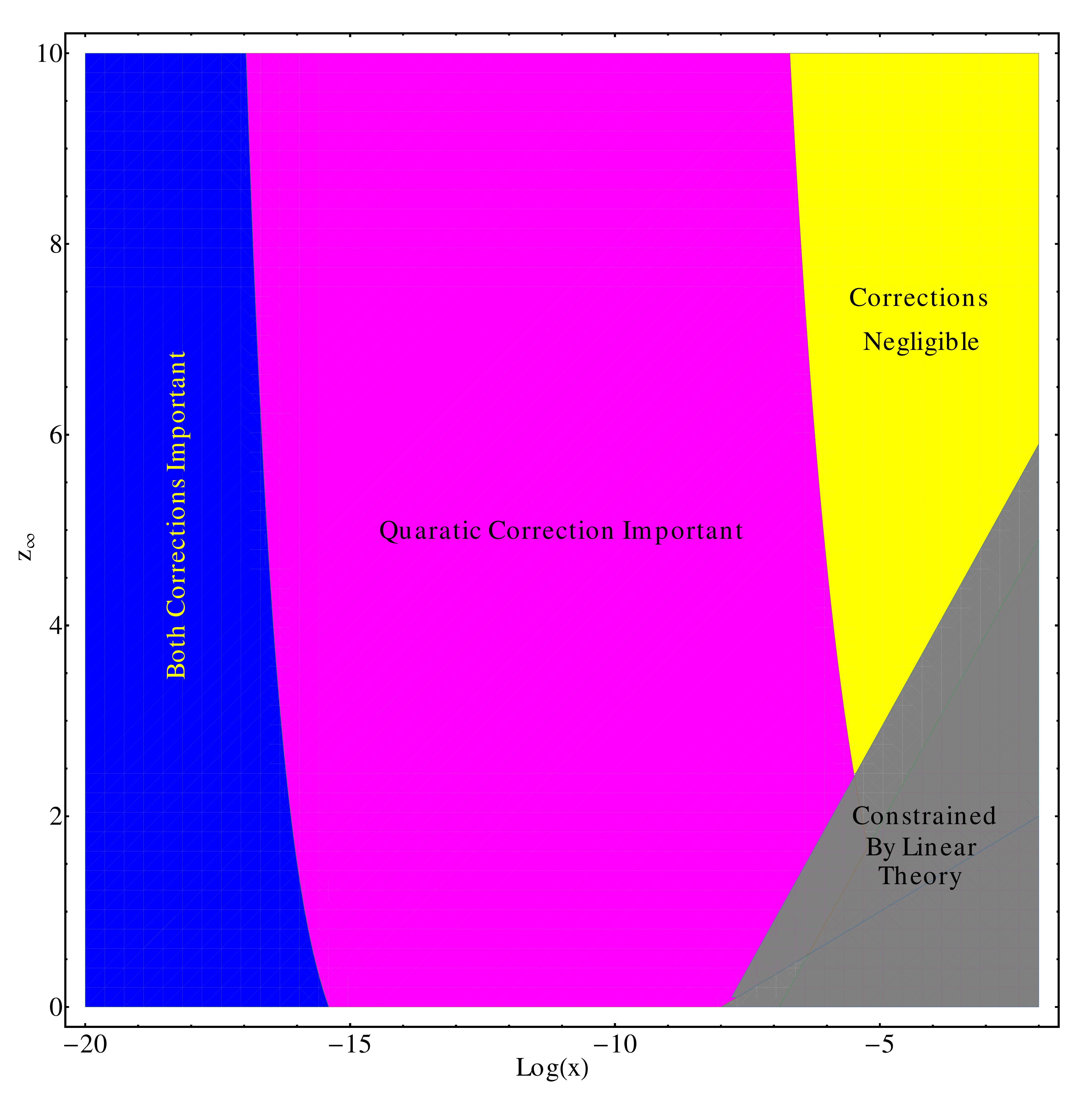}
\caption{The various regions in the $z_\infty$--$\log(x)$ plane with $n=\delta=2$, $\mu=10^3$ TeV and $\mathcal{G}=10^{-2}$. The
yellow
region shows the parameter range where the corrections are negligible. The magenta region shows the ranges where the quadratic
correction is important, the dark blue region where both corrections are important. The grey region corresponds to parameters
where the model deviates from $\Lambda$CDM at the level of linear perturbations.}\label{fig:zx}
\end{figure}

Let us begin with the indices. $n$\footnote{Note that $n=2(\gamma-\alpha)$ and $n$ can only assume even values.} and $\delta$ have
very similar effects:
if their value is increased whilst fixing the other five parameters then the
region of parameter space where the corrections can be neglected will increase.
Being indices, these should not stray too far from $\mathcal{O}(1)$ and so their
effects are far less pronounced than the other parameters, some of which can
vary over many orders of magnitude. Hence, from here on we will fix $n=\delta=2$.
In figure \ref{fig:zx} we plot
the $z_\infty$--$\log(x)$ plane with $\mu=10^3$ TeV and $\mathcal{G}=10^{-2}$ corresponding
to what we have argued above are sensible values. For the sake of brevity we
will set $z_\infty=5$ from here on. This choice is completely arbitrary and
different choices may give rise to very different cross-sections of parameter
space, however, the region where the corrections are negligible is both
ubiquitous and generically large when $z_\infty\gsim0$ and so it is not
necessary to scan this parameter in great detail in order to narrow down a
viable region.

\begin{figure}[ht]\centering
 \includegraphics[width=0.75\textwidth]{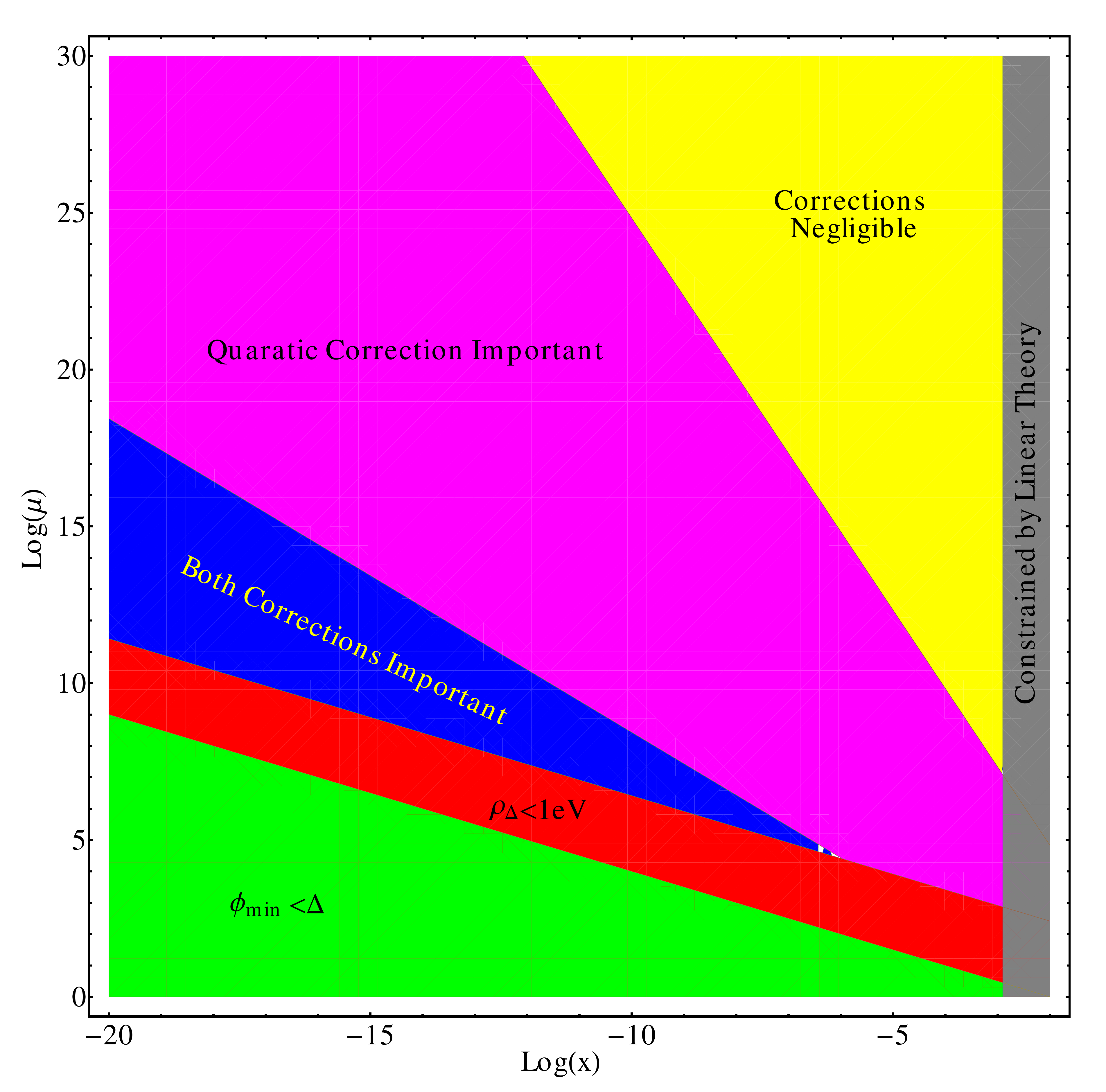}
\caption{The various regions in the $\log(\mu)$--$\log(x)$ plane with $n=\delta=2$, $z_\infty=5$ and $\mathcal{G}=10^{-2}$. The
yellow
region shows the parameter range where the corrections are negligible. The magenta region shows the ranges where the quadratic
correction is important and the dark blue region where both corrections are important. The red region corresponds to models where
a cosmological constant is generated after last scattering and are therefore excluded and the green region corresponds to models
where $\pmi<\Delta$ and a cosmological constant is only generated at some time in the future. The grey region corresponds to
parameters where the model deviates from $\Lambda$CDM at the level of linear perturbations.}\label{fig:mx}
\end{figure}

\begin{figure}[ht]\centering
 \includegraphics[width=0.75\textwidth]{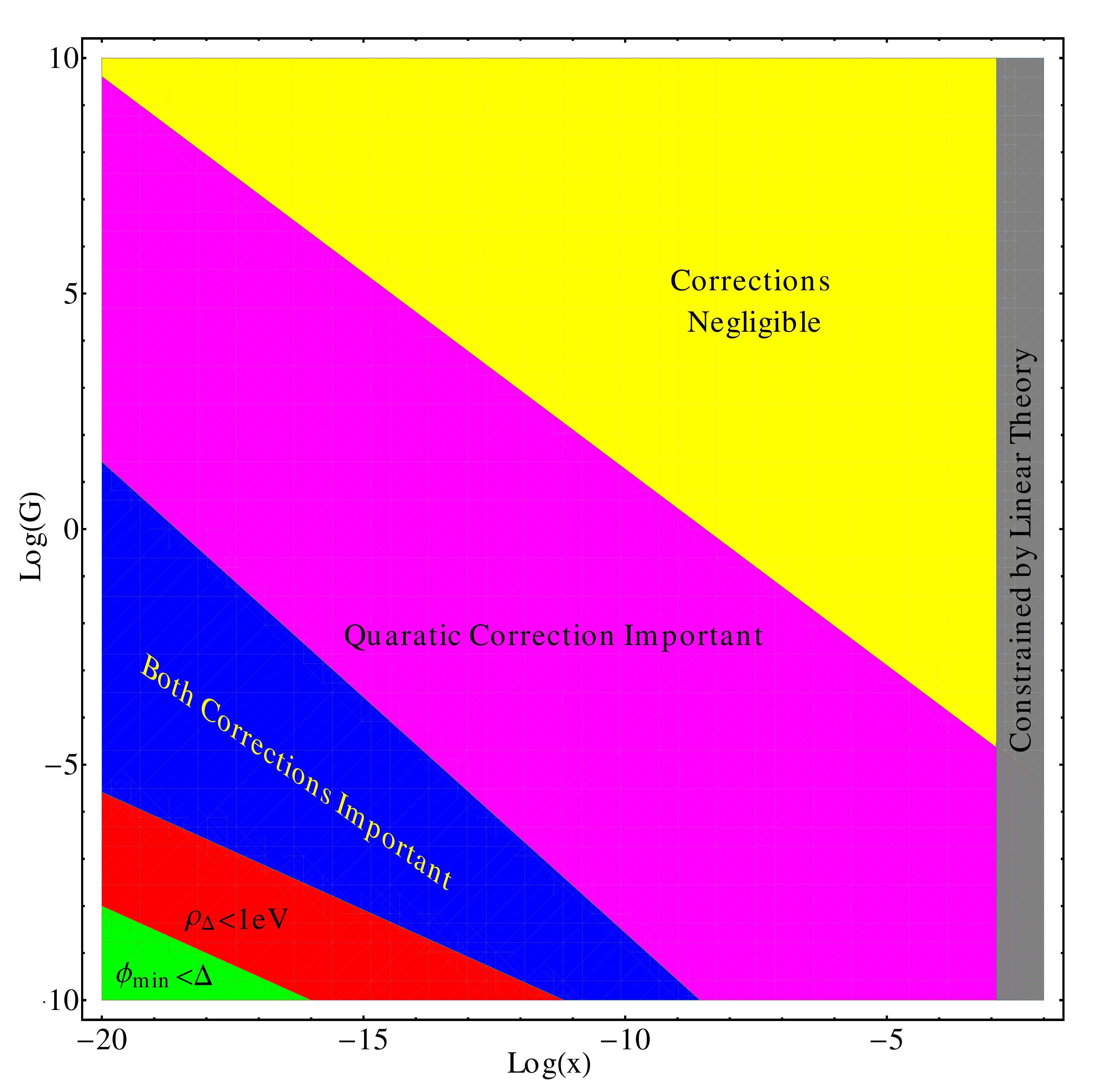}
\caption{The various regions in the $\log(\mathcal{G})$-$\log(x)$ plane with $n=\delta=2$, $z_\infty=5$ TeV and $\mu=10^{3}$ TeV.
The yellow
region shows the parameter range where the corrections are negligible. The magenta region shows the ranges where the quadratic
correction is important and the dark blue region where both corrections are important. The red region corresponds to models where
a cosmological constant is generated after last scattering and are therefore excluded and the green region corresponds to models
where $\pmi<\Delta$ and a cosmological constant is only generated at some time in the future. The grey region corresponds to
parameters where the model deviates from $\Lambda$CDM at the level of linear perturbations.}\label{fig:gx}
\end{figure}

Next, we plot the $\log(\mu)$--$\log(x)$ plane with $n=\delta=2$ and $\mathcal{G}=10^{-2}$ in figure \ref{fig:mx} to investigate
the effects
of varying $\mu$ on the viable region. It is evident from the figure that large ($\!\gsim\!\mathcal{O}(\textrm{TeV})$) values of
$\mu$
are required for there to be a large region with negligible corrections; in fact, if one steadily increases $\mu$ one finds that
this region grows, replacing the regions where the corrections are important. This behaviour can be traced back to equations
(\ref{eq:rhoi}) and (\ref{eq:M4n}) in appendix \ref{app:dcor} where it is shown that $M^{4+n}\propto \mu^n$ and therefore the
density at which the corrections disappear increases slightly faster with $\mu$ than the densities at which the corrections become
important.
Finally, now that we have some idea of the viable values of $z_\infty$ and $\mu$ we plot the $\log(\mathcal{G})$-$\log(x)$ plane
with
$n=\delta=2$ and $\mu=10^{3}$ TeV in figure \ref{fig:gx} in order to investigate the values of $\mathcal{G}$ where the corrections
are
negligible. One can see that when $\mathcal{G}\gsim\mathcal{O}(1)$ the corrections are generally negligible, which is a result of
(\ref{eq:rhoi}) in appendix \ref{app:dcor}, which show that the density at which the corrections disappear generally grows faster
with $\mathcal{G}$ than the density at which they become important. The density at which the corrections are important both
include an
explicit factor of $x^{-1}$ which is absent from the density at which the corrections disappear (there are other factors of $x$
coming from the scale $M$ though these vary with a far smaller power). This is the reason that the region where the corrections
are negligible is larger when $x$ is closer to $1$. The plot clearly shows that there is a large region around $\mathcal{G}\approx
10^{-2}$,
which we have argued above is a sensible range where there is no excessive fine-tuning or strong $\mathrm{U}(1)$ coupling. With
the parameters we have chosen this only exists when $x\gsim10^{-10}$, however this does not really constrain $x$. If one were to
increase either $n$, $\delta$, $z_\infty$ or $\mu$ this region would extend further in the direction of decreasing $x$.
Thus, we have found that when the model parameters assume sensible values there is a large region of parameter space where the
corrections to the effective potential are negligible before they vanish and an FI cosmological constant appears at late times.
When the FI cosmological constant is generated, the background cosmology is indistinguishable from $\Lambda$CDM and so one must
look at the growth of structure on linear scales in order to probe this theory. This will be examined in detail in the next
section. 

\section{Linear Perturbations}\label{sec:perts}
%
In this section we calculate the modified linear CDM power spectrum predicted by these models. This allows us to rule out
parameters that give deviations in tension with
current experiments such as WiggleZ \cite{Contreras:2013bol} and identify those regions that could be probed by
upcoming experiments such as Euclid \cite{Amendola:2012ys}.  

The equation governing the evolution of CDM perturbations in the conformal Newtonian gauge was given in equation
(\ref{eq:pertfull}). Given (\ref{eq:minfbound}), we have $am_{3/2}\gg2.5\times10^{28}a\; \textrm{Mpc}^{-1}$ and so on the scales
of interest
we are always in the limit $k\ll m_\varphi a$. In this limit we can linearise (\ref{eq:pertfull}) to find
\begin{equation}\label{eq:pertlin}
 \ddot{\dc}+2H\dot{\dc}-\frac{3}{2}\Omega_{\rm c}(a)H^2\left(1+2k^2\frac{\beta^2(\varphi)}{m_{\rm eff}^2a^2}\right)\dc\approx0.
\end{equation}
Equation (\ref{eq:pertlin}) will be our starting point in what follows; it has solutions that can be written in terms of
\textit{modified Bessel functions}. A short introduction to
these functions, including how equations of the form (\ref{eq:pertlin}) can be transformed into the modified Bessel equation is
provided in appendix \ref{app:bessel}. The quantity $\dc$ is gauge-dependent and so is not a physical observable. Previous works
have put
constraints on their models by looking for the parameters where the final term in (\ref{eq:pertlin}) is small and the general
relativity result is recovered. ``Small'', is a gauge-dependent statement and  it is the linear power spectrum that is the
physical observable so here we shall
study its deviations from the $\Lambda$CDM predictions as a potential observable probe of supersymmetric chameleons.

In the radiation era, we have
\begin{equation}\label{eq:radpert}
 \ddot{\dc}+2H\dot{\dc}-\frac{3}{2}H^2 \left(\frac{G_{\rm eff}(k)}{G}\right)\frac{\bar{\rc}}{\bar{\rho_c}+\bar{\rho_{\rm
r}}}\dc=0,
\end{equation}
where $G_{\rm eff}(k)$ is defined in (\ref{eq:effG}) and $\bar{\rc}/(\bar{\rho_c}+\bar{\rho_{\rm r}})\approx(1+z_{\rm
eq})/(1+z)$. In general
relativity, $G_{\rm eff}=G$ and this final term is negligible compared with the time derivatives of $\dc$, which scale as
$H^2\dc$, and
can be neglected to give a logarithmic growth of the density contrast, $\dc\propto\ln(t)$. With the inclusion of modified gravity,
one has
\begin{equation}\label{eq:radGeff}
\frac{G_{\rm eff}-G}{G}\left(\frac{1+z_{\rm eq}}{1+z}\right)\approx \mathcal{O}(1)\,x\left(\frac{k}{10^{-5}\textrm{
Mpc}^{-1}}\right)^2\frac{1}{(1+z)^{\frac{2n+5\delta}{n+\delta}}},
\end{equation}
where $H_0^2\sim 10^{-5}h^2$ Mpc$^{-1}$. Now modes deep inside the horizon can have arbitrarily large values of $k$, however, we
know that, in general
relativity at least, modes with $k>0.1h$ Mpc$^{-1}$ will be non-linear today even if they were not in the past so that we do
not need to keep track of their evolution. In modified gravity, we expect this number to be smaller, however here we will use the
general
relativity value in order to be conservative. In this case, the largest mode which is linear today satisfies $k/10^{-5}h\textrm{
Mpc}^{-1}\sim10^4$.
Now $z\gg z_{\rm eq}\sim 10^3$ and the minimum possible value of $(2n+5\delta)/(n+\delta)$ is 2 so that the maximum deviation from
general
relativity satisfies
\begin{equation}
\left.\frac{G_{\rm eff}-G}{G}\left(\frac{1+z_{\rm eq}}{1+z}\right)\right\vert_{\textrm{max}}\ll 10^2x.
\end{equation}
Unless $x$ assumes unrealistically large values $x\sim\mathcal{O}(1)$, which we shall see below gives large deviations from the
CDM power spectrum today in tension with current observations, the final term in (\ref{eq:radpert}) is negligible and the modes
evolve in an identical manner to general relativity.

We can use the standard general
relativity result and for simplicity we will not use the full logarithmic form but will treat the modes as constant
inside the radiation era\footnote{This approximation may be relaxed with little effort and indeed should be if one wishes to
compare with observational data. Given that we shall not do so here there is little to be gained by including the
logarithmic term.}.
Outside the horizon, both during the matter and radiation eras, we have $G_{\rm eff}\approx G$ since the modifications of general
relativity are suppressed and so the perturbations do not evolve. We hence treat the modes as constant until the time of horizon
re-entry. Finally, we shall treat the change from radiation to matter domination as a sharp transition. Whilst not strictly
necessary, this allows us to compute the power spectrum analytically and there are no subtleties associated with modified
gravity in treating the full transition period. We make this approximation in the interest of discerning the new features
introduced by supersymmetric chameleons as simply as possible. 
 
On the scalar field side, we assume that the field settles into its supersymmetric minimum instantaneously at $\rc=\rho_\infty$,
thereby ignoring the short-lived transition period when equation (\ref{eq:phimineq}) has no analytic solution and any
oscillations around the minimum. This was studied in a specific case in \cite{Brax:2004qh}, where it was found that the sharp
settling is a very good approximation and so we do not expect the short-lived transition period to impact upon the power spectrum.
Our power spectrum therefore exhibits unphysical sharp discontinuities at the scales which enter the horizon at matter-radiation
equality and (as we shall see momentarily) at $z_\infty$, which would be found to be smooth curves had we solved the full
equations numerically. We will primarily be concerned with the power spectrum on large scales, since this is where it is
least constrained and so these unphysical features will play no part in our conclusions.

In what follows we shall consider two distinct cases. In the previous subsection we found that we can account for dark energy
without a cosmological constant by imposing $z_\infty<0$, $\rc\gg\rho_\infty$ and $\delta\gg n $ and so we shall first investigate
this case. We will find that this gives unacceptably large deviations from the general relativity prediction and so we then
proceed to investigate the general case $z_\infty\gsim0$. The field is assumed to reach its
supersymmetric minimum sometime around the current epoch i.e. $z_\infty\lsim10$ although our treatment will be valid for
$z_\infty\le z_{\rm eq}$.

In the following we focus on the power spectrum in the matter era. After the end of the matter era around a redshift $z\sim 1$,
the growth of structure is slowed down by the presence of dark energy. Hence deviations of the power spectrum from $\Lambda$CDM
are maximally dependent on the features of modified gravity when calculated at the end of the matter era. Here we shall only make
sure that the deviations from general relativity are no larger than around 10\% to comply with recent observations
\cite{Contreras:2013bol}. Upcoming large scale surveys like Euclid will measure the linear CDM power spectrum to the
percent level \cite{Amendola:2012ys} and so have the power to place further constraints on superchameleon models. The results of
these calculations have been used in figures \ref{fig:zx}--\ref{fig:gx}
in section \ref{subsec:susycons} to indicate where exactly in the parameter space these deviations occur.

\subsection{$z_\infty<0$}

This is the case where the field is displaced from the supersymmetric minimum at all times prior to the present epoch and will
settle into it at some point in the future. We have seen in section \ref{sec:backcos} that when $\delta\gg n$ and
$x\delta(1+z_\infty)^3\approx3n\Omega_{\rm c}^0$ we can have
$w_\varphi\approx-1$ and $V_{\rm eff}\sim \mpl^2H_0^2$, consistent with the current dark energy observations. Since this case is
of particular interest to us we will enforce these conditions below and refer to them as the \textit{dark energy conditions}.
Assuming a matter dominated era we can use equations (\ref{eq:betaphi}) and (\ref{eq:mphi}) in (\ref{eq:pertlin}) to find
\begin{equation}\label{eq:pertvia}
 t^2\ddot{\dc}+\frac{4}{3}t\dot{\dc}-\left[\frac{2}{3}\Omega_{\rm
c}^0+9x(1+z_\infty)^3(kt_0)^2\left(\frac{t}{t_0}\right)^{\frac{8}{3}}\right]\dc.
\end{equation}
Following appendix \ref{app:bessel}, the growing mode solution is
\begin{equation}\label{eq:d>npert}
 \dc(t)=C_{\rm MG}(k)t^{-\frac{1}{6}}I_\nu\left[\sigma kt_0\left(\frac{t}{t_0}\right)^{\frac{4}{3}}\right],
\end{equation}
where
\begin{equation}
 \nu^2=\frac{1}{8}\left(\frac{1}{8}+3\Omega_{\rm c}\right)\quad\textrm{and}\quad \sigma^2\equiv
\frac{81x(1+z_\infty)^3}{16}\approx \frac{243}{16}\frac{n\Omega_{\rm c}^0}{\delta},
\end{equation}
where the last equality for $\sigma$ holds when we impose the dark energy conditions conditions. This should be
compared with the general relativity prediction
\begin{equation}
 \dc(t)=C_{\rm GR}(K)t^{n};\quad\textrm{where}\quad n=-\frac{1}{6}+\frac{1}{2}\sqrt{\frac{1}{9}+\frac{8}{3}\Omega_{\rm c}^0},
\end{equation}
where $\Omega_{\rm c}^0\sim 1$ at the end of the matter era.
For small $x$, we have $I_\nu(x)\sim x^\nu\left[1+\mathcal{O}(x^2)\right]$ to leading order (see appendix \ref{app:bessel}) and
noting that $8\nu=6n+1$ we can see that these expressions agree for small $k$ as indeed they should since $G_{\rm eff}\approx G$
in this limit. 

Given the solution (\ref{eq:d>npert}) we are now in
a position to calculate the power spectrum. We start by noting that the time at which a given mode crosses the horizon ($k=2\pi a
H$) is
\begin{equation}
 t_{\rm H}=t_0\left(\frac{4\pi}{3t_0k}\right)^3
\end{equation}
and assume that the modes are constant during the radiation era and outside the horizon in the matter era as discussed above. In
this case, the contrast during the radiation era (and outside the horizon in the matter era) is given by the primordial
fluctuations from inflation, $\dc^{\rm I}$. Modes that enter the horizon during the matter era, that is modes with $k<k_{\rm
eq}=0.01 h$ Mpc$^{-1}$, will begin evolving according to (\ref{eq:d>npert}) and so we have the boundary condition $\dc(\h)=C_{\rm
MG}(k)\h^{-1/6}I_\nu[\sigma kt_0(\h/t_0)^{4/3}]$, which allows us to find $C_{\rm MG}(k)$ and hence the power spectrum
\begin{equation}\label{eq:ps_dn}
 P(k)=\langle\left\vert\dc(t_0)\right\vert^2\rangle=\langle\left\vert\dc^{\rm I}\right\vert^2\rangle\left\{\begin{array}{l l}
\left(\frac{\h}{t_0}\right)^{\frac{1}{3}}\frac{I^2_\nu\left(\sigma kt_0\right)}{I^2_\nu\left[\sigma
kt_0\left(\frac{\h}{t_0}\right)^{\frac{4}{3}}\right]}& k<k_{\rm eq}\\
\left(\frac{t_{\rm eq}}{t_0}\right)^{\frac{1}{3}}\frac{I^2_\nu\left(\sigma kt_0\right)}{I^2_\nu\left[\sigma kt_0\left(\frac{t_{\rm
eq}}{t_0}\right)^{\frac{4}{3}}\right]}& k>k_{\rm eq}
\end{array}\right. .
\end{equation}
Modified Bessel functions diverge from their leading order general relativity behaviour very rapidly and so if the modified power
spectrum is not to deviate from the general relativity prediction too greatly the arguments of both functions in (\ref{eq:ps_dn})
must be small enough such that the leading order behaviour is a good approximation, at least over the entire range of $k$ where
linear theory is valid. Since $t_0>\h$ this is equivalent to demanding $\sigma k t_0\ll1$. When this is satisfied the power
spectrum will show no deviations from general relativity and when this begins to break down we expect to see small deviations. One
can verify using equation (\ref{eq:seriesI}) in appendix \ref{app:bessel} that the leading order dependence is $k^4$, the same as
predicted by general relativity. Alternatively, according to (\ref{eq:effG}) we should recover the general relativity prediction
whenever $k\ll m_{\rm eff}/\beta(\varphi)$. Using equations (\ref{eq:betaphi}) and (\ref{eq:mphi}) one finds an equivalent
condition up to an order unity coefficient. 

One can then define scale $\tilde{k}$ below which no gravitational enhancement is felt:
\begin{equation}
 \tilde{k}^2\simeq \frac{H_0^2}{x(1+z_\infty)^3}=\frac{\delta H_0^2}{3n\Omega_{\rm c}^0},
\end{equation}
where the last equality holds when we impose the dark energy conditions. Since $k\sim H_0$ corresponds to modes which enter the
horizon today we expect deviations from general relativity on smaller scales. Unless $\delta\gg10^6n$, the linear CDM power
spectrum differs from the general relativity one by several orders of magnitude. Since $\delta$ appears as an index in the
superpotential such a large value seems highly unnatural. With this in mind, we will abandon this limit and proceed to study the
general case where we allow $n$ and $\delta$ to vary independently and $\varphi$ to converge to its supersymmetric minimum at some
point in the recent past.

\subsection{$z_\infty>0$}\label{sec:pertg}

This is the case where the field has settled into its supersymmetric minimum sometime in the past. We start by noting that when
$z_\infty>0$ the final term in (\ref{eq:pertlin}) will exhibit a different time dependence after the
field has converged to its supersymmetric minimum, so we must keep track of modes that enter the horizon before and after this and
match the time evolution appropriately. We therefore begin by solving (\ref{eq:pertlin}) for the case where $z<z_\infty$ so that
$\varphi\approx\varphi_{\rm min}$ and $m_\varphi\approx m_\infty$. Assuming a matter dominated epoch and defining $k_\infty=2\pi
a(t_\infty)H(t_\infty)$ to be the mode which enters the horizon when $z=z_\infty$ ($t_\infty=t_0(1+z_\infty)^{-3/2}$) we have
\begin{equation}\label{eq:pertzlzi}
 t^2\ddot{\dc}+\frac{4}{3}t\dot{\dc}-\left[\frac{2}{3}\Omega_{\rm
c}^0+\frac{9x\delta}{n(1+z_\infty)^{3}}(kt_0)^2\left(\frac{t_0}{t}\right)^{\frac{4}{3}}\right]\dc\quad k<k_\infty.
\end{equation}
Following appendix \ref{app:bessel} we can again write down the solution in terms of modified Bessel functions. Since the final
term decreases with increasing $t$ the growing mode is the modified Bessel function of the second kind:
\begin{equation}\label{eq:infty}
 \dc(t)=C_{\rm MG}^{k<k_\infty}(k)t^{-\frac{1}{6}}K_\omega\left[\zeta kt_0\left(\frac{t_0}{t}\right)^{\frac{2}{3}}\right],
\end{equation}
where
\begin{equation}\label{eq:zeta}
 \zeta^2\equiv \frac{27x\delta}{2n(1+z_\infty)^3};\quad\textrm{and}\quad \omega^2=\frac{9}{4}(\frac{1}{36}+\frac{2}{3}\Omega_{\rm
c}^0).
\end{equation}
Next, we must find the solution when $k>k_\infty$. In this case we have (using equations (\ref{eq:betaphi}), (\ref{eq:vphimin})
and (\ref{eq:mphimin}))
\begin{equation}\label{eq:pertgenzg0}
 t^2\ddot{\dc}+\frac{4}{3}t\dot{\dc}-\left[\frac{2}{3}\Omega_{\rm
c}^0+9\frac{x\delta}{n+\delta}(1+z_\infty)^{\frac{3\delta}{n+\delta}}k^2t_0^2\left(\frac{t}{t_0}\right)^{\frac{8\delta+2n}{
3(\delta+n)}}\right]\dc,
\end{equation}
the solution of which is
\begin{equation}\label{eq:generalpert}
 \dc(t)=C_{\rm MG}^{k>k_\infty}(k)I_{\nu}\left[\sigma kt_0\left(\frac{t}{t_0}\right)^r\,\right];\quad
\nu^2=\left(\frac{\delta+n}{4\delta+n}\right)^{2}\left[\frac{1}{4}+6\Omega_{\rm c}^0\right],
\end{equation}
with
\begin{equation}
 \sigma^2\equiv\frac{81x\delta(\delta+n)(1+z_\infty)^{\frac{3\delta}{\delta+n}}}{(4\delta+n)^2}\quad\textrm{and}\quad
r=\frac{4\delta+n}{3(\delta+n)}.
\end{equation}
We can use this to calculate the power spectrum in the general case. Modes that enter the horizon during the radiation dominated
era (i.e. $k>k_{\rm eq}$) are constant until matter radiation equality when they start growing according to equation
(\ref{eq:generalpert}). In this case we have $C_{\rm MG}^{k>k_\infty}(k>k_{\rm eq})I_{\nu}[\sigma(t_{\rm eq}/t_k)^r]=\dc(t_{\rm
eq})$. On the other hand, modes which enter during matter domination are subject to the condition $C_{\rm
MG}^{k>k_\infty}(k<k_{\rm eq})I_{\nu}\left[\sigma(\h/t_k)^r\right]=\dc(\h)$. Thus, modes that enter the horizon before $z_\infty$
evolve according to
\begin{equation}\label{eq:pspecnew}
 \dc(t)= \dc^{\rm I}\left\{
 \begin{array}{l l}
  \left(\frac{t_{\rm eq}}{t}\right)^{\frac{1}{6}}\frac{I_{\nu}\left[\sigma
kt_0\left(\frac{t}{t_0}\right)^r\right]}{I_{\nu}\left[\sigma kt_0\left(\frac{t_{\rm eq}}{t_0}\right)^r\right]}& k>k_{\rm eq} \\
  \left(\frac{\h}{t}\right)^{\frac{1}{6}}\frac{I_{\nu}\left[\sigma kt\left(\frac{t}{t_0}\right)^r\right]}{I_{\nu}\left[\sigma
kt_0\left(\frac{\h}{t_0}\right)^r\right]}&k<k_{\rm eq}\\
 \end{array}\right. .
\end{equation}
Near $z_\infty$, modes inside the horizon (whether they entered during the radiation or matter era) thus evolve according to the
general form $\dc(t)=D(k)t^{-1/6}I_\nu[\sigma kt_0 (t/t_0)^{r}]$ where the form of $D(k)$ varies depending on when the mode
entered the horizon as detailed in (\ref{eq:pspecnew}). When $z=z_\infty$ the field converges to its supersymmetric minimum and
the evolution now
proceeds according to equation (\ref{eq:infty}) and we must again match the two solutions at $z=z_\infty$ so that
$\dc(t_\infty)=D(k)t_\infty^{-1/6}I_\nu[\sigma kt_0 (t_\infty/t_0)^{r}].$ Modes that enter the horizon later than this simply
evolve according to (\ref{eq:infty}), matching at the time when they enter the horizon, in which case we have $\dc(\h)=C_{\rm
MG}^{k<k_\infty}(k)\h^{-1/6}K_\omega[\zeta kt_0(t_0/\h)^{2/3}]$. Thus, the power spectrum taking on different functional
forms in three different regimes (this is to be contrasted with the two predicted in general relativity):
\begin{equation}\label{eq:genps}
 P(k)=\left\langle\left\vert\dc^{\rm I}\right\vert^2\right\rangle\left\{
 \begin{array}{l l}
  \left(\frac{t_{\rm eq}}{t_0}\right)^{\frac{1}{3}}\frac{I^2_{\nu}\left[\sigma
kt_0\left(\frac{t_\infty}{t_0}\right)^r\right]}{I^2_{\nu}\left[\sigma kt_0\left(\frac{t_{\rm
eq}}{t_0}\right)^r\right]}\frac{K_\omega^2\left[\zeta kt_0\right]}{K^2_\omega\left[{\zeta kt_0}{(1+z_\infty)}\right]}& k>k_{\rm
eq}>k_\infty \\
  \left(\frac{\h}{t_0}\right)^{\frac{1}{3}}\frac{I^2_{\nu}\left[\sigma
kt_0\left(\frac{t_\infty}{t_0}\right)^r\right]}{I^2_{\nu}\left[\sigma
kt_0\left(\frac{\h}{t_0}\right)^r\right]}\frac{K_\omega^2\left[\zeta kt_0\right]}{K^2_\omega\left[{\zeta
kt_0}{(1+z_\infty)}\right]}&k_\infty<k<k_{\rm eq}\\ \left(\frac{\h}{t_0}\right)^{
\frac{1}{3}}\frac{K^2_\omega\left[\zeta kt_0\right]}{K_\omega^2\left[\zeta kt_0\left(\frac{t_0}{\h}\right)^{\frac{2}{3}}\right]}
&k<k_\infty
 \end{array}\right. .
\end{equation}

We are now in a position to explore the deviations from the general relativity prediction, but we must first check that general
relativity is indeed recovered on large scales. We know from our analysis in the previous subsection that this requires taking the
argument of all modified Bessel functions of the first kind to be small, however the second kind functions require more thought.
As detailed in appendix \ref{app:bessel}, these grow with decreasing argument and diverge as it approaches zero and so one may be
concerned that taking the argument to be small is not the correct limit. In fact, $K_\omega[y]\sim
y^{-\omega}+\mathcal{O}(y^{2-\omega})$ (see appendix \ref{app:bessel}) and so in this limit one may neglect the higher order
terms. One can indeed check by expanding the functions according to (\ref{eq:seriesI}) that this leading order behaviour coincides
with the general relativity prediction.

\begin{figure}[ht]\centering
 \includegraphics[width=0.9\textwidth]{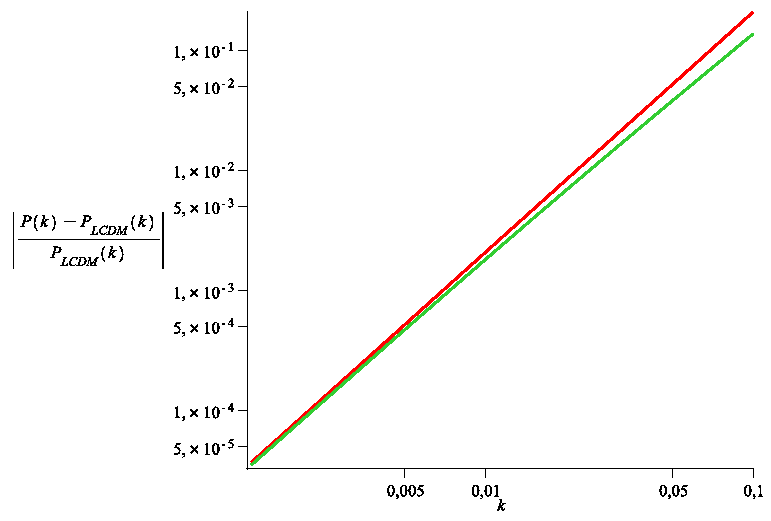}
\caption{The deviation from the $\Lambda$CDM power spectrum when a cosmological constant is present, with $\delta=n=2$, $x=5\times
10^{-7}$ and $z_\infty=5$. In red we show the exact power spectrum deviation. The linearised deviation found using
equation (\ref{eq:genps2}) is shown in green. Smaller values of $x$ result in smaller deviations.}
\end{figure}

We can then go beyond leading order to find the predicted deviations from $\Lambda$CDM. Modified Bessel functions deviate from
their leading order terms very rapidly and so one is interested in the case where the deviations are given by the next-to-leading
order expansion. In this regime, the power spectrum (\ref{eq:genps}) can be expanded (using the power series given
in appendix \ref{app:bessel}) and deviates from the $\Lambda$CDM case by $\Delta P(k)$, which
is scale-dependent:
\begin{equation}\label{eq:genps2}
 \frac{\Delta P(k)}{P_{\rm \Lambda CDM}(k)}=\left\{  \begin{array}{l l}
   \frac{2}{\nu +1} \left(\frac{\sigma k t_0}{2}\right)^{2r}
\left(\frac{t_\infty}{t_0}\right)^{2r}\left[1-\left(\frac{t_{eq}}{t_\infty}\right)^{2r}\right] +2F(\omega)\left(\frac{\zeta k
t_0}{2}\right)^2\left[(1+z_\infty)^2-1\right] & k>k_{\rm eq}>k_\infty \\      \frac{2}{\nu +1} \left(\frac{\sigma k
t_0}{2}\right)^{2r} \left(\frac{t_\infty}{t_0}\right)^{2r}\left[1-\left(\frac{\h}{t_\infty}\right)^ {2r}\right]
+2F(\omega)\left(\frac{\zeta k t_0}{2}\right)^2\left[(1+z_\infty)^2 -1\right]&k_\infty<k<k_{\rm eq}\\ 2F(\omega)\left(\frac{\zeta
k t_0}{2}\right)^2 \left[\left(\frac{t_0}{\h}\right)^{4/3}-1\right]
&k<k_\infty
  \end{array}\right. .
\end{equation}
The factor $F(\omega)$ is given by
\begin{equation}\label{eq:F(w)}
 F(\omega)= \left\{
  \begin{array}{l l}
   \frac{1}{\omega-1}& \omega>1 \\
    \frac{\pi\omega}{\Gamma^2(1+\omega)\sin(\pi\omega)}&\omega<1\\
  \end{array}\right. .
\end{equation}
The case $\omega=1$ corresponds to $\Omega_{\rm c}^0=0.625$, which is not
physically relevant and so is not considered here. This factor arises as a
result of different terms in the expansion of $K_\omega[y]$ becoming sub-leading
when $\omega$ assumes different values. The technical details are given in
appendix \ref{app:bessel}.

The deviation monotonically increases with $k$ in the linear regime. On large
scales entering the horizon after $t_\infty$, the growth of the deviation
$\Delta P/P_{\rm \Lambda CDM}$ is in $k^6$ whereas for smaller scales entering the
horizon before $t_\infty$, the discrepancy grows as $k^2$. Of course, this
result is only valid in the linear regime of perturbation theory. On smaller scales,
typically $k\gsim 0.1 {\rm hMpc}^{-1}$, non-linear effects become important and the screening mechanism is active. We therefore
expect that the deviations
from $\Lambda$CDM will be rapidly suppressed. As a result, we expect the full power spectrum to show large
deviations from $\Lambda$CDM in the linear regime and only mild deviations in
the non-linear one. A complete analysis of this phenomenon requires N-body simulations which are well beyond the scope of this
work.

The CDM power spectrum predicted by the superchameleon is very different from other chameleons that have been previously studied.
Indeed, in the vast majority of models linear scales are slightly outside
the Compton wavelength of the scalar field whose range is around $1$ Mpc (see the discussion in chapter \ref{chap:two}).
Deviations from $\Lambda$CDM are small until the quasi-linear regime for scales
around $1$ Mpc where the effects of the scalar peaks before being damped in the
non-linear regime by the screening mechanism. Here, scales are always outside
the Compton wavelength of the scalar field and the deviation from $\Lambda$CDM
is due to the large coupling $\beta(\varphi)$ to matter. This can be
strong enough to lead to $k$-dependent effects in the linear regime before
being heavily Yukawa suppressed in the non-linear one. Hence we have found that
superchameleons leave a drastically different signature on large scale
structure formation than the majority of previously studied, non-supersymmetric models.

\section{Other Models}

Before concluding this chapter, we pause to discuss how generic the phenomenology of the specific model studied here really is.
This model is by no means the most general or even the simplest. Indeed, the index $\gamma$ appears in the K\"{a}hler potential
and the superpotential but this was not necessary and we could have used two distinct indices. From this point of view, our model
here is a special point in the parameter space of all indices. The phenomenology of these models does not differ drastically from
the model presented here. The scalar potential (\ref{eq:Fpot}) is locally run-away when $\phi\lsim\pmi$ and tends to a constant
when $\phi\gg\pmi$. Generalising these models does not affect the small-$\phi$ behaviour of the potential but changes the
behaviour when $\phi\gg\pmi$. Depending which index is larger the potential either decays to zero tends to $\infty$ as
$\phi\rightarrow\infty$.  The large-field behaviour is irrelevant since the field never passes $\pmi$. It is hardly surprising
then that these models do not alleviate the need for a cosmological constant.

One may also wonder why two terms in the superpotential are necessary? Indeed, if one is ignored then we obtain an F-term
potential of the form $V_{\rm F}\sim \tilde{M}^{4+p}/\phi^p$ provided the index appearing in the K\"{a}hler potential higher than
the one appearing in the superpotential. In this case the potential is more similar to the well-studied potentials found in many
non-supersymmetric models. One also finds that a cosmological constant is required in these models but the absence of any
supersymmetric minimum ensures that supersymmetry is broken at all densities and so these are somewhat less appealing than the
more complicated models investigated here given our original motivations for studying supersymmetric completions.

We can therefore conclude that the qualitative features found in this chapter are ubiquitous in many classes on supersymmetric
models that are not of the no-scale type. 

\section{Summary of Main Results}

In this chapter and the previous one we have presented a bottom-up approach for constructing globally supersymmetric theories
that include chameleon-like screening. In the previous chapter, we presented the general framework and discussed the new
supersymmetric features that arise. In particular, we showed that when one accounts for supergravity corrections, any
supersymmetric model (with the exception of $n=3$ no-scale models) necessarily has $\chi_0\ll10^{-8}$ and so every object in the
universe is self-screened, thereby precluding the possibility of probing these theories using laboratory or astrophysical
methods\footnote{Binary pulsar tests evade this theorem since the observational signatures of modified gravity result from
time-dependent effects not included in the $\chi_0$--$\alpha$ parametrisation \cite{Brax:2013uh}.}. Furthermore, we showed that
the symmetron mechanism cannot operate at all in supersymmetric theories that include an underlying supergravity.

In this chapter we used the general framework to construct a class of supersymmetric chameleon models and examined their
cosmology. We found that they cannot account for the cosmic acceleration and, rather than appeal to an external cosmological
constant or supergravity breaking corrections, we introduced a novel mechanism where the coupling of the superchameleon to two
$\mathrm{U}(1)$ charged chiral scalars induces the appearance of a cosmological constant at late times in the form of an FI
term. This mechanism requires one to set all other contributions to the cosmological constant to zero\footnote{I.e. assume the
old cosmological constant problem in all other sectors is solved.} and so is not a solution to the cosmological constant problem.
However, we have argued that the value of the FI term is more robust to quantum corrections than scalar VEVs (it runs
logarithmically at most) and so a small value, which must be set by hand in our framework, is largely robust to the effects of
particle loops in the matter sector. The no-go theorem of the previous chapter does not apply on linear scales and so next we
derived and solved the equations governing CDM perturbations and calculated the modified form of the power spectrum. We identified
regions of parameter space that are excluded by current measurements and other regions that can be probed using upcoming
experiments. Finally, we explored the full parameter space and identified regions where the FI cosmological
constant can be generated dynamically without interfering with the chameleon mechanism and the theory is not yet ruled out by
current observations of linear cosmological probes. Such regions are ubiquitous and so there is no need to fine-tune the model
parameters. We identified a small region that can be probed by future surveys but there are large viable regions that will remain
unexplored, at least for the foreseeable future.

\thispagestyle{empty}
\thispagestyle{empty}
\newpage

\thispagestyle{empty}
\begin{savequote}[30pc]
A lot of the cosmologists and astrophysicists clearly had been reading science fiction.
\qauthor{Frederik Pohl}
\end{savequote}

\chapter{Discussion and Conclusions}\label{chap:eight}

This thesis has been concerned with several different aspects of modified theories of gravity that include screening mechanisms.
Our primary focus has been on those theories that can be embedded into the conformal scalar-tensor class ---
including the chameleon and symmetron models as well as those that screen using the environment-dependent Damour-Polyakov effect
--- but many of the results presented apply equally to theories that screen using the Vainshtein mechanism. Below we will
summarise some outstanding problems this thesis has attempted to address and review the progress that has been made. We
finish by discussing the prospects for using the work presented here in light of future experimental surveys.

\section{Problems Addressed in this Thesis}

Modified theories of gravity have attracted a renewed interest ever since the acceleration of the universe was first discovered
in 1998. Many theories that can account for the cosmic acceleration fail to satisfy the solar system tests of gravity and this
has prompted an effort towards finding theories that include screening mechanisms. These allow for interesting effects on
intergalactic and cosmological scales but are indistinguishable from general relativity in our own galaxy. Although they cannot
account for the cosmic acceleration, conformal scalar-tensor theories can show interesting effects on intergalactic scales and
may have important consequences for outstanding problems in structure formation. One popular method of probing modified theories
of gravity is to look at cosmological probes such as the luminosity distance-redshift relation or baryon acoustic oscillations
but any effects here are not smoking-gun signals because they are degenerate with other theories of modified gravity and dark
energy. Another testing ground is the laboratory. This has provided some constraints but these tests suffer from the
drawback that they are testing in the heavily screened regime and so one is looking for small deviations from the Newtonian force
law. 

In this thesis we have identified a new regime for testing modified theories of gravity. Astrophysical tests probe scales between
the Milky Way and the Hubble flow. These scales are mildly non-linear and so some degree of screening is present. This allows for
smoking-gun tests since the environment-dependence of this theory predicts that the magnitude of any deviation from general
relativity varies depending on where in the universe one looks. Since the screening is not totally efficient, the theory predicts
novel features in the structure and dynamics of certain stars and galaxies whose magnitude are order-unity compared with the
general relativity predictions. This means that the theories make quantitative predictions of large new signals in the
experimental data. The drawback is that some of these effects can be degenerate with other, non-gravitational, astrophysical
phenomena and it is important to understand the systematic uncertainties in any astrophysical data-set. Furthermore, unlike
laboratory and cosmological probes, there are currently no dedicated surveys searching for signals of modified gravity on
astrophysical scales. This is not as large an issue as one may initially think. Since the signals are large, one only needs a
small sample size (of order 100 galaxies or so) in order to place quantitative constraints and so many of the proposed tests can
piggyback on existing astrophysical surveys with no need for any specific targeting strategy\footnote{We have seen that dwarf
galaxies are particularly useful testing grounds for modified gravity and so this statement assumes that the survey does not
target galaxies with any specific morphologies.}.

The theories this thesis is concerned with screen according to the Newtonian potential; objects with lower Newtonian potentials
are more unscreened provided that they are not screened by any larger neighbours. The current constraints\footnote{Current
meaning prior to the original work presented in this thesis.} rule out theories which self-screen in Newtonian potentials
$\pn\ge10^{-6}$ and so one must look for objects with lower Newtonian potentials if one wishes to place stronger constraints. In
practice, this means one is interested in post-main-sequence stars and dwarf galaxies situated in cosmological voids. This thesis
has used both of these objects to place new constraints on the model parameters.

Recently, several investigations into the quantum stability of these models has provided mounting evidence that they suffer from
quantum instabilities. One may hope to alleviate this problem by imposing some sort of symmetry on the theory so that it is
technically natural. Supersymmetry is a natural choice for this since it enjoys powerful non-renormalisation theorems and it does
not predict Goldstone bosons. This allows the field to vary over many orders of magnitude in field space whilst retaining the
benefits of the underlying symmetry. It is also a natural framework for making contact with more fundamental theories such as
string theory. The second part of this thesis has investigated possible supersymmetric completions of conformal-scalar tensor
theories and investigated the new features and the cosmological dynamics.

\section{Summary of Original Results}

\subsection{Astrophysical Tests of Conformal Scalar-Tensor Theories}

In chapter \ref{chap:two} we presented a model-independent framework for dealing with conformal-scalar tensor theories with
screening mechanisms and introduced two parameters that fully specify the theory\footnote{In fact, we specified these parameters
as values of the fundamental functions $V(\phi)$ and $A(\phi)$ evaluated at the present time. One can reconstruct their entire
cosmic history from these two numbers alone for any individual theory using tomography \cite{Brax:2011aw,Brax:2012gr}. If one
wishes to look at these functions inside screened objects such as the Milky Way then these functions must be specified on a
model-by-model basis.}. These are $\chi_0$, whose value determines how efficient and object is at screening itself, and
$\alpha$, which sets the strength of the fifth-force relative to the Newtonian one in unscreened regions. Prior to the work
presented in this thesis, $\alpha$ was constrained by laboratory experiments on a model-by-model basis and $\mathcal{O}(1)$
values were not ruled out for a large class of models. Cluster counting surveys placed the constraint $\chi_0<10^{-4}$ and it
was often assumed that $\chi_0\lsim\mathcal{O}(10^{-6})$ so that the Milky Way is self-screening. This latter assumption had been
heavily debated in the field since the Milky Way could instead be screened by the local group and no formal analysis had been
performed.

In chapter \ref{chap:three} we derived the equations of modified gravity hydrodynamics and used them to find the
equations governing stellar structure in theories of modified gravity. We then presented a general framework for solving these
equations using the Lane-Emden approximation. Lane-Emden models are particularly useful for studying the effects of modified
gravity
since they decouple the gravitational physics from other physics such as nuclear burning and radiative transfer. We argued that
an unscreened star is brighter, hotter and more ephemeral than its screened counterpart. In order to verify this, we used the
Eddington standard model in conjunction with the modified Lane-Emden equation to find the altered structure of main-sequence
stars. Whereas these are screened given current constraints, their study is still useful since they are far simpler object than
post-main-sequence stars and the effects of modified gravity are unambiguous\footnote{In addition to this, the parameter range
where main-sequence stars are unscreened had not yet been ruled out at the time this work was initiated.}. The criterion that
$\chi_0<\pn$ implies screening
was derived by assuming that the object in question is a static source in the field equations. In fact, stars are dynamical
objects whose equilibrium configuration is set by their interaction with gravity. Our model included this back-reaction and we
subsequently found that main-sequence stars are slightly more unscreened than one may initially believe. Our semi-analytic
treatment confirmed that low-mass, gas pressure-supported main-sequence stars are far brighter than their screened equivalents.
High-mass, radiation supported stars also show an enhancement but this is somewhat reduced since the extra radiation is absorbed
to provide the extra pressure needed to combat the inward gravitational force. We also found that unscreened stars in $f(R)$
gravity have a main-sequence lifetime that can be up to three times shorter than that of a screened star.

Our simple prescription applied strictly to main-sequence stars and neglected many important effects such as convection and
nuclear burning. It is also completely static and so it is somewhat ambiguous as to which point in a star's life these models
correspond. If one wishes to compare with observational data and investigate the structure of post-main-sequence stars a full
numerical treatment is needed. To this end, we presented a modified version of the publicly available code MESA that
can predict the structure and evolution of stars of any initial mass and metallicity including all the absent effects in the
simple Lane-Emden model. This means that its predictions can be compared with experimental data and it is a powerful and
versatile tool for making quantitative predictions of the new effects of modified gravity.

Since the degree of screening varies between different galaxies, the laws of gravitational physics are different in screened
and unscreened galaxies. This means that any theoretical or empirical\footnote{Calibrated on Milky Way or local group objects.}
formulae used to infer the distance to an unscreened galaxy will give an incorrect measurement if they are sensitive to the laws
of gravity. Hence, screened and unscreened distance estimates will agree if the galaxy is screened but will disagree if not. In
chapter \ref{chap:four} we exploited this feature to place new constraints on the model-independent parameters.
Cepheid variable stars are unscreened distance indicators, tip of the red giant branch stars are unscreened when $\chi_0>10^{-6}$
and screened otherwise. Water masers provided a geometric distance estimate and so are screened. 

Using MESA models,
we predicted the change in the inferred distance using screened and unscreened indicators for both TRGB and Cepheid measurements.
By comparing waster maser and TRGB distances to the spiral galaxy NGC 4258 we were able to place a new and independent constraint
$\chi_0<10^{-6}$, resolving the debate in the literature. We then compared Cepheid and TRGB distances to sub-samples of screened
and unscreened galaxies taken from the screening map of \cite{Cabre:2012tq} to place new constraints in the $\chi_0$--$\alpha$
place for chameleon models. We ruled out self-screening parameters $\chi_0>4\times10^{-7}$ for $f(R)$ theories and
$\chi_0>3\times10^{-7}$
for symmetron models. These constraints are currently the strongest in the literature and are three orders of magnitude tighter
than the previous bounds from cluster statistics. These bounds imply that the only unscreened objects in the universe are dwarf
galaxies and heavy post-main-sequence stars with $M\gsim 8M_\odot$.

Chapter \ref{chap:four} used several assumptions about the period-luminosity relation for Cepheid stars in modified gravity. In
particular, the new relation was found using the general relativity formula, which only accounts of the modified equilibrium
structure and ignores the effects of modified gravity on the perturbations about this configuration. In order to investigate
these assumptions, in chapter \ref{chap:five} we perturbed the equations of modified gravity hydrodynamics to first-order and
found the new equation governing radial, adiabatic oscillations, the modified linear adiabatic wave equation. Using the
Sturm-Liouville nature of this equation we were able to predict two new effects. First, an unscreened star is more stable to
linear perturbations than an equivalent screened star and second, the oscillation period of an unscreened star is far shorter
than previously predicted using the assumptions above. We investigated the first effect using Lane-Emden models and found that
the change in the critical value of the first adiabatic index for the onset of the instability is of order $10^{-1}\alpha$. We
then investigated the second effect using both Lane-Emden models of convective stars and MESA models of Cepheid
stars at the blue edge of the instability strip. We found that the period is indeed shorter than one would predict by simply
perturbing the general relativity relation. Using this in the period-luminosity relation, we investigated how the predictions
found in chapter \ref{chap:four} were altered by this new effect. When one neglects the effect of modified gravity perturbations
but retains the altered equilibrium structure we found that the approximations hold very well. When the perturbations are
included we found that the difference in the inferred distance to an unscreened galaxy between screened and unscreened indicators
can be up to three times as large. Hence, the constraints we found in chapter \ref{chap:four} are conservative and it is possible
to improve them using the same data-sets. We estimated that one could probe into the region $\chi_0\sim\mathcal{O}(10^{-8})$ but
the exact value depends on how many unscreened galaxies are left at these small values and this is beyond the scope of this work.

Many of the general results derived here apply equally to Vainshtein screened theories, however in practice the new effects are
not present since these theories are far more efficient at screening an object and the Vainshtein radius is often far larger
than the object's radius.

\subsection{Supersymmetric Completions of Conformal-Scalar Tensor Theories}

In chapter \ref{chap:six} we presented a general framework for embedding conformal scalar-tensor theories into global
supersymmetry. This bottom-up approach is useful because chameleon-like models are infra-red modifications of general
relativity and one can examine their basic properties using global supersymmetry without the technical complications of
supergravity. For simplicity, we only coupled the gravitational scalar to two dark matter fermions but the extension to the
standard model is straightforward. Furthermore, many of the new results including environment-dependent supersymmetry breaking and
efficient screening are independent of the form of the matter coupling.

Using this framework, we were able to show that supersymmetry is always broken at non-zero dark matter densities and showed how
the
scale of the breaking is set by the ambient density and the model parameters. Including supergravity corrections, we were able to
prove a general no-go theorem showing that the mass of the scalar in any supersymmetric theory of modified gravity is at least as
large as the gravitino mass and hence mediates a force whose range is less than $10^{-6}$ m. Using this, we were able to show
that $\chi_0\ll10^{-66}$ and hence astrophysical and laboratory experiments cannot be used to probe these theories. The exception
to the theorem is pure no-scale models where the gravitino mass is undetermined at tree-level. We argued that these models do not
generically include screening mechanisms but this is far from a proof and it remains to be seen if the mechanisms can be realised
using more complicated models. We also showed that supersymmetric symmetrons cannot exist once supergravity corrections are
accounted for unless the model parameters are heavily fine-tuned.

On cosmological scales, $\chi_0$ is not enough to determine whether deviations from general relativity are present. At the level
of the background cosmology, $\chi_0$ has no influence on the homogeneous dynamics of the scalar and on linear scales the
screening mechanism cannot operate efficiently and the large coupling to matter allows for deviations in the cold dark matter
power
spectrum. This is in contrast to non-supersymmetric models with $\mathcal{O}(1)$ couplings that are constrained to affect
non-linear scales only. On non-linear scales the screening mechanism can operate and any deviations form general relativity are
highly suppressed. 

In order to investigate this regime, we presented a class of supersymmetric chameleon models and studied their
cosmological dynamics. We found that the field could not account for dark energy without giving deviations in the CDM power
spectrum that are in tension with current experiments and so a cosmological constant is needed. This presented a new problem in
that supersymmetry is broken if the vacuum energy is positive and it is not possible to include a cosmological constant at the
level of the action.
We presented a novel mechanism where the coupling of the field to two charged scalars can drive their VEV's to zero at late times
leaving a cosmological constant in the form of a Fayet-Illiopoulos term. This is more robust to quantum corrections that scalar
VEVs since it is not overly sensitive to loop corrections and runs logarithmically at most. This is by no means a solution to the
cosmological constant problem since it does not explain why the contribution to the cosmological constant coming from quantum
effects in the other sectors are absent. We investigated the model parameter space where this mechanism is viable and found that
regions where this is the case are ubiquitous.

Finally, we calculated the CDM power spectrum in closed form and found that it deviates from the general relativity prediction on
intermediate scales by an amount dependent on the model parameters. We were able to exclude parameters that are in tension with
current measurements and indicated the region that can be probed using upcoming surveys.

\section{Outlook}

These are exciting times for modified gravity. The past decade has been theoretically dominated with many new theories being
investigated and a plethora of observational tests proposed. The next decade and beyond will see a shift in focus towards more
data-oriented aspects. Upcoming data releases such as SDSS IV MaNGA will provide spectral measurements of nearly
10000 galaxies. DES saw first light in September 2012 and scientific measurements have begun to be made as of
September 2013. This will provide data pertaining to weak lensing, galaxy cluster counts, supernovae distance measurements and
baryon acoustic oscillations, all of which can be used as cosmological probes of dark energy and modified gravity. The megamaser
cosmology project is under way and has already detected water masers in 62 galaxies. The aim of this project is to determine the
Hubble constant geometrically with a precision of 3\% in order to constrain the properties of dark energy. Spitzer
\cite{Werner:2004zk} provides infra-red data pertaining to stars and galaxies and will significantly improve the uncertainties in
the slope of the period-luminosity relation for Cepheid variable stars in the infra-red. 

Looking to the future, missions such as LSST and WFIRST will provide spectroscopic data for a large
number of galaxies and Euclid will provide complementary probes using phenomena such as weak lensing. GAIA \cite{EAS:9131523}
will provide spectroscopic measurements of about one billion stars in the Milky Way and the local group.

The prospects are also good for testing gravity in the strong field regime. Future gravitational wave interferometers such as
Advanced-LIGO and eLISA will provide data relating to black hole mergers and in-spirals and very
recently \cite{Ransom:2014xla} a millisecond pulsar has been observed in stellar triple system. This offers new possibilities to
for testing gravity using the energy radiated into gravitational waves.

What does this mean for the results presented in this thesis and future astrophysical tests of modified gravity? We have already
alluded to the fact that these tests can piggyback on other missions and many of the surveys described above could potentially
improve the current constraints. 

In chapters \ref{chap:six} and \ref{chap:seven} we investigated supersymmetric models of screened modified gravity and we argued
that Euclid could constrain these by measuring the cold dark matter power spectrum to a precision greater than 10\%.

In chapter \ref{chap:three} we argued that the combined effects of the increased luminosity and reduced lifetime of stars should
alter the colour, spectra and luminosity of dwarf galaxies in voids relative to those in clusters and that looking for
systematic offsets could constitute a new observational test. Our simple estimate is by no means accurate enough to compare with
observational data but the tools to make a more quantitative prediction are available. Using MESA to generate mock
stellar populations it is possible to create theoretical isochrones that can be used in conjunction with existing galaxy
synthesiser codes to make predictions that are accurate enough to compare with data. This is a complicated task because the
galactic properties are the result of cumulative effects that begin far in the past when the galaxy was more screened. In
practice, this requires one to account for the time-evolution of the self-screening parameter inside haloes using non-linear
spherical collapse models but this is possible \cite{Li:2011qda}. Progress towards this goal has been made but any quantitative
results are still a long way off. 

In chapter \ref{chap:four} we presented new constraints on chameleon-like models using Cepheid
variable stars in unscreened galaxies. In chapter \ref{chap:five} we argued that these constraints could be greatly improved using
the same data-analysis. We were careful to remark that the small sample of unscreened galaxies is
responsible for the jaggedness of the contours and that testing values of $\chi_0\lsim\mathcal{O}(10^{-8})$ is very data-limited
due to the small number of unscreened galaxies. Surveys such as LSST will provide spectroscopic data for variable
stars in a variety of different environments and help to combat this issue. Furthermore, reducing the uncertainty of the slope in
the period-luminosity relation using Spitzer data has the potential to improve the systematics. Furthermore, there
are other tests that have been alluded to that are not yet possible. Updating the MLAWE to include non-adiabatic processes will
allow for a prediction of the slope of the period-luminosity relation as well as the location of the blue edge of the instability
strip. Data from the surveys mentioned above could then be compared with these predictions therefore providing further tests. 

Recently, a Cepheid variable star in an eclipsing binary system has been discovered for the first time
\cite{2010Natur.468..542P,Marconi:2013tta}. This has allowed an accurate measurement of the Cepheid's mass using both the orbital
properties and the period-luminosity-mass relation and the two agree at the 1\% level. This provides a unique testing ground for
modified theories of gravity. The agreement of the mass measurements can probe chameleon models once the MLAWE is updated to
include non-adiabatic driving processes since the orbit is screened but the pulsation is not. For Vainshtein screened theories,
this test can be reversed since the star is screened but its finite extent may lead to deviations in the mass inferred from its
orbit \cite{Hiramatsu:2012xj}.

RR Lyrae variable stars are another class of pulsating object. They have smaller masses and temperatures than Cepheids and
pulsate due to the same mechanism. Currently, they are not as powerful as distance indicators. This will change with data coming
from GAIA, which will provide new measurements allowing them to become competitive with Cepheid distances. GAIA will not probe
stars outside the local group and so it is unlikely that these will be useful for testing chameleon-like
models. We argued in chapter \ref{chap:five} that Vainshtein screened theories may give rise to scalar radiation from pulsating
stars and this may be detectable by looking for beating effects in the period-luminosity relation or decays in its amplitude.
Since this effect is insensitive to the degree of screening it is possible that Cepheid and RR Lyrae stars in the local group
could provide a possible testing ground.

Another distance indicator that has yet to be exploited is the planetary nebula luminosity function (PNLF). Planetary nebulae are
the remnants of asymptotic giant branch (AGB) stars that blow off their convective envelopes leaving a hot, dense core that
emits ionising radiation. This radiation causes transitions in [OIII] atoms, which is detectable as emission lines at
approximately 507 nm. The luminosity function for these lines has a universal turn-off and the PNLF is hence a distance
indicator \cite{Ciardullo:2012vc}. Since the core of AGB stars is screened but the envelope is not, it is possible that the extra
centripetal force felt by the envelope could move it away from the core so that the PLNF is greatly diminished. The PLNF may
therefore be an unscreened distance indicator and could be used to place new constraints. This effect certainly merits further
investigation.

Chameleon-like models are already heavily constrained by much of the work presented in this thesis and it is only a matter of
time before the theory is rendered astrophysically uninteresting but what about Vainshtein screened theories? These are far less
constrained due to the lack of astrophysical effects. We have already alluded to some potential tests above but are there other
possible signatures? \cite{Hui:2012jb} have argued that the central black holes in under-dense galaxies should be offset from the
galactic centre by an amount of order 0.1 kpc, which presents several possible avenues for testing these theories. The
black hole has a sphere of influence on the surrounding stellar matter and will drag a disk of stars away from the galactic
centre. One may expect an offset between the optical and
dark matter centroids in under-dense galaxies. This is a smoking-gun signal since the direction of the offset should be
correlated with the direction of motion of the galaxy. Work on these tests is already under way and optical, near-infra-red and X-ray data from many of the
surveys above could be used to place new constraints.

In conclusion, in this thesis we have presented a thorough theoretical, numerical and observational investigation into the
properties of various astrophysical objects when the theory of gravity includes chameleon-like screening. We have already placed
the most stringent constraints to date but there is more work to be done and many potential new features to be exploited.
The next decade will see numerous astrophysical surveys, all of which could constrain the theory of gravity with
unprecedented precision. The future is bright, and if gravity really is described by these theories, it may be brighter than we
expect!
\newpage
\thispagestyle{empty}
\newpage
\clearpage

\newpage

\thispagestyle{empty}

 \vspace*{\fill}

\begin{quote}
Mit der Dummheit k\"{a}mpfen G\"{o}tter selbst vergebens.

Against stupidity the gods themselves contend in vain.

\qauthor{Friedrich Schiller, \textit{The Maid of Orleans} }
\end{quote}
 \vspace*{\fill}

\newpage
\thispagestyle{empty}

\newpage
\bibliographystyle{JHEPmodplain}
\bibliography{ref2}

\providecommand{\href}[2]{#2}\begingroup\raggedright\begin{thebibliography}{100}

\bibitem{Einstein:1916vd}
A.~Einstein, {\it {The Foundation of the General Theory of Relativity}},  {\sl
  Annalen Phys.} {\bf 49} (1916) 769--822,
  [\href{http://dx.doi.org/10.1002/andp.200590044}{{\sf
  doi:10.1002/andp.200590044}}].

\bibitem{Will:2004nx}
C.~Will, {\it {The confrontation between general relativity and experiment}},
  {\sl Pramana} {\bf 63} (2004) 731--740,
  [\href{http://dx.doi.org/10.1007/BF02705195}{{\sf doi:10.1007/BF02705195}}].

\bibitem{Riess:1998cb}
{\bf Supernova Search Team} Collaboration, A.~G. Riess {\em et~al.}, {\it
  {Observational evidence from supernovae for an accelerating universe and a
  cosmological constant}},  {\sl Astron.J.} {\bf 116} (1998) 1009--1038,
  [\href{http://arxiv.org/abs/astro-ph/9805201}{{\sf arXiv:astro-ph/9805201}}],
  [\href{http://dx.doi.org/10.1086/300499}{{\sf doi:10.1086/300499}}].

\bibitem{Perlmutter:1998np}
{\bf Supernova Cosmology Project} Collaboration, S.~Perlmutter {\em et~al.},
  {\it {Measurements of Omega and Lambda from 42 high redshift supernovae}},
  {\sl Astrophys.J.} {\bf 517} (1999) 565--586,
  [\href{http://arxiv.org/abs/astro-ph/9812133}{{\sf arXiv:astro-ph/9812133}}],
  [\href{http://dx.doi.org/10.1086/307221}{{\sf doi:10.1086/307221}}].

\bibitem{Copeland:2006wr}
E.~J. Copeland, M.~Sami, and S.~Tsujikawa, {\it {Dynamics of dark energy}},
  {\sl Int.J.Mod.Phys.} {\bf D15} (2006) 1753--1936,
  [\href{http://arxiv.org/abs/hep-th/0603057}{{\sf arXiv:hep-th/0603057}}],
  [\href{http://dx.doi.org/10.1142/S021827180600942X}{{\sf
  doi:10.1142/S021827180600942X}}].

\bibitem{Clifton:2011jh}
T.~Clifton, P.~G. Ferreira, A.~Padilla, and C.~Skordis, {\it {Modified Gravity
  and Cosmology}},  {\sl Phys.Rept.} {\bf 513} (2012) 1--189,
  [\href{http://arxiv.org/abs/1106.2476}{{\sf arXiv:1106.2476}}],
  [\href{http://dx.doi.org/10.1016/j.physrep.2012.01.001}{{\sf
  doi:10.1016/j.physrep.2012.01.001}}].

\bibitem{Ade:2013ktc}
{\bf Planck Collaboration} Collaboration, P.~Ade {\em et~al.}, {\it {Planck
  2013 results. I. Overview of products and scientific results}},
  \href{http://arxiv.org/abs/1303.5062}{{\sf arXiv:1303.5062}},
  \href{http://dx.doi.org/10.1051/0004-6361/201321529}{{\sf
  doi:10.1051/0004-6361/201321529}}.

\bibitem{Bean:2010zq}
R.~Bean and M.~Tangmatitham, {\it {Current constraints on the cosmic growth
  history}},  {\sl Phys.Rev.} {\bf D81} (2010) 083534,
  [\href{http://arxiv.org/abs/1002.4197}{{\sf arXiv:1002.4197}}],
  [\href{http://dx.doi.org/10.1103/PhysRevD.81.083534}{{\sf
  doi:10.1103/PhysRevD.81.083534}}].

\bibitem{Amendola:2012ys}
{\bf Euclid Theory Working Group} Collaboration, L.~Amendola {\em et~al.}, {\it
  {Cosmology and fundamental physics with the Euclid satellite}},  {\sl Living
  Rev.Rel.} {\bf 16} (2013) 6, [\href{http://arxiv.org/abs/1206.1225}{{\sf
  arXiv:1206.1225}}].

\bibitem{Ivezic:2008fe}
{\bf LSST Collaboration} Collaboration, Z.~Ivezic, J.~Tyson, R.~Allsman,
  J.~Andrew, and R.~Angel, {\it {LSST: from Science Drivers to Reference Design
  and Anticipated Data Products}},  \href{http://arxiv.org/abs/0805.2366}{{\sf
  arXiv:0805.2366}}.

\bibitem{Albrecht:2006um}
A.~Albrecht, G.~Bernstein, R.~Cahn, W.~L. Freedman, J.~Hewitt, {\em et~al.},
  {\it {Report of the Dark Energy Task Force}},
  \href{http://arxiv.org/abs/astro-ph/0609591}{{\sf arXiv:astro-ph/0609591}}.

\bibitem{Spergel:2013tha}
D.~Spergel, N.~Gehrels, J.~Breckinridge, M.~Donahue, A.~Dressler, {\em et~al.},
  {\it {Wide-Field InfraRed Survey Telescope-Astrophysics Focused Telescope
  Assets WFIRST-AFTA Final Report}},
  \href{http://arxiv.org/abs/1305.5422}{{\sf arXiv:1305.5422}}.

\bibitem{Abbott:2007kv}
{\bf LIGO Scientific Collaboration} Collaboration, B.~Abbott {\em et~al.}, {\it
  {LIGO: The Laser interferometer gravitational-wave observatory}},  {\sl
  Rept.Prog.Phys.} {\bf 72} (2009) 076901,
  [\href{http://arxiv.org/abs/0711.3041}{{\sf arXiv:0711.3041}}],
  [\href{http://dx.doi.org/10.1088/0034-4885/72/7/076901}{{\sf
  doi:10.1088/0034-4885/72/7/076901}}].

\bibitem{AmaroSeoane:2012km}
P.~Amaro-Seoane, S.~Aoudia, S.~Babak, P.~Binetruy, E.~Berti, {\em et~al.}, {\it
  {eLISA/NGO: Astrophysics and cosmology in the gravitational-wave millihertz
  regime}},  {\sl GW Notes} {\bf 6} (2013) 4--110,
  [\href{http://arxiv.org/abs/1201.3621}{{\sf arXiv:1201.3621}}].

\bibitem{Weinberg:1965rz}
S.~Weinberg, {\it {Photons and gravitons in perturbation theory: Derivation of
  Maxwell's and Einstein's equations}},  {\sl Phys.Rev.} {\bf 138} (1965)
  B988--B1002, [\href{http://dx.doi.org/10.1103/PhysRev.138.B988}{{\sf
  doi:10.1103/PhysRev.138.B988}}].

\bibitem{Damour:1993hw}
T.~Damour and G.~Esposito-Farese, {\it {Nonperturbative strong field effects in
  tensor - scalar theories of gravitation}},  {\sl Phys.Rev.Lett.} {\bf 70}
  (1993) 2220--2223, [\href{http://dx.doi.org/10.1103/PhysRevLett.70.2220}{{\sf
  doi:10.1103/PhysRevLett.70.2220}}].

\bibitem{Khoury:2003aq}
J.~Khoury and A.~Weltman, {\it {Chameleon fields: Awaiting surprises for tests
  of gravity in space}},  {\sl Phys.Rev.Lett.} {\bf 93} (2004) 171104,
  [\href{http://arxiv.org/abs/astro-ph/0309300}{{\sf arXiv:astro-ph/0309300}}],
  [\href{http://dx.doi.org/10.1103/PhysRevLett.93.171104}{{\sf
  doi:10.1103/PhysRevLett.93.171104}}].

\bibitem{Khoury:2003rn}
J.~Khoury and A.~Weltman, {\it {Chameleon cosmology}},  {\sl Phys.Rev.} {\bf
  D69} (2004) 044026, [\href{http://arxiv.org/abs/astro-ph/0309411}{{\sf
  arXiv:astro-ph/0309411}}],
  [\href{http://dx.doi.org/10.1103/PhysRevD.69.044026}{{\sf
  doi:10.1103/PhysRevD.69.044026}}].

\bibitem{Hinterbichler:2010es}
K.~Hinterbichler and J.~Khoury, {\it {Symmetron Fields: Screening Long-Range
  Forces Through Local Symmetry Restoration}},  {\sl Phys.Rev.Lett.} {\bf 104}
  (2010) 231301, [\href{http://arxiv.org/abs/1001.4525}{{\sf
  arXiv:1001.4525}}],
  [\href{http://dx.doi.org/10.1103/PhysRevLett.104.231301}{{\sf
  doi:10.1103/PhysRevLett.104.231301}}].

\bibitem{Brax:2010gi}
P.~Brax, C.~van~de Bruck, A.-C. Davis, and D.~Shaw, {\it {The Dilaton and
  Modified Gravity}},  {\sl Phys.Rev.} {\bf D82} (2010) 063519,
  [\href{http://arxiv.org/abs/1005.3735}{{\sf arXiv:1005.3735}}],
  [\href{http://dx.doi.org/10.1103/PhysRevD.82.063519}{{\sf
  doi:10.1103/PhysRevD.82.063519}}].

\bibitem{Vainshtein:1972sx}
A.~Vainshtein, {\it {To the problem of nonvanishing gravitation mass}},  {\sl
  Phys.Lett.} {\bf B39} (1972) 393--394,
  [\href{http://dx.doi.org/10.1016/0370-2693(72)90147-5}{{\sf
  doi:10.1016/0370-2693(72)90147-5}}].

\bibitem{Sakstein:2014isa}
J.~Sakstein, {\it {Disformal Theories of Gravity: From the Solar System to
  Cosmology}},  {\sl JCAP} {\bf 1412} (2014), no.~12 012,
  [\href{http://arxiv.org/abs/1409.1734}{{\sf arXiv:1409.1734}}],
  [\href{http://dx.doi.org/10.1088/1475-7516/2014/12/012}{{\sf
  doi:10.1088/1475-7516/2014/12/012}}].

\bibitem{Sakstein:2014aca}
J.~Sakstein, {\it {Towards Viable Cosmological Models of Disformal Theories of
  Gravity}},  {\sl Phys.Rev.} {\bf D91} (2015), no.~2 024036,
  [\href{http://arxiv.org/abs/1409.7296}{{\sf arXiv:1409.7296}}],
  [\href{http://dx.doi.org/10.1103/PhysRevD.91.024036}{{\sf
  doi:10.1103/PhysRevD.91.024036}}].

\bibitem{Hui:2009kc}
L.~Hui, A.~Nicolis, and C.~Stubbs, {\it {Equivalence Principle Implications of
  Modified Gravity Models}},  {\sl Phys.Rev.} {\bf D80} (2009) 104002,
  [\href{http://arxiv.org/abs/0905.2966}{{\sf arXiv:0905.2966}}],
  [\href{http://dx.doi.org/10.1103/PhysRevD.80.104002}{{\sf
  doi:10.1103/PhysRevD.80.104002}}].

\bibitem{Jain:2010ka}
B.~Jain and J.~Khoury, {\it {Cosmological Tests of Gravity}},  {\sl Annals
  Phys.} {\bf 325} (2010) 1479--1516,
  [\href{http://arxiv.org/abs/1004.3294}{{\sf arXiv:1004.3294}}],
  [\href{http://dx.doi.org/10.1016/j.aop.2010.04.002}{{\sf
  doi:10.1016/j.aop.2010.04.002}}].

\bibitem{Khoury:2010xi}
J.~Khoury, {\it {Theories of Dark Energy with Screening Mechanisms}},
  \href{http://arxiv.org/abs/1011.5909}{{\sf arXiv:1011.5909}}.

\bibitem{Davis:2011qf}
A.-C. Davis, E.~A. Lim, J.~Sakstein, and D.~Shaw, {\it {Modified Gravity Makes
  Galaxies Brighter}},  {\sl Phys.Rev.} {\bf D85} (2012) 123006,
  [\href{http://arxiv.org/abs/1102.5278}{{\sf arXiv:1102.5278}}],
  [\href{http://dx.doi.org/10.1103/PhysRevD.85.123006}{{\sf
  doi:10.1103/PhysRevD.85.123006}}].

\bibitem{Brax:2012gr}
P.~Brax, A.-C. Davis, B.~Li, and H.~A. Winther, {\it {A Unified Description of
  Screened Modified Gravity}},  {\sl Phys.Rev.} {\bf D86} (2012) 044015,
  [\href{http://arxiv.org/abs/1203.4812}{{\sf arXiv:1203.4812}}],
  [\href{http://dx.doi.org/10.1103/PhysRevD.86.044015}{{\sf
  doi:10.1103/PhysRevD.86.044015}}].

\bibitem{Sakstein:2013pda}
J.~Sakstein, {\it {Stellar Oscillations in Modified Gravity}},  {\sl Phys.Rev.}
  {\bf D88} (2013) 124013, [\href{http://arxiv.org/abs/1309.0495}{{\sf
  arXiv:1309.0495}}], [\href{http://dx.doi.org/10.1103/PhysRevD.88.124013}{{\sf
  doi:10.1103/PhysRevD.88.124013}}].

\bibitem{Sakstein:2014nfa}
J.~Sakstein, B.~Jain, and V.~Vikram, {\it {Testing Gravity Theories Using
  Stars}},  \href{http://arxiv.org/abs/1409.3708}{{\sf arXiv:1409.3708}},
  \href{http://dx.doi.org/10.1142/S0218271814420024}{{\sf
  doi:10.1142/S0218271814420024}}.

\bibitem{Mota:2006ed}
D.~F. Mota and D.~J. Shaw, {\it {Strongly coupled chameleon fields: New
  horizons in scalar field theory}},  {\sl Phys.Rev.Lett.} {\bf 97} (2006)
  151102, [\href{http://arxiv.org/abs/hep-ph/0606204}{{\sf
  arXiv:hep-ph/0606204}}],
  [\href{http://dx.doi.org/10.1103/PhysRevLett.97.151102}{{\sf
  doi:10.1103/PhysRevLett.97.151102}}].

\bibitem{Mota:2006fz}
D.~F. Mota and D.~J. Shaw, {\it {Evading Equivalence Principle Violations,
  Cosmological and other Experimental Constraints in Scalar Field Theories with
  a Strong Coupling to Matter}},  {\sl Phys.Rev.} {\bf D75} (2007) 063501,
  [\href{http://arxiv.org/abs/hep-ph/0608078}{{\sf arXiv:hep-ph/0608078}}],
  [\href{http://dx.doi.org/10.1103/PhysRevD.75.063501}{{\sf
  doi:10.1103/PhysRevD.75.063501}}].

\bibitem{Brax:2007hi}
P.~Brax, C.~van~de Bruck, A.-C. Davis, D.~F. Mota, and D.~J. Shaw, {\it
  {Testing Chameleon Theories with Light Propagating through a Magnetic
  Field}},  {\sl Phys.Rev.} {\bf D76} (2007) 085010,
  [\href{http://arxiv.org/abs/0707.2801}{{\sf arXiv:0707.2801}}],
  [\href{http://dx.doi.org/10.1103/PhysRevD.76.085010}{{\sf
  doi:10.1103/PhysRevD.76.085010}}].

\bibitem{Gies:2007su}
H.~Gies, D.~F. Mota, and D.~J. Shaw, {\it {Hidden in the Light: Magnetically
  Induced Afterglow from Trapped Chameleon Fields}},  {\sl Phys.Rev.} {\bf D77}
  (2008) 025016, [\href{http://arxiv.org/abs/0710.1556}{{\sf
  arXiv:0710.1556}}], [\href{http://dx.doi.org/10.1103/PhysRevD.77.025016}{{\sf
  doi:10.1103/PhysRevD.77.025016}}].

\bibitem{Brax:2007vm}
P.~Brax, C.~van~de Bruck, A.-C. Davis, D.~F. Mota, and D.~J. Shaw, {\it
  {Detecting chameleons through Casimir force measurements}},  {\sl Phys.Rev.}
  {\bf D76} (2007) 124034, [\href{http://arxiv.org/abs/0709.2075}{{\sf
  arXiv:0709.2075}}], [\href{http://dx.doi.org/10.1103/PhysRevD.76.124034}{{\sf
  doi:10.1103/PhysRevD.76.124034}}].

\bibitem{Brax:2009bk}
P.~Brax, C.~van~de Bruck, A.-C. Davis, and D.~Shaw, {\it {Laboratory Tests of
  Chameleon Models}},  \href{http://arxiv.org/abs/0911.1086}{{\sf
  arXiv:0911.1086}}.

\bibitem{Brax:2010xx}
P.~Brax, C.~van~de Bruck, A.~Davis, D.~Shaw, and D.~Iannuzzi, {\it {Tuning the
  Mass of Chameleon Fields in Casimir Force Experiments}},  {\sl
  Phys.Rev.Lett.} {\bf 104} (2010) 241101,
  [\href{http://arxiv.org/abs/1003.1605}{{\sf arXiv:1003.1605}}],
  [\href{http://dx.doi.org/10.1103/PhysRevLett.104.241101}{{\sf
  doi:10.1103/PhysRevLett.104.241101}}].

\bibitem{Brax:2010gp}
P.~Brax and C.~Burrage, {\it {Atomic Precision Tests and Light Scalar
  Couplings}},  {\sl Phys.Rev.} {\bf D83} (2011) 035020,
  [\href{http://arxiv.org/abs/1010.5108}{{\sf arXiv:1010.5108}}],
  [\href{http://dx.doi.org/10.1103/PhysRevD.83.035020}{{\sf
  doi:10.1103/PhysRevD.83.035020}}].

\bibitem{Brax:2011wp}
P.~Brax, A.~Lindner, and K.~Zioutas, {\it {Detection prospects for solar and
  terrestrial chameleons}},  {\sl Phys.Rev.} {\bf D85} (2012) 043014,
  [\href{http://arxiv.org/abs/1110.2583}{{\sf arXiv:1110.2583}}],
  [\href{http://dx.doi.org/10.1103/PhysRevD.85.043014}{{\sf
  doi:10.1103/PhysRevD.85.043014}}].

\bibitem{Upadhye:2012fz}
A.~Upadhye, {\it {Particles and forces from chameleon dark energy}},
  \href{http://arxiv.org/abs/1211.7066}{{\sf arXiv:1211.7066}}.

\bibitem{Brax:2013uh}
P.~Brax, A.-C. Davis, and J.~Sakstein, {\it {Pulsar Constraints on Screened
  Modified Gravity}},  \href{http://arxiv.org/abs/1301.5587}{{\sf
  arXiv:1301.5587}}.

\bibitem{Brax:2004qh}
P.~Brax, C.~van~de Bruck, A.-C. Davis, J.~Khoury, and A.~Weltman, {\it
  {Detecting dark energy in orbit - The Cosmological chameleon}},  {\sl
  Phys.Rev.} {\bf D70} (2004) 123518,
  [\href{http://arxiv.org/abs/astro-ph/0408415}{{\sf arXiv:astro-ph/0408415}}],
  [\href{http://dx.doi.org/10.1103/PhysRevD.70.123518}{{\sf
  doi:10.1103/PhysRevD.70.123518}}].

\bibitem{Brax:2005ew}
P.~Brax, C.~van~de Bruck, A.-C. Davis, and A.~M. Green, {\it {Small scale
  structure formation in chameleon cosmology}},  {\sl Phys.Lett.} {\bf B633}
  (2006) 441--446, [\href{http://arxiv.org/abs/astro-ph/0509878}{{\sf
  arXiv:astro-ph/0509878}}],
  [\href{http://dx.doi.org/10.1016/j.physletb.2005.12.055}{{\sf
  doi:10.1016/j.physletb.2005.12.055}}].

\bibitem{Hu:2007nk}
W.~Hu and I.~Sawicki, {\it {Models of f(R) Cosmic Acceleration that Evade
  Solar-System Tests}},  {\sl Phys.Rev.} {\bf D76} (2007) 064004,
  [\href{http://arxiv.org/abs/0705.1158}{{\sf arXiv:0705.1158}}],
  [\href{http://dx.doi.org/10.1103/PhysRevD.76.064004}{{\sf
  doi:10.1103/PhysRevD.76.064004}}].

\bibitem{Davis:2009vk}
A.-C. Davis, C.~A. Schelpe, and D.~J. Shaw, {\it {The Effect of a Chameleon
  Scalar Field on the Cosmic Microwave Background}},  {\sl Phys.Rev.} {\bf D80}
  (2009) 064016, [\href{http://arxiv.org/abs/0907.2672}{{\sf
  arXiv:0907.2672}}], [\href{http://dx.doi.org/10.1103/PhysRevD.80.064016}{{\sf
  doi:10.1103/PhysRevD.80.064016}}].

\bibitem{Hinterbichler:2011ca}
K.~Hinterbichler, J.~Khoury, A.~Levy, and A.~Matas, {\it {Symmetron
  Cosmology}},  {\sl Phys.Rev.} {\bf D84} (2011) 103521,
  [\href{http://arxiv.org/abs/1107.2112}{{\sf arXiv:1107.2112}}],
  [\href{http://dx.doi.org/10.1103/PhysRevD.84.103521}{{\sf
  doi:10.1103/PhysRevD.84.103521}}].

\bibitem{Li:2011pj}
B.~Li, G.-B. Zhao, and K.~Koyama, {\it {Halos and Voids in f(R) Gravity}},
  {\sl Mon.Not.Roy.Astron.Soc.} {\bf 421} (2012) 3481,
  [\href{http://arxiv.org/abs/1111.2602}{{\sf arXiv:1111.2602}}],
  [\href{http://dx.doi.org/10.1111/j.1365-2966.2012.20573.x}{{\sf
  doi:10.1111/j.1365-2966.2012.20573.x}}].

\bibitem{Li:2011qda}
B.~Li and G.~Efstathiou, {\it {An Extended Excursion Set Approach to Structure
  Formation in Chameleon Models}},  {\sl Mon.Not.Roy.Astron.Soc.} {\bf 421}
  (2012) 1431, [\href{http://arxiv.org/abs/1110.6440}{{\sf arXiv:1110.6440}}],
  [\href{http://dx.doi.org/10.1111/j.1365-2966.2011.20404.x}{{\sf
  doi:10.1111/j.1365-2966.2011.20404.x}}].

\bibitem{Brax:2012sy}
P.~Brax and P.~Valageas, {\it {Structure Formation in Modified Gravity
  Scenarios}},  {\sl Phys.Rev.} {\bf D86} (2012) 063512,
  [\href{http://arxiv.org/abs/1205.6583}{{\sf arXiv:1205.6583}}],
  [\href{http://dx.doi.org/10.1103/PhysRevD.86.063512}{{\sf
  doi:10.1103/PhysRevD.86.063512}}].

\bibitem{Li:2012by}
B.~Li, W.~A. Hellwing, K.~Koyama, G.-B. Zhao, E.~Jennings, {\em et~al.}, {\it
  {The nonlinear matter and velocity power spectra in f(R) gravity}},  {\sl
  Mon.Not.Roy.Astron.Soc.} {\bf 428} (2013) 743--755,
  [\href{http://arxiv.org/abs/1206.4317}{{\sf arXiv:1206.4317}}],
  [\href{http://dx.doi.org/10.1093/mnras/sts072}{{\sf
  doi:10.1093/mnras/sts072}}].

\bibitem{Brax:2012nk}
P.~Brax, A.-C. Davis, B.~Li, H.~A. Winther, and G.-B. Zhao, {\it {Systematic
  Simulations of Modified Gravity: Symmetron and Dilaton Models}},  {\sl JCAP}
  {\bf 1210} (2012) 002, [\href{http://arxiv.org/abs/1206.3568}{{\sf
  arXiv:1206.3568}}],
  [\href{http://dx.doi.org/10.1088/1475-7516/2012/10/002}{{\sf
  doi:10.1088/1475-7516/2012/10/002}}].

\bibitem{Llinares:2012ds}
C.~Llinares and D.~F. Mota, {\it {Shape of Clusters of Galaxies as a Probe of
  Screening Mechanisms in Modified Gravity}},  {\sl Phys.Rev.Lett.} {\bf 110}
  (2013), no.~15 151104, [\href{http://arxiv.org/abs/1205.5775}{{\sf
  arXiv:1205.5775}}],
  [\href{http://dx.doi.org/10.1103/PhysRevLett.110.151104}{{\sf
  doi:10.1103/PhysRevLett.110.151104}}].

\bibitem{Jennings:2012pt}
E.~Jennings, C.~M. Baugh, B.~Li, G.-B. Zhao, and K.~Koyama, {\it {Redshift
  space distortions in f(R) gravity}},  {\sl Mon.Not.Roy.Astron.Soc.} {\bf 425}
  (2012) 2128--2143, [\href{http://arxiv.org/abs/1205.2698}{{\sf
  arXiv:1205.2698}}],
  [\href{http://dx.doi.org/10.1111/j.1365-2966.2012.21567.x}{{\sf
  doi:10.1111/j.1365-2966.2012.21567.x}}].

\bibitem{Lee:2012bm}
J.~Lee, G.-B. Zhao, B.~Li, and K.~Koyama, {\it {Modified Gravity Spins Up
  Galactic Halos}},  {\sl Astrophys.J.} {\bf 763} (2013) 28,
  [\href{http://arxiv.org/abs/1204.6608}{{\sf arXiv:1204.6608}}],
  [\href{http://dx.doi.org/10.1088/0004-637X/763/1/28}{{\sf
  doi:10.1088/0004-637X/763/1/28}}].

\bibitem{Brax:2013mua}
P.~Brax, A.-C. Davis, B.~Li, H.~A. Winther, and G.-B. Zhao, {\it {Systematic
  simulations of modified gravity: chameleon models}},  {\sl JCAP} {\bf 1304}
  (2013) 029, [\href{http://arxiv.org/abs/1303.0007}{{\sf arXiv:1303.0007}}],
  [\href{http://dx.doi.org/10.1088/1475-7516/2013/04/029}{{\sf
  doi:10.1088/1475-7516/2013/04/029}}].

\bibitem{Upadhye:2012vh}
A.~Upadhye, W.~Hu, and J.~Khoury, {\it {Quantum Stability of Chameleon Field
  Theories}},  {\sl Phys.Rev.Lett.} {\bf 109} (2012) 041301,
  [\href{http://arxiv.org/abs/1204.3906}{{\sf arXiv:1204.3906}}],
  [\href{http://dx.doi.org/10.1103/PhysRevLett.109.041301}{{\sf
  doi:10.1103/PhysRevLett.109.041301}}].

\bibitem{Erickcek:2013dea}
A.~L. Erickcek, N.~Barnaby, C.~Burrage, and Z.~Huang, {\it {Chameleons in the
  Early Universe: Kicks, Rebounds, and Particle Production}},  {\sl Phys.Rev.}
  {\bf D89} (2014) 084074, [\href{http://arxiv.org/abs/1310.5149}{{\sf
  arXiv:1310.5149}}].

\bibitem{Wang:2012kj}
J.~Wang, L.~Hui, and J.~Khoury, {\it {No-Go Theorems for Generalized Chameleon
  Field Theories}},  {\sl Phys.Rev.Lett.} {\bf 109} (2012) 241301,
  [\href{http://arxiv.org/abs/1208.4612}{{\sf arXiv:1208.4612}}],
  [\href{http://dx.doi.org/10.1103/PhysRevLett.109.241301}{{\sf
  doi:10.1103/PhysRevLett.109.241301}}].

\bibitem{Brax:2011ta}
P.~Brax and A.-C. Davis, {\it {Modified Gravity and the CMB}},  {\sl Phys.Rev.}
  {\bf D85} (2012) 023513, [\href{http://arxiv.org/abs/1109.5862}{{\sf
  arXiv:1109.5862}}], [\href{http://dx.doi.org/10.1103/PhysRevD.85.023513}{{\sf
  doi:10.1103/PhysRevD.85.023513}}].

\bibitem{Ade:2013zuv}
{\bf Planck Collaboration} Collaboration, P.~Ade {\em et~al.}, {\it {Planck
  2013 results. XVI. Cosmological parameters}},
  \href{http://arxiv.org/abs/1303.5076}{{\sf arXiv:1303.5076}}.

\bibitem{Sherwin:2011gv}
B.~D. Sherwin, J.~Dunkley, S.~Das, J.~W. Appel, J.~R. Bond, {\em et~al.}, {\it
  {Evidence for dark energy from the cosmic microwave background alone using
  the Atacama Cosmology Telescope lensing measurements}},  {\sl Phys.Rev.Lett.}
  {\bf 107} (2011) 021302, [\href{http://arxiv.org/abs/1105.0419}{{\sf
  arXiv:1105.0419}}],
  [\href{http://dx.doi.org/10.1103/PhysRevLett.107.021302}{{\sf
  doi:10.1103/PhysRevLett.107.021302}}].

\bibitem{prialnik2000introduction}
D.~Prialnik, {\em An Introduction to the Theory of Stellar Structure and
  Evolution}.
\newblock Cambridge University Press, 2000.

\bibitem{kippenhahn1990stellar}
R.~Kippenhahn and A.~Weigert, {\em Stellar structure and evolution}.
\newblock Astronomy and astrophysics library. Springer, 1990.

\bibitem{oai:arXiv.org:1011.1491}
F.~Quevedo, S.~Krippendorf, and O.~Schlotterer, {\it {Cambridge Lectures on
  Supersymmetry and Extra Dimensions}},
  \href{http://arxiv.org/abs/1011.1491}{{\sf arXiv:1011.1491}}.

\bibitem{wess1992supersymmetry}
J.~Wess and J.~Bagger, {\em Supersymmetry and Supergravity}.
\newblock Princeton series in physics. Princeton University Press, 1992.

\bibitem{Jacobson:2000xp}
T.~Jacobson and D.~Mattingly, {\it {Gravity with a dynamical preferred frame}},
   {\sl Phys.Rev.} {\bf D64} (2001) 024028,
  [\href{http://arxiv.org/abs/gr-qc/0007031}{{\sf arXiv:gr-qc/0007031}}],
  [\href{http://dx.doi.org/10.1103/PhysRevD.64.024028}{{\sf
  doi:10.1103/PhysRevD.64.024028}}].

\bibitem{Horava:2009uw}
P.~Horava, {\it {Quantum Gravity at a Lifshitz Point}},  {\sl Phys.Rev.} {\bf
  D79} (2009) 084008, [\href{http://arxiv.org/abs/0901.3775}{{\sf
  arXiv:0901.3775}}], [\href{http://dx.doi.org/10.1103/PhysRevD.79.084008}{{\sf
  doi:10.1103/PhysRevD.79.084008}}].

\bibitem{Horndeski:1974wa}
G.~W. Horndeski, {\it {Second-order scalar-tensor field equations in a
  four-dimensional space}},  {\sl Int.J.Theor.Phys.} {\bf 10} (1974) 363--384,
  [\href{http://dx.doi.org/10.1007/BF01807638}{{\sf doi:10.1007/BF01807638}}].

\bibitem{Deffayet:2009wt}
C.~Deffayet, G.~Esposito-Farese, and A.~Vikman, {\it {Covariant Galileon}},
  {\sl Phys.Rev.} {\bf D79} (2009) 084003,
  [\href{http://arxiv.org/abs/0901.1314}{{\sf arXiv:0901.1314}}],
  [\href{http://dx.doi.org/10.1103/PhysRevD.79.084003}{{\sf
  doi:10.1103/PhysRevD.79.084003}}].

\bibitem{Fierz:1939ix}
M.~Fierz and W.~Pauli, {\it {On relativistic wave equations for particles of
  arbitrary spin in an electromagnetic field}},  {\sl Proc.Roy.Soc.Lond.} {\bf
  A173} (1939) 211--232, [\href{http://dx.doi.org/10.1098/rspa.1939.0140}{{\sf
  doi:10.1098/rspa.1939.0140}}].

\bibitem{Boulware:1973my}
D.~Boulware and S.~Deser, {\it {Can gravitation have a finite range?}},  {\sl
  Phys.Rev.} {\bf D6} (1972) 3368--3382,
  [\href{http://dx.doi.org/10.1103/PhysRevD.6.3368}{{\sf
  doi:10.1103/PhysRevD.6.3368}}].

\bibitem{deRham:2010kj}
C.~de~Rham, G.~Gabadadze, and A.~J. Tolley, {\it {Resummation of Massive
  Gravity}},  {\sl Phys.Rev.Lett.} {\bf 106} (2011) 231101,
  [\href{http://arxiv.org/abs/1011.1232}{{\sf arXiv:1011.1232}}],
  [\href{http://dx.doi.org/10.1103/PhysRevLett.106.231101}{{\sf
  doi:10.1103/PhysRevLett.106.231101}}].

\bibitem{Hinterbichler:2011tt}
K.~Hinterbichler, {\it {Theoretical Aspects of Massive Gravity}},  {\sl
  Rev.Mod.Phys.} {\bf 84} (2012) 671--710,
  [\href{http://arxiv.org/abs/1105.3735}{{\sf arXiv:1105.3735}}],
  [\href{http://dx.doi.org/10.1103/RevModPhys.84.671}{{\sf
  doi:10.1103/RevModPhys.84.671}}].

\bibitem{Woodard:2014iga}
R.~Woodard, {\it {Nonlocal Models of Cosmic Acceleration}},  {\sl Found.Phys.}
  {\bf 44} (2014) 213--233, [\href{http://arxiv.org/abs/1401.0254}{{\sf
  arXiv:1401.0254}}], [\href{http://dx.doi.org/10.1007/s10701-014-9780-6}{{\sf
  doi:10.1007/s10701-014-9780-6}}].

\bibitem{Clemson:2012im}
T.~Clemson and K.~Koyama, {\it {The Distinguishability of Interacting Dark
  Energy from Modified Gravity}},  {\sl JCAP} {\bf 1301} (2013) 010,
  [\href{http://arxiv.org/abs/1209.2618}{{\sf arXiv:1209.2618}}],
  [\href{http://dx.doi.org/10.1088/1475-7516/2013/01/010}{{\sf
  doi:10.1088/1475-7516/2013/01/010}}].

\bibitem{Kapner:2006si}
D.~Kapner, T.~Cook, E.~Adelberger, J.~Gundlach, B.~R. Heckel, {\em et~al.},
  {\it {Tests of the gravitational inverse-square law below the dark-energy
  length scale}},  {\sl Phys.Rev.Lett.} {\bf 98} (2007) 021101,
  [\href{http://arxiv.org/abs/hep-ph/0611184}{{\sf arXiv:hep-ph/0611184}}],
  [\href{http://dx.doi.org/10.1103/PhysRevLett.98.021101}{{\sf
  doi:10.1103/PhysRevLett.98.021101}}].

\bibitem{Brax:2010kv}
P.~Brax, C.~van~de Bruck, D.~F. Mota, N.~J. Nunes, and H.~A. Winther, {\it
  {Chameleons with Field Dependent Couplings}},  {\sl Phys.Rev.} {\bf D82}
  (2010) 083503, [\href{http://arxiv.org/abs/1006.2796}{{\sf
  arXiv:1006.2796}}], [\href{http://dx.doi.org/10.1103/PhysRevD.82.083503}{{\sf
  doi:10.1103/PhysRevD.82.083503}}].

\bibitem{Hinterbichler:2010wu}
K.~Hinterbichler, J.~Khoury, and H.~Nastase, {\it {Towards a UV Completion for
  Chameleon Scalar Theories}},  {\sl JHEP} {\bf 1103} (2011) 061,
  [\href{http://arxiv.org/abs/1012.4462}{{\sf arXiv:1012.4462}}],
  [\href{http://dx.doi.org/10.1007/JHEP06(2011)072,
  10.1007/JHEP03(2011)061}{{\sf doi:10.1007/JHEP06(2011)072,
  10.1007/JHEP03(2011)061}}].

\bibitem{Hinterbichler:2013we}
K.~Hinterbichler, J.~Khoury, H.~Nastase, and R.~Rosenfeld, {\it {Chameleonic
  inflation}},  {\sl JHEP} {\bf 1308} (2013) 053,
  [\href{http://arxiv.org/abs/1301.6756}{{\sf arXiv:1301.6756}}],
  [\href{http://dx.doi.org/10.1007/JHEP08(2013)053}{{\sf
  doi:10.1007/JHEP08(2013)053}}].

\bibitem{Brax:2012mq}
P.~Brax, A.-C. Davis, and J.~Sakstein, {\it {SUPER-Screening}},  {\sl
  Phys.Lett.B} {\bf 719} (2013) 210--217,
  [\href{http://arxiv.org/abs/1212.4392}{{\sf arXiv:1212.4392}}],
  [\href{http://dx.doi.org/10.1016/j.physletb.2013.01.044}{{\sf
  doi:10.1016/j.physletb.2013.01.044}}].

\bibitem{Brax:2013yja}
P.~Brax, A.-C. Davis, and J.~Sakstein, {\it {Dynamics of Supersymmetric
  Chameleons}},  {\sl JCAP} {\bf 1310} (2013) 007,
  [\href{http://arxiv.org/abs/1302.3080}{{\sf arXiv:1302.3080}}],
  [\href{http://dx.doi.org/10.1088/1475-7516/2013/10/007}{{\sf
  doi:10.1088/1475-7516/2013/10/007}}].

\bibitem{Brax:2008hh}
P.~Brax, C.~van~de Bruck, A.-C. Davis, and D.~J. Shaw, {\it {f(R) Gravity and
  Chameleon Theories}},  {\sl Phys.Rev.} {\bf D78} (2008) 104021,
  [\href{http://arxiv.org/abs/0806.3415}{{\sf arXiv:0806.3415}}],
  [\href{http://dx.doi.org/10.1103/PhysRevD.78.104021}{{\sf
  doi:10.1103/PhysRevD.78.104021}}].

\bibitem{Davis:2011pj}
A.-C. Davis, B.~Li, D.~F. Mota, and H.~A. Winther, {\it {Structure Formation in
  the Symmetron Model}},  {\sl Astrophys.J.} {\bf 748} (2012) 61,
  [\href{http://arxiv.org/abs/1108.3081}{{\sf arXiv:1108.3081}}],
  [\href{http://dx.doi.org/10.1088/0004-637X/748/1/61}{{\sf
  doi:10.1088/0004-637X/748/1/61}}].

\bibitem{Damour:1994zq}
T.~Damour and A.~M. Polyakov, {\it {The String dilaton and a least coupling
  principle}},  {\sl Nucl.Phys.} {\bf B423} (1994) 532--558,
  [\href{http://arxiv.org/abs/hep-th/9401069}{{\sf arXiv:hep-th/9401069}}],
  [\href{http://dx.doi.org/10.1016/0550-3213(94)90143-0}{{\sf
  doi:10.1016/0550-3213(94)90143-0}}].

\bibitem{Koivisto:2012za}
T.~S. Koivisto, D.~F. Mota, and M.~Zumalacarregui, {\it {Screening
  Modifications of Gravity through Disformally Coupled Fields}},  {\sl
  Phys.Rev.Lett.} {\bf 109} (2012) 241102,
  [\href{http://arxiv.org/abs/1205.3167}{{\sf arXiv:1205.3167}}],
  [\href{http://dx.doi.org/10.1103/PhysRevLett.109.241102}{{\sf
  doi:10.1103/PhysRevLett.109.241102}}].

\bibitem{Zumalacarregui:2012us}
M.~Zumalacarregui, T.~S. Koivisto, and D.~F. Mota, {\it {DBI Galileons in the
  Einstein Frame: Local Gravity and Cosmology}},  {\sl Phys.Rev.} {\bf D87}
  (2013) 083010, [\href{http://arxiv.org/abs/1210.8016}{{\sf
  arXiv:1210.8016}}], [\href{http://dx.doi.org/10.1103/PhysRevD.87.083010}{{\sf
  doi:10.1103/PhysRevD.87.083010}}].

\bibitem{Gasperini:2001pc}
M.~Gasperini, F.~Piazza, and G.~Veneziano, {\it {Quintessence as a runaway
  dilaton}},  {\sl Phys.Rev.} {\bf D65} (2002) 023508,
  [\href{http://arxiv.org/abs/gr-qc/0108016}{{\sf arXiv:gr-qc/0108016}}],
  [\href{http://dx.doi.org/10.1103/PhysRevD.65.023508}{{\sf
  doi:10.1103/PhysRevD.65.023508}}].

\bibitem{Brax:2011aw}
P.~Brax, A.-C. Davis, and B.~Li, {\it {Modified Gravity Tomography}},  {\sl
  Phys.Lett.} {\bf B715} (2012) 38--43,
  [\href{http://arxiv.org/abs/1111.6613}{{\sf arXiv:1111.6613}}],
  [\href{http://dx.doi.org/10.1016/j.physletb.2012.08.002}{{\sf
  doi:10.1016/j.physletb.2012.08.002}}].

\bibitem{Jain:2012tn}
B.~Jain, V.~Vikram, and J.~Sakstein, {\it {Astrophysical Tests of Modified
  Gravity: Constraints from Distance Indicators in the Nearby Universe}},  {\sl
  Astrophys.J.} {\bf 779} (2013) 39,
  [\href{http://arxiv.org/abs/1204.6044}{{\sf arXiv:1204.6044}}],
  [\href{http://dx.doi.org/10.1088/0004-637X/779/1/39}{{\sf
  doi:10.1088/0004-637X/779/1/39}}].

\bibitem{Schmidt:2008tn}
F.~Schmidt, M.~V. Lima, H.~Oyaizu, and W.~Hu, {\it {Non-linear Evolution of
  f(R) Cosmologies III: Halo Statistics}},  {\sl Phys.Rev.} {\bf D79} (2009)
  083518, [\href{http://arxiv.org/abs/0812.0545}{{\sf arXiv:0812.0545}}],
  [\href{http://dx.doi.org/10.1103/PhysRevD.79.083518}{{\sf
  doi:10.1103/PhysRevD.79.083518}}].

\bibitem{Jain:2011ji}
B.~Jain and J.~VanderPlas, {\it {Tests of Modified Gravity with Dwarf
  Galaxies}},  {\sl JCAP} {\bf 1110} (2011) 032,
  [\href{http://arxiv.org/abs/1106.0065}{{\sf arXiv:1106.0065}}],
  [\href{http://dx.doi.org/10.1088/1475-7516/2011/10/032}{{\sf
  doi:10.1088/1475-7516/2011/10/032}}].

\bibitem{Vikram:2013uba}
V.~Vikram, A.~Cabré, B.~Jain, and J.~VanderPlas, {\it {Astrophysical Tests of
  Modified Gravity: the Morphology and Kinematics of Dwarf Galaxies}},  {\sl
  JCAP} {\bf 1308} (2013) 020, [\href{http://arxiv.org/abs/1303.0295}{{\sf
  arXiv:1303.0295}}],
  [\href{http://dx.doi.org/10.1088/1475-7516/2013/08/020}{{\sf
  doi:10.1088/1475-7516/2013/08/020}}].

\bibitem{Vikram:2014uza}
V.~Vikram, J.~Sakstein, C.~Davis, and A.~Neil, {\it {Astrophysical Tests of
  Modified Gravity: Stellar and Gaseous Rotation Curves in Dwarf Galaxies}},
  \href{http://arxiv.org/abs/1407.6044}{{\sf arXiv:1407.6044}}.

\bibitem{Haas:2011mt}
M.~R. Haas, J.~Schaye, and A.~Jeeson-Daniel, {\it {Disentangling galaxy
  environment and host halo mass}},  {\sl Mon.Not.Roy.Astron.Soc.} {\bf 419}
  (2012) 2133, [\href{http://arxiv.org/abs/1103.0547}{{\sf arXiv:1103.0547}}],
  [\href{http://dx.doi.org/10.1111/j.1365-2966.2011.19863.x}{{\sf
  doi:10.1111/j.1365-2966.2011.19863.x}}].

\bibitem{Zhao:2011cu}
G.-B. Zhao, B.~Li, and K.~Koyama, {\it {Testing General Relativity using the
  Environmental Dependence of Dark Matter Halos}},  {\sl Phys.Rev.Lett.} {\bf
  107} (2011) 071303, [\href{http://arxiv.org/abs/1105.0922}{{\sf
  arXiv:1105.0922}}],
  [\href{http://dx.doi.org/10.1103/PhysRevLett.107.071303}{{\sf
  doi:10.1103/PhysRevLett.107.071303}}].

\bibitem{Cabre:2012tq}
A.~Cabre, V.~Vikram, G.-B. Zhao, B.~Jain, and K.~Koyama, {\it {Astrophysical
  Tests of Modified Gravity: A Screening Map of the Nearby Universe}},  {\sl
  JCAP} {\bf 1207} (2012) 034, [\href{http://arxiv.org/abs/1204.6046}{{\sf
  arXiv:1204.6046}}],
  [\href{http://dx.doi.org/10.1088/1475-7516/2012/07/034}{{\sf
  doi:10.1088/1475-7516/2012/07/034}}].

\bibitem{Erickcek:2013oma}
A.~L. Erickcek, N.~Barnaby, C.~Burrage, and Z.~Huang, {\it {Catastrophic
  Consequences of Kicking the Chameleon}},  {\sl Phys. Rev. Lett.} {\bf
  110:171101} (2013) [\href{http://arxiv.org/abs/1304.0009}{{\sf
  arXiv:1304.0009}}].

\bibitem{Brax:2011qs}
P.~Brax and A.-C. Davis, {\it {Supersymmetron}},  {\sl Phys.Lett.} {\bf B707}
  (2012) 1--7, [\href{http://arxiv.org/abs/1109.0468}{{\sf arXiv:1109.0468}}],
  [\href{http://dx.doi.org/10.1016/j.physletb.2011.11.060}{{\sf
  doi:10.1016/j.physletb.2011.11.060}}].

\bibitem{Babichev:2009us}
E.~Babichev, C.~Deffayet, and R.~Ziour, {\it {The Vainshtein mechanism in the
  Decoupling Limit of massive gravity}},  {\sl JHEP} {\bf 0905} (2009) 098,
  [\href{http://arxiv.org/abs/0901.0393}{{\sf arXiv:0901.0393}}],
  [\href{http://dx.doi.org/10.1088/1126-6708/2009/05/098}{{\sf
  doi:10.1088/1126-6708/2009/05/098}}].

\bibitem{Nordtvedt:2003pj}
K.~Nordtvedt, {\it {Lunar laser ranging: A Comprehensive probe of postNewtonian
  gravity}},  \href{http://arxiv.org/abs/gr-qc/0301024}{{\sf
  arXiv:gr-qc/0301024}}.

\bibitem{Dvali:2007kt}
G.~Dvali, S.~Hofmann, and J.~Khoury, {\it {Degravitation of the cosmological
  constant and graviton width}},  {\sl Phys.Rev.} {\bf D76} (2007) 084006,
  [\href{http://arxiv.org/abs/hep-th/0703027}{{\sf arXiv:hep-th/0703027}}],
  [\href{http://dx.doi.org/10.1103/PhysRevD.76.084006}{{\sf
  doi:10.1103/PhysRevD.76.084006}}].

\bibitem{Afshordi:2008rd}
N.~Afshordi, G.~Geshnizjani, and J.~Khoury, {\it {Do observations offer
  evidence for cosmological-scale extra dimensions?}},  {\sl JCAP} {\bf 0908}
  (2009) 030, [\href{http://arxiv.org/abs/0812.2244}{{\sf arXiv:0812.2244}}],
  [\href{http://dx.doi.org/10.1088/1475-7516/2009/08/030}{{\sf
  doi:10.1088/1475-7516/2009/08/030}}].

\bibitem{Murphy:2012rea}
J.~Murphy, T.W., E.~Adelberger, J.~Battat, C.~Hoyle, N.~Johnson, {\em et~al.},
  {\it {APOLLO: millimeter lunar laser ranging}},  {\sl Class.Quant.Grav.} {\bf
  29} (2012) 184005,
  [\href{http://dx.doi.org/10.1088/0264-9381/29/18/184005}{{\sf
  doi:10.1088/0264-9381/29/18/184005}}].

\bibitem{Hiramatsu:2012xj}
T.~Hiramatsu, W.~Hu, K.~Koyama, and F.~Schmidt, {\it {Equivalence Principle
  Violation in Vainshtein Screened Two-Body Systems}},  {\sl Phys.Rev.} {\bf
  D87} (2013), no.~6 063525, [\href{http://arxiv.org/abs/1209.3364}{{\sf
  arXiv:1209.3364}}], [\href{http://dx.doi.org/10.1103/PhysRevD.87.063525}{{\sf
  doi:10.1103/PhysRevD.87.063525}}].

\bibitem{Andrews:2013qva}
M.~Andrews, Y.-Z. Chu, and M.~Trodden, {\it {Galileon forces in the Solar
  System}},  {\sl Phys.Rev.} {\bf D88} (2013) 084028,
  [\href{http://arxiv.org/abs/1305.2194}{{\sf arXiv:1305.2194}}],
  [\href{http://dx.doi.org/10.1103/PhysRevD.88.084028}{{\sf
  doi:10.1103/PhysRevD.88.084028}}].

\bibitem{Hui:2012jb}
L.~Hui and A.~Nicolis, {\it {Proposal for an Observational Test of the
  Vainshtein Mechanism}},  {\sl Phys.Rev.Lett.} {\bf 109} (2012) 051304,
  [\href{http://arxiv.org/abs/1201.1508}{{\sf arXiv:1201.1508}}],
  [\href{http://dx.doi.org/10.1103/PhysRevLett.109.051304}{{\sf
  doi:10.1103/PhysRevLett.109.051304}}].

\bibitem{Hui:2012qt}
L.~Hui and A.~Nicolis, {\it {No-Hair Theorem for the Galileon}},  {\sl
  Phys.Rev.Lett.} {\bf 110} (2013), no.~24 241104,
  [\href{http://arxiv.org/abs/1202.1296}{{\sf arXiv:1202.1296}}],
  [\href{http://dx.doi.org/10.1103/PhysRevLett.110.241104}{{\sf
  doi:10.1103/PhysRevLett.110.241104}}].

\bibitem{Paxton:2010ji}
B.~Paxton, L.~Bildsten, A.~Dotter, F.~Herwig, P.~Lesaffre, {\em et~al.}, {\it
  {Modules for Experiments in Stellar Astrophysics (MESA)}},  {\sl
  Astrophys.J.Suppl.} {\bf 192} (2011) 3,
  [\href{http://arxiv.org/abs/1009.1622}{{\sf arXiv:1009.1622}}],
  [\href{http://dx.doi.org/10.1088/0067-0049/192/1/3}{{\sf
  doi:10.1088/0067-0049/192/1/3}}].

\bibitem{Paxton:2013pj}
B.~Paxton, M.~Cantiello, P.~Arras, L.~Bildsten, E.~F. Brown, {\em et~al.}, {\it
  {Modules for Experiments in Stellar Astrophysics (MESA): Planets,
  Oscillations, Rotation, and Massive Stars}},  {\sl Astrophys.J.Suppl.} {\bf
  208} (2013) 4, [\href{http://arxiv.org/abs/1301.0319}{{\sf
  arXiv:1301.0319}}], [\href{http://dx.doi.org/10.1088/0067-0049/208/1/4}{{\sf
  doi:10.1088/0067-0049/208/1/4}}].

\bibitem{Chang:2010xh}
P.~Chang and L.~Hui, {\it {Stellar Structure and Tests of Modified Gravity}},
  {\sl Astrophys.J.} {\bf 732} (2011) 25,
  [\href{http://arxiv.org/abs/1011.4107}{{\sf arXiv:1011.4107}}],
  [\href{http://dx.doi.org/10.1088/0004-637X/732/1/25}{{\sf
  doi:10.1088/0004-637X/732/1/25}}].

\bibitem{Adams:2008ad}
F.~C. Adams, {\it {Stars In Other Universes: Stellar structure with different
  fundamental constants}},  {\sl JCAP} {\bf 0808} (2008) 010,
  [\href{http://arxiv.org/abs/0807.3697}{{\sf arXiv:0807.3697}}],
  [\href{http://dx.doi.org/10.1088/1475-7516/2008/08/010}{{\sf
  doi:10.1088/1475-7516/2008/08/010}}].

\bibitem{Salpeter:1955it}
E.~E. Salpeter, {\it {The Luminosity function and stellar evolution}},  {\sl
  Astrophys.J.} {\bf 121} (1955) 161--167,
  [\href{http://dx.doi.org/10.1086/145971}{{\sf doi:10.1086/145971}}].

\bibitem{Freedman:2010xv}
W.~L. Freedman and B.~F. Madore, {\it {The Hubble Constant}},  {\sl
  Ann.Rev.Astron.Astrophys.} {\bf 48} (2010) 673--710,
  [\href{http://arxiv.org/abs/1004.1856}{{\sf arXiv:1004.1856}}],
  [\href{http://dx.doi.org/10.1146/annurev-astro-082708-101829}{{\sf
  doi:10.1146/annurev-astro-082708-101829}}].

\bibitem{Alibert:1999an}
Y.~Alibert, I.~Baraffe, P.~Hauschildt, and F.~Allard, {\it {Period - luminosity
  - color - radius relationships of cepheids as a function of metallicity:
  evolutionary effects}},  {\sl Astron.Astrophys.} {\bf 344} (1999) 551--572,
  [\href{http://arxiv.org/abs/astro-ph/9901294}{{\sf arXiv:astro-ph/9901294}}].

\bibitem{2011PhDT.......205K}
C.-Y. {Kuo}, {\em {The megamaser cosmology project : geometric distances to
  megamaser galaxies and accurate masses of supermassive black holes at their
  centers}}.
\newblock PhD thesis, University of Virginia, 2011.

\bibitem{Herrnstein:1999cw}
J.~Herrnstein, J.~Moran, L.~Greenhill, P.~Diamond, M.~Inoue, {\em et~al.}, {\it
  {A Geometric distance to the galaxy NGC 4258 from orbital motions in a
  nuclear gas disk}},  {\sl Nature} {\bf 400} (1999) 539--541,
  [\href{http://dx.doi.org/10.1038/22972}{{\sf doi:10.1038/22972}}].

\bibitem{doi:10.1146/annurev.astro.41.011802.094927}
K.~Lo, {\it Mega-masers and galaxies},  {\sl Annual Review of Astronomy and
  Astrophysics} {\bf 43} (2005), no.~1 625--676,
  [\href{http://arxiv.org/abs/http://www.annualreviews.org/doi/pdf/10.1146/annurev.astro.41.011802.094927}{{\sf
  http://www.annualreviews.org/doi/pdf/10.1146/annurev.astro.41.011802.094927}}],
  [\href{http://dx.doi.org/10.1146/annurev.astro.41.011802.094927}{{\sf
  doi:10.1146/annurev.astro.41.011802.094927}}].

\bibitem{Moran:2007sm}
J.~Moran, L.~Humphreys, L.~Greenhill, M.~Reid, and A.~Argon, {\it {The
  structure of the accretion disk in NGC 4258 derived from observations of its
  water vapor masers}},  {\sl IAU Symp.} (2007)
  [\href{http://arxiv.org/abs/0707.1032}{{\sf arXiv:0707.1032}}].

\bibitem{Herrnstein:1998mh}
J.~Herrnstein, {\it {Cosmology with high-redshift water masers}},
  \href{http://arxiv.org/abs/astro-ph/9803039}{{\sf arXiv:astro-ph/9803039}}.

\bibitem{1950ApJ...112....6E}
I.~{Epstein}, {\it {Pulsation Properties of Giant-Star Models.}},  {\sl
  Astrophys. J.} {\bf 112} (July, 1950) 6,
  [\href{http://dx.doi.org/10.1086/145316}{{\sf doi:10.1086/145316}}].

\bibitem{Humphreys:2007ir}
E.~Humphreys, M.~Reid, L.~Greenhill, J.~Moran, and A.~Argon, {\it {Toward a New
  Distance to the Active Galaxy NGC 4258: II. Centripetal Accelerations and
  Investigation of Spiral Structure}},  {\sl Astrophys.J.} {\bf 672} (2008)
  800--816, [\href{http://arxiv.org/abs/0709.0925}{{\sf arXiv:0709.0925}}],
  [\href{http://dx.doi.org/10.1086/523637}{{\sf doi:10.1086/523637}}].

\bibitem{Humphreys:2013eja}
E.~Humphreys, M.~J. Reid, J.~M. Moran, L.~J. Greenhill, and A.~L. Argon, {\it
  {Toward a New Geometric Distance to the Active Galaxy NGC 4258. III. Final
  Results and the Hubble Constant}},  {\sl Astrophys.J.} {\bf 775} (2013) 13,
  [\href{http://arxiv.org/abs/1307.6031}{{\sf arXiv:1307.6031}}],
  [\href{http://dx.doi.org/10.1088/0004-637X/775/1/13}{{\sf
  doi:10.1088/0004-637X/775/1/13}}].

\bibitem{Henkel:2012by}
C.~Henkel, J.~Braatz, M.~Reid, J.~Condon, K.~Lo, {\em et~al.}, {\it {Cosmology
  and the Hubble Constant: On the Megamaser Cosmology Project (MCP)}},
  \href{http://arxiv.org/abs/1205.0823}{{\sf arXiv:1205.0823}}.

\bibitem{Lombriser:2010mp}
L.~Lombriser, A.~Slosar, U.~Seljak, and W.~Hu, {\it {Constraints on f(R)
  gravity from probing the large-scale structure}},  {\sl Phys.Rev.} {\bf D85}
  (2012) 124038, [\href{http://arxiv.org/abs/1003.3009}{{\sf
  arXiv:1003.3009}}], [\href{http://dx.doi.org/10.1103/PhysRevD.85.124038}{{\sf
  doi:10.1103/PhysRevD.85.124038}}].

\bibitem{Schmidt:2010jr}
F.~Schmidt, {\it {Dynamical Masses in Modified Gravity}},  {\sl Phys.Rev.} {\bf
  D81} (2010) 103002, [\href{http://arxiv.org/abs/1003.0409}{{\sf
  arXiv:1003.0409}}], [\href{http://dx.doi.org/10.1103/PhysRevD.81.103002}{{\sf
  doi:10.1103/PhysRevD.81.103002}}].

\bibitem{Terukina:2013eqa}
A.~Terukina, L.~Lombriser, K.~Yamamoto, D.~Bacon, K.~Koyama, {\em et~al.}, {\it
  {Testing chameleon gravity with the Coma cluster}},  {\sl JCAP} {\bf 1404}
  (2014) 013, [\href{http://arxiv.org/abs/1312.5083}{{\sf arXiv:1312.5083}}],
  [\href{http://dx.doi.org/10.1088/1475-7516/2014/04/013}{{\sf
  doi:10.1088/1475-7516/2014/04/013}}].

\bibitem{Silvestri:2011ch}
A.~Silvestri, {\it {Scalar radiation from Chameleon-shielded regions}},  {\sl
  Phys.Rev.Lett.} {\bf 106} (2011) 251101,
  [\href{http://arxiv.org/abs/1103.4013}{{\sf arXiv:1103.4013}}],
  [\href{http://dx.doi.org/10.1103/PhysRevLett.106.251101}{{\sf
  doi:10.1103/PhysRevLett.106.251101}}].

\bibitem{Upadhye:2013nfa}
A.~Upadhye and J.~H. Steffen, {\it {Monopole radiation in modified gravity}},
  \href{http://arxiv.org/abs/1306.6113}{{\sf arXiv:1306.6113}}.

\bibitem{cox1980theory}
J.~Cox, {\em The Theory of Stellar Pulsation}.
\newblock Princeton series in astrophysics. Princeton Univversity Press, 1980.

\bibitem{Chandrasekhar:1964zza}
S.~Chandrasekhar, {\it {Dynamical Instability of Gaseous Masses Approaching the
  Schwarzschild Limit in General Relativity}},  {\sl Phys.Rev.Lett.} {\bf 12}
  (1964) 114--116, [\href{http://dx.doi.org/10.1103/PhysRevLett.12.114}{{\sf
  doi:10.1103/PhysRevLett.12.114}}].

\bibitem{Hurley:1966}
M.~Hurley, P.~H. Roberts, and K.~Wright, {\it {The Oscillations of Gas
  Spheres}},  {\sl Astrophys. J.} {\bf 143} (1966) 535.

\bibitem{Brax:2006kg}
P.~Brax and J.~Martin, {\it {The SUGRA Quintessence Model Coupled to the
  MSSM}},  {\sl JCAP} {\bf 0611} (2006) 008,
  [\href{http://arxiv.org/abs/astro-ph/0606306}{{\sf arXiv:astro-ph/0606306}}],
  [\href{http://dx.doi.org/10.1088/1475-7516/2006/11/008}{{\sf
  doi:10.1088/1475-7516/2006/11/008}}].

\bibitem{Brax:2006dc}
P.~Brax and J.~Martin, {\it {Dark Energy and the MSSM}},  {\sl Phys.Rev.} {\bf
  D75} (2007) 083507, [\href{http://arxiv.org/abs/hep-th/0605228}{{\sf
  arXiv:hep-th/0605228}}],
  [\href{http://dx.doi.org/10.1103/PhysRevD.75.083507}{{\sf
  doi:10.1103/PhysRevD.75.083507}}].

\bibitem{Brax:2006np}
P.~Brax and J.~Martin, {\it {Moduli Fields as Quintessence and the Chameleon}},
   {\sl Phys.Lett.} {\bf B647} (2007) 320--329,
  [\href{http://arxiv.org/abs/hep-th/0612208}{{\sf arXiv:hep-th/0612208}}],
  [\href{http://dx.doi.org/10.1016/j.physletb.2007.02.019}{{\sf
  doi:10.1016/j.physletb.2007.02.019}}].

\bibitem{Kachru:2003aw}
S.~Kachru, R.~Kallosh, A.~D. Linde, and S.~P. Trivedi, {\it {De Sitter vacua in
  string theory}},  {\sl Phys.Rev.} {\bf D68} (2003) 046005,
  [\href{http://arxiv.org/abs/hep-th/0301240}{{\sf arXiv:hep-th/0301240}}],
  [\href{http://dx.doi.org/10.1103/PhysRevD.68.046005}{{\sf
  doi:10.1103/PhysRevD.68.046005}}].

\bibitem{Conlon:2010jq}
J.~P. Conlon and F.~G. Pedro, {\it {Moduli-Induced Vacuum Destabilisation}},
  {\sl JHEP} {\bf 1105} (2011) 079, [\href{http://arxiv.org/abs/1010.2665}{{\sf
  arXiv:1010.2665}}], [\href{http://dx.doi.org/10.1007/JHEP05(2011)079}{{\sf
  doi:10.1007/JHEP05(2011)079}}].

\bibitem{Nastase:2013ik}
H.~Nastase and A.~Weltman, {\it {Chameleons on the Racetrack}},  {\sl JHEP}
  {\bf 1308} (2013) 059, [\href{http://arxiv.org/abs/1301.7120}{{\sf
  arXiv:1301.7120}}], [\href{http://dx.doi.org/10.1007/JHEP08(2013)059}{{\sf
  doi:10.1007/JHEP08(2013)059}}].

\bibitem{Nilles:1983ge}
H.~P. Nilles, {\it {Supersymmetry, Supergravity and Particle Physics}},  {\sl
  Phys.Rept.} {\bf 110} (1984) 1--162,
  [\href{http://dx.doi.org/10.1016/0370-1573(84)90008-5}{{\sf
  doi:10.1016/0370-1573(84)90008-5}}].

\bibitem{Brignole:1993dj}
A.~Brignole, L.~E. Ibanez, and C.~Munoz, {\it {Towards a theory of soft terms
  for the supersymmetric Standard Model}},  {\sl Nucl.Phys.} {\bf B422} (1994)
  125--171, [\href{http://arxiv.org/abs/hep-ph/9308271}{{\sf
  arXiv:hep-ph/9308271}}],
  [\href{http://dx.doi.org/10.1016/0550-3213(94)00068-9}{{\sf
  doi:10.1016/0550-3213(94)00068-9}}].

\bibitem{Brignole:1997dp}
A.~Brignole, L.~E. Ibanez, and C.~Munoz, {\it {Soft supersymmetry breaking
  terms from supergravity and superstring models}},
  \href{http://arxiv.org/abs/hep-ph/9707209}{{\sf arXiv:hep-ph/9707209}}.

\bibitem{Farquet:2012cs}
D.~Farquet and C.~A. Scrucca, {\it {Scalar geometry and masses in Calabi-Yau
  string models}},  {\sl JHEP} {\bf 1209} (2012) 025,
  [\href{http://arxiv.org/abs/1205.5728}{{\sf arXiv:1205.5728}}],
  [\href{http://dx.doi.org/10.1007/JHEP09(2012)025}{{\sf
  doi:10.1007/JHEP09(2012)025}}].

\bibitem{Jack:1999zs}
I.~Jack and D.~Jones, {\it {Renormalization of the Fayet-Iliopoulos D term}},
  {\sl Phys.Lett.} {\bf B473} (2000) 102--108,
  [\href{http://arxiv.org/abs/hep-ph/9911491}{{\sf arXiv:hep-ph/9911491}}],
  [\href{http://dx.doi.org/10.1016/S0370-2693(99)01484-7}{{\sf
  doi:10.1016/S0370-2693(99)01484-7}}].

\bibitem{Contreras:2013bol}
{\bf WiggleZ Collaboration} Collaboration, C.~Contreras {\em et~al.}, {\it {The
  WiggleZ Dark Energy Survey: measuring the cosmic growth rate with the
  two-point galaxy correlation function}},
  \href{http://arxiv.org/abs/1302.5178}{{\sf arXiv:1302.5178}},
  \href{http://dx.doi.org/10.1093/mnras/sts608}{{\sf
  doi:10.1093/mnras/sts608}}.

\bibitem{PhysRevD.37.3406}
B.~Ratra and P.~J.~E. Peebles, {\it Cosmological consequences of a rolling
  homogeneous scalar field},  {\sl Phys. Rev. D} {\bf 37} (Jun, 1988)
  3406--3427, [\href{http://dx.doi.org/10.1103/PhysRevD.37.3406}{{\sf
  doi:10.1103/PhysRevD.37.3406}}].

\bibitem{Brax:2011bh}
P.~Brax, A.-C. Davis, and H.~A. Winther, {\it {Cosmological supersymmetric
  model of dark energy}},  {\sl Phys.Rev.} {\bf D85} (2012) 083512,
  [\href{http://arxiv.org/abs/1112.3676}{{\sf arXiv:1112.3676}}],
  [\href{http://dx.doi.org/10.1103/PhysRevD.85.083512}{{\sf
  doi:10.1103/PhysRevD.85.083512}}].

\bibitem{Werner:2004zk}
M.~Werner, T.~L. Roellig, F.~Low, G.~Rieke, M.~Rieke, {\em et~al.}, {\it {The
  Spitzer Space Telescope mission}},  {\sl Astrophys.J.Suppl.} {\bf 154} (2004)
  1--9, [\href{http://arxiv.org/abs/astro-ph/0406223}{{\sf
  arXiv:astro-ph/0406223}}], [\href{http://dx.doi.org/10.1086/422992}{{\sf
  doi:10.1086/422992}}].

\bibitem{EAS:9131523}
F.~Thévenin, {\it Measuring stars with gaia},  {\sl EAS Publications Series}
  {\bf null} (1, 2013) 85--88,
  [\href{http://dx.doi.org/10.1051/eas/1363010}{{\sf
  doi:10.1051/eas/1363010}}].

\bibitem{Ransom:2014xla}
S.~Ransom, I.~Stairs, A.~Archibald, J.~Hessels, D.~Kaplan, {\em et~al.}, {\it
  {A millisecond pulsar in a stellar triple system}},
  \href{http://arxiv.org/abs/1401.0535}{{\sf arXiv:1401.0535}},
  \href{http://dx.doi.org/10.1038/nature12917}{{\sf doi:10.1038/nature12917}}.

\bibitem{2010Natur.468..542P}
G.~{Pietrzy{\'n}ski}, I.~B. {Thompson}, W.~{Gieren}, D.~{Graczyk}, G.~{Bono},
  A.~{Udalski}, I.~{Soszy{\'n}ski}, D.~{Minniti}, and B.~{Pilecki}, {\it {The
  dynamical mass of a classical Cepheid variable star in an eclipsing binary
  system}},  {\sl nature} {\bf 468} (Nov., 2010) 542--544,
  [\href{http://dx.doi.org/10.1038/nature09598}{{\sf
  doi:10.1038/nature09598}}].

\bibitem{Marconi:2013tta}
M.~Marconi, R.~Molinaro, G.~Bono, G.~Pietrzynski, W.~Gieren, {\em et~al.}, {\it
  {The Eclipsing Binary Cepheid OGLE-LMC-CEP-0227 in the Large Magellanic
  Cloud: pulsation modelling of light and radial velocity curves}},  {\sl
  Astrophys.J.} {\bf 768} (2013) L6,
  [\href{http://arxiv.org/abs/1304.0860}{{\sf arXiv:1304.0860}}],
  [\href{http://dx.doi.org/10.1088/2041-8205/768/1/L6}{{\sf
  doi:10.1088/2041-8205/768/1/L6}}].

\bibitem{Ciardullo:2012vc}
R.~Ciardullo, {\it {The Planetary Nebula Luminosity Function at the Dawn of
  Gaia}},  {\sl Astrophys.Space Sci.} {\bf 341} (2012) 151--161,
  [\href{http://arxiv.org/abs/1203.5551}{{\sf arXiv:1203.5551}}],
  [\href{http://dx.doi.org/10.1007/s10509-012-1061-2}{{\sf
  doi:10.1007/s10509-012-1061-2}}].

\bibitem{wald1984general}
R.~Wald, {\em General Relativity}.
\newblock Physics, astrophysics. University of Chicago Press, 1984.

\bibitem{Pietrzynski:2002pn}
G.~Pietrzynski, W.~Gieren, P.~Fouque, and F.~Pont, {\it {The araucaria project.
  discovery of cepheid variables in ngc 300 from a wide-field imaging survey}},
   \href{http://arxiv.org/abs/astro-ph/0202494}{{\sf arXiv:astro-ph/0202494}}.

\bibitem{Saha:2006ng}
A.~Saha, F.~Thim, G.~Tammann, B.~Reindl, and A.~Sandage, {\it {Cepheid
  distances to sne ia host galaxies based on a revised photometric zero-point
  of the hst-wfpc2 and new p-l relations and metallicity corrections}},  {\sl
  Astrophys.J.Suppl.} {\bf 165} (2006) 108--137,
  [\href{http://arxiv.org/abs/astro-ph/0602572}{{\sf arXiv:astro-ph/0602572}}],
  [\href{http://dx.doi.org/10.1086/503800}{{\sf doi:10.1086/503800}}].

\bibitem{1996ApJ...470....1S}
N.~A. {Silbermann}, P.~{Harding}, B.~F. {Madore}, R.~C. {Kennicutt}, Jr.,
  A.~{Saha}, P.~B. {Stetson}, W.~L. {Freedman}, J.~R. {Mould}, J.~A. {Graham},
  R.~J. {Hill}, A.~{Turner}, F.~{Bresolin}, L.~{Ferrarese}, H.~{Ford}, J.~G.
  {Hoessel}, M.~{Han}, J.~{Huchra}, S.~M.~G. {Hughes}, G.~D. {Illingworth},
  R.~{Phelps}, and S.~{Sakai}, {\it {The Hubble Space Telescope Key Project on
  the Extragalactic Distance Scale. VI. The Cepheids in NGC 925}},  {\sl
  Astrophys. J.} {\bf 470} (Oct., 1996) 1,
  [\href{http://dx.doi.org/10.1086/177845}{{\sf doi:10.1086/177845}}].

\bibitem{Ferrarese:1998tt}
L.~Ferrarese, F.~Bresolin, J.~Kennicutt, Robert~C., A.~Saha, P.~B. Stetson,
  {\em et~al.}, {\it {The HST Key Project on the extragalactic distance scale.
  12. The Discovery of Cepheids and a new distance to NGC 2541}},  {\sl
  Astrophys.J.} (1998) [\href{http://arxiv.org/abs/astro-ph/9805365}{{\sf
  arXiv:astro-ph/9805365}}].

\bibitem{Sakai:1999yn}
S.~Sakai, L.~Ferrarese, R.~C. Kennicutt, J.~A. Graham, N.~Silbermann, {\em
  et~al.}, {\it {The Hubble Space Telescope Extragalactic Distance Scale Key
  Project. 23. The Discovery of Cepheids in NGC 3319}},  {\sl Astrophys.J.}
  (1999) [\href{http://arxiv.org/abs/astro-ph/9906487}{{\sf
  arXiv:astro-ph/9906487}}].

\bibitem{Prosser:1999ym}
C.~F. Prosser, J.~Kennicutt, Robert~C., F.~Bresolin, A.~Saha, S.~Sakai, {\em
  et~al.}, {\it {The hst key project on the extragalactic distance scale. xxii.
  the discovery of cepheids in ngc 1326-a}},
  \href{http://arxiv.org/abs/astro-ph/9906486}{{\sf arXiv:astro-ph/9906486}}.

\bibitem{1998ApJ...500..763P}
R.~L. {Phelps}, S.~{Sakai}, W.~L. {Freedman}, B.~F. {Madore}, A.~{Saha}, P.~B.
  {Stetson}, R.~C. {Kennicutt}, J.~R. {Mould}, L.~{Ferrarese}, H.~C. {Ford},
  B.~K. {Gibson}, J.~A. {Graham}, M.~{Han}, J.~G. {Hoessel}, J.~P. {Huchra},
  S.~M. {Hughes}, G.~D. {Illingworth}, and N.~A. {Silbermann}, {\it {The Hubble
  Space Telescope Extragalactic Distance Scale Key Project. IX. The Discovery
  of Cepheids in NGC 2090}},  {\sl Astrophys. J.} {\bf 500} (June, 1998) 763,
  [\href{http://dx.doi.org/10.1086/305766}{{\sf doi:10.1086/305766}}].

\bibitem{1994ApJ...427..628F}
W.~L. {Freedman}, S.~M. {Hughes}, B.~F. {Madore}, J.~R. {Mould}, M.~G. {Lee},
  P.~{Stetson}, R.~C. {Kennicutt}, A.~{Turner}, L.~{Ferrarese}, H.~{Ford},
  J.~A. {Graham}, R.~{Hill}, J.~G. {Hoessel}, J.~{Huchra}, and G.~D.
  {Illingworth}, {\it {The Hubble Space Telescope Extragalactic Distance Scale
  Key Project. 1: The discovery of Cepheids and a new distance to M81}},  {\sl
  Astrophys. J.} {\bf 427} (June, 1994) 628--655,
  [\href{http://dx.doi.org/10.1086/174172}{{\sf doi:10.1086/174172}}].

\bibitem{1999ApJ...514..614K}
D.~D. {Kelson}, G.~D. {Illingworth}, A.~{Saha}, J.~A. {Graham}, P.~B.
  {Stetson}, W.~L. {Freedman}, R.~C. {Kennicutt}, J.~R. {Mould},
  L.~{Ferrarese}, J.~P. {Huchra}, B.~F. {Madore}, C.~F. {Prosser},
  F.~{Bresolin}, H.~C. {Ford}, B.~K. {Gibson}, J.~G. {Hoessel}, S.~M.~G.
  {Hughes}, L.~M. {Macri}, S.~{Sakai}, and N.~A. {Silbermann}, {\it {The Hubble
  Space Telescope Key Project on the Extragalactic Distance Scale. XIX. The
  Discovery of Cepheids in and a New Distance to NGC 3198}},  {\sl Astrophys.
  J.} {\bf 514} (Apr., 1999) 614--636,
  [\href{http://dx.doi.org/10.1086/306989}{{\sf doi:10.1086/306989}}].

\bibitem{1997ApJ...477..535G}
J.~A. {Graham}, R.~L. {Phelps}, W.~L. {Freedman}, A.~{Saha}, L.~{Ferrarese},
  P.~B. {Stetson}, B.~F. {Madore}, N.~A. {Silbermann}, S.~{Sakai}, R.~C.
  {Kennicutt}, P.~{Harding}, F.~{Bresolin}, A.~{Turner}, J.~R. {Mould}, D.~M.
  {Rawson}, H.~C. {Ford}, J.~G. {Hoessel}, M.~{Han}, J.~P. {Huchra}, L.~M.
  {Macri}, S.~M. {Hughes}, G.~D. {Illingworth}, and D.~D. {Kelson}, {\it {The
  Hubble Space Telescope Extragalactic Distance Scale Key Project. VII. The
  Discovery of Cepheids in the Leo I Group Galaxy NGC 3351}},  {\sl Asteophys.
  J.} {\bf 477} (Mar., 1997) 535, [\href{http://dx.doi.org/10.1086/303740}{{\sf
  doi:10.1086/303740}}].

\bibitem{Rawson:1997hr}
D.~M. Rawson, L.~M. Macri, J.~R. Mould, J.~P. Huchra, W.~L. Freedman, {\em
  et~al.}, {\it {The Extragalactic Distance Scale Key Project. 8. The Discovery
  of Cepheids and a new distance to NGC 3621 using the Hubble Space
  Telescope}},  {\sl Astrophys.J.} (1997)
  [\href{http://arxiv.org/abs/astro-ph/9705259}{{\sf arXiv:astro-ph/9705259}}].

\bibitem{1998ApJ...505..207T}
A.~{Turner}, L.~{Ferrarese}, A.~{Saha}, F.~{Bresolin}, R.~C. {Kennicutt}, Jr.,
  P.~B. {Stetson}, J.~R. {Mould}, W.~L. {Freedman}, B.~K. {Gibson}, J.~A.
  {Graham}, H.~{Ford}, M.~{Han}, P.~{Harding}, J.~G. {Hoessel}, J.~P. {Huchra},
  S.~M.~G. {Hughes}, G.~D. {Illingworth}, D.~D. {Kelson}, L.~{Macri}, B.~F.
  {Madore}, R.~{Phelps}, D.~{Rawson}, S.~{Sakai}, and N.~A. {Silbermann}, {\it
  {The Hubble Space Telescope Key Project on the Extragalactic Distance Scale.
  XI. The Cepheids in NGC 4414}},  {\sl Astrophys. J.} {\bf 505} (Sept., 1998)
  207--229, [\href{http://dx.doi.org/10.1086/306150}{{\sf
  doi:10.1086/306150}}].

\bibitem{Macri:1999kd}
L.~M. Macri, J.~Huchra, P.~Stetson, N.~Silbermann, W.~Freedman, {\em et~al.},
  {\it {The Extragalactic Distance Scale Key Project. 18. The Discovery of
  Cepheids and a new distance to NGC 4535 using the Hubble Space Telescope}},
  \href{http://arxiv.org/abs/astro-ph/9901332}{{\sf arXiv:astro-ph/9901332}}.

\bibitem{1999ApJ...516..626G}
J.~A. {Graham}, L.~{Ferrarese}, W.~L. {Freedman}, R.~C. {Kennicutt}, Jr., J.~R.
  {Mould}, A.~{Saha}, P.~B. {Stetson}, B.~F. {Madore}, F.~{Bresolin}, H.~C.
  {Ford}, B.~K. {Gibson}, M.~{Han}, J.~G. {Hoessel}, J.~{Huchra}, S.~M.
  {Hughes}, G.~D. {Illingworth}, D.~D. {Kelson}, L.~{Macri}, R.~{Phelps},
  S.~{Sakai}, N.~A. {Silbermann}, and A.~{Turner}, {\it {The Hubble Space
  Telescope Key Project on the Extragalactic Distance Scale. XX. The Discovery
  of Cepheids in the Virgo Cluster Galaxy NGC 4548}},  {\sl Astrophys. J.} {\bf
  516} (May, 1999) 626--646, [\href{http://dx.doi.org/10.1086/307151}{{\sf
  doi:10.1086/307151}}].

\bibitem{Gibson:1998qc}
B.~K. Gibson, S.~M. Hughes, P.~B. Stetson, W.~L. Freedman, J.~Kennicutt,
  Robert~C., {\em et~al.}, {\it {The HST Key Project on the extragalactic
  distance scale. 17. The Cepheid distance to NGC 4725}},  {\sl Astrophys.J.}
  (1998) [\href{http://arxiv.org/abs/astro-ph/9810003}{{\sf
  arXiv:astro-ph/9810003}}].

\bibitem{1996ApJ...463...26K}
D.~D. {Kelson}, G.~D. {Illingworth}, W.~F. {Freedman}, J.~A. {Graham},
  R.~{Hill}, B.~F. {Madore}, A.~{Saha}, P.~B. {Stetson}, R.~C. {Kennicutt},
  Jr., J.~R. {Mould}, S.~M. {Hughes}, L.~{Ferrarese}, R.~{Phelps}, A.~{Turner},
  K.~H. {Cook}, H.~{Ford}, J.~G. {Hoessel}, and J.~{Huchra}, {\it {The
  Extragalactic Distance Scale Key Project. III. The Discovery of Cepheids and
  a New Distance to M101 Using the Hubble Space Telescope}},  {\sl Astrophys.
  J.} {\bf 463} (May, 1996) 26, [\href{http://dx.doi.org/10.1086/177221}{{\sf
  doi:10.1086/177221}}].

\bibitem{1998ApJ...501...32H}
S.~M.~G. {Hughes}, M.~{Han}, J.~{Hoessel}, W.~L. {Freedman}, R.~C. {Kennicutt},
  Jr., J.~R. {Mould}, A.~{Saha}, P.~B. {Stetson}, B.~F. {Madore}, N.~A.
  {Silbermann}, P.~{Harding}, L.~{Ferrarese}, H.~{Ford}, B.~K. {Gibson}, J.~A.
  {Graham}, R.~{Hill}, J.~{Huchra}, G.~D. {Illingworth}, R.~{Phelps}, and
  S.~{Sakai}, {\it {The Hubble Space Telescope Extragalactic Distance Scale Key
  Project. X. The Cepheid Distance to NGC 7331}},  {\sl Astrophys. J.} {\bf
  501} (July, 1998) 32, [\href{http://dx.doi.org/10.1086/305808}{{\sf
  doi:10.1086/305808}}].

\bibitem{1981PThPS..70..115S}
D.~{Sugimoto}, K.~{Nomoto}, and Y.~{Eriguchi}, {\it {Stable Numerical Method in
  Computation of Stellar Evolution}},  {\sl Progress of Theoretical Physics
  Supplement} {\bf 70} (1981) 115--131,
  [\href{http://dx.doi.org/10.1143/PTPS.70.115}{{\sf
  doi:10.1143/PTPS.70.115}}].

\end{thebibliography}\endgroup
\appendix
\chapter{Weyl Rescalings of the Metric}
\label{app:weyl}

In this appendix we will study the transformation properties of several important geometric tensors that are used to derive the
results presented in chapter \ref{chap:two}. We will then go on to show how the conservation law for the Einstein frame
energy-momentum tensor can be obtained from the Jordan frame equation. Here we will present only those properties necessary to
derive the results of chapter \ref{chap:two}. A more general account can be found in \cite{wald1984general}, appendix
D\footnote{Note that the method used in our derivation will differ from theirs but identical results are obtained.}.

\section{Transformation of the Ricci Scalar}

We are interested in the transformation of various curvature tensors constructed using the Jordan frame metric $\tg_{\mu\nu}$ in
order to find the equivalent tensors constructed using the Einstein frame metric $g_{\mu\nu}$. The two metric are related via
\begin{equation}\label{eq:gtilde}
 \tg_{\mu\nu}=A^2(\phi)g_{\mu\nu}.
\end{equation}
We are interested in quantities such as the Riemann tensor and the Ricci tensor and scalar, which are constructed using the
metric and the Christoffel symbols and so we must first find how these are related. Defining
\begin{equation}\label{eq:kdef}
 \mathcal{K}_{\mu\nu}^\alpha=\tilde{\Gamma}_{\mu\nu}^\alpha-\Gamma_{\mu\nu}^\alpha,
\end{equation}
which, by direct computation using equation (\ref{eq:gtilde}) is
\begin{equation}\label{eq:kconf}
\mathcal{K}_{\mu\nu}^\alpha=2\delta^\alpha\vphantom{\nabla}_{\left(\mu\vphantom{\nu}\right.}\nabla\vphantom{\delta^\alpha}_{
\left.\vphantom { \mu } \nu\right) } \ln
A-g_{\mu\nu}\nabla^\alpha\ln A.
\end{equation}
Next, using the definition of the Riemann tensor, 
\begin{equation}\label{eq:riemanndef}
 2\tilde{\nabla}_{\left[\mu\right.}\tilde{\nabla}\vphantom{\delta^\alpha}_{\left.\nu\vphantom{\mu}\right]}v^\alpha=\tilde{R}
^\alpha_ {\,\, \beta\mu\nu } v^\beta
\end{equation}
we have 
\begin{equation}\label{eq:riemanntransform}
 \tilde{R}^\alpha_{\beta\mu\nu}=R^\alpha_{\beta\mu\nu}+2\nabla_{\left[\mu\right.}\mathcal{K}^\alpha_{\left.\vphantom{\mu}\nu\right
]\beta}+2\mathcal{K}_{\gamma\left[\mu\right.}^\alpha\mathcal{K}^\gamma_{\left.\vphantom{\mu}\nu\right]\beta}.
\end{equation}
Contracting the $\alpha$ and $\mu$ index\footnote{Note that it does not matter which metric we use for this contraction since we
are contracting with a $\delta$-symbol.} and inserting equation (\ref{eq:kconf}) we can calculate the Ricci tensor:
\begin{equation}\label{eq:rsjf}
\tilde{R}_{\mu\nu}=R_{\mu\nu}-4\nabla_\mu\nabla_\nu\ln A -g_{\mu\nu}\Box\ln A+2\nabla_\mu\ln A\nabla_\nu\ln
A-4g_{\mu\nu}\nabla_\rho\ln A\nabla^\rho\ln A.
\end{equation}
Contracting with $\tg^{\mu\nu}=A^{-2}(\phi)g^{\mu\nu}$ we obtain the Ricci scalar:
\begin{equation}\label{eq:ricciscalarJR}
 \tilde{R}=\frac{1}{A^2}\left[R-6\Box\ln A-6\nabla_\mu\ln A\nabla^\mu\ln A\right].
\end{equation}
This is precisely the transformation we need to find the Einstein
frame expression for a theory whose action is defined in the Jordan frame. Indeed, this is how we obtained equation
(\ref{eq:nmexampleef}) from equation (\ref{eq:nmactionexample}). In section \ref{sec:suppresscharge} we defined our model in
the Einstein frame and so the inverse transformation is needed to calculate the Jordan frame expression in equation
(\ref{eq:chamactJF}). Clearly this can be achieved using equation (\ref{eq:ricciscalarJR}) and treating the tilded Ricci scalar as
the Einstein frame quantity. In this case we have $g_{\mu\nu}=A^{-2}\tg_{\mu\nu}$ and so if we set $A\rightarrow A^{-1}$ in
equation (\ref{eq:ricciscalarJR}) and swap the tilded and untilded Ricci scalars we can find the equivalent expression for the
Einstein frame Ricci scalar. Note that the term proportional to $\Box\ln A$ is a total derivative and so does not contribute to
the equations of motion. For this reason, we have ignored this term when transforming between different frames in chapter
\ref{chap:two}.

\section{The Energy-Momentum Tensor}

Finally, we want to derive equation (\ref{eq:geode2}), which gives the conservation law for the energy-momentum tensor in the
Einstein frame. Now particles move on geodesics of the Jordan frame metric and so it is this frame where the energy momentum is
covariantly conserved:
\begin{equation}\label{eq:emconsjf}
 \tilde{\nabla}_\mu\tilde{T}^{\mu\nu}=0.
\end{equation}
In terms of the Einstein frame energy-momentum tensor, we have
\begin{equation}
 \tilde{T}^{\mu\nu}=\frac{2}{\sqrt{-\tg}}\frac{\delta S\mmm}{\delta \tilde{g}_{\mu\nu}}=\frac{2}{A^4\sqrt{-g}}\frac{\delta
S\mmm}{\delta g_{\alpha\beta}}\frac{\delta g_{\alpha\beta}}{\delta\tg_{\alpha\beta}}=A^{-6}T^{\mu\nu}
\end{equation}
and so the problem is then to calculate the transformation law from the Jordan frame relation
$\tilde{\nabla}_\mu(A^{-6}T^{\mu\nu})=0$.
Using equation (\ref{eq:kdef}) we have
\begin{equation}
 \tilde{\nabla}_\mu\left(A^{-6}T^{\mu\nu}\right)=\nabla_{\mu}\left(A^{-6}T^{\mu\nu}\right)+A^{-6}\left(\mathcal{K}_{\mu\rho}
^\mu+\mathcal { K } ^\nu_ { \
mu\rho} \right)T^ { \rho\mu }.
\end{equation}
Replacing the factors of $\mathcal{K}$ using equation (\ref{eq:kconf}) we then have
\begin{equation}
 \nabla_\mu T^{\mu\nu}=\frac{A^\prime(\phi)}{A(\phi)}T\nabla^\nu\phi,
\end{equation}
which can be used with the definition of $\beta(\phi)$ (equation (\ref{eq:betacham})) to recover equation (\ref{eq:geode2}).
\chapter{Cepheid and TRGB Data}
\label{app:data}

In this appendix we present a list of the galaxies that were used in the distance indicator comparison performed in chapter
\ref{chap:four} along with their journal references. In table \ref{tab:P-L} we list the galaxies used in the period-luminosity
relation and in table \ref{tab:distance} we list the TRGB and Cepheid distances for each galaxy used to obtain the constraints in
figure \ref{fig:confidence-regions}.

\begin{table*}[h]
\centering
\heavyrulewidth=.08em
	\lightrulewidth=.05em
	\cmidrulewidth=.03em
	\belowrulesep=.65ex
	\belowbottomsep=0pt
	\aboverulesep=.4ex
	\abovetopsep=0pt
	\cmidrulesep=\doublerulesep
	\cmidrulekern=.5em
	\defaultaddspace=.5em
	\renewcommand{\arraystretch}{1.6}
\caption{The galaxies used in the period-luminosity relation and their references. The
second column labelled $N$ gives the number of Cepheids observed in each galaxy.
Names that end with * are galaxies with unacceptably large dispersion in period-luminosity relation.}
\begin{tabu}[c]{l|[2pt]c|[2pt]l}\label{tab:P-L}
Name & $N$ & Reference\\
\tabucline[2pt]{-}
NGC300 & 117 & \cite{Pietrzynski:2002pn}\\
NGC5253 & 5 & \cite{Saha:2006ng}\\
IC4182 & 13 & \cite{Saha:2006ng}\\
NGC925 & 79 & \cite{1996ApJ...470....1S}\\
NGC2541 & 28 & \cite{Ferrarese:1998tt}\\
NGC3319 & 28 & \cite{Sakai:1999yn} \\
NGC1326A & 17 & \cite{Prosser:1999ym}\\
NGC 2090 & 34 & \cite{1998ApJ...500..763P}\\
NGC 3031 & 25 & \cite{1994ApJ...427..628F}\\
NGC 3198 & 52 & \cite{1999ApJ...514..614K}\\
NGC 3351* & 49 & \cite{1997ApJ...477..535G}\\
NGC 3621* & 69 & \cite{Rawson:1997hr}\\
NGC 4321* & 52 & \cite{Ferrarese:1998tt}\\
NGC 4414* & 9 & \cite{1998ApJ...505..207T}\\
NGC 4535 & 50 & \cite{Macri:1999kd}\\
NGC 4548* & 24 & \cite{1999ApJ...516..626G}\\
NGC 4725 & 20 & \cite{Gibson:1998qc}\\
NGC 5457 & 29 & \cite{1996ApJ...463...26K}\\
NGC 7331 & 13 & \cite{1998ApJ...501...32H}
\end{tabu}
\end{table*}

\begin{table}[h]
\caption{Cepheid and TRGB based distances to the galaxies used in the
comparison. The final column gives the screening for $\phi = 4 \times 10^{-7}$ as follows: 
 0: Unscreened, 1: environmentally screened, 2: self-screened.}
\centering
\heavyrulewidth=.08em
	\lightrulewidth=.05em
	\cmidrulewidth=.03em
	\belowrulesep=.65ex
	\belowbottomsep=0pt
	\aboverulesep=.4ex
	\abovetopsep=0pt
	\cmidrulesep=\doublerulesep
	\cmidrulekern=.5em
	\defaultaddspace=.5em
	\renewcommand{\arraystretch}{1.6}
\begin{tabu}[c]{l|[2pt]c|[2pt]c|[2pt]c}\label{tab:distance}
Name & Cepheid distance (Mpc) & TRGB distance (Mpc) & Screening\\
\tabucline[2pt]{-}
DDO 069 & $0.71$ $\pm$ $0.01$ & $0.78$ $\pm$ $0.03$ & 0 \\
NGC 3109 &$ 1.15$ $\pm$ $0.03$ & $1.27$ $\pm$ $0.02$ & 0 \\
DDO 216 & $1.27$ $\pm$ $0.27$ & $0.97$ $\pm$ $0.03$ & 0\\
Sextans A &$ 1.31$ $\pm$ $0.03$ & $1.38$ $\pm$ $0.03$ & 0 \\
Sextans B & $1.49$ $\pm$ $0.11$ & $1.34$ $\pm$ $0.02$ & 0 \\
GR8 & $1.80$ $\pm$ $0.06$ & $2.19$ $\pm$ $0.15$ & 0 \\
NGC 0300 & $2.03$ $\pm$ $0.05$ & $1.95$ $\pm$ $0.06$ & 0 \\
NGC 2403 & $3.20$ $\pm$ $0.15$ & $3.18$ $\pm$ $0.35$ & 0\\
NGC 2366 & $3.28$ $\pm$ $0.30$ & $3.19$ $\pm$ $0.41$ & 0 \\
NGC 5253 & $3.43$ $\pm$ $0.08$ & $3.77$ $\pm$ $0.19$ & 0\\
NGC 4395 & $4.45$ $\pm$ $0.37$ & $4.61$ $\pm$ $0.62$ & 0 \\
IC 4182 & $4.68$ $\pm$ $0.04$ & $4.47$ $\pm$ $0.12$ & 0 \\
NGC 3621 & $7.17$ $\pm$ $0.06$& $7.45$ $\pm$ $0.38$ & 0\\
SMC & $0.06$ $\pm$ $0.00$ & $0.06$ $\pm$ $0.00$ & 1\\
NGC 6822 &$ 0.51$ $\pm$ $0.03$ & $0.48$ $\pm$ $0.01$ & 1\\
IC 1613 & $0.69$ $\pm$ $0.01$ & $0.72$ $\pm$ $0.01$ & 1 \\
IC 0010 & $0.72$ $\pm$ $0.05$ & $0.50$ $\pm$ $0.04$ & 1\\
M33 & $0.90$ $\pm$ $0.02$ & $0.88$ $\pm$ $0.02$ & 1 \\
WLM & $0.95$ $\pm$ $0.05$ & $0.91$ $\pm$ $0.02$ & 1 \\
M31 & $0.86$ $\pm$ $0.02$ & $0.78$ $\pm$ $0.02$ & 2 \\
NGC 5128 & $3.44$ $\pm$ $0.19$ & $3.73$ $\pm$ $0.24$ & 2 \\
M81 & $3.84$ $\pm$ $0.06$ & $4.04$ $\pm$ $0.22$ & 2 \\
M83 & $5.01$ $\pm$ $0.23$ & $4.51$ $\pm$ $0.31$ & 2 \\
M101 & $7.13$ $\pm$ $0.14$ & $7.66$ $\pm$ $0.39$ & 2\\
M106 & $8.41$ $\pm$ $0.07$ & $7.31$ $\pm$ $0.30$ & 2 \\
\end{tabu}
\end{table}
\chapter{Numerical Methods}
\label{app:nto}

In this appendix we describe the various numerical methods used throughout this thesis. We first present a brief introduction to
the architecture of MESA and explain how the effects of modified gravity are implemented numerically. Next, we describe the
numerical methods used to obtain the solutions of the MLAWE presented in chapter \ref{chap:five}.

\section{MESA}

In this section we introduce the basic numerical features of MESA. A complete description is well beyond the scope of this
appendix and the reader is referred to \cite{Paxton:2010ji} for the full details; in this appendix we will only present the
features used to obtain the results presented in this thesis. In particular, we have not made use of processes such as convection,
mass-loss, overshooting or semi-convection and so we will not discuss these here.

\subsection{The Architecture of MESA}

MESA is a one-dimensional code (meaning that it assumes spherical symmetry) that solves the stellar structure equations in a
fully consistent manner. A stellar model is defined as a star at any specific time-step. MESA divides the star into $N$ cells of
variable width labelled by an integer $k$ that varies from $1$ at the stellar surface to $N$ at the centre ($r=0$). The number of
cells is determined by MESA according to the complexity of the phase of stellar evolution and can vary from $\mathcal{O}(10^2)$
on the main-sequence to $\mathcal{O}(10^5)$ for more complicated post-main-sequence stars. Note that MESA defines the
stellar surface as the radius at which the pressure falls to zero. When observing real stars, one actually observes light emitted
from the radius at which the optical depth falls to $2/3$ --- the so-called \textit{photosphere} --- and so the surface
temperature is not equal to the effective temperature (or $B-V$ if one is dealing with magnitudes). MESA keeps track of the
location of the photosphere and so it is possible to output data on any variables if one so desires. In particular, when one
refers to the stellar luminosity, one is in fact referring to the luminosity at the location of the photosphere and it is this
that MESA outputs.

\begin{figure}
\centering
\includegraphics[width=15cm]{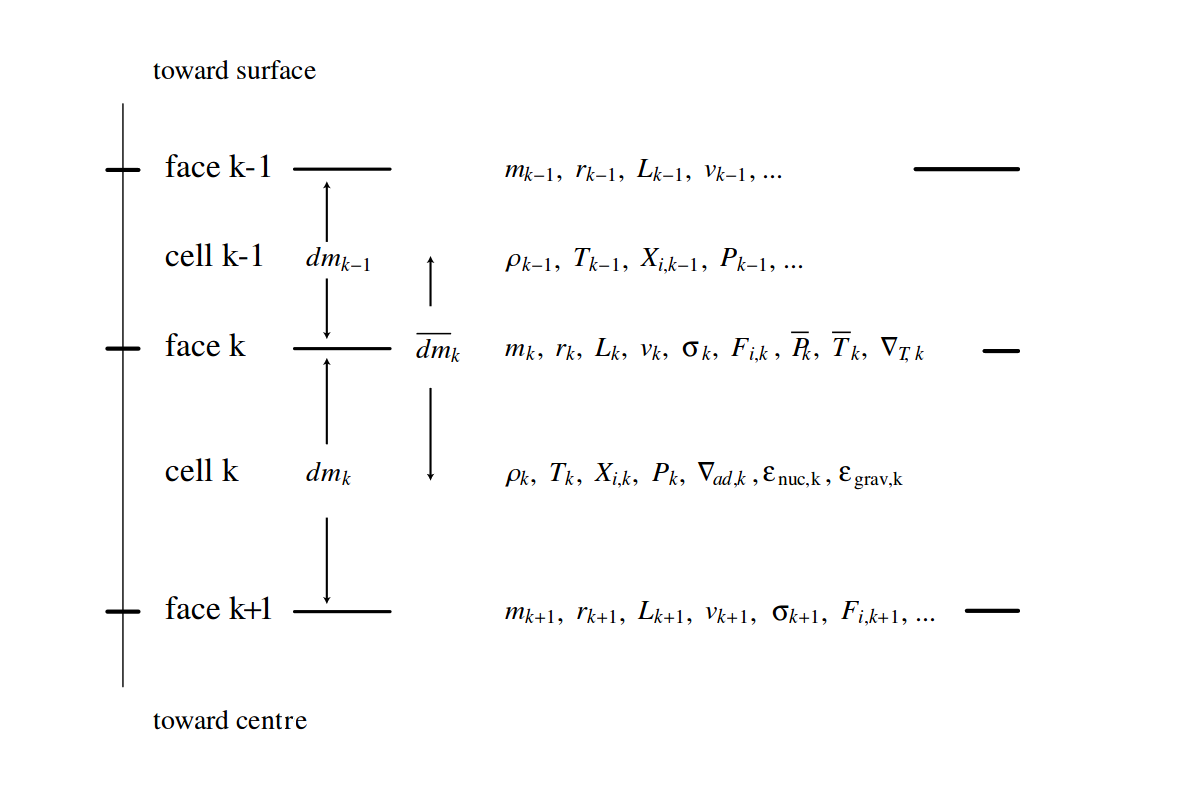}
\caption{The cell structure of stars in MESA and the cell assignments of various stellar quantities. Figure
reproduced from \cite{Paxton:2010ji}. The variables shown that are relevant for this thesis are the temperature
$T$, the density $\rho$ the mass $m$ enclosed by radius $r$ and the luminosity $L$. $\dd m_k$ refers to the mass
enclosed within cell $k$ and $\bar{\dd m}_k$ is the average mass enclosed in cells $k$ and $k-1$. The reader is referred to 
\cite{Paxton:2010ji} for the details of the other quantities shown.}\label{fig:marc}
\end{figure}

MESA works with a variety of different variables such as pressure, temperature, density etc., which may be defined either at the
cell centres or the cell boundaries. As a rule of thumb, extensive variables such as the mass enclosed within the cell or the
luminosity in the cell are defined at the cell boundary whereas intensive variables such as the density and temperature are
defined at the cell centre. A schematic diagram of this is shown in figure \ref{fig:marc} which includes some examples of the cell
assignments of some stellar quantities. Different quantities are defined at different locations for reasons of numerical stability
and efficiency \cite{1981PThPS..70..115S}, however one can always interpolate from cell centres to cell boundaries and vice versa
if necessary. MESA uses $q\equiv M(r)/M(R)$ as the radial variable and not $r$ and so we must interpolate in $m$. As an example,
we will interpolate the pressure --- which is cell-centred and whose value in cell $k$ is $P_k$ --- from the centre of cell $k$ to
its boundary. We start by defining $\dd q_k$ as the ratio of the mass enclosed inside the cell to the stellar mass so that in the
notation of figure \ref{fig:marc} (see the caption) we have $\dd q_k= \dd m_k/M(R)$. Next, we define the interpolation
variable $\delta_k$ for cell $k$ via
\begin{equation}
 \delta_k=\frac{\dd q_{k-1}}{\dd q_k + \dd q_{k-1}}.
\end{equation}
The pressure at the cell boundary $\bar{P}_k$ is then
\begin{equation}
 \bar{P}_k=\delta_kP_k+(1-\delta_k)P_{k-1}.
\end{equation}
If one wishes to interpolate from cell boundaries to cell centres one simply replaces $\dd m_k$ by ${\overline{\dd m}}_k$ in the
definition of $\dd q_k$ (see the caption in figure \ref{fig:marc}).

Given some initial stellar model, MESA uses the values of the various quantities defined at the cell boundaries or centres to
solve the stellar structure equations using the Newton-Raphson method. The initial stellar model is used as the trial solution and
the system is iterated until convergence is achieved. Where necessary, MESA uses opacity tables and reaction networks (including
reaction rates) to close the system. At no point is any equation of state assumed. MESA uses a variable time-step that is
chosen in order to be short enough to give a fast convergence but long enough that the stellar evolution is efficient. The
details of the algorithm are very complicated and are presented in \cite{Paxton:2010ji} section 6.4.

\subsection{Implementation of Modified Gravity}

Modified gravity is implemented into MESA by including the radial variation of $G$ according to equation (\ref{eq:g(r)}). This is
done in two stages: first, the screening radius is found and second, the value of $G$ in each cell exterior to this is updated
according to (\ref{eq:g(r)}). Given input values of $\chi_0$ and $\alpha$, this is achieved numerically using the following
algorithm:
\begin{enumerate}
 \item Starting from an initial stellar model, the cell-centred density is interpolated to the cell boundary, where the radial
coordinate is defined.
\item Using the trapezium rule, the integral (\ref{eq:chiint}) is performed using the density at the cell boundary and the cell
width as the discretised radial coordinate\footnote{MESA uses CGS units and so the appropriate factors of $c$ are included.}. The
integral is performed inwards from the outer cell ($k=1$) to successively deeper cells until (\ref{eq:chiint}) is satisfied. This
cell is designated as the screening cell $k_{\rm s}$. 
\item The total mass of the cells with $k\ge k_{\rm s}$ ($M(\rs)$) is calculated, which is all that is needed to find $G(r)$
according to (\ref{eq:g(r)}).
\item The value of $G$ in the cells with $k\ge k_{\rm s}$ is left unaltered but the value in each cell with $k<k_{\rm s}$ is
updated according to (\ref{eq:g(r)}). 
\item The code is evolved to the next time-step i.e. the stellar structure equations with $G\rightarrow G(r)$ are
solved.
\item The above steps are repeated using the new stellar model.
\end{enumerate}

\cite{Chang:2010xh} have used a different method of implementing modified gravity into MESA using an ansatz for the field profile
and interpolating the screening radius to a cell-centred value. We have compared the output of our code with theirs and the
results are indistinguishable. One may still worry that this method is not self-consistent, i.e. we use a given stellar model to
find $G(r)$ and then evolve the system rather than calculating $G(r)$ at the same time using the Newton-Raphson solver. This is
what \cite{Chang:2010xh} do and since our results match theirs to a high accuracy we conclude that our method does not introduce
any errors due to our numerical procedure\footnote{The advantage of our method is that MESA is constantly developing and changing.
Our implementation can easily be incorporated into successive versions of MESA whereas including $G(r)$ as another variable found
using the Newton-Raphson solver requires a complete re-writing of the code with every new release.}. This is because of the
time-step selection mentioned above. If any large error is introduced by our procedure, the stellar structure equations are not
satisfied and MESA reduces the time-step accordingly until convergence is achieved\footnote{The disadvantage of this is that
the run-time for a given star is longer than if one had included $G(r)$ as a dynamical variable. In fact, the run-time for most
modified gravity simulations using our method is of order minutes and so this trade-off against compatibility with future releases
is perfectly reasonable.}. 

\section{Numerical Methods for Solving the MLAWE}

In this section we describe the numerical procedure used to solve the MLAWE and obtain the results in chapter \ref{chap:five}.

\subsection{First-Order form of the MLAWE}

Whereas the form presented in equation
(\ref{eq:MLAWE}) is useful for extracting the new physical features,
numerically, it is more convenient to work with a first order system. One may
re-write the homogeneous MLAWE in the form
\begin{equation}\label{eq:1storderstart}
 \frac{\dd }{\dd r}\left[\frac{\gao P_0}{r^2}\frac{\dd}{\dd
r}\left(r^3\xi\right)\right]-4\frac{\dd P_0}{\dd
r}-4\pi\alpha G\rho_0^2r\xi+\omega^2r\rho_0\xi=0.
\end{equation}
In order to cast this into first order form we introduce the new variable
$\eta$, defined by
\begin{equation}
 \eta(r) = \frac{1}{r^2}\frac{\dd}{\dd
r}\left(r^3\xi\right),
\end{equation}
which, using equation (\ref{eq:lagpresspert}), is nothing but
$-\delta\rho/\rho_0$. Using this definition, we can cast
(\ref{eq:1storderstart}) into a convenient first-order from for both Lane-Emden
and MESA models.

\subsection{Lane-Emden Form}\label{sec:lawele}

When solving for Lane-Emden profile, we replace $r$ with $y$ using (\ref{eq:rc})
and the pressure and density with $\theta$ to obtain
\begin{align}
 \frac{\dd \eta}{\dd y} &= \frac{1}{\gao\theta}\left[4(n+1)\frac{\dd
\theta}{\dd
y}\xi+\alpha(n+1)y\theta^n\xi\right.\label{eq:leeq1}\\&\left.-(n+1)\gao\frac{\dd
\theta}{\dd y}\eta-\tilde{\omega}^2y\xi\right]\nonumber\\ \frac{\dd \xi}{\dd y}
&= \frac{\eta-3\xi}{y}\label{eq:leeq2},
\end{align}
where $\gao$ is treated as constant in Lane-Emden models, $\tilde{\omega}$
is defined in (\ref{eq:omtild}) and the term proportional to $\alpha$ is
present only when $r>\rs$. The value of $\gao$ may be chosen at will.
These are supplemented by the boundary conditions
\begin{align}
 \eta(0)&=3\xi(0)\label{eq:lecentrebc}\\
\eta(y_{\rm R}) &= \frac{\xi(y_{\rm
R})}{\gao}\left.\left(4+\frac{\tilde{\omega}^2y_{\rm R}}{(n+1)\dd\theta/\dd
y}\right)\right\vert_{y=y_{\rm R}},\label{eq:lesurfacebc}
\end{align}
which are the Lane-Emden equivalents of (\ref{eq:BCC}) and (\ref{eq:BCS}). We
solve this equation using the shooting technique. Using the fourth-order
Runge-Kutta method, we solve equations (\ref{eq:leeq1}) and (\ref{eq:leeq2}) in
conjunction with the modified Lane-Emden equation (\ref{eq:MGLE}) using a trial
value of $\tilde{\omega}^2$ and the centre boundary condition
(\ref{eq:lecentrebc}). We then test the surface condition
(\ref{eq:lesurfacebc}) against our solution, iterating over different
values of $\tilde{\omega}^2$ until the difference between our solution and the
boundary condition is less than some predefined tolerance (exactly how much
depends on the accuracy required for the eigenvalue). Using this method, one can
obtain $\tilde{\omega}^2$ and the corresponding eigenfunction to the desired
accuracy.

\subsection{MESA form}

MESA produces pressure, density, temperature etc. profiles in
physical units, and so it is convenient to use the dimensionless quantities
defined in (\ref{eq:dimlessP}), (\ref{eq:dimlessrho}) and (\ref{eq:dimlessr}) so
that the MLAWE is
\begin{align}
 \frac{\dd \eta}{\dd
x}&=\left(\gao(x)\bar{P}_0(x)\right)^{-1}\left[-\eta\left(\bar{P}_0\frac{\dd
\gao}{\dd x}+\gao\frac{\dd \bar{P}_0}{\dd x} \right)\right.\\&+4\frac{\dd
\bar{P}_0}{\dd x}\left.\vphantom{-\eta\left(\bar{P}_0\frac{\dd
\gao}{\dd x}+\gao\frac{\dd \bar{P}_0}{\dd x} \right)}
+4\pi\alpha
x\bar{\rho}_0^2\xi-\Omega^2x\bar{\rho}_0\xi\right],\\\frac{\dd \xi}{\dd
x}&=\frac{\eta-3\xi}{x},
\end{align}
where, again, the term proportional to $\alpha$ is only present when $x>x_{\rm
s}\equiv \rs/R$ and $\Omega^2$ is the dimensionless eigenfrequency defined in
(\ref{eq:dimOmega}).

The eigenvalue problem is solved in a similar manner to the Lane-Emden models. Using profiles for $\bar{P}_0$ and $\bar{\rho}_0$
at both the cell boundaries
and centres as well as the screening radius from MESA, these equations
are integrated from the stellar centre using the fourth order Runge-Kutta scheme
for a test value of $\Omega^2$. $\Omega^2$ is then iterated until the boundary
condition at the stellar surface,
\begin{equation}
 \eta(1)=\frac{1}{\gao(1)}\left(\Omega^2+4\right)\xi(1),
\end{equation}
is satisfied up to some pre-set tolerance. In this manner, $\Omega^2$ and the
corresponding eigenfunction can be found for any given model and the period
found by inverting (\ref{eq:dimOmega}).

\chapter{Modified Bessel Equations}
\label{app:bessel}

This appendix details some important properties of modified Bessel functions that appear in the solution of the equations
governing linear perturbations in the CDM power spectrum studied in chapter \ref{chap:seven}.

\section{Generalised Modified Bessel Equations}
We have seen in section \ref{sec:perts} that the linearised perturbation equations are all of the form
\begin{equation}
 t^2\ddot{\dc}+at\dot{\dc}-\left(b^2+c^2t^{2r}\right)\dc=0,
\end{equation}
with $a=4/3$. The substitution $\dc=t^{n}\tdc(t)$ with $n=(1-a)/2$ may be used to find the following equation for $\tdc$:
\begin{equation}
 t^2\ddot{\tdc}+t\dot{\tdc}-\left(\frac{(a-1)^2}{4}+b^2+c^2t^{2r}\right)\tdc=0,
\end{equation}
which may further be transformed into the form
\begin{equation}\label{eq:modbes}
 u^2\tdc^{\prime\prime}+u\tdc^\prime-\left(\nu^2+u^2\right)\tdc;\quad \nu^2\equiv \frac{(a-1)^2}{4r^2}+\frac{b^2}{r^2}
\end{equation}
using the substitution $u=ct^r/r$ and the notation $\prime\equiv \dd/\dd u$. Equation (\ref{eq:modbes}) is a \textit{modified
Bessel equation}, the solutions of which are \textit{modified Bessel functions} of the first and second kind, $I_\nu(u)$ and
$K_\nu(u)$. Unlike regular Bessel functions, which are oscillatory in nature, these functions either grow ($I_\nu$) or decay
($K_\nu$) with increasing $u$. The general solution is then
\begin{equation}
 \dc(t)=t^{\frac{1-a}{2}}\left[C_1I_\nu\left(\frac{c}{r}t^r\right)+C_2K_\nu\left(\frac{c}{r}t^r\right)\right],
\end{equation}
where $C_{1,\,2}$ are arbitrary integration constants. In chapter \ref{chap:seven} we shall only be interested in
whichever function is the growing mode (this depends on the
sign of $r$). The modified Bessel function of the first kind has the power series expansion
\begin{equation}\label{eq:seriesI}
 I_\nu(x)=\sum_{k=0}^\infty\frac{1}{k!\Gamma(\nu+k+1)}\left(\frac{x}{2}\right)^{\nu+2k}=\frac{1}{\Gamma(1+\nu)}\left(\frac{x}{2}
\right)^\nu\left(1+\mathcal{O}(x^2)+\ldots\right),
\end{equation}
where $\Gamma(m)$ is the gamma function. The modified Bessel function of the second kind is defined via
\begin{equation}
 K_\nu(x)=\lim_{n\to \nu}\frac{\pi\left[I_{-n}(x)-I_n(x)\right]}{2\sin(n\pi)}.
\end{equation}
Its power series expansion for $\nu\notin\mathbb{Z}$ is as follows:
\begin{equation}\label{eq:seriesK}
 K_{\nu}(x)=\frac{\pi\textrm{csc}(\pi\nu)}{2}\left[\sum_{k=0}^\infty\frac{1}{
\Gamma(k-\nu+1)k!}\left(\frac{x}{2}\right)^{2k-\nu}-\sum_{k=0}^\infty\frac{1}{
\Gamma(k+\nu+1)k!}\left(\frac{x}{2}\right)^{2k+\nu}\right] .
\end{equation}

\section{Second Kind Power Series Expansions for the Power Spectra}
In this section we briefly outline the steps that lead to the factor
$F(\omega)$ given in (\ref{eq:F(w)}) and appearing in the deviation from the
$\Lambda$CDM spectra (\ref{eq:genps2}). The expansion of the first kind modified
Bessel functions (\ref{eq:seriesI}) is a trivial exercise in algebra, however
the second kind expansion (\ref{eq:seriesK}) requires more thought. The three
leading terms are:
\begin{equation}
 K_{\nu}(x)=\frac{\pi\textrm{csc}(\pi\nu)}{2\Gamma(1-\nu)}\left(\frac{2}{x}
\right)^\nu\left[1+\frac{\Gamma(1-\nu)}{\Gamma(2-\nu)}\left(\frac{x}{
2}\right)^2-\frac{\Gamma(1-\nu)}{\Gamma(1+\nu)}\left(\frac{x}{2}\right)^{2\nu}
\right],
\end{equation}
which correspond to the first two terms in the first sum in (\ref{eq:seriesK})
and the first term in the second sum. The next-to-leading order correction to
the leading order term ($\propto x^{-\nu}$) depends on whether $\nu>1$ or
the converse, which gives rise to the factor $F(\omega)$ in (\ref{eq:genps2}).
When $\nu=1$ a different power series is needed, which we do not give here
since it corresponds to an uninteresting cosmological scenario. When $\nu<1$, the
final factor can be evaluated using the relation
\begin{equation}
 \Gamma(1-m)\Gamma(m)=\frac{\pi}{\sin(\pi m)}.
\end{equation}
\chapter{The Global Potential for Supersymmetric Chameleon Models}
\label{app:mgp}

In this appendix we show how the global F-term scalar potential for supersymmetric chameleons coupled to charged scalar fields
studied in chapter \ref{chap:seven} can be minimised to recover the simple form given in equation (\ref{eq:Fpot}). 

\section{Minimisation of the Global Potential}

Ignoring the contribution from the K\"{a}hler potential for now (it
depends on $|\phi|$ only) and setting $\langle\phi_\pm\rangle=0$, which is always a minimum, we have:
\begin{align}\label{eq:dW}
 \left\vert\frac{\dd
W}{\dd\phi}\right\vert^2=&{g^\prime}^2|\pi_+|^2|\pi_-|^2+\frac{\gamma^2}{2}
\left(\frac{|\phi|^{2\alpha-2}}{\Lambda_0^{2\alpha-6}}+\frac{|\phi|^{2\gamma-2}}
{ \Lambda_2^{2\gamma-6}}\right)
+\gamma\Re\left(\frac{\phi^{\alpha-1}{\phi^*}^{\gamma-1}}{\Lambda_0^{\alpha-3}
\Lambda_2^{\gamma-3}}\right)\nonumber\\&+\sqrt{2}{g^\prime}\gamma\Re\left[
\pi_+\pi_-\left(\frac{{\phi^*}^{\alpha-1}}{\Lambda_0^{\alpha-3}}+\frac{{\phi^*}^
{\gamma-1}}{\Lambda_2^{\gamma-3}}\right)\right],
\end{align}
which, as can be seen, simplifies greatly when the negatively charged field has zero VEV. We shall see now that this VEV does
indeed minimise the potential. We begin by writing the charged fields in polar form $\pi_\pm\equiv \pi_\pm e^{i\theta_\pm}$.
Ideally, one would hope to set the three angular fields $\{\theta,\theta_\pm\}$ to constant values in order to give negative signs
in front of the final three terms in (\ref{eq:dW})\footnote{This is the approach taken when the charged fields are absent
\cite{Brax:2011qs}.}, however, this is not possible and instead one must eliminate them in terms of the other fields. In order to
do this, we exploit the local $\mathrm{U}(1)$ symmetry, which acts as $\theta_\pm(x)\rightarrow\pm q\alpha(x)$, to set
$\theta_+(x)=\theta_-(x)=\chi(x)/2$, which reduces the angular fields to the set $\{\theta,\chi\}$. With this in mind, the scalar
potential, including the contribution from the K\"{a}hler potential is
\begin{align}\label{eq:global_V}
 V(\phi,&\theta,\pi_+,\pi_-,\chi)=\frac{2{g^\prime}^2\pi_+^2\pi_-^2}{\gamma^2}
\left(\frac{\Lambda_1}{\phi}\right)^{2\gamma-2}+\left(\frac{\Lambda_1}{\phi}
\right)^{2\gamma-2}\left(\frac{\phi^{2\alpha-2}}{\Lambda_0^{2\alpha-6}}+\frac{
\phi^{2\gamma-2}}{\Lambda_2^{2\gamma-6}}\right)\nonumber\\&
+\frac{2}{\gamma}\left(\frac{\Lambda_1}{\phi}\right)^{2\gamma-2}\frac{\phi^{
\alpha+\gamma-2}}{\Lambda_0^{\alpha-3}\Lambda_2^{\gamma-3}}\cos[
(\alpha-\gamma)\theta]\nonumber\\&+\frac{g^\prime}{\sqrt{2}\gamma}\left(\frac{
\Lambda_1}{\phi}\right)^{2\gamma-2}\pi_+\pi_-\left[\frac{\phi^{\alpha-1}}{
\Lambda_0^{\alpha-3}}\cos[\chi-(\alpha-1)\theta]+\frac{\phi^{\gamma-1}}{
\Lambda_2^{\gamma-3}}\cos[\chi-(\gamma-1)\theta]\right].
\end{align}
Minimising this with respect to $\chi$ one finds
\begin{equation}
 \frac{\phi^{\alpha-\gamma}}{\Lambda_0^{\alpha-3}}\sin[\chi-(\alpha-1)\theta]
+\frac{1}{\Lambda_2^{\gamma-3}}\sin[\chi-(\gamma-1)\theta]=0,
\end{equation}
which may be used in the equation found by minimising equation
(\ref{eq:global_V}) with respect to $\theta$ to find a relation between
$\sin[(\alpha-\gamma)\theta]$ and $\sin[\chi-(\gamma-1)\theta)]$ (or
equivalently $\sin[\chi-(\alpha-1)\theta]$):
\begin{equation}\label{eq:sinsin}
 \frac{2\phi^{\alpha-1}}{\Lambda_0^{\alpha-3}}\sin[(\alpha-\gamma)\theta]=\frac{
g^\prime}{2}\pi_+\pi_-\sin[\chi-(\gamma-1)\theta].
\end{equation}
This may be used to eliminate $\chi$ from the potential to find:
\begin{align}\label{eq:glob_pot}
V(\phi,&\theta,\pi_+,\pi_-)=
\frac{2{g^\prime}^2\pi_+^2\pi_-^2}{\gamma^2}\left(\frac{\Lambda_1}{\phi}\right)^
{2\gamma-2}+\left(\frac{\Lambda_1}{\phi}\right)^{2\gamma-2}\left(\frac{\phi^{
2\alpha-2}}{\Lambda_0^{2\alpha-6}}+\frac{\phi^{2\gamma-2}}{\Lambda_2^{2\gamma-6}
} \right)\nonumber\\&
+\frac{2}{\gamma}\left(\frac{\Lambda_1}{\phi}\right)^{2\gamma-2}\frac{\phi^{
\alpha+\gamma-2}}{\Lambda_0^{\alpha-3}\Lambda_2^{\gamma-3}}\cos[
(\alpha-\gamma)\theta]\nonumber\\&+\frac{g^\prime}{\sqrt{2}\gamma\Lambda_0^{
\alpha-3}}\phi^{\alpha-1}\left(\frac{\Lambda_1}{\phi}\right)^{2\gamma-2}\sqrt{
\pi_+^2\pi_-^2-\frac{8\phi^{2\gamma-2}}{{g^\prime}^2\Lambda_2^{2\gamma-6}}\sin^2
[ (\alpha-\gamma)\theta]}\nonumber\\&+\frac{g^\prime}{\sqrt{2}\gamma\Lambda_2^{
\gamma-3}}\phi^{\gamma-1}\left(\frac{\Lambda_1}{\phi}\right)^{2\gamma-2}\sqrt{
\pi_+^2\pi_-^2-\frac{8\phi^{2\alpha-2}}{{g^\prime}^2\Lambda_0^{2\alpha-6}}\sin^2
[(\alpha-\gamma)\theta]}.
\end{align}
At first glance, one may worry about the square roots. In fact, these are somewhat spurious since the above expression is
only true when $\chi$ is fixed to its minimum. Furthermore, if one examines equation (\ref{eq:sinsin}) then it is evident
that as $\pi_+,\pi_-\rightarrow0$ the second term in the square root has exactly the same behaviour and so there is never a region
in configuration space where the argument is negative. Minimising (\ref{eq:glob_pot}) together with the potential coming from the
D-term and the $\pi_\pm$ terms in the superpotential,
\begin{equation}
 V_D+\left\vert\frac{\dd W}{\dd \pi_+}\right\vert+\left\vert\frac{\dd W}{\dd \pi_-}\right\vert=
\frac{1}{2}\left(q\pi_+^2+-q\pi_-^2-\xi^2\right)^2+{g^{\prime}}^2\phi^2\left(\pi_+^2+\pi_-^2\right),
\end{equation}
with respect to $\pi_-$ one indeed finds that $\langle\pi_-\rangle=0$ is a solution. If one expands the global potential around
this minimum by setting $\pi_-\rightarrow\langle\pi_-\rangle+\delta\pi_-$ then the coefficient of the $\delta\pi_-^2$ term is
\begin{equation}
 q\left(\xi^2-q\pi_+^2\right)+{g^\prime}^2|\phi|^2.
\end{equation}
In theory, this can be negative, however we have not yet finished minimising the
potential. In section \ref{sec:cosgen} we learnt that there are two possible
solutions for $\pi_+$ given by equation (\ref{eq:pi+}) when $\pi_-=0$ and so we
should check that these are indeed stable minima of the global potential. The
case where $\pi_+=0$ is clearly a minimum since the negative term vanishes. The
second case gives the coefficient as $2{g^\prime}^2|\phi|^2$ and so in either
case the coefficient is positive and the stationary point is a stable minimum.
When $\langle\pi_-\rangle=0$ equation (\ref{eq:sinsin}) gives
$\sin[(\alpha-\gamma)\theta]=0$ and hence $\cos[(\alpha-\gamma)\theta]=-1$.
Making this substitution in (\ref{eq:glob_pot}) yields the far simpler form of
the potential given in (\ref{eq:Fpot}).
\chapter{Corrections to the Effective Potential for Supersymmetric Chameleon Models}
\label{app:dcor}

In this appendix we show how one can explore the importance of the corrections to the effective potential for the
supersymmetric chameleon studied in chapter \ref{chap:seven} at early times when the $\mathbb{Z}_2$ symmetry of the potential for
the charged scalars is broken and they acquire non-zero VEVs.

\section{Late Time Importance of the Corrections}

The corrections to the effective potential are of the form
\begin{equation}\label{eq:dcorr2}
 V_{\rm corr}=\frac{{g^\prime}^2\xi^2}{q}\phi^2-\frac{{g^\prime}^4}{2q^2}\phi^4
\end{equation}
and we shall make use of the definition (\ref{eq:Gdef}) and for brevity define $Z\equiv(1+z_\infty)^3$.

We can estimate the density at which each correction becomes important and we can no longer neglect them by equating each one with
the magnitude of the density dependent term in turn. In this case, one finds that the field values $\phi_i$ at which the order $i$
corrections are important are
\begin{align}
\phi_2&=\left(\frac{\mathcal{G}^2\xi^2}{x\rho_{\rm
c}}\right)^{\frac{1}{\delta-2}}\pmi^{\frac{\delta}{\delta-4}}\quad\textrm{and}\\
 \phi_4&=\left(\frac{\mathcal{G}^4}{x\rc}\right)^{\frac{1}{\delta-4}}\pmi^{\frac{\delta}{\delta-4}}
\end{align}
respectively. Now we can always begin the cosmic evolution far enough in the past such that the density is large enough that the
corrections are negligible, in which case the field evolves according to the background cosmology detailed in section
\ref{sec:backcos}. As the field evolves, the coefficient of the density dependent term becomes smaller and the corrections will
eventually become important. If this occurs before the field rolls past $\Delta$ then we must correct the dynamics appropriately.
If, on the other hand, this occurs after the field has passed $\Delta$ then these corrections will no longer be present and we can
neglect them completely. Using equations (\ref{eq:vphimin}) and (\ref{eq:vphi}), we can estimate the densities $\rho_i$ at which
$\phi=\phi_i$ and the density $\rho_\Delta$ at which $\phi=\Delta$:
\begin{align}\label{eq:rhoi}\begin{split}
\rho_2^{\frac{n+2}{n+\delta}}&=\frac{M^2\mathcal{G}^2\xi^2}{x}\left(\frac{M}{\Lambda}\right)^{\frac{4}{n}}\rho_\infty^{-\frac{
\delta-2 } {
n+\delta}}\\\rho_4^{\frac{n+4}{n+\delta}}&=\frac{M^4\mathcal{G}^4}{2x}\left(\frac{M}{\Lambda}\right)^{\frac{16}{n}}\rho_\infty^{
-\frac{
\delta-4}{n+4}}\\
 \rho_\Delta^{\frac{1}{n+\delta}}&=\frac{GM}{\xi}\left(\frac{M}{\Lambda}\right)^{\frac{4}{n}}\rho_\infty^{\frac{1}{n+\delta}}
\end{split}
\end{align}
The condition that the corrections can be neglected is then $\rho_\Delta\gg\rho_i$. In this analysis we shall take ``much greater
than'' to mean an order of magnitude i.e. $\rho_\Delta\ge 10\rho_i$.

\subsection{Mass Scales}

One must be careful that the model parameters are inter-dependent and it is important to keep track of which are fixed by
specific
choices of others. In the analysis of section \ref{sec:cosgen} we consider the low-energy model parameters
$\{n,\delta,x,m,\mathcal{G},z_\infty\}$ independent, which completely fixes the derived scales $M$ and $\Lambda$ via equation
(\ref{eq:scales}). We start by writing equations (\ref{eq:x}) and (\ref{eq:rhoinf}) in the form
\begin{align}
 \left(\frac{\Lambda}{10^{-3}\textrm{eV}}\right)^4&=\frac{\delta x}{n}Z \label{eq:Lambda}\\
\pmi&=x^{\frac{1}{\delta}}\mu\label{eq:pmi},
\end{align}
which can be combined using equation (\ref{eq:scales}) to find
\begin{equation}\label{eq:M4n}
M^{4+n}=10^{-12}\frac{\delta}{n}Zx^{\frac{n+\delta}{\delta}}\mu^n\textrm{eV}^4.
\end{equation}
These relations can then be used to eliminate the quantities $M$ and $\Lambda$ in equation (\ref{eq:rhoi}) in favour of the
low-energy parameters.

\section{The Quadratic Correction}
When $\rho_2\gg \rho_\Delta,\rho_4,\rc$ the effective potential is
\begin{equation}
 V_{\rm eff}(\phi)\approx\Lambda^4\left[1-\left(\frac{\phi_{\rm
min}}{\phi}\right)^{\frac{n}{2}}\right]^2+\mathcal{G}^2\xi^2\phi^2.
\end{equation}
This is minimised at field values satisfying
\begin{equation}
 \left(\frac{\pmi}{\phi}\right)^{n+2}-\left(\frac{\pmi}{\phi}\right)^{\frac{n}{2}+2}=\frac{\mathcal{G}^2\xi^2x^{\frac{2}{\delta}}{
\mu^2}}{
n\Lambda^4}.
\end{equation}
When $\mathcal{G}\xi x^{1/\delta}\mu\ll \Lambda^2$ we have $\phi\approx\pmi$ and so this case is still viable provided that
$\pmi>\Delta$.
If the converse is true then the minimum lies at field values
\begin{equation}
 \phi=\pmi\left(\frac{n\Lambda^4}{\mathcal{G}^2\xi^2x^{\frac{2}{\delta}}\mu^2}\right)^{\frac{1}{n+2}}
\end{equation}
and demanding that this is larger than $\Delta$ we find that the parameters must satisfy
\begin{equation}\label{eq:quadcorr}
 \xi^{n+4}\mathcal{G}^nx^{\frac{n}{\delta}}\mu^n>n\Lambda^4
\end{equation}
in order for the cosmological constant to be generated.

\section{The Quartic Correction}\label{sec:quart}

When the quartic correction is important the potential takes the following form:
\begin{equation}
 V(\phi)=\Lambda^4\left(1-\left(\frac{\phi_{\rm min}}{\phi}\right)^{\frac{n}{2}}\right)^2-\frac{\mathcal{G}^4}{2}\phi^4.
\end{equation}
The minimum, if it exists, is given by the solution of
\begin{equation}
 \left(\frac{\phi_{\rm min}}{\phi}\right)^{\frac{n}{2}+4}-\left(\frac{\phi_{\rm
min}}{\phi}\right)^{n+4}=\frac{2}{n}\mathcal{G}^4\left(\frac{\phi_{\rm min}}{\Lambda}\right)^4
\end{equation}
and so the only possible solutions have $\phi>\phi_{\rm min}$. This means that when this correction only is important the field
will always pass $\Delta$ at some time and generate a cosmological constant. With this in mind, one may wonder if the case
$\pmi<\Delta$ is allowed since in this case the field can still pass $\Delta$ if the minimum lies at large enough field values.
This situation is highly unnatural since once the corrections vanish the field lies at values greater than $\pmi$ and will
subsequently roll backwards, reintroducing the corrections. Hubble friction will eventually reduce the amplitude of the
oscillations, however this leads to a situation that is highly fine-tuned and sensitive to the initial conditions and so we
exclude it.

\section{Simultaneous Corrections}

When both corrections are simultaneously important the effective potential, including the matter coupling, is
\begin{equation}
 V(\phi)=\Lambda^4\left(1-\left(\frac{\phi_{\rm min}}{\phi}\right)^{\frac{n}{2}}\right)^2+x\rc\left(\frac{\phi}{\phi_{\rm
min}}\right)^\delta-\frac{\mathcal{G}^4}{2}\phi^4+\mathcal{G}^2\xi^2\phi^2.
\end{equation}
When including the density term one should technically solve the entire dynamical system in terms of $\varphi$ and $\rc(t)$,
however we can glean all the information we need if we simply set $\rc=1$ eV$^4$. As mentioned above, we require that the field
rolls past $\Delta$ before $\rc=1$ eV$^4$ so that the cosmological constant is generated before last scattering and this will be
the case if the minimum is located at field values greater than this by last scattering. Technically, this condition is not
sufficient since it only guarantees that the minimum is located at values greater than $\Delta$ by last scattering, not that the
field passes $\Delta$ by this time, however the large mass of the field ensures that this approximation is sensible. Far enough in
the past the corrections are unimportant and field tracks its minimum owing to this large mass. Eventually we reach the epoch
where all three terms become important and pass to the regime where the density-dependent term is negligible. This transition is
smooth and so given the large mass we expect that the field should simply remain fixed at the new, density-independent minimum and
therefore the dynamics should not differ largely from the static analysis we will employ here.

The field value at the minimum satisfies
\begin{equation}
 \frac{n\Lambda^4}{\phi_{\rm min}^2}\left[\left(\frac{\phi_{\rm min}}{\phi}\right)^{\frac{n}{2}+2}-\left(\frac{\phi_{\rm
min}}{\phi}\right)^{n+2}\right]+\frac{x\delta \rho_{\rm c}}{\phi_{\rm min}^2}\left(\frac{\phi_{\rm
min}}{\phi}\right)^{2-\delta}+2G^2\xi^2-2G^4\phi^2=0,
\end{equation}
which must be solved numerically for the minimum given a specific set of parameters and for the same reasons given in section
\ref{sec:quart}, we will impose $\pmi>\Delta$. If this has no solutions then the potential is runaway near $\rc=1$ eV$^4$ and the
field will be able to pass $\Delta$. When solutions exist the parameters where the minimum occurs at field values larger than
$\Delta$ are viable and those where the converse is true are not.

\end{document}